\documentclass[twocolumn,aps,prd,amsmath,amssymb,preprintnumbers,longbibliography]{revtex4-1}
\usepackage{amsmath} \usepackage{amsfonts} \usepackage{amssymb}
\usepackage{bbm}
\usepackage{epsfig}
\usepackage{graphics}
\usepackage{graphicx}
\usepackage{titlesec}
\usepackage{mathtools}
\usepackage{environ}
\usepackage{dsfont}
\usepackage{version}
\usepackage[colorlinks=true]{hyperref}
\hypersetup{urlcolor=black,linkcolor=black,citecolor=black}
\usepackage{natbib}
\usepackage[toc,page]{appendix}

\usepackage{dsfont}

\usepackage{caption}
\usepackage{subcaption}
\usepackage{tabularx}
\newcolumntype{C}[1]{>{\centering\arraybackslash}p{#1}}

\makeatletter
\renewcommand*\env@matrix[1][c]{\hskip -\arraycolsep
  \let\@ifnextchar\new@ifnextchar
  \array{*\c@MaxMatrixCols #1}}
\makeatother

\newcommand{\be}{\begin{equation}}
\newcommand{\ee}{\end{equation}}
\newcommand{\bel}[1]{\be\label{#1}}

\newcommand{\ba}{\begin{align}}
\newcommand{\ea}{\end{align}}
\newcommand{\nn}{\nonumber}

\newcommand{\gl}{\big(}
\newcommand{\gr}{\big)}

\titleformat{\subsection}[block]{\normalfont\bfseries}{\thesubsection.}{1ex}{}
\titlespacing{\subsection}{0pt}{10pt}{1pt}[0pt]
\titleformat*{\section}{\large\bfseries}
\renewcommand{\thesubsection}{\arabic{subsection}}

\pdfoutput=1 

\usepackage{todonotes}
\usepackage{braket}

\newcommand{\tr}{\mathrm{tr}}

\newcommand{\exval}[1]{\langle #1 \rangle}

\newcommand{\psibar}{\overline{\psi}}
\newcommand{\epstil}{\tilde{\varepsilon}}
\newcommand{\cD}{\mathcal{D}}
\newcommand{\cR}{\mathcal{R}}
\newcommand{\cK}{\mathcal{K}}
\newcommand{\cA}{\mathcal{A}}
\newcommand{\Ktil}{\widetilde{\mathcal{K}}}
\newcommand{\Kbar}{\overline{\mathcal{K}}}

\newcommand{\Khat}{\widehat{\cK}}
\newcommand{\Bhat}{\widehat{B}}
\newcommand{\Ahat}{\hat{A}}
\newcommand{\gbar}{\bar{g}}

\newcommand{\tbar}{\bar{t}}
\newcommand{\xbar}{\bar{x}}
\newcommand{\xhat}{\hat{x}}
\newcommand{\fhat}{\hat{f}}
\newcommand{\nhat}{\hat{n}}

\newcommand{\mbar}{\bar{m}}
\newcommand{\qtil}{\tilde{q}}

\newcommand{\Stil}{\widetilde{S}}
\newcommand{\stil}{\tilde{s}}
\newcommand{\atil}{\tilde{a}}
\newcommand{\Shat}{\widehat{S}}
\newcommand{\vp}{{\varphi}}
\newcommand{\vpbar}{\bar\vp}
\newcommand{\eps}{\varepsilon}

\newcommand{\taubar}{\bar{\tau}}
\newcommand{\rhobar}{\bar{\rho}}
\newcommand{\cL}{\mathcal{L}}
\newcommand{\Ltil}{\tilde{\cL}}
\newcommand{\Lbar}{\overline{\cL}}
\newcommand{\Lba}{\overline{L}}
\newcommand{\Dbar}{\overline{D}}
\newcommand{\Htil}{\widetilde{H}}
\newcommand{\Hbar}{\overline{H}}
\newcommand{\Etil}{\widetilde{E}}
\newcommand{\Nhat}{\widehat{N}}
\newcommand{\Qtil}{\widetilde{Q}}
\newcommand{\Vbar}{\overline{V}}
\newcommand{\Vtil}{\widetilde{V}}

\newcommand{\subt}[1]{_{\text{#1}}}

\newcommand{\inn}{\subt{in}}
\newcommand{\tf}{t_f}

\newcommand{\g}{_{\gamma}}
\newcommand{\al}{_{\alpha}}
\newcommand{\bet}{_{\beta}}
\newcommand{\del}{_{\delta}}
\newcommand{\gd}{_{\gamma\delta}}
\newcommand{\diag}{\text{diag}}
\newcommand{\Ra}{_{Ra}}
\newcommand{\La}{_{La}}
\newcommand{\Rb}{_{Rb}}
\newcommand{\Lb}{_{Lb}}
\newcommand{\Rc}{_{Rc}}
\newcommand{\Lc}{_{Lc}}
\newcommand{\Rd}{_{Rd}}
\newcommand{\Ld}{_{Ld}}

\newcommand{\opsi}{[\psi]}
\newcommand{\opsib}{[\,\psibar\,]}
\newcommand{\opsit}{[\psi(t)]}

\newcommand{\lopsi}{\cL[\psi,\psibar\,]}

\newcommand{\psig}{\psi\g(x)}
\newcommand{\psibarg}{\psibar\g(x)}

\newcommand{\zetal}{\tilde{\zeta}}
\newcommand{\zetalg}{\zetal\g(x)}
\newcommand{\zetalgg}{\zetal\g^*(x)}
\newcommand{\zetabar}{\overline{\zeta}}

\newcommand{\Kbapsi}[2]{\Kbar[\,\psibar(#1),\psi(#2)]}
\newcommand{\Ktipsi}[2]{\Ktil[\psi(#1),\psibar(#2)\,]}
\newcommand{\Kpsi}[2]{\cK[\psi(#1),\psi(#2)]}
\newcommand{\taur}{_{\tau\rho}}
\newcommand{\psip}{\psi_+}
\newcommand{\psim}{\psi_-}
\newcommand{\te}{t+\eps}
\newcommand{\efr}[1]{\frac{#1\eps}{2}}
\newcommand{\xe}{x+\eps}

\newcommand{\qq}[1]{``#1"}

\newcommand{\dt}{\partial_{t}}
\newcommand{\dpl}{\partial_+}
\newcommand{\dm}{\partial_-}
\newcommand{\herm}{^{\dagger}}
\newcommand{\hateq}{\widehat{=}}
\newcommand{\pmat}[4]{\begin{pmatrix} #1 & #2 \\ #3 & #4 \end{pmatrix}}
\newcommand{\pvec}[2]{\begin{pmatrix} #1 \\ #2 \end{pmatrix}}
\newcommand{\mf}{m_f}
\newcommand{\otx}{(t,x)}
\newcommand{\otex}{(\te,x)}
\newcommand{\otxe}{(t,\xe)}
\newcommand{\otexe}{(\te,\xe)}

\newcommand{\Ptil}{\tilde{P}}
\newcommand{\abs}[1]{|#1|}
\newcommand{\nbar}{\bar{n}}

\newcommand{\wpx}{\widehat{\partial}_{x}}
\newcommand{\wpt}{\widehat{\partial}_{t}}

\definecolor{refkey}{rgb}{0,0,1}
\definecolor{labelkey}{rgb}{0,1,0}

\begin{document}

\title{\LARGE Fermion picture for cellular automata}

\author{C. Wetterich}

\affiliation{Institut  f\"ur Theoretische Physik\\
Universit\"at Heidelberg\\
Philosophenweg 16, D-69120 Heidelberg}

\begin{abstract}

How do cellular automata behave in the limit of a very large number of cells? Is there a continuum limit with simple properties? We attack this problem by mapping certain classes of automata to quantum field theories for which powerful methods exist for this type of problem. Indeed, many cellular automata admit an interpretation in terms of fermionic particles. Reversible automata on space-lattices with a local updating rule can be described by a partition function or Grassmann functional integral for interacting fermions moving in this space. We discuss large classes of automata that are equivalent to discretized fermionic quantum field theories with various types of interactions. Two-dimensional models include relativistic Thirring or Gross-Neveu type models with abelian or non-abelian continuous global symmetries, models with local gauge symmetries, and spinor gravity with local Lorentz symmetry as well as diffeomorphism invariance in the (naive) continuum limit.

The limit of a very large number of cells needs a probabilistic description. Probabilistic cellular automata are characterized by a probability distribution over initial bit-configurations. They can be described by the quantum formalism with wave functions, density matrix and non-commuting operators associated to observables, which are the same for the automata and associated fermionic quantum theories. This formalism is crucial for a discussion of concepts as vacuum states, spontaneous symmetry breaking, coarse graining and the continuum limit for probabilistic cellular automata. In particular, we perform explicitly the continuum limit for an automaton that describes a quantum particle in a potential for one space dimension.

\end{abstract}

\maketitle


\tableofcontents

\bigskip
\noindent

\section{Introduction}
\label{sec: I}

Since the pioneering work by von Neumann~\cite{JVN}, Ulam~\cite{ULA} and Zuse~\cite{ZUS} cellular automata have found applications in many areas of science~\cite{GAR, LIRO, TOOM, DKT, WOLF, VICH, PREDU, TOMA, FLN}. One often describes the states of the automaton by a configuration of bits, or equivalently fermionic occupation numbers or Ising spins. Deterministic classical cellular automata map a bit-configuration at time $t$ to a new bit-configuration at the next discrete time step $t+\eps$. The bits are grouped into cells $x$, and the updating is such that only the state of a few neighboring cells at $t$ influences the values of the bits in the cell $x$ at $\te$. We restrict our discussion here to reversible or invertible cellular automata~\cite{HED, RICH, AMPA, HPP, CREU, TOMA} for which the inverse of the updating map exists.

We concentrate on automata that admit a simple particle interpretation. Bit configurations, with bits $n\al(t)$ taking values of one or zero, can be viewed as the states of fermionic many body systems in the occupation number basis. In this view $n\al(t)=n\g\otx=1,0$ denotes the occupation number of a fermion $\alpha$. The index $\alpha=(x,\gamma)$ involves a discrete position variable $x$, as well as internal properties as \qq{colors} or \qq{flavors} $\gamma$. Local configurations of the bits or occupation numbers $n\g(x)$ at the position $x$ are used to define the state of the cell $x$. We are interested in automata that can be interpreted in terms of propagation and scattering of fermionic particles. They will describe the dynamics of fermionic quantum field theories or many body systems.

We are interested in the continuum limit of a very large number of cells. This requires a probabilistic description, as given by a probability distribution for initial bit configurations. A possible mapping of the probabilistic cellular automaton to a quantum field theory for fermions can be an important help for the continuum limit, both on the conceptual side and more quantitatively by the existence of powerful methods for the investigation of the continuum limit in quantum field theories. In this respect one would like to know which types of automata can be mapped to fermionic quantum field theories, and vice versa.

Generic fermionic models cannot be represented as automata. For the representation as a probabilistic cellular automaton it is necessary that a discretization exists for which the evolution operator for a discrete time step $\eps$ takes the particular form of a \qq{unique jump operator}. One of the aims of this paper is to find out which types of fermionic quantum field theories with interactions can be represented as cellular automata. For this purpose we have to cast candidate automata into the language of Grassmann variables and Grassmann (functional) integrals that are usually employed for fermionic many body systems and quantum field theories. We will employ a general bit-fermion map~\cite{CWFCS, CWFGI} which can be implemented whenever the \qq{global map} or \qq{overall map} for the updating of the whole bit configuration is known explicitly. This general map will be illustrated by a number of rather simple automata that give rise, nevertheless, to interacting discrete quantum field theories with various global or local symmetries. As compared to previous work~\cite{CWPCA, CWFCB, CWNEW} this extends considerably the class of two-dimensional fermionic quantum field theories which have a cellular automaton representation. In particular, they include models with local gauge symmetries as well as a large family of Lorentz-invariant models with non-abelian global continuous symmetries.

The behavior of the investigated automata for a small number of cells remains very simple. We are interested here in a very large number of cells for which interesting collective phenomena can emerge. The fermionic picture for automata constitutes a bridge to the highly developed formalism of quantum field theory which precisely deals with these collective phenomena. Cellular automata are equivalent to particular discretizations of fermionic quantum field theories. The limit of the number of cells going to infinity corresponds to the continuum limit of the discrete quantum field theories. For many discrete models the continuum limit can show new symmetries as Lorentz symmetry. In the continuum limit one often observes universality of the macroscopic behavior which becomes independent of the particular discretization. New phenomena as order parameters, spontaneous symmetry breaking, phase transitions or topological excitations can emerge.

For a very large number of cells only a probabilistic approach is meaningful. We therefore focus on \qq{probabilistic cellular automata}. In our context the probabilistic aspects enter only by a probability distribution over initial conditions, while the deterministic updating remains unchanged. This differs from stochastic cellular automata~\cite{TOOM, DKT, VICH, FLN} for which the updating is described by a Markov process with probabilistic properties itself. Our models preserve the important property that the initial probabilistic information is processed but not lost after arbitrarily many updatings. In contrast, the loss of information is a characteristic feature of Markov chains (except for deterministic limiting cases). Since stochastic cellular automata are sometimes called probabilistic cellular automata as well, a more specific naming for our models could be \qq{unitary probabilistic cellular automata}.

Unitary probabilistic cellular automata share many features of discrete real quantum mechanics, for which time is discretized and the complex wave function or density matrix is written in terms of real quantities. For discrete quantum mechanics the evolution is encoded in a unitary step evolution operator that describes the change for a minimal discrete time step from $t$ to $\te$. A discrete evolution equation replaces the continuous Schrödinger- or von Neumann equation. We introduce the formalism of discrete quantum mechanics for the description of probabilistic cellular automata, based on wave functions, operators for observables, and the density matrix. The \qq{quantum law} for expectation values follows from its standard classical statistical definition. The particular feature of a cellular automaton is the specific form of the step evolution operator which is a \qq{unique jump matrix}.

For unitary probabilistic cellular automata the step evolution operator is an orthogonal matrix acting on a real wave function. For invertible automata is is a \qq{unique jump matrix} with exactly one element equal to $\pm1$ in each row and column, and all other elements zero. The step evolution operator of a fermionic quantum field theory can be derived from a Grassmann functional integral. If this fermionic step evolution operator has the unique jump property, both the wave function and the step evolution operator, as well as the operators for observables, are the same for the probabilistic cellular automaton and its fermionic equivalent. \qq{Classical} probabilistic cellular automata are then formulated as real discrete quantum systems. In the presence of a suitable complex structure, that we associate here to a map between particles and holes, this becomes standard discrete quantum mechanics with a complex Hilbert space.

The construction of a discrete quantum field theory for fermions which is equivalent to a cellular automaton is highly non-trivial. In one direction one needs to compute first the step evolution operator for a given discretized fermionic model. In general, this will be an orthogonal matrix in a real formulation, but it will not be a unique jump matrix. One has to find particular discretizations that realize the automaton property of a unique jump step evolution operator. In the opposite direction, one may translate for a given automaton the updating rule and unique jump step evolution operator to an expression in terms of Grassmann variables. The challenge is now to find out which updating rules can realize the wanted symmetries and other properties of the fermionic quantum field theory that one wants to construct. The present paper describes some general strategies to realize this goal. In addition, an educated guess is often necessary to find a solution.

From the point of view of quantum field theory the quantum mechanics of a single particle (or a few particles) is associated to isolated local \qq{particle excitations} of a given \qq{vacuum state}. The setting is precisely the same for probabilistic automata. Our approach to probabilistic automata includes single-particle quantum mechanics, e.g. a particle in a potential, as a particular limiting case. Typically, the dynamics of a single quantum particle is dictated by the collective properties of the vacuum state. One-particle quantum mechanics refers to a subsystem which is influenced by the properties of its \qq{environment}, associated here to the vacuum. It is not an isolated automaton by itself. Single-particle quantum mechanics arises in the continuum limit of the probabilistic automaton and the associated fermionic quantum field theory.

The focus of this paper concerns the continuum limit for a very large number of cells. The existence of this limit may be expected if the change of the wave function remains small for a \qq{macroscopic time interval} $\Delta t$ which is much larger than the microscopic time step $\eps$. In this case the limit $\eps\to0$ can be formulated which transforms the discrete updating into a continuous differential equation in the form of a Schrödinger or von Neumann equation. The probabilistic setting is crucial for this purpose since it permits a continuous change of the wave function. The argument is simple: If the discrete time evolution is well approximated by a continuous time evolution, the continuum limit has to produce a differential equation for the time evolution. Similarly to the discrete evolution the continuous evolution equation is linear in the wave function or density matrix. If the unitary character of conserved information remains valid in the continuum formulation, a linear time evolution equation has to be of the type of a Schrödinger or von Neumann equation.

For the fermionic quantum field theories discussed here we first establish the naive continuum limit. This corresponds to formally taking $\eps\to0$ in the Grassmann functional integral. In this limit one recovers the continuum quantum field theory that the discretization is thought to represent. It is not guaranteed that the true continuum limit shares the same properties as the naive continuum limit. Both for the true and the naive continuum limit the step evolution operator has no longer the unique jump property of a cellular automaton. The continuum evolution maps no longer a given bit configuration to a unique new bit configuration. It can evolve into many different configurations with certain probabilities. One has to understand how this typical feature of quantum theories arises. The key to this understanding is averaging or coarse graining.

Steps towards the continuum limit need to proceed by some form of coarse graining which preserves the unitary character of the evolution. Using the discrete quantum formulation this coarse graining can be done by choosing suitable subtraces of the density matrix. In this case the orthogonal or unitary form of the step evolution operator can be preserved on the coarse grained level. On the other side, the unique jump property of the step evolution operator is lost on the coarse grained level. After coarse graining we do no longer have an automaton. A given coarse grained state evolves to different coarse grained states with certain probabilities. We do not deal with a Markov chain, however, since the coarse grained evolution still preserves the probabilistic information of the appropriate subsystem specified by the coarse grained density matrix. The off-diagonal elements of this density matrix are needed to formulate a linear evolution law on the coarse grained level. The probability distribution at a given time $t$ alone is not sufficient to determine the coarse grained evolution. We provide first simple examples demonstrating explicitly the loss of the automaton property on the coarse grained level.

Considerable simplifications occur for the continuum limit of one-particle states. We mimic the effect of order parameters in an interacting fermionic quantum field theory by studying a \qq{random cellular automaton} for the one-particle state. For this type of cellular automaton one distributes randomly \qq{disorder points} in the space-time lattice. In one space dimension a particle moving towards increasing or decreasing $x$ reverses its direction if it encounters a disorder point. For a sparse enough distribution of these scattering events we can perform the continuum limit explicitly. The result is a two-dimensional Dirac equation with a potential $V(x)$. This potential is determined by the distribution of the scattering events. In the non-relativistic limit this random cellular automaton yields the Schrödinger equation for a quantum particle in a potential. This realizes a classical statistical system for the quantum particle in one space dimension.

Beyond the one-particle state the continuum limit for automata with a very large number of cells is largely unexplored terrain. The present work is a new step in this direction. This part should be considered only as a beginning, demonstrating directly only a few important conceptual issues. Considerable effort remains necessary for establishing the true continuum limit of the models presented here. Fortunately, the automata introduced in this paper have rather simple updating rules. This will allow to study directly the behavior as the number of cells is increased, as far as this number remains numerically tractable. Already for the simple quantum field theories discussed here the equivalence to cellular automata offers interesting new avenues. This concerns, in particular, non-equilibrium quantum field theory. The automaton can be initialized with an arbitrary wave function or density matrix. Following its evolution either analytically or numerically describes directly the time-dependence in the quantum field theory with the corresponding initial conditions.

\section{Outline}
\label{sec: OL}

The first aim of this paper is a demonstration that classical automata admit a description in terms of fermionic many body systems with a particular form of the evolution operator. In consequence, probabilistic automata will be equivalent to fermionic quantum systems. This general equivalence is sufficient for the use of the powerful concepts of quantum many body theory or quantum field theory for probabilistic automata. The first part of this paper in sects.~\ref{sec: FPGU}-~\ref{sec: DSG} establishes a rich family of fermionic quantum field theories which have a representation as probabilistic cellular automata. These models constitute the basis for the approach to the continuum limit which is discussed in the second part of this paper in sects.~\ref{sec: OO}-~\ref{sec: CSQM}.

Our basic setting applies to general automata. The cellular property will be discussed explicitly for the particular models. We begin with deterministic general automata which are characterized at each step in their evolution by some specific configuration of bits $\rho$. An updating rule maps this bit-configuration to a new bit-configuration $\taubar(\rho)$ at the next evolution step. We associate the bit-configurations with states of a many body model for fermions in the occupation number basis. A bit value one corresponds to a fermion present at the \qq{position} of the bit, while for a bit value zero no fermion is present. Here the general notion of \qq{position} typically (but not necessarily) includes a position in space. It may further refer to \qq{internal properties} for different \qq{species} of bits or fermions. A chain of updatings of the automaton can be associated with a time-evolution with discrete time steps $\eps$.

Deterministic cellular automata have been proposed for a deterministic setting of quantum theory~\cite{GTH, ELZE, HOOFT2, HOOFT3, HOOFT4}. In contrast, we investigate here probabilistic cellular automata for which the initial state is characterized by a probability distribution over the possible bit-configurations. We will find that the probabilistic information and its evolution gives rise to quantum features. The continuity of the probabilistic information accounts for the wave aspect in particle-wave duality. Probabilistic cellular automata can be associated to quantum many body theories or quantum field theories for fermions~\cite{Wetterich:2020kqi, CWPCA, CWFCB, CWNEW}. Our discussion remains within classical statistics. We do not investigate \qq{quantum cellular automata}~\cite{Arrighi2019}.

Simple cellular automata acting on a chain of bits have been shown to be equivalent to specific free or interacting quantum field theories for fermions in $1+1$-dimensions~\cite{CWFCS, CWFGI, CWPCA, CWNEW}. The present paper generalizes this approach to large classes of automata that are equivalent to fermionic models with various forms of interactions. The specific models discussed here have one space and one time dimension. While our approach can be implemented in arbitrary dimensions the focus on $1+1$-dimensional systems helps for a simple presentation of the key points.

The equivalence between cellular automata and fermionic systems is based on a general \qq{bit-fermion} map~\cite{CWFGI, CWNEW} which exists in arbitrary dimensions. This map associates classical bits or Ising spins~\cite{LENZ, ISING, BINDER} to occupation numbers of fermions. Grassmann elements in the fermionic description of Grassmann functional integrals are associated to bit-configurations. This general bit-fermion map differs from other particular maps between fermions and Ising spins in two dimensions~\cite{PLECH, BER1, BER2, SAM, ITS, PLE1} or other forms of fermion-boson equivalence for particular models~\cite{COL, DNS, FUR, NAO}. We exploit the bit-fermion map in order to construct for arbitrary updatings of bit-configurations in a cellular automaton an associated transformation between corresponding Grassmann elements. This transformation can be interpreted as a particular discrete time evolution step of the fermionic model that is equivalent to the cellular automaton. A chain of updatings in the cellular automaton can be represented by a Grassmann (functional) integral.

Instead of a general formal discussion we often demonstrate key features of the fermionic picture for cellular automata by two classes of models. The first are fermionic quantum field theories with interactions of a type of Thirring model~\cite{THI, KLA, AAR, FAIV} or Gross-Neveu model~\cite{GN, WWE, RSHA, RWP, SZKSR}. This includes models with continuous non-abelian symmetries. The second class of models implements local gauge symmetries. In particular, we construct an automaton for a discretization of two-dimensional spinor gravity~\cite{CWLSG, CWSGDI, SG3, SG4}. For this model the discrete fermionic action is invariant under local Lorentz transformations. The continuum limit realizes diffeomorphism symmetry. Together with general construction rules these examples should demonstrate that probabilistic cellular automata can indeed describe a wide variety of fermionic models.

In sect.~\ref{sec: FPGU} we start with a single updating step of an automaton and map it to a product of Grassmann variables at neighboring times $t$ and $\te$. Sect.~\ref{sec: LU} deals with local updatings for which the updating of a cell involves only the cell itself. This will be combined subsequently with other updating steps inducing non-trivial propagation dynamics. In sect.~\ref{sec: CA} we discuss chain automata that involve bits positioned at a chain of discrete locations $x$. We focus on automata described by local elementary processes which only involve bits in a certain neighborhood of positions. They constitute cellular automata. Sect.~\ref{sec: SUG} considers the sequence of updatings and represents the corresponding automaton as a Grassmann (functional) integral.

In sect.~\ref{sec: PCA} we introduce probabilistic cellular automata. The probabilistic aspects arise from a probability distribution over initial bit-configurations for the automaton. This is directly related to the probabilistic features of quantum mechanics. We introduce the wave function for the probabilistic cellular automaton. It is identical to the one for the associated fermionic quantum system. The probabilistic formulation gives access to discrete and continuous symmetries that we discuss in sect.~\ref{sec: S}. This includes abelian and non-abelian continuous global symmetries as well as local gauge symmetries. These symmetry concepts are not easily visible in the usual formulation of automata. As familiar from quantum field theories they constitute powerful tools for an understanding of the dynamics, in particular for the continuum limit of the automata.

In sect.~\ref{sec: AOPS} we turn to the construction of automata that constitute interesting discrete fermionic quantum field theories. We begin by sequences of updatings for propagation and scattering, rather analogous to the Feynman path integral for quantum field theories. In sect.~\ref{sec: ANA} we focus on the automata with non-abelian continuous symmetries. The fermionic quantum field theories corresponding to the automata discussed in sects.~\ref{sec: AOPS},~\ref{sec: ANA} are generalized Thirring or Gross-Neveu models. Sect.~\ref{sec: MP} investigates different forms of the propagation step of the updating. This includes disorder and its possible relation to a mass term for the fermions. Sect.~\ref{sec: USB} discusses an updating by shifted blocks. This is employed for the realization of local gauge symmetries in sect.~\ref{sec: CAGT}. In sect.~\ref{sec: DSG} we present a particular automaton that represents spinor gravity~\cite{CWLSG, CWSGDI, SG3, SG4}. The associated quantum field theory exhibits local Lorentz symmetry. Diffeomorphism symmetry is realized in the naive continuum limit.

The second part of this paper discusses important aspects of the continuum limit. We begin by introducing the operators playing an important role for this limit, as the Hamiltonian, the momentum operator and fermion bilinears that may play the role of order parameters. Subsequently, we discuss the concepts of \qq{vacuum} and spontaneous symmetry breaking. The continuum limit is constructed explicitly for one-particle states which may be considered as excitations of a given vacuum. Finally, we address the general mechanisms how coarse graining leads to a loss of the automaton property on the averaged level. The map to quantum theory is completed by a discussion of the complex structure.

In sect.~\ref{sec: OO} we deepen the quantum formalism by introducing operators for observables. The association of operators to observables and the quantum rule for expectation values follow directly from the classical statistical setting for the automaton. Fermionic operators familiar from quantum field theories are translated to the probabilistic automata in sect.~\ref{sec: FO}. For given observables in the fermionic quantum system we construct the equivalent observables for the probabilistic cellular automaton. The operators associated to these observables need not to commute. In sect.~\ref{sec: VOS} we employ concepts familiar from quantum field theory, as vacua and spontaneous symmetry breaking, for the description of stationary states of the probabilistic cellular automaton.

Sect.~\ref{sec: SPWF} discusses the one-particle excitations for various vacua. We introduce the complex wave function and the Fourier transform to a momentum basis for the corresponding states of the cellular automaton. The explicit continuum limit for the one-particle states in one space dimension is performed in sect.~\ref{sec: QMPP} based on a random cellular automaton. This yields a Dirac equation with a potential. In the non-relativistic limit this results in the Schrödinger equation for a particle in a potential. In sect.~\ref{sec: CLCGS} we turn to conceptual issues of the continuum limit and the associated coarse graining which averages over microscopic states. On the coarse grained level the one-particle state follows a unitary evolution, but is no longer described by a cellular automaton. A given state evolves with various probabilities to different states and the uniqueness of the automaton is lost. With this insight we address the emergence of one-particle quantum mechanics in sect.~\ref{sec: CSQM}. In sect.~\ref{sec: C} we present our conclusions.

\section{Fermion picture for updating}
\label{sec: FPGU}

In this section we translate the updating rule for cellular automata to a fermionic expression in terms of Grassmann variables. We take an overall perspective on the states of all cells at a given \qq{time} of the updating. They are characterized by a configuration of $M$ bits or fermionic occupation numbers. There are therefore $2^M$ possible overall states at a given time. In an overall view the updating is realized by mapping each bit configuration to a new one. We cast this map into the form of a step evolution operator. This setting of general automata is more general than the one for cellular automata where the updating is determined by the states of a well defined set of neighboring cells. In the following we will describe case by case for our examples which are the neighbors that determine the updating in a given cell. This will relate the overall view to the standard view of local updatings of cells.

\subsection*{Grassmann elements for bit configurations}

Our first task is an expression of the updating rule of the automaton in terms of Grassmann variables for fermions. For this purpose our starting point is the association of every configuration of $M$ bits (or Ising spins) to a Grassmann basis element for $M$ Grassmann variables~\cite{CWFCB, CWFGI}. We characterize the bits by fermionic occupation numbers $n_\alpha$ that can take the values zero or one, $\alpha=1\dots M$. Corresponding Ising spins are given by $s_\alpha=2n_\alpha-1$.
On the other hand we consider $M$ anticommuting Grassmann variables $\psi_\alpha$, $\{\psi_\alpha,\psi_\beta\}=0$. A Grassmann basis element $g_\tau$ is a product of these Grassmann variables,
\bel{GU1}
g_\tau\opsi=\stil_\tau\prod_{\alpha=1}^{M}\atil_\alpha\ ,
\ee
with factors $\atil_\alpha$ either given by $\psi_\alpha$ or one, and $\stil_\tau=\pm1$ a conveniently chosen sign factor. We can associate $\atil_\alpha=1$ with a particle $\alpha$ present or $n_\alpha=1$, and $\atil\al=\psi\al$ with no particle $\alpha$ present or $n\al=0$. A bit configuration is a sequence of zeros and ones, or occupation numbers $\{n\al\}=\{n_1,n_2\dots n_M\}$. To every Grassmann basis element $g_\tau$ one associates a unique bit configuration. Every factor $\psi\al$ in $g_\tau$ corresponds to a zero in the bit configuration at the position $\alpha$. All other bits equal one. The number of different bit configurations or Grassmann basis elements is given by $N=2^M$, $\tau=1\dots N$. We can use the sequence of occupation numbers $\{n\al\}$ to label $\tau$. For example, we can order $\tau$ by using binary numbers.

\subsection*{State vector and step evolution operator}

We focus first on deterministic cellular automata with some given specific initial bit-configuration. At every step in the updating chain one therefore has a specific bit-configuration $\taubar$. It is convenient to denote this bit configuration by a state vector $q$. This is an $N$-component unit vector with components $q_\tau=\pm\delta_{\tau,\taubar}$. The only non-vanishing component of $q$ therefore directly indicates a particular bit configuration $\taubar$. The change of the bit configuration according to the updating rule is described by a change of the state vector
\bel{GU2}
q_\tau'=\Shat_{\tau\rho}q_\rho\ .
\ee
The \qq{step evolution operator} $\Shat$ is a real $N\times N$ matrix. We restrict our discussion to reversible or invertible cellular automata for which the inverse process also constitutes an automaton. In this case $\Shat$ is given by a \qq{unique jump matrix} that has only elements $\pm1$ or $0$, with a single element $\pm1$ in each row and column. Unique jump operators are orthogonal matrices
\bel{GU3}
\Shat^T\Shat=1\ ,
\ee
and therefore invertible.

An updating rule $\rho\to\taubar$ or $\taubar=\taubar(\rho)$ is realized by elements
\bel{GU4}
\Shat_{\tau\rho}=\pm\delta_{\tau,\taubar(\rho)}=\pm\delta_{\rho,\rhobar(\tau)}\ ,
\ee
with $\rhobar(\tau)$ the inverse map of $\taubar(\rho)$. This realizes for every configuration $\rhobar$ the jump or updating $q\to q'$, with $q_\tau'=\pm q_{\rhobar(\tau)}$. The association between $g_\tau$ and $\tau$ in eq.~\eqref{GU1} does not depend on the sign $\stil_\tau$. The Grassmann description is redundant in this respect. As a consequence, the signs of the non-zero elements of $\Shat$ do not matter for the association of $\Shat$ to a cellular automaton. Also the sign of the non-zero component of $q$ will not be relevant at the present stage.

The step evolution operator specifies the updating rule completely. It provides for an overall picture for the simultaneous updating of all cells of an automaton. As such it is a very general concept. We will have to specify a local structure of cells and its neighbors in order to realize particular cellular automata.

\subsection*{Grassmann expression for updating rule}

As a basic fermionic expression for the updating rule encoded in $\Shat$ we introduce a \qq{local factor} $\Ktil[\psi,\psibar\,]$ which depends on two sets of Grassmann variables $\psi\al$ and $\psibar\al$. It is defined in terms of the step evolution operator by
\bel{GU5}
\Ktil[\psi,\psibar\,]=g_\tau\opsi\Shat_{\tau\rho}g_\rho'\opsib\ .
\ee
Here we introduce for the convenience of sign-manipulations of Grassmann elements
\bel{GU6}
g_\tau'\opsi=\eps_\tau g_\tau\opsi\ ,\quad \eps_\tau=(-1)^{\tfrac{m_\tau(m_\tau-1)}{2}}\ ,
\ee
with $m_\tau$ the number of $\psi$-factors in $g_\tau$. This allows us to employ the convention of summing over double indices in eq.~\eqref{GU5}, or for the identity
\bel{GU7}
\exp\gl\psi\al\psibar\al\gr=\prod\al\gl1+\psi\al\psibar\al\gr=g_\tau\opsi g_\tau'\opsib\ .
\ee
Every arbitrary $\Ktil[\psi,\psibar]$ defines a matrix $\Shat$ by a double expansion in Grassmann basis elements~\eqref{GU5}. What singles out the fermionic description of an automaton are those $\Ktil$ for which $\Shat$ is a unique jump operator.

As an example we consider three Grassmann variables $\psi_1$, $\psi_2$, $\psi_3$ and a local factor
\bel{GU8}
\begin{split}
\Ktil=&1+\psi_1\psibar_2+\psi_2\psibar_3+\psi_3\psibar_1-\psi_1\psi_2\psibar_2\psibar_3\\
&-\psi_2\psi_3\psibar_3\psibar_1-\psi_3\psi_1\psibar_1\psibar_2-\psi_1\psi_2\psi_3\psibar_1\psibar_2\psibar_3\ .
\end{split}
\ee
The step evolution operator is an $8\times8$ matrix that decays into four blocks. The first term $1$ is the product of two Grassmann elements $g_\tau\opsi=1$ and $g_\tau'\opsi=1$. The totally filled bit configuration $(1,1,1)$ is not changed by the updating rule,
\bel{GU9}
\Shat_{(111),(111)}=1\ .
\ee
Similarly, the last term is a product of the basis elements $\psi_1\psi_2\psi_3$ and $\psibar_1\psibar_2\psibar_3$ for the totally empty configuration. The totally empty state remains invariant under the updating
\bel{GU10}
\Shat_{(000),(000)}=1\ .
\ee

Next we have the three products of single Grassmann variables $\psi\al\psibar_\beta$. With $\psibar_1$, $\psibar_2$, $\psibar_3$ associated to the configurations $(011)$, $(101)$, $(110)$, respectively, we infer the elements
\bel{GU11}
\Shat_{(011),(101)}=\Shat_{(101),(110)}=\Shat_{(110),(011)}=1\ .
\ee
In this sector $\Shat$ is a $3\times3$ matrix
\bel{GU12}
\Shat=\begin{pmatrix}0&1&0\\0&0&1\\1&0&0\end{pmatrix}\ .
\ee
For example, it maps the state vector for the configuration $(101)$ to the one for the configuration $(011)$,
\bel{GU13}
q'=\Shat q_{(101)}=\Shat\begin{pmatrix}0\\1\\0\end{pmatrix}=\begin{pmatrix}1\\0\\0\end{pmatrix}=q'_{(011)}\ .
\ee

We can associate the configurations $(011)$, $(101)$, $(110)$ with a \qq{hole} or \qq{antiparticle} of type $1$, $2$, $3$, or $h_1$, $h_2$, $h_3$ respectively.
The cellular automaton maps $h_2\to h_1$, $h_3\to h_2$, $h_1\to h_3$. Looking at eq.~\eqref{GU8}, we can say that a factor $\psibar\al$ in $\Ktil$ annihilates a hole $h\al$, while $\psi\al$ creates a hole $\alpha$ for the updated configuration. This part of  $\Shat$ therefore describes an automaton for which the colors of single holes are changed in a cyclic way.

Finally, associating the factors $\psibar_1\psibar_2$, $\psibar_2\psibar_3$, $\psibar_3\psibar_1$ with the configurations $(001)$, $(010)$, $(100)$ one infers
\bel{GU14}
\Shat_{(001),(100)}=\Shat_{(100),(010)}=\Shat_{(010),(001)}=1\ .
\ee
This part accounts for a cyclic color rotation of the single particle states, $p_1\to p_3$, $p_2\to p_1$, $p_3\to p_2$. As compared to eq.~\eqref{GU13} this part obtains by exchanging particles and holes, corresponding to an exchange $n\al=0\leftrightarrow n\al=1$ between eqs.~\eqref{GU11} and~\eqref{GU14}. With a similar exchange relating eqs.~\eqref{GU9} and~\eqref{GU10} the automaton exhibits a particle-hole symmetry.

\subsection*{Elementary processes}

The construction of the fermionic local factor $\Ktil$ according to eq.~\eqref{GU5} is possible for arbitrary cellular automata. It becomes, however, rather involved for a high number $M$ of particle species, with $\Ktil$ involving a sum over $2^M$ terms. This issue is rather cumbersome for automata formulated on a space lattice with $M_x^D$ points, for which $M$ is proportional to $M_x^D$ and becomes typically very large. We want to reduce the updating rule to a local \qq{elementary process} that at once gives a rule for configurations at arbitrary lattice points.

Elementary processes can be defined for cellular automata for which $\Ktil$ can be expressed as an exponential
\bel{GU15}
\Ktil[\psi,\psibar\,]=\exp\big\{-\cL[\psi,\psibar\,]\big\}\ .
\ee
The quantity $\cL$ will describe the \qq{elementary processes} for the cellular automaton, from which the updating rule can be computed by exponentiation. For the simple cellular automaton~\eqref{GU8} one has
\bel{GU16}
\cL=-\gl\psi_1\psibar_2+\psi_2\psibar_3+\psi_3\psibar_1\gr\ .
\ee
This expression is simpler than the expression for $\Ktil$ in eq.~\eqref{GU8}. The elementary process in this case is the cyclic permutation of holes.

\subsection*{Clock systems}

This structure generalizes to an arbitrary number $M$ of Grassmann variables $\psi\al$. For this purpose we choose
\bel{GU17}
\cL=-F_{\alpha\beta}\psi\al\psibar_\beta\ ,
\ee
where $F_{\alpha\beta}$ is an $M\times M$-matrix with all elements zero or one, and a single element one in each row and column. The matrix $F$ is again a unique jump matrix, but now in the $M$-dimensional space of Grassmann elements, as compared to $\Shat$ which acts in the $2^M$-dimensional space of Grassmann basis elements. For a non-zero element $F_{38}$, for example, the elementary process changes a hole of type $8$ to a hole of type $3$.

For the construction of the step evolution operator associated to $F$ we expand the exponential
\bel{GU18}
\Ktil=\sum_{n=0}^{\infty}\frac{1}{n!}\gl\psi\al F_{\alpha\beta}\psibar_\beta\,\gr^n=\sum_n\Ktil_n\ .
\ee
For $n=0$ the term $1$ maps the totally filled state to the totally filled state. The term with $n=1$ generates the permutation of single holes encoded in $F$. For
\bel{GU19}
\begin{split}
\Ktil_2=&-\frac{1}{2}F_{\alpha\beta}F_{\gamma\delta}\,\psi\al\psi\g\psibar_\beta\psibar_\delta\\
=&-\sum\al\sum_\beta\sum_{\gamma\neq\alpha}\sum_{\delta>\beta}F_{\alpha\beta}F_{\gamma\delta}\,\psi\al\psi\g\psibar_\beta\psibar_\delta\
\end{split}
\ee
we observe that only pairs of different Grassmann variables contribute since $\psi\al^2=0$, $\psibar_\beta^2=0$. Every bilinear $\psibar_\beta\psibar_\delta$ with given $\bar\beta$ and $\bar\delta$ appears in the sum. It actually appears twice, once for $\beta=\bar\beta$ and a second time for $\delta=\bar\beta$. The two terms in the sum are identical. We restrict the sum to $\delta>\beta$, absorbing in this way the factor $1/2$. The bilinears $\psibar_\beta\psibar_\delta$, $\delta>\beta$, denote bit configurations with precisely two holes, one of type $\beta$ and the other of type $\delta$. The part of the step evolution operator corresponding to $\Ktil_2$ maps every two-hole state $(\beta,\delta)$ to a new two-hole state $(\bar\alpha(\beta),\bar\gamma(\delta))$ according to the non-zero elements of $F$. The change of color of two holes is the same as for the single hole, now applied to each hole separately.

This simple discussion demonstrates once more the correspondence between bit-configurations and Grassmann basis elements. A bit-configuration can be characterized by the position of zeros or holes in the configuration $\{n\al\}$. Correspondingly, Grassmann basis elements can only have a single (or no) factor $\psibar\al$ for every $\alpha$, as guaranteed by $\psibar\al^2=0$. Furthermore, the Grassmann rule $\psibar_\delta\psibar_\beta=-\psibar_\beta\psibar_\delta$ implies that $\psibar_\delta\psibar_\beta$ and $\psibar_\beta\psibar_\delta$ are not independent, such that a full set of basis elements describing two-hole states or bit-configurations with two zeros are $\psibar_\beta\psibar_\delta$, $\delta>\beta$.

The continuation of the expansion series is straightforward. The term $n=3$ accounts for the map of three-hole states to color changed three-hole states, and so on. The term $n=M-1$ involves $M-1$ $\psibar$-factors and $M-1$ $\psi$-factors. At this level each element $g_\tau\opsi$ and $g_\tau'\opsib$ has only a single factor $1$, and therefore corresponds to a single particle of the type for which no $\psibar$ and $\psi$ factor is present. The corresponding part of the step evolution operator induces for the single particles the same color change as for single holes. This cellular automaton exhibits again particle-hole symmetry. Accordingly, the totally empty state for $n=M$ remains invariant.

As a special case, for $F_{\alpha\beta}=\delta_{\alpha\beta}$ one obtains the unit step evolution operator. This coincides with the identity~\eqref{GU7}. In contrast, a local factor $\Ktil=1$ does not correspond to a unique jump matrix. The corresponding step evolution operator has almost all elements vanishing, except a single one that maps the totally filled state onto itself.

In the general case $F$ can describe cycles with different periods. The maximal period is $M$, realized for $F^M=1$. It is sufficient to consider the one-hole states which are permuted by $F$. The same permutation holds for the single-particle states. More complex states are simply composed by the corresponding permutations of particles or holes. If $F$ is block diagonal, the permutation of single holes decays into different sectors, with maximal period in each sector given by the size of the block. The general clock character of the elementary process~\eqref{GU17} can be understood as follows: One can start with a given $\alpha_1$ that is mapped to a color that we denote by $\alpha_2$, which in turn transforms to $\alpha_3$, and so on. At the latest after $M$ steps the cycle is complete by reaching again $\alpha_1$. For a smaller cycle $\alpha_1$ is already reached after $M'$ steps, with $M'<M$. In the subsector of species $\alpha_j$ belonging to this chain the period is $M'$. For the remaining part one can start again with some $\beta_1$ among the colors not belonging to the ones of the first cycle, and repeat the procedure for the remaining $M-M'$ colors. One may call the simple automata of this type \qq{cyclic automata} or \qq{clock systems}. If the system has more than one period it consists of several synchronized clocks.

For clock systems the possible structures of cellular automata are easily identified. One may group a family of bits into a cell, denoting the cell by $x$ and the bits belonging to this cell by $\bar\alpha(x)$. We assume that every bit $\alpha$ belongs to precisely one cell. If we arrange the cells $x$ as positions on a $D$-dimensional lattice we can define which cells $y(x)$ are neighbors of $x$. Since clock systems only exchange bits we can determine from $F_{\alpha\beta}$ which bits $\hat\beta$ at $t$ have moved inside the cell $x$ at $\te$. They are characterized by $F_{\bar\alpha\hat\beta}=\pm1$ for one of the $\bar\alpha$ belonging to the cell. A cellular automaton is realized if all $\hat\beta$ belong to cells $y$ in a suitable neighborhood of $x$. The updating is such that only the bits $\hat\beta$ in the neighboring cells $\{y\}$ (which may include $x$) influence the values of the bits $\bar\alpha$ in the cell $x$ at time $\te$. For each bit $\bar\alpha$ in the cell $x$ the non-zero element of the unique jump matrix $F_{\bar\alpha\beta}$ defines a bit $\hat\beta(\bar\alpha)$. The updating rule is $n_{\bar\alpha}(\te)=1$ if $n_{\hat\beta(\bar\alpha)}(t)=1$, and $n_{\bar\alpha}(\te)=0$ if $n_{\hat\beta(\bar\alpha)}(t)=0$.

\section{Local updatings}
\label{sec: LU}

The clock automata are equivalent to fermionic models for which the action involves only terms that are quadratic in the Grassmann variables. This can describe a generalized propagation of particles, but no scattering. We will discuss in sect.~\ref{sec: MP} several interesting aspects of a generalized propagation, as the propagation in a potential. Here we will turn to the issue of scattering which will involve fermionic models for which the action contains terms with more than two Grassmann variables.

We start with a zero-dimensional automaton. The updating in each cell depends only on the state of the cell - possible other neighbors play no role. This is not yet a cellular automaton in the usual sense. We will turn to one-dimensional automata in the next section. The zero-dimensional automaton is useful, however, to introduce the implementation of updating rules for scattering by maps between elements of a Grassmann algebra as the basis for the fermion picture. In sect.~\ref{sec: AOPS} we will combine the local updatings of the present section which describe scattering with a propagation of particles. This can be realized by sequences of different updating steps. For the present section a given state of a single cell is characterized by a configuration of $M$ bits or fermionic occupation numbers. The number of possible states in a cell is therefore $2^M$.

For an automaton with local updatings the only neighbor of a cell is the cell itself. Every unique jump operator $\Shat$ describes a consistent reversible automaton. It can be translated to a local factor $\Ktil[\psi,\psibar\,]$ by the rule~\eqref{GU5}. What is not guaranteed, however, is an expression of $\Ktil[\psi,\psibar\,]$ by a local process $\cL[\psi,\psibar\,]$ by eq.~\eqref{GU15} such that $\cL$ remains simple. We will proceed first in the inverse direction and investigate which simple forms of $\cL$ lead to a unique jump step evolution operator. In the second part of this section we discuss the construction of $\cL$ for simple scattering rules as encoded in a simple for of $\Ktil$.

On the level of fermion bilinears~\eqref{GU17} for clock automata arbitrary unique jump matrices $F_{\alpha\beta}$ are associated to automata. Such a general feature does no longer hold if $\cL$ contains higher powers of Grassmann variables, as appropriate for scattering. A general form of fermion interaction encoded in $\cL$ will not correspond to an automaton. In this section we discuss recipes for the construction of interactions that render a fermion model equivalent to a cellular automaton. In later sections we will present more concrete examples.

\subsection*{Scattering processes}

As a next more general and more complex class of automata built on elementary processes we take $\cL$ as a sum of products of Grassmann basis elements $g_{\taubar}\opsi$ and $g_{\bar\rho}'\opsib$,
\bel{GU20}
\cL=-B_{\taubar\bar\rho}g_{\taubar}\opsi g_{\bar\rho}'\opsib\ .
\ee
Here we restrict the number of indices $\bar\rho$ for which $B_{\taubar\bar\rho}$ differs from zero to $P$, with typically $P$ much smaller than $N$. Similarly, $B_{\taubar\bar\rho}$ differs from zero only for $P$ values of the index $\taubar$. Within the restricted index spaces $B$ is taken as a $P\times P$ unique jump matrix.

We focus on particle number conserving automata for which $B_{\taubar\bar\rho}$ vanishes whenever $g_{\taubar}\opsi$ and $g_{\bar\rho}\opsib$ have a different number of $\psi$ or $\psibar$ factors.
As a first example we take only Grassmann basis elements with precisely two $\psi$-factors
\bel{GU21}
\begin{split}
g_{\bar\rho}\opsib=&\psibar_\beta\psibar_\delta\ ,\quad \delta \geq\beta\ ,\\
g_{\bar\rho}\opsib=&\psibar\al\psibar\g\ ,\quad \gamma\geq\alpha\ .
\end{split}
\ee
This yields
\bel{GU22}
\cL=-\sum_{\alpha,\beta,\gamma>\alpha,\delta>\beta}B_{\alpha\gamma,\beta\delta}\psi\al\psi\g\psibar_\beta\psibar_\delta\ .
\ee
The first term in the expansion~\eqref{GU18} for $n=0$ maps the totally filled state into itself. The term $n=1$ maps two-hole states onto two-hole states. It is similar to the expression~\eqref{GU19}, which obtains as a special case
\bel{GU23}
B_{\alpha\gamma,\beta\delta}=F_{\alpha\beta}F_{\gamma\delta}\ .
\ee
It is more general, however, since the rule how a pair of holes transforms into another pair of holes is left free, instead of being specified by the combination of single hole transformations in eq.~\eqref{GU19}.

The process~\eqref{GU22} describes an \qq{elementary scattering process} where two holes of type $\beta$ and $\delta$ are changed into two holes of type $\alpha$ and $\gamma$. From this elementary $2\to2$ scattering higher order scatterings follow by higher order terms in the expansion of the exponential.

The term $n=2$ in the exponential expansion maps four-hole states into four-hole states. For the Grassmann basis elements corresponding to four holes $\psibar_\beta\psibar_\delta\psibar_{\beta'}\psibar_{\delta'}$ all colors $\beta$, $\delta$, $\beta'$, $\delta'$ must be different. This arises from the expansion for the products $\psibar_\beta\psibar_\delta$ with $\psibar_{\beta'}\psibar_{\delta'}$ which differ from zero only if no indices coincide.
The cellular automaton is defined only for those four-hole states that appear in the expansion of the exponential. Each independent Grassmann basis element $\psibar_\beta\psibar_\delta\psibar_{\beta'}\psibar_{\delta'}$ must be multiplied with a unique Grassmann basis element $\psi_\alpha\psi_\gamma\psi_{\alpha'}\psi_{\gamma'}$ in $\Ktil$, with a coefficient $\pm1$. Otherwise $\Shat_{\tau\rho}$ is not a unique jump operator in the four-hole sector.

This generalizes to higher orders $n$. First the Grassmann basis elements $g_\tau\opsib$ appearing in the exponential expansion define the space of bit configurations on which the cellular automaton acts. In our case these are configurations with an even number of zeros or holes, and colors of the holes dictated by $B$. Second, each Grassmann basis element $g_\tau\opsib$ appearing in $\Ktil$ must be multiplied with a unique Grassmann basis element $g_\rho\opsi$. Otherwise the uniqueness and completeness of the updating is lost.

\subsection*{Restrictions for cellular automata}

Arbitrary elementary $2\to2$ scatterings do not necessarily define a cellular automaton. There are restrictions on the allowed $B_{\alpha\gamma,\beta\delta}$ in eq.~\eqref{GU22} which result from the requirement that $\Shat_{\tau\rho}$ must be a unique jump operator. Already at the level $n=2$ a problem becomes immediately visible, namely that a given four-hole configuration can be constructed by several distinct combinations of two-hole Grassmann elements. For example, $\psibar_1\psibar_2\psibar_3\psibar_4$ can be obtained from the products $\gl\psibar_1\psibar_2\gr\cdot\gl\psibar_3\psibar_4\gr$, $\gl\psibar_1\psibar_3\gr\cdot\gl\psibar_2\psibar_4\gr$ or $\gl\psibar_1\psibar_4\gr\cdot\gl\psibar_2\psibar_3\gr$. There can therefore be three distinct terms in the sum contributing to $\psibar_1\psibar_2\psibar_3\psibar_4$ for 
\bel{GU24}
\Ktil_2=\frac{1}{2}\Big(B_{\taubar\bar\rho}g_{\taubar}\opsi g_{\bar\rho}\opsib\Big)^2\ .
\ee
The sum of the different contributions could be multiplied by a sum of different basis elements $g_\tau\opsi$, destroying uniqueness and thereby preventing $\Shat$ to be a unique jump operator in this sector.
(In addition there is the standard factor two since each of the three combinations can be realized by exchanging contributions from the two factors in the square~\eqref{GU24}. This removes the prefactor $1/2$, as discussed previously.) As an example we can take $\cL=-\gl\psi_1\psi_2\psibar_3\psibar_4+\psi_3\psi_4\psibar_1\psibar_2+\psi_1\psi_5\psibar_2\psibar_4+\psi_2\psi_6\psibar_1\psibar_3+\text{other terms}\gr$. For $\Ktil_2$ one obtains $\Ktil_2=\gl\psi_1\psi_2\gr\gl\psi_3\psi_4+\psi_5\psi_6\gr\psibar_1\psibar_2\psibar_3\psibar_4+\dots$. While the two-hole states are scattered to unique two-hole states, a particular four-hole state is scattered to two different four-hole states. This is not a cellular automaton.

There are different ways how a valid unique jump step evolution operator can be obtained. In the present section we focus first on the possibility that not all possible combinations $\psibar_\beta\psibar_\delta$ are contained in the list of those contributing non-vanishing elements to $B_{\taubar\bar\rho}$ in eq.~\eqref{GU20} or $B_{\alpha\gamma,\beta,\delta}$ in eq.~\eqref{GU22}. By restricting the space of states the step evolution operator can be a unique jump operator for the remaining states. Instead of eliminating the states not participating in the scattering one can also implement a unit evolution for those states, as discussed below and in sect.~\ref{sec: AOPS}. As a first example for restricted states we can take $\cL=-\gl\psi_1\psi_2\psibar_3\psibar_4+\psi_3\psi_4\psibar_1\psibar_2\gr$. The automaton acts only on the two-hole states corresponding to $\psibar_1\psibar_2$ and $\psibar_3\psibar_4$, as well as $\psi_1\psi_2$ and $\psi_3\psi_4$. On the level of four-hole scattering there is then a unique four-hole state that is left invariant, according to $\Ktil_2=\psi_1\psi_2\psi_3\psi_4\psibar_1\psibar_2\psibar_3\psibar_4$. This can be accompanied by other terms involving sets of Grassmann variables different from $\alpha=1,\dots4$.

For a somewhat more complex example that we will use later, the Grassmann variables $\psi\al$ may be split into two classes $\psi_{+a}$ and $\psi_{-a}$, $a=1\dots M/2$. Suppose that the sum~\eqref{GU20} or~\eqref{GU22} contains only terms with one factor $\psi_+$ and the other $\psi_-$, and similar for $\psibar$,
\bel{GU25}
\cL=-B_{ac,bd}\psi_+^a\psi_-^c\psibar_+^b\psibar_-^d\ .
\ee
A given four-hole combination $\psi_+^1\psi_-^2\psibar_+^3\psibar_-^4$ could still arise from two different products $\gl\psibar_+^1\psibar_-^3\gr\cdot\gl\psibar_+^2\psibar_-^4\gr$ or $\gl\psibar_+^1\psibar_-^4\gr\cdot\gl\psibar_+^2\psibar_-^3\gr$. We may divide the index $a$ into even and odd, and choose $B$ such that it differs from zero only if the pairs $(ac)$ and $(bd)$ contain each one even and one odd index. In this case $\psibar_+^1\psibar_-^2\psibar_+^3\psibar_-^4$ can only be realized by the product $\gl\psibar_+^1\psibar_-^4\gr\cdot\gl\psibar_+^2\psibar_-^3\gr$.

\subsection*{Simple scattering automaton}

We may consider for eight Grassmann variables, $a=1\dots4$, a model based on
\bel{GU26}
\begin{split}
\cL=&-\gl\psi_+^1\psi_-^4\psibar_+^2\psibar_-^3+\psi_+^2\psi_-^3\psibar_+^1\psibar_-^4\\
&+\psi_-^1\psi_+^4\psibar_-^2\psibar_+^3+\psi_-^2\psi_+^3\psibar_-^1\psibar_+^4\gr\ .
\end{split}
\ee
The second term in the expansion of the exponential reads
\bel{GU27}
\begin{split}
\frac{1}{2}\cL^2=\phantom{+}&\psi_+^1\psi_+^2\psi_-^3\psi_-^4\psibar_+^1\psibar_+^2\psibar_-^3\psibar_-^4\\
+&\psi_-^1\psi_-^2\psi_+^3\psi_+^4\psibar_-^1\psibar_-^2\psibar_+^3\psibar_+^4\\
+&\psi_+^1\psi_-^2\psi_+^3\psi_-^4\psibar_-^1\psibar_+^2\psibar_-^3\psibar_+^4\\
+&\psi_-^1\psi_+^2\psi_-^3\psi_+^4\psibar_+^1\psibar_-^2\psibar_+^3\psibar_-^4\\
+&\psi_+^1\psi_-^1\psi_+^4\psi_-^4\psibar_+^2\psibar_-^2\psibar_+^3\psibar_-^3\\
+&\psi_+^2\psi_-^2\psi_+^3\psi_-^3\psibar_+^1\psibar_-^1\psibar_+^4\psibar_-^4\ .
\end{split}
\ee
The six terms are different products of Grassmann basis elements for four-hole states. The step evolution operator in this sector is therefore indeed a unique jump matrix. The four terms in third order,
\bel{GU28}
\begin{split}
-\frac{1}{6}\cL^3=\phantom{+}&\psi_+^2\psi_+^3\psi_+^4\psi_-^1\psi_-^2\psi_-^3\psibar_+^1\psibar_+^3\psibar_+^4\psibar_-^1\psibar_-^2\psibar_-^4\\
+&\psi_+^1\psi_+^3\psi_+^4\psi_-^1\psi_-^2\psi_-^4\psibar_+^2\psibar_+^3\psibar_+^4\psibar_-^1\psibar_-^2\psibar_-^3\\
+&\gl\psi_+\leftrightarrow\psi_-\,,\,\psibar_+\leftrightarrow\psibar_-\gr\ ,
\end{split}
\ee
involve products of Grassmann basis elements corresponding to six-hole states. For eight different Grassmann variables $\psi_\pm^a$ the six-hole states correspond to two-particle states. The four terms~\eqref{GU28} are the particle-hole transformed terms of eq.~\eqref{GU26}. Indeed, for the product $\psibar_+^1\psibar_+^3\psibar_+^4$ the missing $\psibar_+$-factor is $\psibar_+^2$ etc. Finally, the term $\tfrac{1}{24}\cL^4$ is a product of the Grassmann elements containing all eight spinors. The totally empty state is left invariant. With eq.~\eqref{GU27} being invariant under the particle-hole transformation the step evolution operator for the model~\eqref{GU26} is particle-hole invariant.

We conclude that eq.~\eqref{GU26} defines a unique jump step evolution operator acting on the particular states with $0$, $2$, $4$, $6$, $8$ holes that appear in the expressions~\eqref{GU26}\eqref{GU27}\eqref{GU28}, plus totally filled and empty states.
If we restrict the bit-configurations to this subset, eq.~\eqref{GU26} defines a cellular automaton. The strategy followed for this model is to employ only a sparse set of Grassmann basis elements in eq.~\eqref{GU20}, such that for all of the terms appearing in the expansion of the exponential there is a unique product from which it is obtained. In our case, only four out of the $28$ possible bilinears for eight Grassmann variables are used. In this restricted sector the elementary process is a two-hole to two-hole scattering. The restricted cellular automaton involves $1+4+6+4+1=16$ bit-configurations, out of the $2^8=256$ possible bit-configurations for eight bits. The restriction operates by appropriate selection rules for the \qq{allowed} states. The updating rule for this automaton is simple. Pairs of \qq{occupied bits} ($n\al=1$) of the types $+$ and $-$ with appropriate colors are scattered into pairs of occupied bits with associated colors, according to
\bel{28A}
(1,4)\leftrightarrow(2,3)\ ,\quad (4,1)\leftrightarrow(3,2)\ .
\ee
Here the first color stands for $n_+$ and the second for $n_-$. The same rule holds for pairs of \qq{empty bits} or \qq{holes} ($n\al=0$).

It is straightforward to build other automata based on the structure~\eqref{GU26}. For example, the four bilinears $\gl\psi_+^1\psi_-^4\gr$, $\gl\psi_+^2\psi_-^3\gr$, $\gl\psi_-^1\psi_+^4\gr$ and $\gl\psi_-^2\psi_+^3\gr$ can be permuted arbitrarily. The exchange $\gl\psi_+^1\psi_-^4\gr\leftrightarrow\gl\psi_+^2\psi_-^3\gr$ can be achieved by a simple transformation of variables $\psi_+^1\leftrightarrow\psi_+^2$, $\psi_-^4\leftrightarrow\psi_-^3$. This does not change the structure of the automaton, the other two pairs $\gl\psi_-^1\psi_+^4\gr$ and $\gl\psi_-^2\psi_+^3\gr$ remaining unaffected. One also can perform more general variable transformations on $\psi\al$ at the price of changing the space of allowed bit configurations after the updating steps.

\subsection*{Complete local scattering automata}

So far we have achieved the unique jump property of the step evolution operator by restricting the states of the automaton to the ones participating in the scattering. This leads to a restricted local factor $\Ktil\subt{sc}$. It defines the bit-configurations taking part in the scattering by those for which the associated Grassmann elements appear in the double expansion of $\Ktil\subt{sc}$ according to eq.~\eqref{GU5}. The Grassmann elements not appearing in this expansion define the bit-configurations in the complement of the scattering bit-configurations. We can define $\Ktil\subt{com}$ by a unique jump step evolution operator in the sector of the complement states. If it is defined for all states in the complement we can define the \qq{complete local scattering automaton} by a local factor
\bel{CS1}
\Ktil[\psi,\psibar\,]=\Ktil\subt{sc}[\psi,\psibar\,]+\Ktil\subt{com}[\psi,\psibar\,]\ .
\ee
For the example~\eqref{GU26} the $16$ scattering bit-configurations have a complement of $240$ bit-configurations. The factor $\Ktil\subt{com}$ has to involve all $240$ Grassmann elements in the complement for $\psibar$. Each such element has to be multiplied by precisely one element from the complement of $\psi$. The complete local scattering automaton is defined for all $2^8$ bit configurations formed from $8$ bits.

A simple type of complete automaton takes a unit step evolution operator in the complement sector. For this type of updating all bit-configurations in the complement remain invariant, while the other configurations take part in the scattering according to $\Ktil\subt{sc}$. While very simple on the level of the automaton or the step evolution operator, this type of automaton often takes a somewhat more complex form of $\cL$ in the exponential expression~\eqref{GU15}. One has to realize the restricted unit operator by a suitable local factor
\bel{CS3}
\begin{split}
\cK\subt{com}=&\exp\big\{\psi\g\psibar\g\big\}\big|\subt{restr.}\\
=&\exp\big\{\psi\g\psibar\g\big\}-\Ktil\subt{cor}\ .
\end{split}
\ee
The correction factor $\Ktil\subt{cor}$ becomes necessary since the expansion of $\exp\{\psi\g,\psibar\g\}$ also involves products of Grassmann elements $\psibar$ that belong to the scattering configurations. These products have to be subtracted in order to restrict the unit step evolution operator to the complement.

Subsequently, one has to find $\cL$ according to
\bel{CS3}
\exp\big\{\psi\g\psibar\g\big\}-\Ktil\subt{cor}+\Ktil\subt{sc}=\exp\big\{-\cL\big\}=\exp\big\{\psi\g\psibar\g-\cL\subt{int}\big\}\ .
\ee
If $\Ktil\subt{sc}$ contains a constant term, $\Ktil\subt{sc}=1+\dots$, we subtract it such that $\Ktil\subt{cor}$ does not contain a constant term either. Assume next that the remaining $\Ktil\subt{sc}$ contains as leading terms in an expansion in $\psibar$ an expression $\Ktil\subt{sc}=\frac12A_{\alpha\beta}[\psi]\psibar\al\psibar_\beta$, $A_{\alpha\beta}=-A_{\beta\alpha}$. We then have to subtract the identity contribution in the sector $\psibar\al\psibar_\beta$ for those pairs $(\alpha,\beta)$ for which $A_{\alpha\beta}[\psi]$ does not vanish,
\bel{CS4}
\Ktil\subt{cor}=-\frac12\sum_{(\alpha,\beta)}\psi\al\psi_\beta\psibar\al\psibar_\beta+\dots\ .
\ee
This yields a leading term
\bel{CS5}
\cL\subt{int}^{(0)}=-\frac12\sum_{(\alpha,\beta)}\gl A_{\alpha\beta}[\psi]-\psi\al\psi_\beta\gr\psibar\al\psibar_\beta\ ,
\ee
where the sum extends only over those pairs $(\alpha,\beta)$ for which $A_{\alpha\beta}\neq0$. One next expands $\exp\{\psi\g\psibar\g-\cL\subt{int}^{(0)}\big\}$ in the order $\psibar\al\psibar_\beta\psibar\g$. The difference to the expansion of the l.h.s. of eq.~\eqref{CS3} in the same order yields the first subleading term $\cL\subt{int}^{(1)}$. This procedure can be repeated in order to obtain an expansion $\cL\subt{int}=\cL\subt{int}^{(0)}+\cL\subt{int}^{(1)}+\cL\subt{int}^{(2)}+\dots$ in powers of $\psibar$. Typically the expansion terminates because only a finite number of different $\psibar\al$ is available. While the procedure is systematic, the result for $\cL\subt{int}$ can be somewhat involved. Examples are discussed in sects.~\ref{sec: AOPS},~\ref{sec: ANA},~\ref{sec: USB}. We observe that $\Ktil\subt{sc}$ needs not be in an exponential form. For example, we can describe with $\Ktil\subt{sc}$ a pure two-hole to two-hole scattering without introducing the multi-hole scattering which would be induced by an exponential expression.

\subsection*{Automata without particle number conservation}

For particle number conserving automata both $\cL$ and $\Ktil$ have only terms with an equal number of factors $\psi$ and $\psibar$. The fermion picture is not limited to this case. Often a simple scattering process generates a somewhat more complex form of $\cL$. Consider four bits, $\alpha=1\dots4$, and a $2$ to $4$ scattering together with its inverse. For definiteness assume that species $1$ and $2$ are scattered to four particles. Together with the inverse process this amounts to
\bel{CS6}
\Ktil\subt{sc}=\psi_1\psi_2\psi_3\psi_4\psibar_1\psibar_2+\psi_1\psi_2\psibar_1\psibar_2\psibar_3\psibar_4\ .
\ee
With
\bel{CS7}
\Ktil\subt{cor}=-\psi_1\psi_2\psibar_1\psibar_2+\psi_1\psi_2\psi_3\psi_4\psibar_1\psibar_2\psibar_3\psibar_4
\ee
this yields
\begin{align}
\label{CS8}
&\Ktil=1+\psi\al\psibar\al-\frac12\psi\al\psi\bet\psibar\al\psibar\bet+\psi_1\psi_2\psibar_1\psibar_2\\
&-\frac16\psi\al\psi\bet\psi\g\psibar\al\psibar\bet\psibar\g+\psi_1\psi_2\psi_3\psi_4\psibar_1\psibar_2+\psi_1\psi_2\psibar_1\psibar_2\psibar_3\psibar_4\ .\nn
\end{align}
Following the procedure outlined before one obtains for the interaction part in eq.~\eqref{CS3}
\begin{align}
\label{CS9}
\cL\subt{int}=-\Big\{&\psi_1\psi_2\psibar_1\psibar_2+\psi_1\psi_2\psi_3\psi_4\psibar_1\psibar_2+\psi_1\psi_2\psibar_1\psibar_2\psibar_3\psibar_4\nn\\
-&\psi_1\psi_2\psi_3\psibar_1\psibar_2\psibar_3-\psi_1\psi_2\psi_4\psibar_1\psibar_2\psibar_4\nn\\
-&\psi_1\psi_2\psi_3\psi_4\psibar_1\psibar_2\psibar_3\psibar_4\Big\}\ .
\end{align}

\section{Chain automata}
\label{sec: CA}

Chain automata are defined by an ordering of bits on a chain of \qq{space points} $x=m_x\eps$, $m_x$ integer. The lattice distance $\eps$ can be used to introduce units of length for $x$. The integers $m_x$ or the positions $x$ label the cells of the automaton. Neighbors are defined in an obvious way. For example, the next neighbors of the cell $m_x$ are the cells $m_x+1$ and $m_x-1$. We will often take the chain to be periodic, with a total number of space points $M_x$. On each point $x$ one may have several species, denoted by an index $\gamma=1\dots M$. The occupation numbers or bits are therefor $n\g(x)$, and the associated Grassmann variables are $\psi\g(x)$. There are $2^M$ possible states of a cell, and a total of $2^{M_xM}$ different possible bit configurations for the overall state of the automaton at any given time $t$.

We focus on chain automata which are based on \qq{local elementary processes}. This means that $\cL$ is a sum of local terms
\bel{C1}
\cL=\sum_x L(x)\ ,
\ee
with $L$ involving only Grassmann variables in the neighborhood of $x$. As a result, the local factor $\Ktil$ becomes a product
\bel{C2}
\Ktil=\prod_x\exp\{-L(x)\}\ .
\ee
We assume that $L(x)$ is chosen such that $\Ktil$ is associated to a unique jump step evolution operator and therefore to an automaton.

In appendix~\ref{app: A1} we establish that every local chain automaton defined by eq.~\eqref{C2} is a cellular automaton. This general establishment of the cellular property of all local chain automata may seem somewhat involved. In practice, it will be very straightforwards to establish for our concrete models from which positions $y$ a fermion of a given species $\gamma$ at position $x$ can have originated, and to classify the different possibilities for $\eta$ to lead to a given $\sigma$. The proof that local chains are cellular automata relies on the property that $\Ktil$ generates an overall unique jump operator $\Shat$. We will have to understand the conditions on $L(x)$ for $\Ktil$ to generate a unique jump step evolution operator.

\subsection*{Transport automata}

If $L(x)$ contains only one factor $\psi$ and one factor $\psibar$ the updating retains the structure~\eqref{GU17}, provided every factor $\psibar\g(x)$ and every factor $\psi\g(x)$ appears in the sum $\sum_x L(x)$. A simple example is the \qq{right-transport automaton}
\bel{C3}
L(x)=-\psi\g(x+\eps)\psibar\g(x)\ .
\ee
The local elementary process transports a hole of type $\gamma$ at $x$ to a hole of type $\gamma$ at $x+\eps$. This is done for all colors $\gamma$ and all positions $x$. By exponentiation every hole in the bit configuration is transported by one unit $\eps$ to the right (i.e. to increasing $x$). As a consequence, the whole bit configuration is displaced by one unit to the right. The right-transport automaton is particle-hole symmetric. The corresponding left-transport operator replaces in eq.~\eqref{C3} $x+\eps$ by $x-\eps$.

One can combine the right-transport with a color rotation by defining
\bel{C4}
L(x)=-F_{\gamma\delta}(x)\psi\g(x+\eps)\psibar_\delta(x)\ ,
\ee
with $F_{\gamma\delta}(x)$ a unique jump matrix for every $x$. The simple bilinear form~\eqref{C4} offers a rather rich variety of transport operators. One can combine right-transport and left-transport by having them act on different species of holes or particles, that may be named right-movers or left-movers. While pure right-transport or left-transport has for the updating of the cell $x$ a single neighbor $x-\eps$ or $x+\eps$, a combined right- and left-transport describes a cellular automaton with two neighbors of $x$, namely $x-\eps$ and $x+\eps$.

\subsection*{Local scattering automata}

One may combine transport with local scattering in a next step of the updating, cf. sect.~\ref{sec: AOPS}. Rather interesting automata can be constructed in this way. In the remaining parts of this section we discuss different local scattering automata. They may or may not be combined with transport or other local scattering automata in subsequent steps.

A particularly simple form of local chain automata is realized for \qq{strictly local} scattering for which $L(x)$ in eq.~\eqref{C1} involves only $\psi\g(x)$ and $\psibar_\delta(x)$ at the point $x$. The updating proceeds for every point $x$ separately. The only cell neighboring $x$ is the cell $x$ itself. If $L$ defines a unique jump operator, $\Ktil$ leads to a unique jump operator.
For the local scattering automata we can employ the zero-dimensional automata discussed in the preceding section.

\subsection*{Next neighbor scattering}

Instead of local scattering one can consider \qq{next-neighbor scattering}. The elementary process maps two neighboring holes at $x$ and $x+\eps$ to two neighboring holes with different colors. This is encoded in
\bel{C5}
L(x)=-B_{\gamma\delta,\eta\varphi}(x)\psi\g(x+\eps)\psi_\delta(x)\psibar_\eta(x+\eps)\psibar_\varphi(x)\ ,
\ee
with $B_{\gamma\delta,\eta\varphi}$ a unique jump operator in the space spanned by the index pairs $(\eta\varphi)$ and $(\gamma\delta)$. We have to find forms of $B_{\gamma\delta,\eta\vp}$ for which $L(x)$ generates a unique jump step evolution operator.

Consider first the case that $B_{\gamma\delta,\eta\varphi}(x)$ vanishes for all odd $x$ $(x=m_x\eps\,,\ m_x\ \text{odd})$. In this case the sum over $x$ in eq.~\eqref{C1} extends effectively only over even $x$. The factors $\exp\gl-L(x)\gr$ contain all fermions at different positions, such that
\bel{C6}
\Ktil=\prod_{x\,\text{even}}\exp\gl-L(x)\gr
\ee
involves independent local automata for all even $x$. From the point of view of updating the neighbors of the cell $x$ are the cells $x$ and $x+\eps$. The same holds for the cell $x+\eps$. We could also consider extended cells $(x,x+\eps)$ for which the updating is strictly local.

The local automata are of the type of the two-hole scattering~\eqref{GU25}, with $\gl\psi_+,\psi_-\gr$ are replaced by $\gl\psi(x+\eps),\psi(x)\gr$, colors $b,d$ corresponding here to $\eta,\varphi$, and similar for $\psibar$. We can realize local automata of the type~\eqref{GU26} by
\begin{align}
\label{C7}
L(x)=&-\big{\{}\big{[}\psi_1(x+\eps)\psi_4(x)\psibar_2(x+\eps)\psibar_3(x)+(\psi\leftrightarrow\psibar)\big{]}\nn\\
&+\big{[}x+\eps\leftrightarrow x\big{]}\big{\}}\ .
\end{align}
The corresponding automaton acts on all bit configurations for which all zeros can be grouped into neighboring pairs at $x+\eps$ and $x$ (for $x$ even), with an additional selection rule for the allowed color combinations of the pairs.

Adding non-zero $B(x)$ at odd $x$ the structure becomes more complicated. The terms $L(x)$ and $L(x+\eps)$ both contain Grassmann variables $\psibar\g(x+\eps)$. Similarly, the variables $\psi\g(x)$ appear in $L(x)$ and $L(x-\eps)$. For the updating of the cell $x$ the neighbors are $(x-\eps,x,x+\eps)$. As a result, the factors in $\Ktil=\prod_x\exp\{-L(x)\}$ are no longer disconnected, and the automaton looses its (almost) strictly local structure. Propagation can now be implemented by the next-neighbor scattering alone, even if no additional transport steps are implemented.

One may guess that the fermion model~\eqref{C7} describes an automaton for which the allowed states are pairs of holes or bits in two neighboring cells, in the color combinations $(1,4)$, $(4,1)$, $(2,3)$, $(3,2)$. The updating changes $(1,4)\leftrightarrow(2,3)$, $(4,1)\leftrightarrow(3,2)$. We will see that this is indeed the case. For a proof one needs to establish that $\Ktil$ generates a unique jump step evolution operator with these properties. The exponential expansion~\eqref{C6} indicates which overall bit configurations take part in the scattering.

The connection of different cells raises the question if every Grassmann basis element $g_\tau\opsib$ appearing in the double expansion of $\Ktil$ emerges in a unique way from factors in the exponential expansion for $\exp(-\cL)$, being then multiplied by a unique $g_\rho\opsi$. For the local automaton~\eqref{C7} this is the case, such that a cellular automaton is realized according to our general argument.

A proof of this statement is rather instructive for the construction of cellular automata and is therefore given in some detail. (We consider here open boundary conditions.) The automaton property of the first term in the expansion which describes the scattering of a pair of holes is rather obvious. For the second term in the expansion of the exponential the new types of terms induced by $L$-factors at neighboring $x$ correspond to
\bel{C8}
\begin{split}
\frac{1}{2}L(x+\eps)&L(x)=\psi_1(x+2\eps)\psi_4(x+\eps)\psi_1(x+\eps)\psi_4(x)\\
\times&\psibar_2(x+2\eps)\psibar_3(x+\eps)\psibar_2(x+\eps)\psibar_3(x)+\dots
\end{split}
\ee
They describe the transformation of states with two holes at $x+\eps$, one hole at $x$ and another hole at $x+2\eps$, to similar states with different colors. The fermion bilinear at $x+\eps$ does not arise in a unique way from the expansion of the exponential. For example, the factor $\psibar_4(x+\eps)\psibar_1(x+\eps)$ could be generated by the product of $\psibar_1(x+2\eps)\psibar_4(x+\eps)$ from $L(x+\eps)$ and $\psibar_1(x+\eps)\psibar_4(x)$ from $L(x)$. Alternatively, it could arise from $\psibar_4(x+2\eps)\psibar_1(\xe)$ in $L(\xe)$ and $\psibar_4(\xe)\psibar_1(x)$ in $L(x)$. The two possibilities are distinguished, however, by different Grassmann variables at $x+2\eps$ and $x$. The Grassmann basis element $\psibar_1(x+2\eps)\psibar_1(x+\eps)\psibar_4(x+\eps)\psibar_4(x)$ has a unique origin from the expansion.

This generalizes to all other terms in second order in the expansion. Indicating by $(n_1,n_2,n_3)$ the number of holes at neighboring points $x+2\eps$, $x+\eps$, $x$, the mixed terms have the structure $(1,2,1)$. The origin of a particular color combination is dictated by the colors of the \qq{exterior sites} with one hole.
The color at $x$ contributes a hole with a definite color at $x+\eps$ according to the pairs appearing in $L(x)$. Also the color at $x+2\eps$ contributes another hole at $x+\eps$ with color dictated by $L(x+\eps)$. Non-vanishing contributions require that the two colors at $x+\eps$ are different.

At higher levels of the expansion the mixed terms from products of neighboring $L(x)$ generate hole-configurations of the type $(1,2,2,1)$, $(1,2,2,2,1)$, $(1,3,3,1)$, $(2,3,3,2)$, $(2,3,4,3)$ etc. The unique origin of a given color combination from the various factors of $L(x)$ can be inferred stepwise. One starts at $x$. If there is a single hole at $x$ the factor $L(x)$ contributes a particular color for the holes at $x+\eps$. If there are several holes at $x$, several colors at $x+\eps$ are fixed by $L(x)$. Subtracting the colors of the holes at $x+\eps$ that arise from these factors $L(x)$ one remains at $x+\eps$ with a certain combination of colors for holes. The subtraction of the contribution from $L(x)$ leads to reduced hole configurations for $x+\eps$, $x+2\eps$ etc. They are given for the examples above by $(1,2,1)$, $(1,2,2,1)$, $(1,3,2)$, $(2,3,1)$ and $(2,3,1)$. We can now repeat the procedure in a second step. The remaining colors of holes at $x+\eps$ contribute particular colors for the holes at $x+2\eps$, according to the pairs allowed for $L(x+\eps)$. Subtracting those again one obtains reduced configurations for $x+2\eps$, $x+3\eps\dots$, which are given for our examples by $(1,1)$, $(1,2,1)$, $(1,1)$, $(2,2)$, $(2,2)$. If only pairs with equal numbers remain, as $(1,1)$ or $(2,2)$ in our case, the possible color combinations are dictated by $L(x+2\eps)$. In this case the color-assignment process ends, with the color content for all factors of $L(x)$, $L(x+\eps)$ and $L(x+2\eps)$ fixed uniquely. If not, one proceeds to a further reduction step. If the assignment procedure ends before all holes are accounted for one starts a new cluster at the next value for $x$ for which a hole is present.

In summary, if $L(x)$ is of a type where for every $\psibar\g(x)$ the accompanying color of $\psibar_\delta(x+\eps)$ is uniquely fixed, we can define $\cL(x)=\sum_xL(x)$ by a sum over all $x$, and $\exp\gl-\cL(x)\gr$ defines a local factor for which the step evolution operator is a unique jump operator. This is realized by a matrix $B$ in eq.~\eqref{C5} which differs from zero only if every index $\eta$ and $\varphi$ belongs uniquely to a given pair, and similar for $\gamma$ and $\delta$. The bit-configuration on which the cellular automaton acts have an even number of zeros. The colors of the bit-configurations obey selection rules that correspond to the pairs appearing in $L(x)$. A large variety of cellular automata can be constructed in this way.

For the particular local chain automaton~\eqref{C7} the updating rule is finally very simple. Once one restricts the allowed overall configurations to the ones where all holes or bits with value zero can be grouped into neighboring pairs with color combinations $(1,4)$, $(4,1)$, $(2,3)$ and $(3,2)$, the updating proceeds by a simple exchange $(1\leftrightarrow2)$ and $(3\leftrightarrow4)$ in each cell $x$. For the restricted configurations the neighbors at $x-\eps$ and $x+\eps$ actually do not influence this updating. The only point beyond this almost trivial updating is the restriction to a subset of allowed configurations which remains preserved by the updating. We could \qq{complete} the updating by a unit operator for the remaining configurations not belonging to the restricted set. Then the neighbors at $x-\eps$ and $x+\eps$ are needed in order to decide if the holes at $x$ belong to one of the restricted configurations or not, and therefore if colors are exchanged or not.

We can further combine the next-neighbor scattering with transport. This is achieved by replacing $L(x)$ in eq.~\eqref{C5} by
\bel{C9}
L_R(x)=-B_{\gamma\delta,\eta\varphi}^{(R)}(x)\psi\g(x+2\eps)\psi_\delta(x+\eps)\psibar_\eta(x+\eps)\psibar_\varphi(x)\ .
\ee
The resulting step evolution operator $\Shat$ is a product of the operator for the next-neighboring scattering~\eqref{C5} with the right-transport operator restricted to the allowed configurations. The two operators commute. The neighbors of the cell $x$ for the updating are shifted by one position to the left, e.g. $y_i=(x-2\eps,x-\eps,x)$. Left-transport is realized in a similar way
\bel{C10}
L_L(x)=-B_{\gamma\delta,\eta\varphi}^{(L)}(x)\psi\g(x)\psi_\delta(x-\eps)\psibar_\eta(x+\eps)\psibar_\varphi(x)\ .
\ee
In general, $L_R(x)+L_L(x)$ will not generate an automaton. One can, however, assign some of the indices $\varphi$, $\eta$ to right-movers, and the other to left-movers. In this case cellular automata based on $L_R(x)+L_L(x)$ become possible. The neighbors for updating $x$ are the five cells $(x-2\eps,x-\eps,x,x+\eps,x+2\eps)$.

\subsection*{Next-neighbor double-hole scattering}

As a last example for chain automata we discuss next-neighbor double-hole scattering. A process of this type will be employed for spinor gravity in two dimensions. The elementary process maps four holes situated at two neighboring $x$ and $x+\eps$ to similar four-hole configurations with different colors. On each site one has a double-hole consisting of one hole of type $+$ , and a second of type $-$. The elementary process is described by a product of four Grassmann variables $\psi_\pm^a(x)$ and four Grassmann variables $\psibar_\pm^a(x)$,
\bel{C11}
\begin{split}
L(x)=&-B_{abcd,efgh}(x)\psi_+^a(x+\eps)\psi_-^b(x+\eps)\psi_+^c(x)\psi_-^d(x)\\
&\times\psibar_+^e(x+\eps)\psibar_-^f(x+\eps)\psibar_+^g(x)\psibar_-^h(x)\ ,
\end{split}
\ee
with $B$ an appropriate unique jump matrix in the allowed color index space. We choose non-vanishing elements of $B$ such that every index pair $(gh)$ is uniquely associated to an index pair $(ef)$, and similarly for the pairs $(ab)$ and $(cd)$. The structures for the neighboring cells for the updating is the same as for next-neighbor scattering.

As an example we take for the non-zero elements
\bel{C12}
B_{1234,1342}=B_{4321,3124}=B_{3412,4213}=B_{2143,2431}=1\ .
\ee
The elementary four hole scattering process maps the pairs
\be
\begin{tabular}{cccc}
(12)(34)&(43)(21)&(34)(12)&(21)(43)\\
$\uparrow$&$\uparrow$&$\uparrow$&$\uparrow$\\
(13)(42)&(31)(24)&(42)(13)&(24)(31)
\end{tabular}\ .
\ee
Here the first pair is at $x$, the second at $x+\eps$. Within each pair the first color is for $\psi_+$ and the second for $\psi_-$. Summing first only over even $x$ the scattering of pairs at $(x\,,\ x+\eps)$ gets disconnected from $(x+2\eps\,,\ x+3\eps)$ etc., and $\Ktil$ is composed of factors $\Ktil(x)$ without common Grassmann variables. For the local scattering factors $\Ktil(x)$ we observe that any combination of $0$, $4$, $8$, $12$ or $16$ holes that can be formed from the color pairs $(13)$, $(31)$, $(24)$ or $(42)$ at $x$ and $x+\eps$ has a unique origin from the expansion of $\exp\gl-L(x)\gr$ in the sense that the product of factors which yields a given hole configuration is unique. We observe the selection rule that every color $1$, $2$, $3$, $4$ appears an equal number of times in the allowed hole-configurations. For example, an allowed eight-hole configuration is $(1+2\,,\ 3+4)(3+4\,,\ 1+2)$. The unique product that yields this configuration is $(13)(42)\cdot(24)(31)$. The corresponding product of $\psibar$-variables multiplies a unique color combination for $\psi$-variables.

Adding the sum over odd $x$ we can proceed as for the next-neighbor scattering. For products of factors $L(x+\eps)$ and $L(x)$ the allowed color combinations at $x$ indicate uniquely which colors at $x+\eps$ originate from factors $L(x)$. Subtracting them yields again reduced hole configurations, for which the remaining colors at $x+\eps$ have to arise from $L(x+\eps)$, and so on. We conclude that $\cL(x)=\sum_xL(x)$, with $L(x)$ given by eqs.~\eqref{C11}\eqref{C12}, yields indeed a cellular automaton.

\subsection*{Spinor gravity automaton}

An interesting automaton that we will discuss later in more detail combines the index combinations $(13)(42)$ and $(24)(31)$ with right-transport, and the index combinations $(31)(24)$ and $(42)(13)$ with left-transport. We place the right-transport and the left-transport on different sublattices such that these two parts do not interfere and can be treated separately. For the double-holes taking part in the right transport the variables $\psibar$ are placed at even $x$ and $x+2\eps$, and the variables $\psi$ at $x+\eps$ and $x+3\eps$. On the other hand, for the double-holes that move to the left we place $\psibar$ at $x+\eps$ and $x+3\eps$, and $\psi$ at $x$ and $x+2\eps$. Together with the color content of the double-holes this assignment is shown if Fig.~\ref{fig: C1}. We choose periodic boundary conditions with a number of $x$-points $M_x=4\ \text{mod}~4$. The corresponding model is given by terms with eight Grassmann variables according to
\bel{C14}
\begin{split}
\cL=\sum_{x\ \text{even}}&\phantom{\times}\psi_+^1(x+\eps)\psi_-^2(x+\eps)\psi_+^3(x+3\eps)\psi_-^4(x+3\eps)\\
&\times\psibar_+^1(x)\psibar_-^3(x)\psibar_+^4(x+2\eps)\psibar_-^2(x+2\eps)\\
&+\ \text{three more terms.}
\end{split}
\ee

The updating for the cellular automaton can be directly read off from Fig.~\ref{fig: C1}. On the restricted set of configurations for which the holes can be grouped into the appropriate double pairs one needs for the updating of the right-movers in cell $x+\eps$ only the configurations in the cell $x$ to its left. For the species $+$ the colors are exchanged according to $(1\to1, 2\to2, 3\to4, 4\to3)$, while for the species $-$ one has $(1\to3, 2\to4, 3\to2, 4\to1)$. The left-movers obey the same rule of color changes. We observe that the double-pairs after the updating are not the same as before the updating. We therefore will need a second step of updating with different incoming pairs. For the spinor gravity automaton this will be described in sec.~\ref{sec: DSG}.
\begin{figure}
\centering
\renewcommand{\arraystretch}{1.2}

\begin{tabular}{C{0.07\textwidth}*{5}{C{0.07\textwidth}}}

$x$ & $x+\eps$ & $x+2\eps$ & $x+3\eps$ & & \\
\noalign{\hrule height 1pt}
-- & $(12)$ & -- & $(34)$ & $\psi$ & $t+2\eps$ \\
$(13)$ & -- & $(42)$ & - & $\psibar$ & $t+\eps$ \\
\hline
-- & $(21)$ & -- & $(43)$ & $\psi$ & $t+2\eps$ \\
$(24)$ & -- & $(31)$ & -- & $\psibar$ & $t+\eps$ \\
\hline
(34) & -- & $(12)$ & -- & $\psi$ & $t+2\eps$ \\
-- & $(42)$ & -- & $(13)$ & $\psibar$ & $t+\eps$ \\
\hline
$(43)$ & -- & $(21)$ & -- & $\psi$ & $t+2\eps$ \\
-- & $(31)$ & -- & $(24)$ & $\psibar$ & $t+\eps$ \\
\hline

\end{tabular}

\caption{Elementary local process for next-neighbor double-hole scattering with right and left transport. The pairs indicate the colors of first $\psi_+$ and second $\psi_-$ and we place the double-hole pairs at the positions indicated. The time labels will be discussed in sect.~\ref{sec: DSG}.}
\label{fig: C1}

\end{figure}

The updating rule is invariant under the exchange of particles and holes. In order to see this we begin with the first two elementary processes in Fig.~\ref{fig: C1} which realize right-handed transport. This part maps particles and holes at even $x$ to particles and holes at odd $x$. We first construct the totally empty state on the sublattice of even $x$ by products of $\psibar$-variables arising from appropriate products of $L(x)$.
For this purpose we place on every even $x$ two pairs of double-holes as indicated in Fig.~\ref{fig: C2}. The assignment of factors from $L(x)$ and $L(x+2\eps)$ is unique. The totally empty state on the even sublattice obtains in the expansion of the exponential in order $n=M_x$, with assignment of colors for $\psibar$ shown in Fig.~\ref{fig: C2a}. This state is uniquely mapped to the totally empty state on the odd sublattice, with colors for $\psi$ shown in Fig.~\ref{fig: C2b}. The scattering of two neighboring two-particle pairs with the same colors as the scattering for the neighboring two-hole pairs obtains by omitting the corresponding factor $L(x)$ out of the $M_x$ factors for the empty state. In this subsector the next-neighbor double-particle scattering for four particles involves $M_x-1$ factors in the expansion of the exponential.
\begin{figure}
\centering
\renewcommand{\arraystretch}{1.2}

\vspace{0.25cm}

\begin{subfigure}[b]{\linewidth}

\begin{tabular}{C{0.14\textwidth}*{5}{C{0.14\textwidth}}}

$x$ & $x+2\eps$ & $x+4\eps$ & $x+6\eps$ & $x+8\eps$ & $x+10\eps$ \\
\noalign{\hrule height 1pt}
$(13)$ & $(42)$ & $(13)$ & $(42)$ & $(13)$ & $(42)$ \\
$(24)$ & $(31)$ & $(24)$ & $(31)$ & $(24)$ & $(31)$ \\
$(42)$ & $(13)$ & $(42)$ & $(13)$ & $(42)$ & $(13)$ \\
$(31)$ & $(24)$ & $(31)$ & $(24)$ & $(31)$ & $(24)$ \\
\hline

\end{tabular}

\caption{}
\label{fig: C2a}

\end{subfigure}


\begin{subfigure}[b]{\linewidth}

\begin{tabular}{C{0.14\textwidth}*{5}{C{0.14\textwidth}}}

$x-\eps$ & $x+\eps$ & $x+3\eps$ & $x+5\eps$ & $x+7\eps$ & $x+9\eps$ \\
\noalign{\hrule height 1pt}
$(34)$ & $(12)$ & $(34)$ & $(12)$ & $(34)$ & $(12)$ \\
$(43)$ & $(21)$ & $(43)$ & $(21)$ & $(43)$ & $(21)$ \\
$(12)$ & $(34)$ & $(12)$ & $(34)$ & $(12)$ & $(34)$ \\
$(21)$ & $(43)$ & $(21)$ & $(43)$ & $(21)$ & $(43)$ \\
\hline

\end{tabular}

\caption{}
\label{fig: C2b}

\end{subfigure}

\caption{Totally empty state for even $x$ for next-neighbor double-hole scattering. We show the assignment of $\psibar$ in part (a). This is mapped to the totally empty state for odd $x$, for which we show the assignment in part (b). For part (a) the first two lines arise from $L(x)$ according to the first two processes in Fig.~\ref{fig: C1}, and the last two lines emerge from $L(x+2\eps)$. Periodic boundary conditions are assumed.}
\label{fig: C2}

\end{figure}

The discussion of the complementary lattice for the left-moving pairs of double-holes proceeds in complete analogy to the right-moving part. Now the totally empty state in the odd sublattice in $x$ is transported to the totally empty state in the even sublattice. Altogether, the scattering processes for $4p$ particles omit $p$-factors $L(x)$ from the $2M_x$ factors of the totally empty state for the whole lattice.

In summary, we have provided in this section a toolbox of local chain automata for which a formulation in terms of Grassmann variables for fermions is possible. Different updatings can be combined sequentially in order to construct fermionic quantum field theories that are equivalent to cellular automata. Most of the automata discussed in this section show particle-hole symmetry, which will be an important symmetry for the fermion models.

\section{Sequence of updatings and\\Grassmann functional integral}
\label{sec: SUG}

In this section we discuss sequences of updatings of cellular automata. This provides for an overall view of the automaton for all steps of its evolution. The correspondence in the fermion picture is a Grassmann functional integral. This integral is over variables at all times, as familiar in the functional integral formulation of quantum field theory.

For cellular automata updating rules are applied in a sequential way. We may label the steps by a time $t=m_t\eps$ with integer $m_t$. The updating at $t$ changes the bit configuration $\{n\al(t)\}$ to a new bit configuration $\{n\al(t+\eps)\}$. This change is encoded in the rotation of the state vector $q(t)$ to $q(t+\eps)$ according to
\bel{S1}
q_\tau(t+\eps)=\Shat_{\tau\rho}(t)q_\rho(t)\ ,
\ee
with step evolution operator $\Shat(t)$. The next step of the updating from $\{n\al(\te)\}$ to $\{n\al(t+2\eps)\}$ is encoded in the step evolution operator $\Shat(t+\eps)$. The sequence of the two evolution steps obtains as a matrix product
\bel{S2}
U(t+2\eps,t)=\Shat(t+\eps)\Shat(t)\ .
\ee
Both $\Shat(t)$ and $\Shat(t+\eps)$ have to be unique jump matrices. Furthermore, the ranges of allowed bit-configurations have to match. For $\te$ the bit-configurations $\tau$ for which $\Shat_{\tau\rho}(t)$ has non-zero entries have to be the same as the ones for which $\Shat_{\sigma\tau}(t+\eps)$ has non-zero entries. There is no need that $\Shat(t+\eps)$ and $\Shat(t)$ are identical.

Rather complex automata can be constructed by a sequence of different step evolution operators, for example a sequence of transport operations and scattering operations. The updating can be continued to large $t$ by continuing $U(t+4\eps,t)=\Shat(t+3\eps)\Shat(t+2\eps)\Shat(t+\eps)\Shat(t)$ etc. For many automata one requires repetitivity. In this case the matrix $U(t+q\eps,t)=U$ will be repeated, such that ($p$, $q$ integer)
\bel{S3}
U(t+pq\eps,t)=U^p\ ,\quad q_\tau(t+pq\eps)=\gl U^p\gr\taur q_\rho(t)\ .
\ee
One may treat $U$ as a $t$-independent combined updating operator.

\subsection*{Light cones and causality}

If every step in a sequence of updating is a cellular automaton, the hole sequence can be considered as a new cellular automaton, typically with a larger number of neighbors. Consider a cell $x$ at the last step of the updating. Its neighbors are $y_i(x)$. In turn, the neighbors of $y_i$ in the second to last steps are $z_{j_i}(y_i)$. Taking the two last steps together the neighbors of $x$ are $\gl z_{j_1}(y_1(x)),z_{j_2}(y_2(x)),\dots z_{j_k}(y_k(x))\gr$. Continuing this procedure spans the \qq{past light cone} of the cell $x$. Only bit-configurations (\qq{events}) within the past light cone can influence the configuration in the cell $x$ at the end of the sequence.

In the other direction towards the future the configurations in the cell $x$ can only influence configurations in \qq{future neighboring cells} $\bar{y}_i(x)$. Repeating this constructs the \qq{future light cone} of the events in the cell $x$. This is a causal structure, as familiar from Lorentz-invariant quantum field theories. The precise shape of the light cones depends on the particular updating mechanism, but the causal structure is a built in feature for sequences of local chain automata. The border of the light cones defines the \qq{light-velocity}.

\subsection*{Modulo two property of Grassmann functional\\integrals}

On the level of step evolution operators the sequence of updatings proceeds by simple matrix multiplication. We want to transfer this to the formulation with Grassmann variables. This is slightly more complicated due to the modulo two property of Grassmann functional integrals~\cite{CWFGI}.

For the combined updating $U(t+2\eps,t)$ we would like to construct a combined local factor $\cK(t+2\eps,t)$. This should depend on Grassmann variables $\psi\al(t)$ and $\psi\al(t+2\eps)$. The idea is to use a product of local factors $\Kbapsi{t+\eps}{t}$ and $\Ktipsi{t+2\eps}{t+\eps}$, and to integrate over the intermediate Grassmann variables $\psibar(t+\eps)$,
\bel{S4}
\begin{split}
\cK&(t+2\eps,t)=\Kpsi{t+2\eps}{t}\\
&=\int\cD\psibar(t+\eps)\Ktil(t+\eps)\Kbar(t)\\
&=\int\cD\psibar(t+\eps)\Ktipsi{t+2\eps}{t+\eps}\Kbapsi{t+\eps}{t}\ .
\end{split}
\ee
We will employ the definition~\eqref{GU5} for odd $t+\eps$ (or even $t$), where $\Ktil[\psi,\,\psibar\,]$ corresponds to $\Ktil(t+\eps)$, $\psi$ to $\psi(t+2\eps)$, $\psibar$ to $\psibar(t+\eps)$, and therefore $\Shat$ to $\Shat(t+\eps)$. We have to find out the form of $\Kbar(t)$ for even $t$ that reproduces the appropriate $\cK(t+2\eps,t)$.

In general, employing the definition~\eqref{GU5} also for even $t$ will not work since the integral $\int\cD\psibar g_\rho'\opsib g_\tau\opsib$ has no simple properties. For the relation between Grassmann elements and the step evolution operator we therefore chose for even $t$ a different definition in terms of conjugate Grassmann basis elements
\bel{S5}
\Kbar(t)=\gbar_\tau'(t+\eps)\Shat\taur\gbar_\rho(t)\ .
\ee
The conjugate Grassmann basis elements are defined by the relation
\bel{S6}
\int\cD\psi\gbar_\tau\opsi g_\rho\opsi=\delta\taur\ .
\ee
This ensures the appearance of the matrix product in the expression
\bel{S7}
\begin{split}
\int&\cD\psi(t+2\eps)\Kbar(t+2\eps)\Ktil(t+\eps)\\
&=\gbar_\tau'(t+3\eps)\gl\Shat(t+2\eps)\Shat(t+\eps)\gr\taur g_\rho'(t+\eps)\ .
\end{split}
\ee

The elements $\gbar_\tau'$ are defined similar to eq.~\eqref{GU6},
\bel{S8}
\gbar_\tau'=\eps_\tau'\gbar_\tau=(-1)^{\tfrac{m_\tau'(m_\tau'-1)}{2}}\gbar_\tau\ ,
\ee
with $m_\tau'$ the number of $\psi$-factors in $\gbar_\tau$. They obey
\bel{S9}
\int\cD\psi g_\tau'\opsi \gbar_\rho'\opsi=\delta\taur\ ,
\ee
where we omit an additional minus sign~\cite{CWNEW} appearing if the total number of Grassmann variables equals $2$, $3$ $\text{mod}~4$. Eq.~\eqref{S9} ensures the expression
\bel{S10}
\int\psibar(t+\eps)\Ktil(t+\eps)\Kbar(t)=g_\tau(t+2\eps)\gl\Shat(t+\eps)\Shat(t)\gr\taur\gbar_\rho(t)\ .
\ee
We can produce in this way sequences of arbitrary length by chains of local factors, for example
\bel{S11}
\begin{split}
g_\tau(t+4\eps)&U\taur(t+4\eps,t)\gbar_\rho(t)\\=&\int\cD\psibar(t+3\eps)\cD\psi(t+2\eps)\cD\psibar(t+\eps)\Ktil(t+3\eps)\\&\times\Kbar(t+2\eps)\Ktil(t+\eps)\Kbar(t)\ .
\end{split}
\ee

\subsection*{Grassmann functional integrals}

Let us define the expression
\bel{S12}
\Khat(t)=\Ktil(t+\eps)\Kbar(t)=\Khat[\psi(t+2\eps),\,\psibar(t+\eps)\,,\psi(t)]\ .
\ee
Its integration over $\psibar(t+\eps)$ yields $\cK(t)$ in eq.~\eqref{S4}. We further encode the initial bit-configuration in the initial state vector $q(t\inn)$, and define the Grassmann wave function for even $t$ by
\bel{S13}
g(t)=q_\tau(t)g_\tau\opsi\ .
\ee
Setting initial conditions at $t\inn=0$ we obtain for even $\mf$ the relation
\bel{S14}
\begin{split}
q_\tau&(\mf\eps)=U\taur(\mf\eps,0)q_\rho(0)\\
=&\int\cD\psibar\cD\psi\,\gbar_\tau(\mf\eps)\Khat\gl(\mf-2)\eps\gr\Khat\gl(\mf-4)\eps\gr\dots\\
&\times\Khat(2\eps)\Khat(0)g(0)\ .
\end{split}
\ee
For the \qq{Grassmann functional integral} on the r.h.s. of eq.~\eqref{S14} the integration $\int\cD\psibar\cD\psi$ involves an integration over all Grassmann variables $\psi(t')$ for even $t'$ between $0$ and $\mf$, and over Grassmann variables $\psibar(t'+\eps)$ for odd $t'+\eps$ between $\eps$ and $(\mf-1)\eps$. The l.h.s. of eq.~\eqref{S14} expresses directly the bit-configuration of the automaton after $\mf$ updatings. The relation~\eqref{S14} constitutes the general Grassmann functional integral expression for the cellular automaton. One can find a similar expression for odd $\mf$, which will not be needed here, however.

We can further employ the exponential form
\bel{S15}
\Khat(t)=\exp\big\{-\gl\cL(t+\eps)+\Lbar(t)\gr\big\}\ ,
\ee
where we define for $t$ even
\bel{S16}
\Kbar(t)=\exp\big\{-\Lbar(t)\big\}\ .
\ee
This yields for the state vector after $\mf$ updatings
\bel{S17}
q_\tau(\mf\eps)=\int\cD\psibar\cD\psi\gbar_\tau(\mf\eps)e^{-S}g(0)\ ,
\ee
with action
\bel{S18}
S=\sum_t\gl\cL(t+\eps)+\Lbar(t)\gr\ .
\ee
Here the sum is over all even $t$ from zero to $(\mf-2)\eps$. If $\cL(t+\eps)$ and $\Lbar(t)$ involve simple sums of polynomials of Grassmann variables, as for the simple elementary processes discussed above, eq.~\eqref{S17} is a rather standard form of a Grassmann functional integral. It is tailored here for an initial value problem, as encoded in $g(0)$. The \qq{readout} of the state after $\mf$ steps is achieved by multiplication of the integral for $g(\mf\eps)$ with $\gbar_\tau(\mf\eps)$ and integration over $\psi(\mf\eps)$.

For chain automata obeying eq.~\eqref{C1}, and similarly $\Lbar(t)=\sum_xL(t,x)$, the Grassmann functional integral is formulated on a two-dimensional quadratic lattice, with variables $\psi\g(t,x)$ and $\psibar(t+\eps,x)$ and local action
\bel{S19}
S=\sum_t\sum_x\gl L(t+\eps,x)+\overline{L}(t,x)\gr\ .
\ee
This is a standard expression for quantum field theories for fermions.

\subsection*{Particle-hole conjugation}

Up to signs the conjugate Grassmann basis elements $\gbar_\tau\opsi$ obtain from $g_\tau\opsi$ by replacing every factor $\psi\al$ by $1$, and every $1$ by $\psi\al$. In this way the product $\gbar_\tau\opsi g_\tau\opsi$ (no sum here) contains all Grassmann variables $\psi\al$, and the Grassmann integral over this product equals one.
In contrast, for $\rho\neq\tau$ the product $\gbar_\tau\opsi g_\rho\opsi$ either vanishes because at least one Grassmann variable appears twice or it does not contain all variables $\psi\al$ such that the integral~\eqref{S6} vanishes. The map from $g_\tau$ to $\gbar_\tau$ replaces particles by holes and vice versa. It can be associated to particle-hole conjugation.

The alternating relations between the step evolution operator and the local factors $\Kbar$ and $\Ktil$ imply that for the same choice of $\Ktil$ and $\Kbar$ (up to time translation of the variables) the step evolution operator switches between particles and holes. More precisely, we denote the particle-hole conjugate of $\Shat$ by $\Shat^c$, and the switch is between $\Shat^c$ and $\Shat$. The particle-hole conjugate $\Shat^c$ obtains from $\Shat$ by exchanging particles and holes for each species $\gamma$ and location $x$ of the fermions or bits. For a particle-hole invariant updating, as encountered in many of the examples discussed so far, this alternation is not effective. In this case it does not matter if we evaluate $\Shat$ from eq.~\eqref{GU5} or~\eqref{S5}.

For the automaton the particle-hole transform of a given state at $t$ replaces in the bit-configuration all occupation numbers $n\g(x)=1$ by $n\g(x)=0$, and vice versa. Expressed in terms of configurations of Ising spins $s\g(x)=2n\g(x)-1$ the particle-hole transformation switches the signs of all Ising spins. Correspondingly, the particle-hole transform $\Shat^c(t)$ of the step evolution operator $\Shat(t)$ corresponds to a switch of sign of all Ising spins at $t$ and $\te$.

More formally, we define the particle-hole conjugate Grassmann basis elements $g_\tau^c$ by
\bel{S20}
g_\tau^c=\eps_\tau^c\gbar_\tau\ ,\quad \eps_\tau^c=\pm1\ ,
\ee
where the sign is chosen such that $g_\tau^c$ is one of the Grassmann elements $g_\rho$ without any additional minus sign.
Writing
\bel{S21}
g_\tau^c=K\taur g_\rho\ ,
\ee
the matrix $K$ is a unique jump operator with only positive elements, obeying
\bel{S22}
K^2=1\ ,\quad K^T=K\ .
\ee
We will later use the involution $K$ for the definition of complex conjugation.

Assume now that we use for $\Kbar(t)$ the same form as for $\Ktil(t+\eps)$, with the replacements $\psibar(t+\eps)\to\psi(t)$, $\psi(t+2\eps)\to\psibar(t+\eps)$,
\bel{S23}
\begin{split}
\Kbar(t)=&g_\tau[\,\psibar(t+\eps)]\Shat\taur(t+\eps)g_\rho'[\psi(t)]\\
=&\gbar_\tau'[\,\psibar(t+\eps)]\Shat\taur(t)g_\rho[\psi(t)]\ .
\end{split}
\ee
For the relation between $\Shat(t)$ and $\Shat(t+\eps)$ we write
\bel{S24}
g_\tau=\gl KD_1\gr_{\tau\alpha}\gbar_\alpha'\ ,\quad g_\rho'=\gl KD_2\gr_{\rho\beta}\gbar_\beta\ ,
\ee
where the \qq{sign matrices} $D_1$, $D_2$ are diagonal matrices with elements $\pm1$. They account for possible relative minus signs. Insertion into eq.~\eqref{S23} yields
\bel{S25}
\Shat(t)=D_1K\Shat(t+\eps)KD_2=D_1\Shat^c(t+\eps)D_2\ .
\ee
The matrix $\Shat^c(\te)=K\Shat(t+\eps)K$ is the particle-hole conjugate of the matrix $\Shat(t+\eps)$, for which all holes are transformed to particles and vice versa. For a particle-hole symmetric cellular automaton one has $\Shat^c(t+\eps)=D_3\Shat(t+\eps)D_4$, where $D_3$ and $D_4$ are again sign matrices. The signs of the elements of $\Shat$ do not play a role for the association to an updating rule. We conclude that for particle-hole symmetric automata it does not play a role for the updating rule if we extract $\Shat$ according to the rule~\eqref{GU5} or~\eqref{S5}. If the automaton is not particle-hole invariant and we want to use for $\Kbar$ the same dependence on $\psi$ up to a shift of the variables in $t$, the matrix $\Shat(t)$ for even $t$ is related to the matrix $\Shat(t+\eps)$ by an exchange of particles and holes, up to irrelevant signs.

\section{Probabilistic cellular automata}
\label{sec: PCA}

Quantum theories are probabilistic theories. This holds for the quantum field theories for fermions described here. Typical statements are about expectation values of observables that often take discrete values. An example is the occupation number $n\g(t,x)$ for a fermion of type $\gamma$ located at $x$ at time $t$. The expectation value for this observable is a real number between zero and one.

For a deterministic cellular automaton $n\g(t,x)$ either takes the sharp value one or zero, depending on the precise initial condition. For any initial bit-configuration $n\g(t,x)$ can be computed by following the updating rule of the automaton. The probabilistic aspects of a quantum theory emerge from probabilistic initial conditions. Instead of a unique initial bit configuration one specifies a probability distribution over all possible initial bit configurations. We call this setting a \qq{probabilistic cellular automaton}. We will see that probabilistic initial conditions are sufficient for an implementation of the probabilistic quantum features. They are crucial for the existence of a continuum limit.

In the present section we introduce the wave function. It is identical for the automaton and its fermionic counterpart. At this stage we deal with a real formulation of quantum systems in a setting with discrete time steps. Further basic concepts of the quantum formalism as the density matrix, operators for observables, a complex structure and continuous time evolution will be described in later sections.

\subsection*{Wave function for cellular automata}

We have described in sect.~\ref{sec: FPGU} the deterministic initial condition of a given bit-configuration by a state vector $q(t\inn)$. For this \qq{sharp initial condition} $q(t\inn)$ is a vector with only one particular component equal to $\pm1$, and all other components zero, $q_\tau=\pm\delta_{\tau,\taubar}$. A probabilistic cellular automaton is described by a more general initial state vector $q(t\inn)$. It only needs to be a real unit vector,
\bel{P1}
q_\tau(t\inn)q_\tau(t\inn)=1\ .
\ee
The initial probabilities are defined by
\bel{P2}
p_\tau(t\inn)=q_\tau^2(t\inn)\ .
\ee
The positivity of the probabilities is ensured by the square of a real quantity in eq.~\eqref{P2}, and eq.~\eqref{P1} guarantees the normalization.

The time evolution \eqref{GU2} of the state vector according to the updating rule of the cellular automaton generalizes to probabilistic initial conditions,
\bel{P3}
q_\tau(t+\eps)=\Shat\taur(t)q_\rho(t)\ .
\ee
This reproduces indeed the updating of the probability distribution for the cellular automaton. If we know at a given time $t$ the probability $p_\rho(t)$ for every bit-configuration $\rho$, we can infer the probabilities $p_\tau(t+\eps)$ for all bit configurations $\tau$ at $t+\eps$. If the updating rule maps a given configuration $\rho$ to $\taubar(\rho)$, the probability for this updated configuration is the same as the one for the original configuration,
\bel{P4}
p_{\taubar(\rho)}(t+\eps)=p_\rho(t)\ ,\quad p_\tau(t+\eps)=p_{\bar\rho(\tau)}(t)\ .
\ee
Generalizing eq.~\eqref{P2} to arbitrary $t$,
\bel{P5}
p_\tau(t)=q_\tau^2(t)\ ,
\ee
and employing the evolution law~\eqref{P3} yields
\bel{P6}
\begin{split}
p_\tau(t+\eps)=&\sum_{\rho,\rho'}\Shat\taur(t)q_\rho(t)\Shat_{\tau\rho'}(t)q_{\rho'}(t)\\
=&\sum_{\rho,\rho'}\delta_{\tau,\taubar(\rho)}q_\rho(t)\delta_{\tau,\taubar(\rho')}q_{\rho'}(t)\\
=&\sum_\rho\delta_{\tau,\taubar(\rho)}q_\rho^2(t)=\sum_\rho\delta_{\tau,\taubar(\rho)}p_\rho(t)\\
=&\sum_\rho\delta_{\bar\rho(\tau),\rho}p_\rho(t)=p_{\bar\rho(\tau)}(t)\ ,
\end{split}
\ee
in accordance with eq.~\eqref{P4}. Here we employ the fact that $\Shat$ is a unique jump operator $\Shat\taur=\pm\delta_{\tau,\taubar(\rho)}=\pm\delta_{\rho,\bar\rho(\tau)}$.

Since $\Shat$ is an orthogonal matrix the length of the vector $q(t+\eps)$ is the same as for $q(t)$, such that the normalization condition holds for all $t$,
\bel{P7}
q_\tau(t)q_\tau(t)=1\ .
\ee
The time evolution finds a simple expression as a rotation of the state vector. We will associate the state vector $q(t)$ with the wave function of quantum mechanics in a real formulation. In general, a complex quantum wave function can be written as a real unit vector with twice the number of components. In this real representation the unitary evolution becomes an orthogonal evolution or rotation of the real wave function. Furthermore, for particular cases as Majorana-Weyl spinors the wave function can be taken real anyhow. The conditions~\eqref{P7},~\eqref{P3} account for the properties of the real representation of quantum mechanics. The state vector $q(t)$ for the probabilistic cellular automaton is an example for a \qq{classical wave function}~\cite{CWQPCS}, or the appearance of concepts of quantum mechanics in classical statistics~\cite{CWQMCS, CWEM}. In presence of a suitable complex structure it can be written as a complex wave function with half the number of components, see later.

\subsection*{Wave function for fermionic quantum field theory}

So far we have found Grassmann functionals for fermions that have the same step evolution operator as a cellular automaton. We will next establish that also the state vector of the probabilistic automaton can be identified with the wave function of an associated fermionic quantum model. For these models the quantum wave function $q(t)$ is the same as for the probabilistic cellular automaton.

For a given wave function $q(t)$ of the cellular automaton we can define for even $t$ the Grassmann wave function
\bel{P8}
g(t)=q_\tau(t)g_\tau\opsit\ .
\ee
Inversely, every Grassmann element $g(t)$ formed with the Grassmann variables $\psi\al(t)$ can be expanded in terms of the Grassmann basis elements $g_\tau\opsit$. The expansion coefficients $q_\tau(t)$ are associated with the components of the wave function. The evolution of the Grassmann wave function from $t$ to $t+\eps$ is given by multiplication with the local factor $\Kbar(t)$, and integration over the Grassmann variables $\psi(t)$,
\bel{P9}
\begin{split}
g(t+\eps)=&\int\cD\psi(t)\Kbar(t)g(t)\\
=&\int\cD\psi(t)\gbar_\tau'(t+\eps)\Shat\taur(t)\gbar_\rho(t)q_\sigma(t)g_\sigma(t)\\
=&\Shat\taur(t)q_\rho(t)\gbar_\tau'(t+\eps)=q_\tau(t+\eps)\gbar_\tau'(t+\eps)\ .
\end{split}
\ee
We therefore use for odd $t+\eps$ a different expansion of the Grassmann wave function
\bel{P10}
g(t+\eps)=q_\tau(t+\eps)\gbar_\tau'[\psi(t+\eps)]\ .
\ee
In the next step the evolution is obtained by multiplication with $\Ktil(t+\eps)$ and integration over $\psi(t+\eps)$,
\bel{P11}
\begin{split}
&g(t+2\eps)=\int\cD\psi(t+\eps)\Ktil(t+\eps)g(t+\eps)\\
=&\int\cD\psi(t+\eps)g_\tau(t+2\eps)\Shat\taur(t+\eps)g_\rho'(t+\eps)\\
&\times q_\sigma(t+\eps)\gbar_\sigma'(t+\eps)\\
=&\Shat\taur(t+\eps)q_\rho(t+\eps)g_\tau(t+2\eps)=q_\tau(t+2\eps)g_\tau(t+2\eps)\ ,
\end{split}
\ee
in accordance with eq.~\eqref{P8} for even $t+2\eps$. The modulo two property of Grassmann functional integrals is reflected in the alternating expansions~\eqref{P8} and~\eqref{P10} for even and odd $t$.

For a given probabilistic cellular automaton we can construct the Grassmann wave function $g(t)$ directly from the wave function $q(t)$. On the other hand the Grassmann wave function can be extracted from the Grassmann functional integral with a suitable boundary term. In eq.~\eqref{S17} we choose for $t\inn=0$
\bel{P12}
g(0)=q_\tau(0)g_\tau[\psi(0)]\ .
\ee
The Grassmann wave function obtains by restricting the range of integration to Grassmann variables $\psi(t')$ with $t'\leq t-\eps$, with corresponding restriction of the action
\bel{P13}
\begin{split}
g(t)=&\int\cD\psi(t'\leq t-\eps)e^{-S_<}g(0)\ ,\\
S_<=&\sum_{t'=0}^{t-2\eps}\gl\cL(t'+\eps)+\Lbar(t')\gr\ .
\end{split}
\ee
Here we have taken $t$ even, with a simple extension to $t+\eps$ odd given by eq.~\eqref{P9}. The Grassmann wave function $g(t)$ is a Grassmann element formed from the variables $\psi(t)$, as it should be. Indeed, $\cL(t-\eps)$ involves $\psi(t)$, which is not included in the integration variables.

The identity~\eqref{P13} is found easily by expressing the local factors $\Ktil(t+\eps)$ and $\Kbar(t)$ by $\cL(t+\eps)$ and $\Lbar(t)$, and using a chain of identities~\eqref{P9}~\eqref{P11} involving integrations over products of local factors, starting with $g(0)$. We can extract the wave function by
\bel{P14}
q_\tau(t)=\int\cD\psi(t)\gbar_\tau(t)g(t)\ ,
\ee
which coincides with eq.~\eqref{S17} for $\mf\eps=t$, $S_<=S$. The derivation of the identities~\eqref{P13}~\eqref{S17} makes no assumption on the form of the wave function $q(t)$ and is therefore not restricted to \qq{sharp} or deterministic initial conditions.

In conclusion, the wave function extracted from the Grassmann functional integral by eqs.~\eqref{P13},~\eqref{P14} follows the same discrete time evolution as the wave function of the probabilistic automaton. We choose the initial wave function~\eqref{P12} to be identical with the one that defines the probabilistic initial condition of the automaton. The fermionic wave function is then identical to the wave function of the automaton for all times.

\section{Symmetries}
\label{sec: S}

Having mapped the time history of probabilistic automata to the functional integral for a fermionic quantum field theory many powerful concepts of quantum field theory become applicable to cellular automata. Some of them are very familiar for quantum field theories, as the momentum of particles or Fourier transforms, while they are not commonly employed for the description of automata. In the present section we start with the concept of symmetries.

Symmetries are key concepts for an understanding of quantum field theories or many body quantum theories. They are less frequently used for the description of cellular automata. In this section we discuss symmetries acting on the Grassmann variables for the fermion picture of the updating and indicate consequences for the properties of the cellular automaton. More precisely, we focus on symmetries that act as variable transformations on $\psi\g(x)$, $\psibar\g(x)$ which leave $\cL[\psi,\psibar\,]$ invariant. These symmetries include continuous global symmetries as well as local gauge symmetries. We will explicitly construct automata that realize such symmetries.

\subsection*{Discrete symmetries}

Simple discrete symmetries are permutations of the \qq{colors} $\gamma$ of $\psi\g(x)$, $\gamma\to\delta(\gamma)$, and similarly for $\psibar_{\bar\gamma}$, $\bar\gamma\to\bar\delta(\bar\gamma)$, or
\bel{SY1}
\psi\g(x)\to\psi_{\delta(\gamma)}(x)\ ,\quad\psibar_{\bar\gamma}(x)\to\psibar_{\bar\delta(\bar\gamma)}(x)\ .
\ee
These permutations have a direct analogue on the level of bit-configurations for the automaton. A bit configuration is transformed to a new bit configuration for which the colors of all bits are permuted correspondingly. Invariance of $\cL[\psi,\psibar\,]$ implies invariance of the updating under the corresponding color permutation of the ingoing and outgoing bit configuration. It is a simple way of stating this permutation symmetry. In view of the freedom of choice of signs the invariance of $\cL[\psi,\psibar\,]$ under the discrete transformation
\bel{SY2}
\psi\g(x)\to s_{\delta(\gamma)}\psi_{\delta(\gamma)}(x)\ ,\quad\psibar_{\bar\gamma}(x)\to s_{\bar\delta(\bar\gamma)}\psibar_{\bar\delta(\bar\gamma)}(x)\ ,
\ee
with $s_{\delta(\gamma)}=\pm1$, $s_{\bar\delta(\bar\gamma)}=\pm1$, leads to the same invariance of the updating under color permutations of the bits. Discrete symmetries of this type can be extended to transformations that combine color permutations and changes of position by generalizing on the r.h.s. of eqs.~\eqref{SY1},~\eqref{SY2} $\psi_{\delta(\gamma)}(x)$ to $\psi_{\delta(\gamma)}\gl y(x)\gr$, or more general permutations among all the Grassmann variables $\psi\g(x)$, and similar for $\psibar\g(x)$.

There are further discrete symmetries of $\cL[\psi,\psibar\,]$ that have no direct correspondence as transformations among bit configurations of the automaton. As a simple example consider the simultaneous sign reflection for a particular $\gamma$,
\bel{SY3}
\psi\g(x)\to-\psi\g(x)\ ,\quad \psibar\g(x)\to-\psibar\g(x)\ ,
\ee
applied simultaneously to all $x$. If $\cL[\psi,\psibar\,]$ is invariant under this discrete transformation it has to contain only terms with an even number of $\psi\g(x)$ or $\psibar\g(x)$. Any term with an odd number of factors $\psi\g(x)$ has to be mutiplied by an odd number of factors $\psibar\g(x)$, while even numbers of $\psi\g(x)$ come with even numbers of $\psibar\g(x)$. This extends to the local factor $\Ktil[\psi,\psibar\,]=\exp\big\{-\cL[\psi,\psibar\,]\big\}$. As a consequence, the number of particles or holes of the species $\gamma$ can change only by an even number - it is conserved modulo two. As an example, a term $\sim\psi\g(x)\psibar\g(y)$ in $\Ktil$ changes a hole of type $\gamma$ at $y$ to one at $x$, while $\psi\g(x_1)\psi\g(x_2)\psi\g(x_3)\psibar\g(y)$ produces instead three holes at $x_1$, $x_2$ and $x_3$.

There is no transformation among bit configurations which reflects the discrete symmetry~\eqref{SY3}. Nevertheless, this symmetry encodes a simple rule that the updating obeys. There are many other discrete symmetries of a similar type. If $\cL$ is invariant under the simultaneous reflection of a family $(\gamma_1, \gamma_2\dots \gamma_N)$ of Grassmann variables similar to eq.~\eqref{SY3}, the updating rule conserves the number of particles or holes belonging to one of the species $(\gamma_1\dots\gamma_N)$ modulo two. If the number of Grassmann variables is even for all terms in $\cL[\psi,\psibar\,]$ one has a reflection symmetry~\eqref{SY3} acting simultaneously on all $\gamma$. In consequence, the total particle number is conserved modulo two.

\subsection*{Global continuous symmetries}

Continuous transformations among the Grassmann variables correspond to well defined transformations within the Grassmann algebra. In particular, we may consider infinitesimal global transformations
\begin{align}
\label{SY4}
\psi\g(x)\to\psi\g(x)+\delta\psi\g(x)\ ,&\quad\psibar\g(x)\to\psibar\g(x)+\delta\psibar\g(x)\ ,\nn\\
\delta\psi\g(x)=\eps_{\gamma\delta}\psi_\delta(x)\ ,&\quad\delta\psibar\g(x)=\bar\eps_{\gamma\delta}\psibar_{\delta}(x)\ .
\end{align}
If $\lopsi$ remains invariant under the infinitesimal transformation~\eqref{SY4}, it is invariant under a global continuous transformation generated by integrating the infinitesimal transformation. The specification \qq{global} indicates that the transformation~\eqref{SY4} acts simultaneously for all $t,x$ with $\eps_{\gamma\delta}$, $\bar\eps_{\gamma\delta}$ independent of $t$ and $x$.

The right transport automaton~\eqref{C3} (and similarly for left transport) is invariant under global $SO(M)$ transformations for $M$ species $\gamma=1\dots M$. The corresponding $M(M-1)/2$ infinitesimal transformations are realized by antisymmetric infinitesimal parameters,
\bel{SY5}
\eps_{\gamma\delta}=\bar\eps_{\gamma\delta}=-\eps_{\delta\gamma}\ ,
\ee
for arbitrary pairs $(\gamma,\delta)$. In turn, the $SO(M)$-symmetry restricts the possible terms appearing in $\lopsi$. This induces restrictions or selection rules for the updating consistent with $SO(M)$-symmetry. We can combine transport automata for $N$ right-movers and $N$ left-movers, $M=2N$. Automata with $SO(N)$-symmetry for fermion models with interactions will be discussed in sect.~\ref{sec: ANA}.

A particularly simple transformation is a global scaling of all $\psi\g(x)$ and $\psibar\g(x)$ in opposite directions
\bel{SY6}
\delta\psi=\eps\psig\ ,\quad\delta\psibarg=-\eps\psibarg\ .
\ee
Invariance of $\lopsi$ under this transformation implies that every term in $\cL$ has to involve an equal number of factors $\psi$ and $\psibar$. This extends to the local factor $\Ktil$, implying that the total particle number is conserved. The updating of the automaton conserves the total number of bits having the value zero. This is a simple example how a continuous symmetry leads to a conserved quantity. Of course, one can state the conservation law directly for the updating rule. A relation to a continuous symmetry is usually not given for cellular automata, however. The formulation of rules and restrictions for updating rules in terms of continuous symmetries can become a rather useful tool for more complex situations, where the power of group theory for Lie groups can be exploited.

Another simple global symmetry is realized if two species of Grassmann variables $\psi_a$ and $\psi_b$ appear in $\cL$ only in the combination $\psi_a(x)\psi_b(y)-\psi_b(x)\psi_a(y)$. Indeed, for the infinitesimal transformation
\bel{SY7}
\delta\psi_a(x)=\eps\psi_b(x)\ ,\quad\delta\psi_b(x)=-\eps\psi_a(x)
\ee
one finds
\bel{SY8}
\delta\big[\psi_a(x)\psi_b(y)-\psi_b(x)\psi_a(y)\big]=0\ .
\ee
The corresponding continuous transformations are abelian $SO(2)$-transformations.

In particular, $\cL$ could contain only pairs $\psi_a(x)\psi_b(x)$, e.g. $x=y$. In this case the $SO(2)$-symmetry is local, see below. The consequence of this symmetry for the updating rule is that holes of the colors $a$ and $b$ can only be produced in pairs. A similar symmetry acing on $\psibar$ implies that the automaton is defined only for configurations where the number of bits of type $a$ being zero equals the corresponding number of bits of type $b$.

As an example, a local scattering automaton based on eq.~\eqref{GU26} is invariant under several abelian symmetries of this type, based on the pairs $(\psi_a,\psi_b)=(\psi_{+}^1,\psi_{-}^4), (\psi_{-}^1,\psi_{+}^4), (\psi_{+}^2,\psi_{-}^3),\dots$ etc. Another example is realized by suitable pairs for the spinor graviton automaton in eq.~\eqref{C14}.

Finally, global symmetries could mix the variables $\psig$ and $\psibarg$. An example is the infinitesimal transformation
\bel{SY9}
\delta\psig=\eps\psibarg\ ,\quad\delta\psibarg=-\eps\psig\ ,
\ee
acting simultaneously on all $\psig$, $\psibarg$. This symmetry is realized if $\lopsi$ only involves terms with an equal number of $\psi$-factors and $\psibar$-factors in the combination $\psig\psibar_\delta(y)+\psi_\delta(y)\psibarg$, or factors $\psig\psi_\delta(y)+\psibarg\psibar_\delta(y)$., according to
\begin{align}
\label{SY10}
\delta\big[\psig\psibar_\delta(y)+\psi_\delta(y)\psibarg\big]=0\ ,\nn\\
\delta\big[\psig\psi_\delta(y)+\psibarg\psibar_\delta(y)\big]=0\ .
\end{align}
We briefly describe in appendix~\ref{app: A} how the symmetry~\eqref{SY9} can be expressed in terms of complex Grassmann variables.

\subsection*{Local gauge symmetries}

Local gauge symmetries are a central ingredient for the fundamental interactions in particle physics. Using the concepts of ref~\cite{CWSLGT} one can implement these local continuous symmetries for the fermion picture of cellular automata. For each local gauge symmetry the parameters $\eps_{\gamma\delta}$ and $\bar\eps_{\gamma\delta}$ in eq.~\eqref{SY4} depend on the position $x$. We will later associate $\psi$ and $\psibar$ to different times. Independent transformations of $\psi$ and $\psibar$ (i.e. independent $\eps_{\gamma\delta}$ and $\bar\eps_{\gamma\delta}$) can be viewed as infinitesimal transformations depending on spacetime - they are local infinitesimal gauge transformations.

The building blocks for cellular automata with local gauge invariance are local composite invariants $b_i(x)=b_i[\psig]$ and $\bar b_i(x)=\bar b_i[\psibarg]$ formed from $\psig$ and $\psibarg$ respectively. We construct $\cL(t)$ from building blocks involving products $b_i(x)\bar b_j(x)$ or similar. The unique jump character of the step evolution operator will be violated if $b_i(x)$ and $\bar b_j(x)$ consist of sums of several terms. For a simple straightforward possibility each $b_i$ and $\bar b_j$ involves only one Grassmann element.

As an example we consider the gauge group $SO(4)$ with four Grassmann variables $\psig$, $\gamma=1\dots4$, in the four-dimensional vector representation. An invariant involving only a single Grassmann element is 
\begin{align}
\label{SY17}
b(x)=&\psi_1(x)\psi_2(x)\psi_3(x)\psi_4(x)\nn\\
=&\frac1{24}\eps_{\gamma_1\gamma_2\gamma_3\gamma_4}\psi_{\gamma_1}(x)\psi_{\gamma_2}(x)\psi_{\gamma_3}(x)\psi_{\gamma_4}(x)\ .
\end{align}
We base the automaton on independent blocks of two neighboring cells of sites,
\bel{SY18}
\cL(t)=\sum_{x\ \text{even}}L(x)\ ,
\ee
with
\bel{SY19}
L(x)=-\big[b(\xe)\bar b(x)+b(x)\bar b(\xe)\big]\ ,
\ee
and
\begin{align}
\label{SY20}
K(x)=&\exp\big\{-L(x)\big\}\nn\\
=&1+b(\xe)\bar b(x)+b(x)\bar b(\xe)\nn\\
&\phantom1+b(x)b(\xe)\bar b(x)\bar b(\xe)\ .
\end{align}

The corresponding automaton is very simple. It acts only on bit-configurations for which at each position $x$ all bits take equal values, either all zero or all one. The combined two sites $x$ and $\xe$ can therefore be in the four states $(0,0)$, $(0,1)$, $(1,0)$ and $(1,1)$, where $(0,1)$ denotes $n_1(x)=n_2(x)=n_3(x)=n_4(x)=0$, $n_1(\xe)=n_2(\xe)=n_3(\xe)=n_4(\xe)=1$, and similar for the other three possibilities. The updating rule for a single block is
\begin{alignat}{2}
(0,0)&\to(0,0)\ ,\quad&&(1,1)\to(1,1)\ ,\nn\\
(1,0)&\to(0,1)\ ,&&(0,1)\to(1,0)\ .
\end{alignat}
The updating is particle-hole symmetric.

In sect.~\ref{sec: USB} we will discuss more complex models with local gauge symmetries. At the present stage it should have become clear that the fermion picture for cellular automata can accommodate a rich variety of symmetries, including local gauge symmetries.

\subsection*{Symmetries of functional integral}

If a symmetry is shared by each updating step, it is a symmetry of the Grassmann functional integral. In this case the action is invariant under symmetry transformations that act on the Grassmann variables. This is a symmetry in the usual sense of quantum field theory or many body quantum theory for fermions. A given symmetry may not be shared by the initial conditions and therefore the overall state of the system. This situation represents \qq{spontaneous symmetry breaking}, for which the symmetry of the (ground-) state is smaller than the symmetry of the action.

To be more precise, the updating at $t$ encoded in $\Lbar(t)$ involves the same Grassmann variables $\psibar\al(\te)$ as for the updating at $\te$ encoded in $\cL(\te)$. A symmetry of the Grassmann functional integral requires that the same transformation is applied on $\psibar\al(\te)$ in $\Lbar(t)$ and $\cL(\te)$. In case of global symmetries this typically links the transformations of $\psi\al(t)$ and $\psi\al(t+2\eps)$. The global character of the symmetry therefore has not only a global character in space but also in time.

For local gauge symmetries the transformation of $\psi\al(t+2\eps)$ is independent of the transformation of $\psibar\al(t+\eps)$. Also the transformation of $\psi\al(t)$ is independent of $\psibar\al(\te)$. For $\cL(\te)$ we associate the transformation parameters $\eps_{\gamma\delta}(x)$ in eq.~\eqref{SY4} with $\eps_{\gamma\delta}(t+2\eps,x)$ and $\bar\eps_{\gamma\delta}(x)$ with $\eps_{\gamma\delta}(\te,x)$. Similarly, the transformation of $\psi\g(x)$ in $\Lbar(t)$ is given by local parameters $\eps_{\gamma\delta}(t,x)$. Taking things together, and identifying $\psibar\g(t+\eps,x)=\psi\g(t+\eps,x)$, a local gauge transformation transforms all Grassman variables at different space-time points $(t,x)$ independently
\bel{67A}
\delta\psi\g(t,x)=\eps_{\gamma\delta}\otx\psi_\delta(t,x)\ .
\ee
This is the usual meaning of a local gauge symmetry in quantum field theory.

\subsection*{Continuous symmetries for probabilistic cellular automata}

Probabilistic automata allow for a direct implementation of continuous symmetries. Continuous transformations of bit-configurations are not defined. One may formally define a continuous transformation of the step evolution operator, but the transformed object has no longer the unique jump form for an automaton. In contrast, a continuous wave function admits continuous transformations. Those do not respect the form of a deterministic sharp wave function, such that a probabilistic setting is mandatory for this purpose. In short, two wave functions that are related by a symmetry transformation remain related by the same symmetry transformation after the updating. In section~\ref{sec: OO} we will introduce operators for observables. They can be related by continuous symmetry transformations as well. Two operators related by a symmetry transformation lead to the same expectation value for the associated observables. All these features are well known from quantum mechanics. In our probabilistic setting they appear directly in the description of cellular automata.

Let us start with the local factor $\Kbar(t)=\Kbapsi tt$ which is related to the step evolution operator by eq.~\eqref{S5}. Continuous transformations of Grassmann variables, or more generally of Grassmann elements, are well defined within the Grassmann algebra. We take this fermionic representation of continuous symmetries as a starting point and translate it here to the wave function which has a direct interpretation for cellular automata. Consider the transformation of the Grassmann basis elements
\bel{CSP1}
\bar h_\tau\opsit=D_{\tau\rho}^{-1}\gbar_\rho\opsit\ .
\ee
These transformations may be generated by transformations of the Grassmann variables $\psi\al(t)$ from which $\gbar_\tau$ is constructed, but our setting here is more general. Similarly, one has a transformation
\bel{CSP2}
\bar h'_\tau[\psibar(\te)]=\bar g'_\rho[\psibar(\te)]\Dbar_{\rho\tau}
\ee
which may be generated by the transformation of $\psibar\al(\te)$.

These transformations induce a transformation of the local factor
\begin{align}
\label{CSP3}
\Kbar'=&\bar h'_\tau\opsib\Shat\taur\bar h_\rho\opsi\nn\\
=&\gbar'_\tau\opsib\Dbar_{\tau\sigma}\Shat_{\sigma\mu}D_{\mu\rho}^{-1}\gbar_\rho\opsi\nn\\
=&\gbar'_\tau\Shat'\taur\gbar_\rho\opsib\ ,
\end{align}
resulting in a transformed step evolution operator
\bel{CSP4}
\Shat'=\Dbar\Shat D^{-1}\ .
\ee
By definition, a symmetry transformation does not change the local factor and therefore leaves the step evolution operator invariant,
\bel{CSP5}
\Kbar'=\Kbar\ ,\quad \Shat'=\Shat\ ,
\ee
or
\bel{CSP6}
\Dbar(\te)\Shat(t)=\Shat(t)D(t)\ .
\ee

Let us define the symmetry transformation of the wave function by
\bel{CSP7}
q'(t)=D(t)q(t)\ ,\quad q'(\te)=\Dbar(\te)q(\te)\ .
\ee
The evolution of $q'$ obeys
\begin{align}
\label{CSP8}
q'(\te)=&\Shat(t)q'(t)=\Shat(t)D(t)q(t)=\Shat'(t)D(t)q(t)\nn\\
=&\Dbar(\te)\Shat(t)q(t)=\Dbar(\te)q(\te)\ .
\end{align}
If $q'(t)$ and $q(t)$ are related by a symmetry transformation, then also $q'(\te)$ and $q(\te)$ are related by this symmetry transformation. The evolution is compatible with the symmetry. In particular, if $\Dbar(\te)=D(t)=D$, the transformation matrix commutes with the step evolution operator
\bel{CSP9}
\big[D,\Shat(t)\big]=0\ .
\ee
This is typically realized for global symmetry transformations. Typical symmetry transformations result in orthogonal transformations, $D^T(t)D(t)=1$, such that the norm of the wave function is not changed.

The symmetry transformation of a Grassmann element $g(t)=q_\tau g_\tau\opsit$ can be realized equivalently either by a transformation of the wave function or by a transformation of the Grassmann basis elements
\bel{CSP10}
g'(t)=g_\tau\opsit q'_\tau(t)=g_\tau\opsit D\taur(t)q_\rho(t)=h_\rho\opsit q_\rho(t)\ ,
\ee
with
\bel{CSP11}
h_\rho\opsit=g_\tau\opsit D\taur(t)\ .
\ee
This is compatible with the relation~\eqref{S6}
\begin{align}
\label{CSP12}
\int\cD\psi\bar h_\tau\opsi h_\rho\opsi=&\int\cD\psi D^{-1}_{\tau\sigma}\gbar_\sigma\opsi g_\mu\opsi D_{\mu\rho}\nn\\
=&D^{-1}_{\tau\sigma}D_{\sigma\rho}=\delta\taur\ .
\end{align}
In summary, we can realize symmetry transformations either by a transformation of the Grassmann variables or, more generally, Grassmann basis elements at fixed $\Shat$ and $q$. Alternatively, we can employ an equivalent transformation of the wave function, keeping now the Grassmann variables fixed. It is the second version that applies directly to probabilistic cellular automata.

The concept of a continuous wave function for probabilistic cellular automata constitutes an important advantage for the realization of continuous symmetries. This extends to other important ingredients of quantum mechanics, as a change of basis~\cite{CWIT, CWQF}. For a sufficiently smooth wave function the discrete evolution equation can be cast in the form of a continuous Schrödinger equation. One then obtains the usual realization of symmetries for quantum systems with symmetry operators commuting with the Hamiltonian. All this is not possible for the sharp wave functions of deterministic automata.



\section{Alternation of propagation and\\scattering}
\label{sec: AOPS}

In the next part of this work we discuss various quantum field theories that are equivalent to probabilistic cellular automata. This includes models with global and local non-abelian continuous symmetries. We also indicate general strategies how cellular automata describing fermion models with interactions can be constructed.

A simple construction for automata describing fermion systems with interactions alternates a step evolution operator for propagation at even $t$ and another one for the interaction at odd $t$. In other words, for $t$ even the local factor $\Kbar(t)$ accounts for the propagation, while $\Ktil(t+\epstil)$ describes the interaction. For a simple example of a propagation a particle will move by $\pm\eps$ in the $x$-direction for a time step $\epstil$. The interaction entails no motion such that the particle has moved by $\pm\eps$ in the combined time interval $2\epstil$ for propagation and interaction. For automata based on the alternation of propagation and interaction steps we choose time units $\epstil=\eps/2$ such that the overall motion for the two combined steps amounts to $\Delta x=\pm\Delta t$. More complex settings with several interaction steps are possible as well.

This setting is appropriate if we want to construct automata for which bilinear terms in the fermionic action describe the propagation, while terms involving products of more than two Grassmann variables account for the interaction. If we want to realize step evolution operators that act on arbitrary bit configurations, the interaction term may be a complete local scattering in the sense of sect.~\ref{sec: LU}. In consequence, $\cL$ will be somewhat more complicated than for the examples discussed so far. In sects.~\ref{sec: USB},~\ref{sec: DSG} we will give an example for an alternative construction for which the fermionic action contains no fermion bilinear.

For the propagation we consider two types of bits or fermions, namely right-movers with occupation numbers $n_{R,a}$ or associated Grassmann variables $\psi_{R,a}$, and left-movers with $n_{L,a}$ or $\psi_{L,a}$. The index $a$ distinguishes between possible additional internal properties as \qq{colors}. At even $t$ the step evolution operator for a free propagation acts independently on the right-movers and left-movers. Right-movers move one position in $x$ to the right, such that a particle at $(t,x)$ will be found at $(t+\epstil,x+\eps)$. Similarly, all left-movers move one position to the left. The step evolution operator for this propagation is very simple. All bits (and holes) of right-movers move one position to the right, such that the whole configuration of right-moving bits is displaced by one unit to larger $x$, and similarly for the left-movers to lower $x$. On the fermionic level we deal with the structure~\eqref{GU17}, leading to
\bel{TS1}
\begin{split}
\Lbar(t)=-\sum_x\sum_a\Big\{&\psibar_{R,a}(t+\epstil,x+\eps)\psi_{R,a}(t,x)\\
+&\psibar_{L,a}(t+\epstil,x-\eps)\psi_{L,a}(t,x)\Big\}\ .
\end{split}
\ee

\subsection*{Thirring scattering}

The alternation of propagation and scattering is a very general recipe for the construction of automata representing fermion models with interactions. For the interaction step we can take any of the strictly local interactions, as the examples discussed in sect.~\ref{sec: LU}. One can also employ the scattering steps discussed in sect.~\ref{sec: CA}, provided they are promoted to complete scattering. In the present section we recall one example~\cite{CWNEW} for which the fermionic picture is a type of discretized Thirring or Gross-Neveu model. It is based on two colors of right- and left-movers and a simple updating rule: whenever a single right-mover meets a single left-mover the colors are exchanged.

For odd $t+\epstil$ (even $t$) we realize local scattering by independent local factors for every position $x$
\bel{TS2}
\Shat(t+\epstil)=\Shat(x=1)\otimes\Shat(x=2)\otimes\dots\otimes\Shat(x=M_x)\ .
\ee
We focus on two colors $a=1,2=\text{red, green}$. The local step evolution operator is therefore a $16\times16$-matrix acting on the states denoted by $(n_{R,1}, n_{R,2}, n_{L,1}, n_{L,2})$. The updating rule for the Thirring automaton~\cite{CWNEW} exchanges the colors whenever a single right-mover encounters a single left-mover. This is realized for
\bel{TS2*}
\begin{split}
&\Shat_{(1001),(0110)}=\Shat_{(0110),(1001)}\\
=&\Shat_{(1010),(0101)}=\Shat_{(0101),(1010)}=1\ .
\end{split}
\ee
For all other sectors $\Shat(x)$ is the unit matrix.

The translation to the fermionic picture reads~\cite{CWNEW}
\bel{TS3}
\begin{split}
\Ltil(t+\epstil)=&\sum_xL(t+\epstil,x)\ ,\\
L(t+\epstil,x)=&-\big[\sum_{\gamma=1}^{4}\psi\g\psibar\g+\Qtil\big](1-\Qtil)\ ,
\end{split}
\ee
with
\bel{TS4}
\begin{split}
\Qtil=&\gl\psi_1\psi_4-\psi_2\psi_3\gr\gl\psibar_1\psibar_4-\psibar_2\psibar_3\gr\\
+&\gl\psi_1\psi_3+\psi_2\psi_4\gr\gl\psibar_1\psibar_3+\psibar_2\psibar_4\gr\ .
\end{split}
\ee
Here the shorthands $(\psi_1,\psi_2,\psi_3,\psi_4)$ stand for $(\psi_{R1},\psi_{R2},\psi_{L1},\psi_{L2})$ taken at $(t+2\epstil,x)$ and $(\psibar_1,\psibar_2,\psibar_3,\psibar_4)$ are the corresponding Grassmann variables taken at $(t+\epstil,x)$. The first term in the square bracket in eq.~\eqref{TS3} accounts for the unit step evolution operator in the sector of configurations not involved in the scattering~\eqref{TS2*}. The second term $\Qtil$ entails the scattering and subtracts the unit matrix in the sector of the scattered particles. The multiplication with the factor $(1-\Qtil)$ yields the expression~\eqref{TS3} somewhat more complicated than the elementary processes discussed before. It is necessary to ensure for the local factor the simple form
\bel{TS5}
\Ktil(t+\epstil)=\exp\big\{-\Ltil(t+\epstil)\big\}=\prod_x\big[\exp(\psi\g\psibar\g+\Qtil\big]\ .
\ee

\begin{figure}

\includegraphics{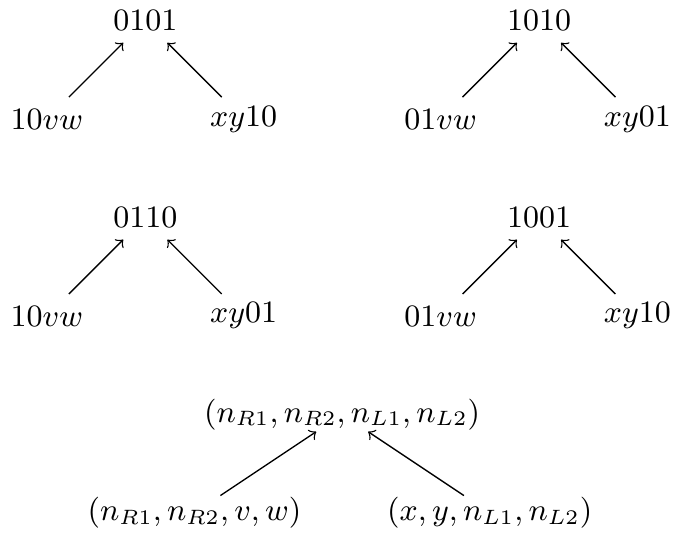}
\caption{Updating for the Thirring automaton}
\label{fig: ACA}

\end{figure}

The updating for the cellular automaton which combines propagation and Thirring scattering involves for the cell $x$ the two neighbors $x-\eps$ and $x+\eps$. The four scattering processes are shown in Fig.~\ref{fig: ACA}. Here $v,w,x,y$ stand for arbitrary values of the bits $0$ or $1$. We also indicate in this figure the remaining $12$ processes without scattering, with $(n_{R1},n_{R2},n_{L1},n_{L2})$ the values of bits in the cell $x$ in the combinations not appearing in the first four scattering processes.

We observe that the relation between a simple $\Ktil(t+\eps)$ and a somewhat more complex $\Ltil(t+\eps)$ is an alternative construction to the choice of restricted states discussed in sect.~\ref{sec: CA}. It can combine the scattering of a subclass of states with a unit step evolution operator for the other states. In this way a restriction to a subclass of states is avoided. The price to pay is a slightly more complicated form of the fermionic action for the scattering process. Arbitrary forms of invertible $2\to2$ scattering can be implemented in this way.

Combining eqs.~\eqref{TS1} and~\eqref{TS3} we can write
\bel{TS6}
\begin{split}
&\Lbar(t)+\Ltil(t+\epstil)\\
&=\eps\sum_x\sum_a\Big[\psibar_{R,a}(t+\epstil,x+\eps)\gl\dt+\partial_x\gr\psi_{R,a}(t+\epstil,x+\epstil)\\
&+\psibar_{L,a}(t+\epstil,x-\eps)\gl\dt-\partial_x\gr\psi_{L,a}(t+\epstil,x-\epstil)\Big]+\cL_\text{int}\ ,
\end{split}
\ee
where the interaction term $\cL_\text{int}$ involves all terms with more than two Grassmann variables and we employ the lattice derivatives$(\epstil=\eps/2)$
\bel{TS7}
\psi(t+\eps,x\pm\eps)-\psi\otx=\eps\gl\dt\pm\partial_x\gr\psi(t+\epstil,x\pm\epstil)\ .
\ee

\subsection*{Continuum model and Lorentz symmetry}

We can consider our discrete setting as a particular discretization of a continuum model. This continuum model obtains by neglecting terms in subleading order in an expansion in $\eps$. In the continuum formulation the lattice derivatives~\eqref{TS7} become partial derivatives with respect to continuous time $t$ and position $x$. Correspondingly, we can drop the differences between $\psi(t+n\eps,x+m\eps)$ and $\psi\otx$ beyond leading order for small $n$ and $m$. Furthermore, we replace sums by integrals $\sum_{t,x}=1/(2\eps^2)\int_{t,x}$, taking into account that the distance between two positions $x$ at fixed $t$ amounts to $2\eps$.
We finally renormalize the Grassmann variables $\psi\to\sqrt{2\eps}\psi$. Then the factor $(1-\Qtil)$ in eq.~\eqref{TS3} can be replaced by one. One arrives at the continuum action for fermions
\begin{align}
\label{TS8}
S=\int_{t,x}\Big\{&\sum_a\gl\psibar_{R,a}\gl\dt+\partial_x\gr\psi_{R,a}+\psibar_{L,a}\gl\dt-\partial_x\gr\psi_{L,a}\gr\nn\\
&+L_\text{int}\Big\}\ ,
\end{align}
where the Grassmann variables are taken at $\otx$ and we keep first the distinction between even and odd $t$, such that
\begin{align}
\label{TS9}
L_\text{int}=&-2\eps^2\Qtil \\
=&-2\big\{\gl\psibar_{R1}\psibar_{L2}-\psibar_{R2}\psibar_{L1}\gr\gl\psi_{R1}\psi_{L2}-\psi_{R2}\psi_{L1}\gr\nn\\
&+\gl\psibar_{R1}\psibar_{L1}+\psibar_{R2}\psibar_{L2}\gr\gl\psi_{R1}\psi_{L1}+\psi_{R2}\psi_{L2}\gr\big\}\ .\nn
\end{align}

The action~\eqref{TS8} is invariant under Lorentz transformations. We combine $\psi_{Ra}$ and $\psi_{La}$ into a two-component spinor
\bel{TS10}
\psi_a=\pvec{\psi_{Ra}}{\psi_{La}}\ ,\quad \psibar_a=\gl\psibar_{La},-\psibar_{Ra}\gr\ .
\ee
It transforms under infinitesimal global Lorentz transformations
$$\delta\psi_a=-\eta\Sigma^{01}\psi_a\ ,\quad \delta\psibar_a=\eta\psibar_a\Sigma^{01}\ ,\quad \Sigma^{mn}=\frac14\big[\gamma^m,\gamma^n\big]\ ,$$
with two-dimensional Dirac matrices obeying
\bel{S31}
\{\gamma^m,\gamma^n\}=2\eta^{mn}\ ,\quad \eta_{mn}=\eta^{mn}=\text{diag}(-1,1)\ ,
\ee
given explicitly by the Pauli-matrices
\bel{S32}
\gamma^0=-i\tau_2\ ,\quad \gamma^1=\tau_1\ ,\quad \Sigma^{01}=\frac{1}{2}\tau_3\ .
\ee
Here $\eta_{mn}$ is the invariant signature tensor or Lorentz-metric.
Using for flat space $\gamma^\mu=\gamma^m$ the action~\eqref{TS8} takes a manifestly Lorentz-invariant form
\bel{TS11}
\begin{split}
S=\int_{t,x}\Big\{&-\psibar_a\gamma^\mu\partial_\mu\psi_a-\frac12\psibar_a\gamma^\mu\psi_a\psibar_b\gamma_\mu\psi_b\\
&+\psibar_a\gamma^\mu\psi_b\eps_{ab}\psibar_c\gamma_\mu\psi_d\eps_{cd}\Big\}\ ,
\end{split}
\ee
with antisymmetric tensor $\eps_{12}=-\eps_{21}=1$ and summation over repeated indices.

In the continuum limit one may identify for certain purposes
\bel{TS12}
\psibar\g\otx=\psi\g(t+\epstil,x)=\psi\g\otx\ ,\quad \psibar=\psi^T\gamma^0\ .
\ee
The action~\eqref{TS11} describes a type of generalized Thirring model~\cite{THI, KLA, AAR, FAIV} for Majorana fermions with two colors. It is invariant under $SO(2)$-color rotations. Introducing a suitable complex structure this is the Thirring- or Gross-Neveu-model with $U(1)$-symmetry.

\section{Automata with non-abelian or abelian continuous symmetries}
\label{sec: ANA}

In this section we discuss cellular automata for which the fermion picture describes generalized Gross-Neveu or Thirring models with non-abelian or abelian continuous symmetry. This will serve as a basis for a subsequent investigation in sect.~\ref{sec: OO} how many features of the fermionic quantum field theories, as conserved charges corresponding to the non-commuting generators of the group, find a useful application for the dynamics of the  probabilistic automaton. A discussion of the possible continuum limit of the fermionic model sheds light on the questions that arise for the continuum limit for cellular automata with a very large number of cells.

\subsection*{$\boldsymbol{SO(N)}$-symmetric scattering}

As an example for a continuous symmetry we discuss models with global $SO(N)$-symmetry. For $N>2$ the symmetry group is non-abelian. The degrees of freedom are right-moving and left-moving bits with $N$ colors each, $n\g(x)=\gl n_{Ra}(x), n_{La}(x)\gr$, $a=1\dots N$. They are represented by Grassmann variables $\psi_{Ra}$, $\psi_{La}$, $\psibar_{Ra}$, $\psibar_{La}$. For the local factor we take $\Ktil(t+\epstil)=\prod_x K(x)$,
\bel{NY1}
K(x)=\exp\big\{-\gl L\subt{int}+L_0+\Delta L\gr\big\}\ ,
\ee
with
\bel{NY2}
-L\subt{int}=\psi_{La}\psi_{Rb}\psibar_{Ra}\psibar_{Lb}+\psi_{Ra}\psi_{Lb}\psibar_{Ra}\psibar_{Lb}\ ,
\ee
and
\bel{NY3}
-L_0=\psi_{Ra}\psibar_{Ra}+\psi_{La}\psibar_{La}\ .
\ee
Here we use the shorthands $\psi\g=\psi\g(\te,x)$, $\psibar\g=\psi\g(t+\epstil,x)$.
The term $\Delta L$ contains \qq{corrections} involving six or more Grassmann variables that are needed to ensure the unique jump property of the step evolution operator. The setting is similar to the Thirring type model~\eqref{TS3}, with $L\subt{int}$ playing the role of $-\Qtil$. The global $SO(N)$-invariance of $L_0$ and $L\subt{int}$ is manifest, such that $K$ is $SO(N)$-invariant for $SO(N)$-invariant $\Delta L$.
Furthermore, the interaction term is invariant under the exchange $\psi\g\leftrightarrow\psibar\g$, and it is $L\leftrightarrow R$ symmetric, $\psi_{Ra}\leftrightarrow\psi_{La}$, $\psibar_{Ra}\leftrightarrow\psibar_{La}$.

The updating rule for the automaton can be obtained by expanding the exponential in eq.~\eqref{NY1}. We start with the two-particle sector, noting that one has a unit evolution in the zero- and one-particle sectors. In the order $\psi^2\psibar^2$ one finds a two-to-two scattering
\bel{NY4}
K_{2,2}=-\psi\Rb\psi\La\psibar\Ra\psibar\Lb\ .
\ee
Whenever a single right-mover encounters a single left-mover they exchange colors. This includes the unit evolution if the colors are identical. The second term in eq.~\eqref{NY2} cancels a similar contribution in $L_0^2/2$. The remaining part of $L_0^2/2$ leads to a unit step evolution operator in the sector of two left-movers or two right-movers. Taking the various contributions together we have a unique jump evolution in the two-particle sector.

For the four-particle sector we consider in the sector with one pair of right-movers and one pair of left-movers the four-to-four scattering
\begin{align}
\label{NY5}
K_{4,4}=&\frac14\psi_{Rc}\psi_{Rd}\psi\La\psi\Lb\psibar\Ra\psibar\Rb\psibar_{Lc}\psibar_{Ld}\nn\\
=&\frac14\gl K_{2,2}\gr^2\ .
\end{align}
This corresponds to an updating for which colors are exchanged if a pair of right-movers meets a pair of left-movers. Equivalently, right-movers change to left-movers with the same color, and vice versa. We observe that the expansion of $\exp\gl K_{2,2}\gr$ from eq.~\eqref{NY1} produces a factor $\gl K_{2,2}\gr^2/2$ instead of  $\gl K_{2,2}\gr^2/4$. In order to guarantee the unique jump property we include in the correction term $\Delta L$ terms with a similar structure. This generalizes to higher order terms.

To be specific, we concentrate on $N=4$ where
\bel{NY6}
\Delta L=\frac14K_{2,2}^2-\frac19K_{2,2}^3-\frac1{192}K_{2,2}^4+\Delta L'\ .
\ee
With the addition of these corrections one has
\begin{align}
\label{NY7}
K_1=&\exp\big\{K_{2,2}-\frac14K_{2,2}^2+\frac19K_{2,2}^3-\frac{11}{192}K_{2,2}^4\big\}\nn\\
=&1+K_{2,2}+K_{4,4}+K_{6,6}+K_{8,8}\ ,
\end{align}
where $K_{2,2}$, $K_{4,4}$, $K_{6,6}$, $K_{8,8}$ generate unique jump step evolution operators in the sectors with one, two, three and four pairs of right- and left-movers. In these sectors all colors are exchanged, or equivalently, right-movers become left-movers with the same colors, and vice versa. We note the relations
\bel{NY8}
K_{6,6}=\frac1{36}K_{2,2}^3\ ,\quad K_{8,8}=\frac1{576}K_{2,2}^4\ .
\ee

\subsection*{Particle-hole symmetry}

The scattering in the sector spanned by $K_1$ is particle-hole symmetric. Indeed, we can write
\bel{NY9}
K_{6,6}=-\psi\Rb^c\psi\La^c\psibar\Ra^c\psibar\La^c\ ,
\ee
where we define
\bel{NY10}
\psi\Ra^c=\frac16\eps_{abcd}\psi\Rb\psi_{Rc}\psi_{Rd}\ ,
\ee
and similar for $\psi\La^c$, $\psibar\Ra^c$, $\psibar\La^c$. The expression $\psi\Rb^c\psi\La^c$ corresponds to a right-moving hole with color $b$ and and left-moving hole with color $a$, and similar for $\psibar\Ra^c\psibar\Lb^c$ for the incoming holes.
Here holes are defined with respect to the fully occupied state. Thus $K_{6,6}$ accounts for an exchange of colors if a right-moving hole encounters a left-moving hole. This scattering is the particle-hole transform of the scattering described by $K_{2,2}$. The double-pair-double-pair scattering $K_{4,4}$ is particle hole invariant by itself. Finally, one has
\bel{NY11}
K_{8,8}=b_Rb_L\bar b_R\bar b_L\ ,
\ee
with
\bel{NY12}
b_R=\frac1{24}\eps_{abcd}\psi\Ra\psi\Rb\psi_{Rc}\psi_{Rd}=\psi_{R1}\psi_{R2}\psi_{R3}\psi_{R4}\ ,
\ee
and similarly for $b_L$, $\bar b_R$, $\bar b_L$. By virtue of the $SO(4)$-invariant totally antisymmetric tensor $\eps_{abcd}$ the quantities $b_L$, $b_R$, $\bar b_L$, $\bar b_R$ are $SO(4)$-invariant. The term $K_{8,8}$ accounts for the invariance of the totally occupied state. It is the particle-hole conjugate of the factor $1$ which encodes the invariance of the totally empty state.

We choose the remaining correction terms such that 
\bel{NY13}
K(x)=K_1+K'\ ,
\ee
where $K'$ only contributes to the step evolution operator in the sector not spanned by $K_1$. We also require particle-hole symmetry for $K'$.
For one possible choice of the correction terms $K'$ generates the unit evolution in all sectors not covered by $K_1$. For this choice the updating rule of the automaton is very simple. Scattering occurs only if one, two or three pairs consisting of one right-mover and one left-mover meet at a given position $x$. Then all colors are exchanged between right-movers and left-movers. Equivalently, right-movers become left-movers and vice versa, keeping in this picture their colors. A given updating rule of the automaton fixes the higher order terms $\Delta L$ in eq.~\eqref{NY1}. They can be constructed systematically, as outlined in appendix~\ref{app: B}.

\subsection*{Naive continuum limit}

The naive continuum limit can be taken in complete analogy to the Thirring-type model with two colors discussed before. This yields again a fermionic action~\eqref{TS8}, with indices in the kinetic term running now from one to $N$. We can interpret the discrete fermion model for the $SO(N)$-symmetric automaton as a discrete lattice formulation of the continuum quantum field theory obtained in the naive continuum limit. In this naive continuum limit the higher order terms in $\Delta L$ in eq.~\eqref{NY1} do not contribute. Different possible forms of $\Delta L$ which are compatible with $SO(N)$-symmetry correspond to different discretizations of the continuum quantum field theory. Omitting $\Delta L$ the interaction term $L\subt{int}$ in eq.~\eqref{TS8} is given by eq.~\eqref{NY2}. This interaction is invariant under Lorentz transformations,
\begin{align}
\label{NY22}
\delta\psi\Ra=&-\frac\eta2\psi\Ra\ ,\quad \delta\psi\La=\frac\eta2\psi\La\ ,\nn\\
\delta\psibar\Ra=&-\frac\eta2\psibar\Ra\ ,\quad\delta\psibar\La=\frac\eta2\psibar\La\ .
\end{align}
We can again write it as a Thirring type model, now with four colors, with
\bel{NY23}
L\subt{int}=\frac12\Big\{\gl\psibar_a\gamma^\mu\psi_b\gr\gl\psibar_b\gamma_\mu\psi_a\gr-\gl\psibar_a\gamma^\mu\psi_a\gr\gl\psibar_b\gamma_\mu\psi_b\gr\Big\}\ .
\ee
Here an additional factor of two arises for $L\subt{int}$ from the continuum normalization of the Grassmann variables~\cite{CWNEW}. The continuum action
\bel{154A}
S=\int_{t,x}\big\{-\psibar_a\gamma^\mu\partial_\mu\psi_a+L\subt{int}\big\}
\ee
is Lorentz invariant.

This structure holds for an arbitrary number of colors $N$. The $SO(N)$-symmetry is manifest since all contractions are done with the $SO(N)$-invariant tensor $\delta_{ab}$. For $N=2$ we obtain a Thirring-type model different from eq.~\eqref{TS11}. The reason is a different updating rule for the automaton. Now a pair with an incoming right-mover and left-mover with the same color remains unchanged, instead of changing color as for the model~\eqref{TS11}. For $N>4$ the series of scattering terms~\eqref{NY7} gets now extended by additional higher order terms $K_{m,m}$, with $m\leq2N$. Also the series of correction terms in $\Delta L$ comprises additional higher order terms. For $N<4$ the series are shorter, since for $N\leq3$ one has $K_{8,8}=0$, and for $N=2$ also $K_{6,6}=0$. For $N=1$ there is no scattering since $K_{2,2}=0$. The updating rule remains the same for all $N$. Whenever a number of pairs of right-movers and left-movers meet at $x$, and no other particles are present, the colors are exchanged.

By a different ordering of the Grassmann variables we can also write $L\subt{int}$ in the form
\begin{align}
\label{NY23A}
L\subt{int}=-\frac12\Big\{&\gl\psibar_a\psi_a\gr\gl\psibar_b\psi_b\gr-\gl\psibar_a\bar\gamma\psi_a\gr\gl\psibar_b\bar\gamma\psi_b\gr\nn\\
-&\gl\psibar_a\psi_b\gr\gl\psibar_b\psi_a\gr+\gl\psibar_a\bar\gamma\psi_b\gr\gl\psibar_b\bar\gamma\psi_a\gr\Big\}\ ,
\end{align}
with $\bar\gamma=-\gamma^0\gamma^1=\tau_3$. This is a type of generalized Gross-Neveu model~\cite{GN, WWE, RSHA, RWP, SZKSR}. We observe that for $N=1$ we have no interaction since $L\subt{int}=0$. The different terms in eq.~\eqref{NY23A} are necessary in order to ensure the unique jump property of the step evolution operator for the automaton. For $L=-\gl\psibar_a\psi_a\gr\gl\psibar_b\psi_b\gr/2$ the unique jump property is not realized.

\subsection*{Majorana and Dirac fermions}

There are two different ways to treat the relation between $\psibarg$ and $\psig$ in the naive continuum limit. One may either keep them as distinct Grassmann variables, as for the original formulation for which $\psibar$ corresponds to Grassmann variables at $\te$, and $\psi$ denotes Grassmann variables at $t$ (with $t$ even). Alternatively, one may encounter situations for which the difference between $t$ and $\te$ is no longer resolved. Therefore one identifies in the continuum limit
\bel{MF1}
\psibar\g\otx=\psi\g\otx\ .
\ee
In this case the fermionic scattering describes a quantum field theory for $N$ Majorana fermions.

For Majorana fermions the interaction term simplifies further and eq.~\eqref{NY2} becomes
\begin{align}
\label{MF2}
-L\subt{int}=&2\psi\Ra\psi\La\psi\Rb\psi\Lb\nn\\
=&-\frac12\gl\psi\Ra\psi\Rb-\psi\Rb\psi\Ra\gr\gl\psi\La\psi\Lb-\psi\Lb\psi\La\gr\ .
\end{align}
In particular, for $N=2$ this yields
\bel{MF3}
-L\subt{int}=4\psi_{R1}\psi_{L1}\psi_{R2}\psi_{L2}\ .
\ee
This is the most general local interaction for Majorana fermions with two colors. The right-movers and left-movers are Majorana-Weyl fermions.

For even $N$ we can introduce complex Grassmann variables by
\bel{MF4}
\zeta_a=\psi_a+i\psi_{\frac{N}{2}+a}\ ,\quad a=1\dots \frac{N}2\ .
\ee
They will describe $N/2$ Dirac fermions. In terms of the complex Grassmann variables the action of the fermionic quantum field theory reads in the naive continuum limit
\bel{MF5}
S=\int_{t,x}\Big\{\zeta\Ra^*\gl\partial_t+\partial_x\gr\zeta\Ra+\zeta\La^*\gl\partial_t-\partial_x\gr\zeta\La+L\subt{int}\Big\}\ .
\ee
Using the standard two-component notation for the Dirac spinors
\bel{MF6}
\zeta=\pvec{\zeta_L}{\zeta_R}\ ,\quad\zetabar=\zeta\herm\gamma^0=\gl\zeta_L^*,-\zeta_R^*\gr\ ,
\ee
we can write the continuum action in a manifestly Lorentz-invariant form
\bel{MF7}
S=-\int_{t,x}\Big\{\zetabar_a\gamma^\mu\partial_\mu\zeta_a+\frac12\gl\zetabar_a\zeta_a\gr\gl\zetabar_b\zeta_b\gr\Big\}\ .
\ee
Here we employ the identity
\bel{MF8}
\sum_{a=1}^{N/2}\zetabar_a\zeta_a=-2\sum_{a=1}^N\psi\Ra\psi\La\ .
\ee

In particular, for $N=2$ the interaction term takes the form of the Gross-Neveu model
\bel{MF9}
L\subt{int}=-\frac12\gl\zetabar\zeta\gr^2\ .
\ee
With the choice~\eqref{MF1} the automaton for $N=2$ can be interpreted as a discretization of the Gross-Neveu model with a particular strength of the interaction. With the continuum identification $\psibar=\psi$ also the model~\eqref{TS11} of the previous section is a discretization of the Gross-Neveu model, now with a different strength of the interaction, larger by a factor of two.

\subsection*{Continuum limit in fermionic quantum field theory}

The true continuum limit for the $SO(N)$-invariant automaton is more complex than the naive continuum limit. It is best discussed in the fermion language where we deal with a discrete quantum field theory. For the true continuum limit one has to include the effects of fluctuations. This is done conceptually by computing the quantum effective action, from which exact field equations can be derived by variation. These field equations correspond to macroscopic evolution equations. Furthermore, the quantum effective action generates the one-particle irreducible correlation functions for the quantum field theory. For small momenta or large distances the effective action is typically a continuum object, even of one starts at short distances with a discrete setting. The transition from the microscopic or short-distance action to the effective action is typically given by flow equations for running couplings, which now depend on a renormalization scale $k$. A suitable approach can proceed by functional renormalization, starting at $k\sim\pi/\eps$ with the discrete fermion model for the cellular automaton.  In this setting $k$ is an effective infrared cutoff which can be lowered continuously. For $k\to0$ all fluctuations are included and one obtains the quantum effective action.

For a continuum formulation the dimension of the renormalized Grassmann variables $\psi$, $\psibar$ is $(\text{mass})^{1/2}$. The terms appearing in eq.~\eqref{NY23} correspond to (naively) renormalizable interactions, that may be multiplied by dimensionless running couplings. Due to the flow the macroscopic couplings can deviate from the value $1/4$. There is no symmetry reason why the two terms in eq.~\eqref{NY23} need to have equal couplings. (We assume here a scale-dependent renormalization of the Grassmann variables such that the kinetic term retains its canonical form.) The full symmetries do not allow further quartic couplings, however. Besides Lorentz invariance, translation symmetry in space and time and $SO(N)$-symmetry, these symmetries include a chiral symmetry that guarantees that the numbers of right-movers and left-movers are conserved separately. There are further discrete symmetries as parity or the transformation $\psi\leftrightarrow\psibar$.

The terms omitted for the naive continuum limit correspond to (naively) non-renormalizable interactions. For example, the couplings of terms $\sim\psi^4\psibar^4$ have dimension $(\text{mass})^{-2}$. In the short distance limit they are $\sim\eps^2$. If they would retain values of this size for the effective action, they would be indeed negligible in the continuum limit. The flow of these couplings typically leads, however, to an increase of their absolute values as the (renormalization) momentum scale $k$ decreases. For low $k\eps$ the effective couplings will be fluctuation dominated, rather than being given by the \qq{classical} or microscopic values. For example, typical couplings for terms $\sim\psi^4\psibar^4$ are $\sim k^{-2}$ instead of $\eps^2$. If the higher order couplings correspond to irrelevant parameters for the renormalization flow their values become predictable for $k\eps\ll1$.

The chiral symmetry together with the Lorentz-symmetry forbids a mass term for the fermions. The question arises if the quantum effective action becomes Lorentz-invariant, as suggested by the naive continuum limit.
The interaction part of the step evolution operator is not Lorentz-invariant if we include the correction terms in $\Delta L$. Some correction terms (e.g. the ones canceling the contribution~\eqref{NY18} in appendix~\ref{app: B}) contain different powers of right-movers and left-movers, thereby violating Lorentz symmetry. Also the kinetic term with lattice-derivatives instead of partial derivatives breaks the Lorentz-symmetry. Lorentz-invariance of the continuum limit requires that the Lorentz-invariant partial fixed point for the flow of couplings is approached for $k\to0$.

The chiral symmetry requires an equal number of factors $\psi_R$ and $\psibar_R$ in each term, as well as an equal number of factor $\psi_L$ and $\psibar_L$. This symmetry is respected in the discrete setting. With $\Delta L$ being chiral invariant, the local factor $K(x)$ is chiral invariant, and also the sequence of step evolution operators is chiral invariant. We conclude that a Lorentz-invariant continuum limit, together with the absence of spontaneous chiral symmetry breaking, leads in the continuum limit to an interesting quantum field theory with massless fermions.

\subsection*{Universality classes}

The fermion interactions in the quantum effective action need no longer to obey a unique jump condition for the associated evolution operator. Including fluctuations amounts effectively to some sort of coarse graining or averaging. We will see explicitly in sect.~\ref{sec: CLCGS} how the unique jump property is lost in the course of coarse graining.

The continuum limit of a quantum field theory is typically related to a fixed point (sometimes approximate fixed point) in the flow of couplings. These fixed points define universality classes. In general, the universality classes depend only on the species of massless (or very light) particles and the symmetries (and dimension) of a model. A quantum field theory with the same symmetries as our $SO(N)$-invariant automaton is the $N$-component Gross-Neveu model for which only the first term in eq.~\eqref{NY23A} is included in the microscopic action. It is an interesting question if our automaton belongs to the same universality class as these Gross-Neveu models.

The answer to this question is not trivial. For the case $N=1$ the answer is negative. Our automaton describes a free-theory, in contrast to the Gross-Neveu model. A free theory is indeed a local fixed point and defines its own universality class. For $N=2$ the naive continuum limit with the identification $\psibar=\psi$ is the Gross-Neveu model. This case could belong to the same universality class as the asymptotically free Gross-Neveu model.
Nevertheless, it is possible that the particular discretization corresponding to an automaton constitutes a new universality class different from the Gross-Neveu model. If not, and if Lorentz-symmetry is realized in the continuum limit, one expects that the continuum limit of the $SO(2)$-invariant automaton is given by the Gross-Neveu model. For $N>2$ the universality class of the $SO(N)$-invariant automata may be generalizations of the Gross-Neveu model with non-abelian continuous $SO(N)$-symmetry. For $N$ even an interesting subgroup of $SO(N)$ is $SU(N/2)$. The $SU(N/2)$-transformations are compatible with a complex structure.

\section{Modified propagation and particles in a potential}
\label{sec: MP}

So far we have only considered a very simple propagation where at each updating step the right-movers jump one position to the right and left-movers one position to the left. This propagation can be modified in many ways. One may further combine the modified propagation with interaction steps. In this section we concentrate on the propagation. This will lead to an action that is quadratic in the fermion fields. Higher order terms for interactions can be added by interaction steps. We also take for simplicity a single color, with straightforward generalization.

The simple propagation for right-movers and left-movers corresponds to massless fermions. A modified propagation could induce a mass term such that massive fermions can be described. Spatial inhomogeneities could describe a massive particle in a potential. This shows the way towards an automaton that realizes the motion of a massive particle in a potential, as described by the standard Schrödinger equation. Characteristic quantum effects as tunneling through a potential barrier or dispersion of the wave function should become visible in this way.

The evolution of a quantum particle in a potential is no longer described by an automaton for which a particle at a position $x$ is moved in the next step to a definite position. The one-particle step evolution operator is no longer a unique jump matrix. This loss of the automaton property is related to the coarse graining associated with the one-particle subsystem in the continuum limit. We argue that the motion of a quantum particle in a potential is of a random nature, where the randomness may be due to expectation values of fermion composites.

\subsection*{Reflection and delay layers}

One possibility to modify the propagation is the introduction of particular \qq{layers} at certain values of $t$ for which the propagation differs from the \qq{standard propagation} for right-movers and left-movers. For such a layer a particle at a certain position $x$ may propagate \qq{normally} while a particle at a neighboring position $\xe$ propagates \qq{exceptionally}. For a coarse graining corresponding to an averaging over space the distinction between the two positions is lost. On the coarse grained level both a normal and an exceptional propagation become possible, similar to the behavior of a massive quantum particle. In this section we start by discussing homogeneous layers, and turn to inhomogeneities later.

A \emph{reflection layer} at a particular (even) $\tbar$ is realized by replacing the step evolution operator for the standard propagation by a modified operator. In the step from $\tbar$ to $\tbar+\eps$ the right-movers move one position to the left and become left-movers, and the left-movers move one position to the right and become right-movers. Every particle changes the direction of its trajectory when it hits the reflection layer.

In the fermion language one has for $\tbar$
\bel{MP1}
\Lbar(\tbar)=\sum_x\Big[\psibar_R(\tbar+\epstil,x+\eps)\psi_L(\tbar,x)-\psibar_L(\tbar+\epstil,x-\eps)\psi_R(\tbar,x)\Big]\ .
\ee
The choice of sign is, in principle, a matter of convention. A given choice, as the one in eq.~\eqref{MP1}, may permit a simple formulation of certain structures, in particular the continuum limit, as compared to alternative choices.

If we combine the reflection layer with a standard propagation $(c=1)$ or unit step evolution operator $(c=0)$ at the next evolution step from $\tbar+\epstil$ to $\tbar+2\epstil$ one adds
\begin{align}
\label{MP2}
\Ltil(t+\epstil)=-&\sum_x\Big[\psi_R\gl t+2\epstil,x+(1+c)\eps\gr\psibar_R(\tbar+\epstil,x+\eps)\nn\\
+&\psi_L\gl t+2\epstil,x-(1+c)\eps\gr\psibar_L(\tbar+\epstil,x-\eps)\Big]\ .
\end{align}
The sum
\begin{align}
\label{MP3}
&\Lbar(\tbar)+\Ltil(\tbar+\epstil)=\\
\sum_x\Big[&\psibar_R(\tbar+\epstil,x+\eps)\gl\psi_R\gl t+2\epstil,x+(1+c)\eps\gr+\psi_L(\tbar,x)\gr\nn\\
+&\psibar_L(t+\epstil,x-\eps)\gl\psi_L\gl t+2\epstil,x-(1+c)\eps\gr-\psi_R(\tbar,x)\gr\Big]\nn
\end{align}
can no longer be written as a lattice derivative. Also the naive continuum limit is not Lorentz invariant.

For a \emph{delay double-layer} we combine two reflection layers at neighboring $\tbar$ and $\tbar+\epstil$. The trajectory for every particle changes first direction at $\tbar$, and again subsequently at $\tbar+\epstil$. As a result, it will not change its position for the combination of the two evolution steps. After the two evolution steps the right-movers are again right-movers and the left-movers are again left-movers. As compared to the standard evolution at $\tbar$ and unit-evolution at $\tbar+\epstil$, for which particles would have moved by one position to the right or left, the particles have not moved in the presence of the delay double-layer. Their trajectories are \qq{delayed} in this sense. We will use this simple system as an example that the naive continuum limit can fail to reproduce the correct behavior.

In the fermion picture one adds to $\Lbar(\tbar)$ in eq.~\eqref{MP1} the term
\begin{align}
\label{MP4}
\Ltil(\tbar+\epstil)=&\sum_x\Big[\psi_R(\tbar+2\epstil,x)\psibar_L(\tbar+\epstil,x-\eps)\nn\\
&-\psi_L(\tbar+2\epstil,x)\psibar_R(\tbar+\epstil,x+\eps)\Big]\ .
\end{align}
For the combination of the two evolution steps one obtains
\begin{align}
\label{MP5}
&\Lbar(\tbar)+\Ltil(\tbar+\epstil)=\nn\\
\sum_x\Big[&\psibar_R(\tbar+\epstil,x+\eps)\gl\psi_L(\tbar+2\epstil,x)+\psi_L(\tbar,x)\gr\nn\\
-&\psibar_L(\tbar+\epstil,x-\eps)\gl\psi_R(\tbar+2\epstil,x)+\psi_R(\tbar,x)\gr\Big]\ .
\end{align}

For an automaton with double layers for all $t$ the naive continuum action becomes
\bel{MP6}
S=\frac{1}{\eps^2}\int_{t,x}\gl\psibar_R\psi_L-\psibar_L\psi_R\gr\ .
\ee
For a continuum normalization $\psi\to\sqrt{a\eps}\psi$ similar to sect.~\ref{sec: AOPS} this reads
\bel{MP7}
S=-\int_{t,x}\mbar\gl\psibar_L\psi_R-\psibar_R\psi_L\gr=-\int_{t,x}\mbar\psibar\psi\ ,
\ee
with
\bel{MP8}
\mbar=\frac a\eps\ .
\ee
For the last term in eq.~\eqref{MP7} we use the two-component spinors~\eqref{TS10}. This makes the Lorentz-invariance manifest and identifies $\mbar$ with a particle mass up to multiplicative normalization. There is no propagation in this case.

We can combine a delay double-layer at $\tbar$ and $\tbar+\epstil$ with a standard propagation at $\tbar+2\epstil$ and unit evolution at $\tbar+3\epstil$. This leads to a type of delayed propagation. Instead of moving two positions to the right or left for the combined evolution from $\tbar$ to $\tbar+4\epstil=\tbar+2\eps$ the particle now moves by only one position. In the fermion picture this adds to eq.~\eqref{MP5} the terms
\begin{align}
\label{MP9}
&\Lbar(\tbar+2\eps)+\Ltil(\tbar+3\epstil)=\nn\\
&\sum_x\Big[\psibar_R(\tbar+3\epstil,x+\eps)\gl\psi_R(\tbar+4\epstil,x+\eps)-\psi_R(\tbar+2\epstil,x)\gr\nn\\
&\quad+\psibar_L(\tbar+3\epstil,x-\eps)\gl\psi_L(\tbar+4\epstil,x-\eps)-\psi_L(\tbar+2\epstil,x)\Big]\nn\\
&=\eps\sum_x\Big[\psibar_R(\tbar+3\epstil,x+\eps)\gl\dt+\partial_x\gr\psi_R(\tbar+3\epstil,x+\epstil)\nn\\
&\quad+\psibar_L(\tbar+3\epstil,x-\eps)\gl(\dt-\partial_x\gr\psi_L(\tbar+3\epstil,x-\epstil)\Big]\ .
\end{align}
A naive continuum limit neglects the detailed positions of the various Grassmann variables after transmuting the lattice derivatives in eq.~\eqref{MP9} to partial derivatives. We may build an automaton by iteration of the four layers for the delayed evolution. For a continuum normalization $\psi\to2\sqrt{\eps}\psi$ the combination of eqs.~\eqref{MP5} and~\eqref{MP9} becomes in terms of the two-component spinors
\bel{MP10}
S=-\int_{t,x}\gl\psibar\gamma^\mu\partial_\mu\psi+m\psibar\psi\gr\ ,
\ee
with
\bel{MP11}
m=\frac2\eps\ .
\ee
This is the Lorentz-invariant action for a free massive particle. The particle mass is proportional to the inverse lattice distance and diverges for $\eps\to0$.

The naive continuum limit is not a valid true continuum limit for this setting, however. The reason is that the mass term does not act as a small change. In contrast to a massive quantum particle the displacement after four evolution steps remains unique. The overall effect of the delay double layer is simply a change of the velocity. For realizing a quantum particle the lack of commutativity of the step evolution operators for the double delay layers and the normal propagation should become a negligible effect. In our case the non-commutativity is important, the order of the operators matters and invalidates the naive continuum limit. At the present stage there should only be a warning that a naive continuum limit can give a misleading picture.

\subsection*{Inhomogeneous propagation and potential}

It is possible to conceive automata that are not homogeneous in the space position $x$. For the example of a \emph{filter layer} one employs at even $t$ a reflection layer for which eq.~\eqref{MP1} is used only for even $x$. For odd $x$ one employs instead a standard propagation~\eqref{TS1} for right- and left-movers. Only half of the particles are reflected, while the other half passes this layer without reflection. For odd $\tbar+\epstil$ we may add a reflection layer only for odd $x$, while now for even $x$ a standard propagation is chosen. As a result of the combination of the two layers one half of the particles is reflected to the right and left and vice versa, and therefore effectively stopped. The other half propagates freely.

If for all $t$ one has sequences of these two layers the particles are filtered into freely moving particles and particles that are not displaced in the average. The filter acts according to odd or even $x$ at even $t$. We can describe this as a homogeneous cellular automaton with larger cells. If a coarse graining averages over the positions $x$ and $\xe$ both propagation and stop become possible for the coarse grained particle.

If we consider only two of these subsequent partial reflection layers the particles are filtered into freely propagating particles and delayed particles. In the naive continuum limit this filtering can be seen as the effect of some periodic potential $V(x)$ that causes the delay of those particles that at even $\tbar$ are situated at even $x$.

This concept of a potential can be generalized by placing the \qq{delay points} at positions that differ from the simple location at every even $x$. In general, the homogeneity of the automaton is lost in this case. The potential can be seen as a type of $x$-dependent mass term $V(x)\hateq\,\mbar(x)$. For a simple filtering the potential has only two possible values, zero and $\sim\eps^{-1}$. The positions of the non-zero values can be chosen arbitrarily. A larger range of different \qq{delays} can be obtained by placing different inhomogeneous layers at different $t$. While these simple examples do not yet describe a quantum particle they provide for simple illustrations how coarse graining can induce the property of different possible motions.

\subsection*{Disorder and mass term}

Consider a particular point $(\bar t,\bar x)$ with even $\bar t$ at which the updating rule is modified such that the right-movers move to the left and become left-movers, and the left-movers move to the right and become right-movers. Every particle whose trajectory hits this point changes its direction. For this automaton with a disorder point $(\bar t,\bar x)$ the step evolution for the propagation is modified to
\bel{DM1}
\Shat_\text{free}(\tbar)\to\Shat_\text{free}(\tbar)\Shat_{LR}(\tbar,\xbar)\ ,
\ee
where $\Shat\subt{free}$ describes the free propagation of right-movers and left-movers, and $\Shat_{LR}(\tbar,\xbar)$ exchanges right-movers and left-movers at the particular position $\xbar$. In the fermionic language we still have the structure~\eqref{GU17}, but $\Lbar(\tbar)$ obtains an additional piece $\Delta\Lbar(\tbar)$
\bel{DM2}
\begin{split}
&\Delta\Lbar(\tbar)=\psibar_R(\tbar+\epstil,\xbar+\eps)\psi_L(\tbar,\xbar)-\psibar_L(\tbar+\epstil,\xbar-\eps)\psi_R(\tbar,\xbar)\\
&+\psibar_R(\tbar+\epstil,\xbar+\eps)\psi_R(\tbar,\xbar)+\psibar_L(\tbar+\epstil,\xbar-\eps)\psi_L(\tbar,\xbar)\ .
\end{split}
\ee
The second term in eq.~\eqref{DM2} cancels a similar term in eq.~\eqref{TS1} at the position $\xbar$, such that the matrix $F$ in eq.~\eqref{GU17} and the step evolution operator indeed remain unique jump matrices. We did not write here color indices. At the disorder point the direction change may act on all colors or only on specific colors.

A single disorder point makes the cellular automaton somewhat more complicated, but it remains still rather straightforward to follow the evolution of a given spin configuration. This changes in the presence of many disorder points that are distributed at places $(\tbar_i,\xbar_i)$ in the two dimensional lattice without a particular order. We have in mind very large lattices with a random distribution of disorder points. The trajectories change now randomly the directions. If the disorder points are rare and the time interval $\Delta t$ for the evolution is not too large, the right-movers at most points $(\Delta t,x)$ will originate from initial right-movers at $(0,x-\Delta t)$. Those particles have average velocity $\Delta x/\Delta t=1$. For some positions $\xhat$ at $\Delta t$ the right-movers may originate from particles that have hit disorder points where they have changed direction. Those particles originate from initial particles within the past light cone of the point $(\Delta t,\xhat)$, i.e. from within the interval $[\xhat-\Delta t,\xhat+\Delta t]$ at $t=0$. For those particles the absolute value of the average velocity is smaller than one.

One would like to make some averaged statements for this situation, working with a probability distribution for the initial bit-configurations. We will discuss this issue in sect.~\ref{sec: SPWF} in the context of the one-particle wave function.
In section~\ref{sec: QMPP} we will construct a cellular automaton for a quantum particle in a potential based on this idea. There we will also make a connection between disorder and possible expectation values for fermion composites. In a theory with interactions a term quartic in the fermion fields may reduce effectively to a quadratic term in the presence of expectation values of fermion bilinears. The propagation in the background of the expectation values can induce the propagation with disorder.

If the effect or disorder is in some sense small the naive continuum limit may become a more reliable guide. This will be shown in sect.~\ref{sec: QMPP} in case of one -particle states. In the continuum formulation a single disorder point at $(\tbar,\xbar)$ adds to the free action a term proportional to a $\delta$-distribution
\bel{DM3}
\begin{split}
S=\int_{t,x}\Big\{\gl&\psibar_R\gl\dt+\partial_x\gr\psi_R+\psibar_L\gl\dt-\partial_x\gr\psi_L\gr\\
\times&\gl1-2\eps^2\delta(t-\tbar)\delta(x-\xbar)\gr\\
+2\eps\gl&\psibar_R\psi_L-\psibar_L\psi_R\gr\delta(t-\tbar)\delta(x-\xbar)\Big\}\ .
\end{split}
\ee
If the disorder points are distributed randomly and we consider averages over a large enough volume $\Omega(x)$ around $x$ we may replace $2\eps^2\delta(t-\tbar)\delta(x-\xbar)$ by a function $\bar p(t,x)$ which is proportional to the average number of disorder points per site. This results in
\bel{DM4}
\begin{split}
S=\int_{t,x}\Big\{Z\gl&\psibar_R\gl(\dt+\partial_x\gr\psi_R+\psibar_L\gl\dt-\partial_x\gr\psi_L\gr\\
+\bar m\gl&\psibar_R\psi_L-\psibar_L\psi_R\gr\Big\}\ ,
\end{split}
\ee
with
\bel{DM5}
\bar m=\frac{\bar p\otx}\eps\ ,\quad Z=1-\bar p\otx=1-\eps\bar m\ .
\ee
For rare disorder points or small $\bar p\otx$ the \qq{mass parameter} $\bar m$ can be small as compared to the inverse lattice distance $1/\eps$.

For $\bar p$ independent of $x$ the interpretation of $m=\bar m/Z$ as a particle mass becomes apparent if we cast eq.~\eqref{DM4} into a standard Lorentz-invariant form. We can renormalize $\psi\to Z^{-1/2}\psi$, $\psibar\to Z^{-1/2}\psibar$ and obtain for the two component spinors~\eqref{TS10}
\bel{DM6}
S=-\int_{t,x}\Big\{\psibar\gamma^\mu\partial_\mu\psi+m\psibar\psi\Big\}\ ,\quad m=\frac{\bar m}{Z}\ .
\ee
This is the standard form of the Lorentz-invariant action for a free massive Dirac fermion in $1+1$-dimensions. For $\psibar$ identified with $\psi$ we deal with a Majorana fermion. An interaction term can be added by using the step evolution for odd $t+\epstil$, as discussed above. A perhaps more familiar form of the Grassmann functional uses $e^{-S}=e^{iS_M}$, with \qq{Minkowski action} $S_M=iS$.

The naive continuum limit leading to eq.~\eqref{DM6} is not guaranteed to provide for a correct description of the automaton. One has to investigate under which circumstances it applies. The formal issue is related to non-commuting pieces of the Hamiltonian, see sect.~\ref{sec: OO}. In any case a probabilistic description will be necessary for a continuum description. It is the small change of the wave function for changes of spacetime locations of the order $\eps$ that allows for a formal limit $\eps\to0$.

If a valid continuum limit allows for a mass term for fermions, the generalization to particles in a potential is straightforward. A potential is simply an inhomogeneous part of the mass term, as realized by an average number of disorder points $~\bar p(x)$ varying over space-distances much larger than $\eps$. We will return to these issues in sects.~\ref{sec: SPWF},~\ref{sec: QMPP} for the restricted setting of one-particle states.

\section{Updating by shifted blocks}
\label{sec: USB}

So far we have constructed automata corresponding to fermionic quantum field theories with interactions by a sequence of propagation and interaction steps. In view of the possibilities of modified propagation and large classes of local interactions a rather rich variety of fermionic quantum field theories are equivalent to probabilistic cellular automata. In the following we will introduce a second efficient method for the construction of invertible cellular automata, namely updating by shifted blocks. This will be particularly useful for the implementation of models with local gauge symmetries for which no gauge invariant terms bilinear in the fermion fields exist.

We have encountered in sect.~\ref{sec: CA} simple updating rules that operate within a block of neighboring positions. As an example, the updating in two neighboring cells $x$ and $x-\eps$ may only depend on the same cells at $x$ and $x-\eps$. We can group the two cells of the block into a larger common cell at $x-\eps$. A simple updating of the block-cell can be strictly local, not being influenced by neighboring block cells. This allows for a rather trivial construction of fermion models which are equivalent to such \qq{block automata}, for example realizing particular symmetries. No spread of information through the lattice occurs, however, for such \qq{ultralocal} models.

Assume that we employ in the next updating step again a similar block structure, but now grouping the positions $x$ and $\xe$ into a common block-cell at $x$. In this case the automaton cannot be divided into local pieces anymore, since the neighbors of the cells at $x$ and $\xe$ two updating steps before are $x-\eps$, $x$, $\xe$, $x+2\eps$, and so on. Both updating steps can have a simple fermionic equivalence, while a rather rich dynamical behavior can emerge. The naive continuum limit is the same for shifted blocks and the ultralocal setting. This points again to the importance of understanding the relation between the true continuum limit and the naive continuum limit.

For shifted blocks one has to guarantee that the possible outgoing bit configurations at the first step match precisely all ingoing configurations at the second step. This condition is not always trivial to implement, restricting somewhat the possibilities. Shifted blocks are an interesting alternative to the sequence of propagation and scattering for constructing cellular automata with interesting dynamics. In the next section we will establish in this way a fermionic quantum field theory with local $SO(M)$-gauge symmetry.

\subsection*{Shifted blocks}

Assume that the updating from $t$ to $\te$ (even $t$) involves blocks combining the cells $x$ and $x-\eps$ (even $x$). In the fermion language we write
\bel{SB1}
\Lbar(t)=\sum_{x\ \text{even}}\Lba(t,x-\eps)\ ,
\ee
where $\Lba(t,x-\eps)$ involves the Grassmann variables $\psi\g(t,x-\eps)$, $\psi\g\otx$, $\psibar\g(\te,x-\eps)$ and $\psibar\g(\te,x)$. The cell $x-\eps$ has two neighbors at $x-\eps$ and $x$ which matter for its updating. Also the cell $x$ has the two neighbors at $x-\eps$ an $x$. The updating in the block denoted by even $x$ is independent of the state of all other blocks at $y\neq x$. It is rather straightforwards to construct an invertible updating for the blocks. If the state of each cell involves $M$ bits, the composite states of the two neighboring sites at $x-\eps$ and $x$ are configurations of $2M$ bits. The updating within a block is given by a unique jump $2M\times2M$-matrix. We have encountered already several automata of this type.

\begin{figure*}

\includegraphics{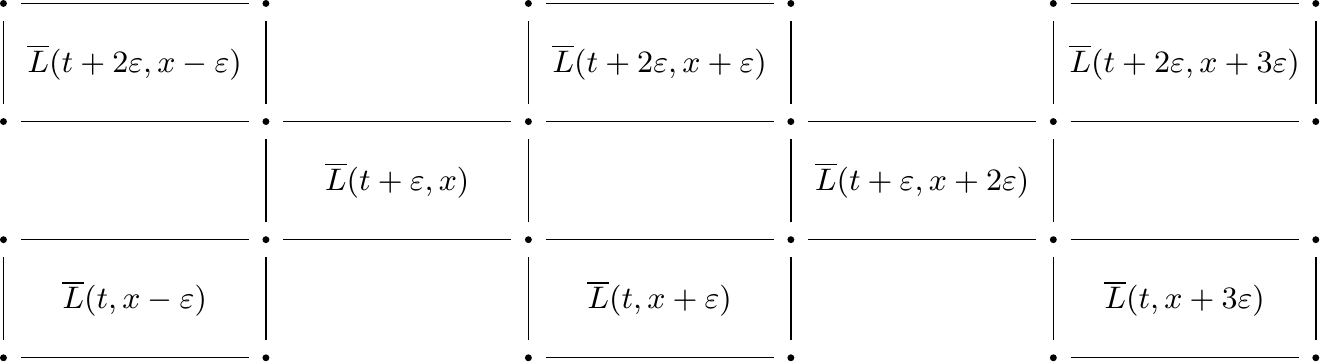}

\caption{Updating with shifted blocks. The dots denote the locations of the different cells, and the lines indicate which cells belong to a given block-cell for a given updating step. The past light cone of the two cells in the center of the top time-layer at $t+3\eps$ comprises all six cells in the lowest time layer at $t$. Similarly, the future light cone of the two cells in the center at $t$ consists of all six cells in the top layer. For the layers at $\te$ and $t+2\eps$ it includes the four cells shown in the figure.}
\label{fig: SBA}

\end{figure*}

For the updating from $\te$ to $t+2\eps$ we use a similar block structure, but with blocks shifted by one unit,
\bel{SB2}
\cL(\te)=\sum_{x\ \text{even}}L(\te,x)\ ,
\ee
where $L(\te,x)$ involves the Grassmann variables $\psibar\g(\te,x)$, $\psibar\g(\te,\xe)$, $\psi\g(t+2\eps,x)$ and $\psi\g(t+2\eps,\xe)$. The block updates the cells at $x$ and $\xe$ from their two neighbors at $x$ and $\xe$. For the sequence of the two steps different positions are now connected over a larger range, since the updating of the cell $\xe$ involves the cell $x$, which in turn involves the cell $x-\eps$ in the previous step. The overall automaton can no longer be decomposed into separated local parts, as visible from Fig.~\ref{fig: SBA}. Similar to the sequence of propagation and scattering each cell is influenced by a non-trivial past light cone, and influences itself a non-trivial future light cone.

In general, the updating rules for the steps at $t$ and $\te$ can be chosen differently. The only thing required is that all allowed ingoing bit configurations for the step at $\te$ can be generated as outgoing bit configurations for the step at $t$, and that the bit configurations generated by the updating at $t$ do not include bit-configurations for which the updating at $\te$ is not defined. This matters for automata with constrained bit configurations. Similar properties have to hold for the updating from $\te$ to $t+2\eps$. A simple structure uses for all even $t$ the same $\Lba(t,x-\eps)$ with variables shifted from $t$ to $t+2\eps$ and so on, while for all odd $\te$ one employs the same $\Lba(\te,x)$ in this sense. Already this simple setting can generate rather rich dynamics, for which the updating at $\otx$ is influenced by a past light-cone, and the state at $\otx$ influences the cells in the future light-cone. The light cones are easily visible from Fig.~\ref{fig: SBA}.

\subsection*{Conditional propagation}

Shifted blocks are convenient for a description of conditional propagation. Suppose we want to describe a propagation of particles that is influenced by their environment. This propagation is conditional in the sense that it depends on the state of the environment. An example is a particle that moves to the right if certain conditions for the environment are met, and to the left or not at all if other conditions are obeyed. We can describe this situation by a set of system variables $\psig$ and the associated occupation numbers and additional variables $\vp_\delta(x)$ accounting for the environment. The automaton acts on the combined configuration of the system- and environment-bits. The configuration of the environment-bits influences the updating of the system bits.

As a simple example we consider blocks with two sites at $x$ and $\xe$. We choose a single system variable and use shorthands $\psi(x)=\psi$, $\psi(\xe)=\psi'$. We also restrict the discussion to a single environment variable $\vp(x)=\vp$, $\vp(\xe)=\vp'$, and use similar shorthands for $\psibar$ and $\vpbar$. For our model the particles associated to $\psi$ move for most configurations of the environment from $x$ before the updating to $\xe$ after the updating, and similarly from $\xe$ to $x$. If this would be the updating for all states of the environment, the shifted blocks would describe free particles moving on straight trajectories corresponding to the diagonals. We will, however, select specific combinations of the environment-bits for which the particles do not move. This realizes a conditional propagation.

The updating rule for our model is shown in table~\ref{tab: CPA}. For our notation $(n_\psi,n_\vp,n_{\psi'},n_{\vp'})$ the first two entries indicate the value of the system-bit and environment-bit at $x$, the second two entries at $\xe$. The lower line specifies the state of the block before the updating, the upper line after the updating. We specify only the updating for eight out of the sixteen configurations. We assume particle-hole symmetry, such that the updating of the other eight configurations can be extracted from the table by exchanging ones and zeros. Inspecting the change in the first bit of each pair (the system-bit) we observe \qq{normal propagation} of ones or zeros along the diagonals of the block for most entries in the table.
\begin{table}

\centering
\renewcommand{\arraystretch}{1.2}

\begin{tabular}{|C{0.11\textwidth}|C{0.11\textwidth}|C{0.11\textwidth}|C{0.11\textwidth}|}

\hline
$11,11$ & $10,11$ & $11,10$ & $10,10$\\
$11,11$ & $11,10$ & $10,11$ & $10,10$\\
\noalign{\hrule height 1pt}
$00,11$ & $01,10$ & $11,01$ & $10,00$\\
$10,01$ & $11,00$ & $01,11$ & $11,01$\\
\hline

\end{tabular}

\caption{Updating rule for conditional propagation. The lower configuration is updated to the upper configuration.}
\label{tab: CPA}

\end{table}
There is only one entry for an \qq{exceptional propagation} for which a single particle at $x$ does not move. It corresponds to the last entry of the table. If the environment-bits (second entries in the pairs) take the value one at both positions a single particle at $x$ does not change its position. By particle-hole symmetry a single particle at $\xe$ does not move if the environment-bits are zero at both positions. In a wider sense this model describes the motion of particles in the presence of \qq{impurities}. The updating rule in the table also specifies the updating of the environment, which depends in turn on the state of the system-particle.

The fermion picture for this model is given by
\begin{align}
\label{SB3}
-L=&\vp\vpbar'+\vp'\vpbar+\gl1-\vp\vpbar-\vp'\vpbar'\gr\psi\psi'\psibar\,\psibar'\nn\\
&+\gl1+\vp\vpbar+\vp'\vpbar'-\vp\vpbar'-\vp'\vpbar+\vp\vp'\vpbar\vpbar'\gr\psi'\psibar\nn\\
&+\gl\vp\vpbar+\vp'\vpbar'+\vp\vp'\vpbar\vpbar'\gr\psi\psibar'\nn\\
&+\vp\vp'\psi'\psibar'+\vpbar\vpbar'\psi\psibar\ .
\end{align}
The exponential $K=\exp(-L)$ indeed describes the updating in the block according to the rules of sect.~\ref{sec: FPGU}. Reading out the motion of the particle we find normal propagation with factors $\psi'\psibar$ or $\psi\psibar'$ for most terms. The exceptional propagation arises from the last two terms.

The corresponding Grassmann functional for the arrangement of shifted blocks is already a rather complicated discrete many fermion model with a variety of interactions. Looking at the action it is a priori not directly visible that such a model can find an exact description in terms of a cellular automaton and becomes \qq{solvable} in this sense. With several system-bits and several environment-bits rather complex updating rules can be implemented. In turn, the functional integral description may provide methods for an investigation of the continuum limit for complex automata with a large number of cells.

\section{Cellular automata for local gauge theories}
\label{sec: CAGT}

The basic fundamental interactions in elementary particle physics are all described by local gauge theories. Pioneered by electromagnetism and general relativity the gauge principle governs today's understanding of the electro-weak and strong interactions. This raises the question if cellular automata can realize local gauge symmetries as well. The answer is positive. Yang-Mills theories of the type of quantum chromodynamics can be based on gauge invariant objects constructed from fermions. The unique jump property of the step evolution operator requires that these gauge invariant objects correspond to a single Grassmann element, rather than a sum of different Grassmann elements. This seems, at first sight, a strong restriction. We will, however, find an equivalent formulation in terms of composite link variables, rather close to lattice gauge theories. In the continuum limit the unique jump property is no longer required such that many different gauge invariant terms in the effective action become possible.

\subsection*{Non-abelian local gauge symmetry}

The aim of this section is an existence proof of automata realizing a non-abelian local continuous symmetry. This is done by constructing a simple example. We will employ the updating with shifted blocks for the construction of an automaton for which the fermion picture is invariant under local $SO(4)$-gauge transformations. For this purpose we associate $L(x)$ in eq.~\eqref{SY19} with $L(\te,x)$ and choose a similar form for $\Lba(t,x-\eps)$, with $t$ and $x$ even,
\begin{align}
\label{SB4}
\Lba(t,x-\eps)=-\big[&\bar b(\te,x)b(t,x-\eps)\nn\\
&+\bar b(\te,x-\eps)b\otx\big]\ ,\nn\\
L(\te,x)=-\big[&b(t+2\eps,\xe)\bar b(\te,x)\nn\\
&+b(t+2\eps,x)\bar b(\te,\xe)\big]\ .
\end{align}
Here $b$ and $\bar b$ are the $SO(4)$-invariants defined in eq.~\eqref{SY17}.
For this automaton the states \qq{$0$} with four bits zero at a given position $x$ and \qq{$1$} with all four bits at $x$ equal to one propagate on straight lines spanned by the diagonals of the blocks. This amounts to constrained scattering in the sense of sect.~\ref{sec: CA}. The automaton acts only on the gauge invariant four-fermion configurations. Due to the shifting of the blocks the gauge invariant four-particle states propagate on diagonals through the whole space-time lattice.

This very simple setting can easily be generalized to $SO(N)$-gauge theories with arbitrary $N$ where $b$ involves $N$ Grassmann variables
\bel{SB5}
b\otx=\psi_1\otx\psi_2\otx\dots\psi_N\otx\ .
\ee
(We omit the bars on objects at odd $\te$ since the time label suffices to designate them.) For $N$ odd we can take $b\otx$ as a collective Grassmann variable. More complex gauge invariant automata can be constructed by introducing further Grassmann variables that are invariant under the gauge transformations, or different gauge singlets $b_i(x)$ built from different variables. These extensions are also possible for even $N$. The only difference is that the collective variables $b_i(x)$ commute with all other variables.

The formulation of local gauge theories in terms of gauge singlets seems almost trivial. In quantum chromodynamics (QCD) this would correspond to a formulation in terms of microscopic gauge invariant objects as plaquettes. For a large range of scales the continuum limit of QCD is better described by the dynamics of gluons, the gauge fields of the gauge group $SU(3)$. The gluons are not gauge invariant objects. They are related in a lattice formulation to link variables. We discuss next that composite link variables can also be used for a description of cellular automata with local gauge symmetry.

\subsection*{Link variables}

Composite link variables are Grassmann elements built from Grassmann variables at neighboring sites in $x$ or $t$. If the number of Grassmann variables is even, the link variables are commuting objects. In their simplest form the link variables are bilinears in the Grassmann variables. For a block or plaquette with four sites $\otx$, $(t,\xe)$, $(\te,x)$, $(\te,\xe)$ we may define the link variables
\begin{align}
\label{SB6}
l\gd^{+1}\otx=&\,\psi\g\otx\vp\del\otxe\ ,\nn\\
l\gd^{-1}\otexe=&\,\psi\g\otexe\vp\del\otex\ ,\nn\\
l\gd^{+0}\otxe=&\,\psi\g\otxe\vp\del\otexe\ ,\nn\\
l\gd^{-0}\otex=&\,\psi\g\otex\vp\del\otx\ .
\end{align}
In this notation the upper indices indicate the direction of the link, i.e. $+1$ corresponds to the positive $x$-direction and $-0$ the the negative $t$-direction. The position argument denotes the \qq{origin} of the link, while the color indices indicate the colors of the variables $\psi$ and $\vp$ involved. One may view the link variables as matrices in color space associated to the directed links of the lattice.

In general, the variables $\vp\g(x)$ may be composites of the variables $\psig$. They may also be Grassmann variables of a different species. For an implementation of local gauge transformations we suppose the infinitesimal transformation
\begin{align}
\label{SB7}
\delta\psi\g\otx=&\,\eps\gd\otx\psi\del\otx\ ,\nn\\
\delta\vp\g\otx=&-\vp\del\otx\eps_{\delta\gamma}\otx\ .
\end{align}
In consequence, the link variables transform with two different transformation parameters
\bel{SB8}
\delta l\gd^{+1}\otx=\eps_{\gamma\eta}\otx l_{\eta\delta}^{+1}\otx-l_{\gamma\eta}^{+1}\otx\eps_{\eta\delta}\otxe\ ,
\ee
as familiar from lattice gauge theories. We may write this in a matrix form, with $\gl A\otx\gr\gd=\eps\gd\otx$,
\bel{SB9}
\delta l^{+1}=A\otx l^{+1}\otx-l^{+1}\otx A\otxe\ .
\ee
The plaquette product of four link variables is gauge invariant,
\begin{align}
\label{SB10}
p\otx=l^{+1}\otx &l^{+0}\otxe l^{-1}\otexe l^{-0}\otex\ ,\nn\\
&\delta p\otx=0\ .
\end{align}

We cannot always use the plaquette product directly for the construction of a cellular automaton with local gauge symmetry. The reason is that the matrix product contains sums,
\begin{align}
\label{SB11}
p\otx=&-\psi_{\gamma_1}\otx\vp_{\gamma_1}\otx\psi_{\gamma_2}\otxe\vp_{\gamma_2}\otxe\nn\\
&\times\psi_{\gamma_3}\otexe\vp_{\gamma_3}\otexe\nn\\
&\times\psi_{\gamma_4}\otex\vp_{\gamma_4}\otex\ ,
\end{align}
where the minus sign is absent if $\vp$ commutes with $\psi$. This may not be compatible with the unique jump property of the step evolution operator. For $\vp\g(x)=\psig$ the plaquette product vanishes due to $\psi\g^2(x)=0$. There is a simple particular choice for which
\bel{SB12}
\psig\vp\g(x)=\psi_1(x)\psi_2(x)\dots\psi_N(x)=b(x)\ ,
\ee
e.g. $\vp_1(x)=\psi_2(x)\dots\psi_N(x)$ etc. In this case one has
\bel{SB13}
p\otx=-N^4b\otx b\otxe b\otexe b\otex\ ,
\ee
such that $\tilde p\otx=p\otx/N^4$ is a single Grassmann basis element. One can therefore use $\tilde p\otx$ together with $b\otx$ as gauge invariant building blocks for gauge invariant cellular automata. In particular, we could choose an action which is a sum over plaquettes, $S=\sum_{t,x}\tilde p\otx$, similar to lattice gauge theories. The gauge coupling takes a fixed value due to the required unique jump property. As before, more complex gauge invariant automata can involve several $\tilde p_i\otx$ and $b_i(x)$ combined with Grassmann variables that are gauge singlets.

At first sight one may wonder what is the use of link variables since the identity~\eqref{SB13} seems much simpler than the expression~\eqref{SB10}. This issue changes if one considers a possible continuum limit. We will see in sect.~\ref{sec: CLCGS} that the coarse graining involved in this limit leads to step evolution operators that are no longer unique jump operators. Once the unique jump property is no longer required there are a larger number of possibilities to construct gauge invariant objects from link variables. It is well conceivable that for suitable models the link variables and associated gauge fields play a dominant role for the continuum limit, similar to QCD.

On a coarse grained level the variables $\psi$ and $\vp$ constituting the composite link variable may be combinations of several objects with the same transformation property~\eqref{SB7}. The gauge coupling is no longer fixed - it can become a scale-dependent \qq{running coupling}. There is still a long way to go in order to decide if it is possible to construct an automaton that realizes in four dimensions a unified gauge group as $SO(10)$ with propagating fermions in chiral representations. Our first investigation here seems to indicate that there is no barrier in principle.

\section{Discrete spinor gravity in two\\dimensions}
\label{sec: DSG}

In this section we further explore cellular automata for which the action does not contain terms quadratic in the Grassmann variables. In continuous spacetime this type of models is needed for spinor gravity~\cite{CWLSG, CWSGDI, SG3, SG4} for which local Lorentz symmetry and diffeomorphism symmetry are realized in a purely fermionic setting. The vierbein and the metric emerge then as composite objects. In the absence of quadratic terms the single fermion excitations of a vacuum without condensates of fermion composites do not have a well defined propagator. Condensates are therefore needed for a realistic model. The setting without quadratic terms for the fermions still allows for a well defined Grassmann functional integral. What is missing are appropriate methods to deal with this situation. A formulation in terms of an equivalent cellular automaton may provide for an interesting starting point for particular models of this type. We will focus here on the structure of the next-neighbor double-hole scattering discussed in sect.~\ref{sec: CA}. The action~\eqref{C11} involves only terms with eight Grassmann variables. We will show that the fermionic model associated to this cellular automaton is a discrete version of two-dimensional spinor gravity.

\subsection*{Next-neighbor double-particle scattering}

For a sequence of updating steps the allowed configurations after the updating of the first step have to match the starting configurations of the second step. As an example, the outgoing configurations in Fig.~\ref{fig: C1} do not coincide with the allowed ingoing configurations. We want to use the updating of Fig.~\ref{fig: C1} for odd $t+\eps$ (or even $t$), say from $t+\eps$ to $t+2\eps$ as indicated in the last column in Fig.~\ref{fig: C1}. The updating at even $t$ has to be different in order to avoid a mismatch. We indicate a possible updating rule for even $t$ in Fig.~\ref{fig: S1}. With this rule the configurations match at $t+\eps$. The outgoing configurations for $\psibar$ or $\psi(t+\eps)$ in Fig.~\ref{fig: S1} match the ingoing configurations for Fig. ~\ref{fig: C1}.

The elementary processes for the updating shown in Fig.~\ref{fig: S1} have the interpretation of scattering of neighboring pairs of two particles. For the first line a particle $(+,2)$ and a particle $(-,1)$ at $x-\eps$, together with particles $(+,4)$ and $(-,3)$ at $x+\eps$, are scattered to particles $(+,1)$ and $(-,3)$ at $x$ and particles $(+,4)$ and $(-,2)$ at $x+2\eps$. Due to the modulo-two property with respect to particles and holes a Grassmann variable $\psi\g(t,x)$ gives now a non-vanishing contribution if at $t$ and $x$ a particle of type $\gamma$ is present. The elementary processes described by $\Lbar(t)$ are scatterings of four particles into four particles.
\begin{figure}
\centering
\renewcommand{\arraystretch}{1.2}

\begin{tabular}{C{0.07\textwidth}*{5}{C{0.07\textwidth}}}

$x-\eps$ & $x$ & $x+\eps$ & $x+2\eps$ & & \\
\noalign{\hrule height 1pt}
-- & $(13)$ & -- & $(42)$ & $\psibar$ & $t+\eps$ \\
$(21)$ & -- & $(43)$ & - & $\psi$ & $t$ \\
\hline
-- & $(24)$ & -- & $(31)$ & $\psibar$ & $t+\eps$ \\
$(12)$ & -- & $(34)$ & -- & $\psi$ & $t$ \\
\hline
(42) & -- & $(13)$ & -- & $\psibar$ & $t+\eps$ \\
-- & $(43)$ & -- & $(21)$ & $\psi$ & $t$ \\
\hline
$(31)$ & -- & $(24)$ & -- & $\psibar$ & $t+\eps$ \\
-- & $(34)$ & -- & $(12)$ & $\psi$ & $t$ \\
\hline

\end{tabular}

\caption{Updating for next-neighbor double-hole scattering for even $t$. This updating rule can be combined with the one for Fig. \ref{fig: C1} for odd $t$.}
\label{fig: S1}

\end{figure}
This seems to contrast with the elementary process of four-hole scattering associated to $\cL(t+\eps)$, which appears together with $\Lbar(t)$ in the combined local factor $\Khat(t)$ in eq.~\eqref{S15}. Due to particle-hole symmetry of the particular next-neighbor double-hole scattering chosen in sec.~\ref{sec: CA} we can equally interpret the elementary processes at $t+\eps$ as four-particle scattering, with the same color-content as for the holes. Interpreting all elementary processes as particle scatterings facilitates the discussion.

\begin{figure}

\includegraphics{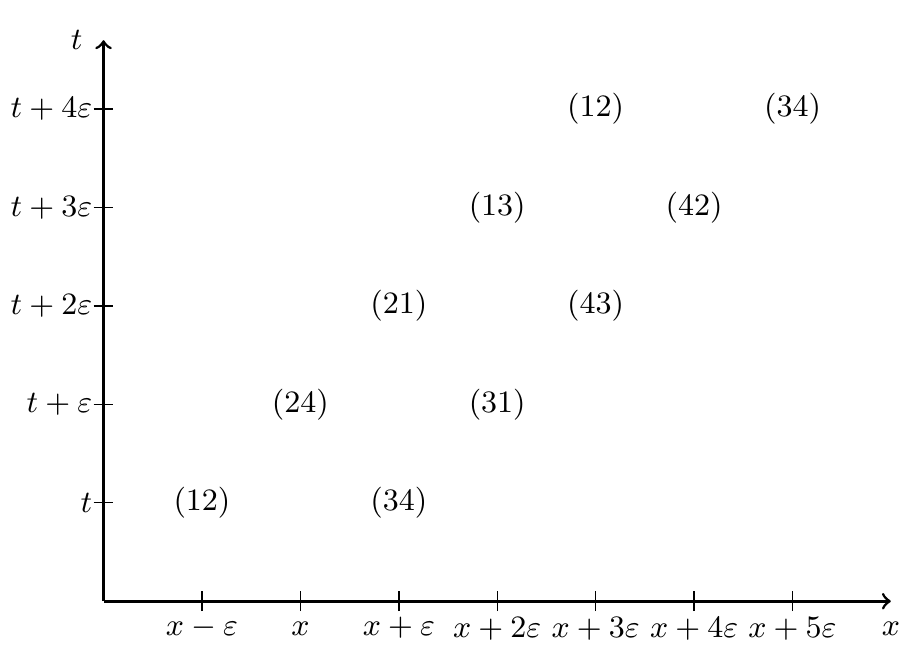}

\caption{Sequential updating for a four-particle state for next-neighbor double-particle scattering. The motion is periodic in $t\to t+4\eps$, $x\to x+4\eps$.}
\label{fig: S2}

\end{figure}

We can now draw the time history of a four-particle state at $t\inn$ by a sequence of updatings according to $\Lbar(t)$ for $t$ even and $\cL(t+\eps)$ for $t+\eps$ odd. An example is shown in Fig.~\ref{fig: S2}. The whole four-particle bloc moves to the right, with changing colors. The same color combinations are found again at $t+4\eps$, with $x$ displaced by four units to the right. 
This state can be characterized by a wave function $q(t,x)$, where $x-\eps$ denotes the position of the particle pair on the left, and $x+\eps$ the particle pair on the right, and we omit color indices. This wave function is periodic with period $4\eps$,
\bel{S26}
q(t+4\eps,x+4\eps)=q(t,x)\ .
\ee
The evolution of this particular four-particle state finds a rather simple description for this cellular automaton. We will describe later the evolution of more complex bit-configurations.

The updating of the spinor gravity automaton in the second step of the sequence can be read from Fig.~\ref{fig: S1}. On the space of restricted spin configurations the updating of the cell $x$ for right-movers involves only the cell $x-\eps$, and similarly $\xe$ for left-movers. Combining two steps in the sequence amounts to the simple exchange of colors $1\leftrightarrow 2$, $3\leftrightarrow 4$ for the appropriate neighbors. This is also visible in fig.~\ref{fig: S2}. We conclude that the spinor gravity automaton is rather simple. Nevertheless, we can discuss non-trivial structures for the associated discrete quantum field theory for fermions.

The Grassmann functional integral~\eqref{S19} for this automaton is specified by
\begin{align}\label{S27}
&\Lba(t,x)+L(t+\eps,x)\nn\\
=&-\Big\{\psi_+^1(t+\eps,x)\psi_-^3(t+\eps,x)\psi_+^4(t+\eps,x+2\eps)\nn\\
&\times\psi_-^2(t+\eps,x+2\eps)\psi_+^2(t,x-\eps)\psi_-^1(t,x-\eps)\nn\\
&\times\psi_+^4(t,x+\eps)\psi_-^3(t,x+\eps)+\text{three more terms}\Big\}\nn\\
&-\Big\{\psi_+^1(t+2\eps,x+\eps)\psi_-^2(t+2\eps,x+\eps)\psi_+^3(t+2\eps,x+3\eps)\nn\\
&\times\psi_-^4(t+2\eps,x+3\eps)\psi_+^1(t+\eps,x)\psi_-^3(t+\eps,x)\nn\\
&\times\psi_+^4(t+\eps,x+2\eps)\psi_-^2(t+\eps,x+2\eps)\nn\\
&+\text{three more terms}\Big\}\ .
\end{align}
The first four terms (first curly bracket) arise from $\Lba(t,x)$, while the second four terms (second curly bracket) emerge from $L(t+\eps,x)$. The color assignments of the three terms not indicated explicitly can be inferred from Figs.~\ref{fig: C1},~\ref{fig: S1}. Eq.~\eqref{S27} constitutes a discrete functional integral for fermions for which the action contains terms with eight Grassmann variables. We will below understand its rather simple structure, which describes a lattice discretization for a continuum model which is invariant under local Lorentz transformations.

\subsection*{Local Lorentz transformations}

The action for the model~\eqref{S27} is invariant under local infinitesimal transformations
\bel{S28}
\begin{split}
\delta\psi_+^a(t,x)&=-\frac{1}{2}\eta(t,x)\psi_+^a(t,x)\ ,\\
\delta\psi_-^a(t,x)&=\phantom{-}\frac{1}{2}\eta(t,x)\psi_-^a(t,x)\ .
\end{split}
\ee
The invariance under this opposite multiplicative rescaling of $\psi_+$ and $\psi_-$ at every position $(t,x)$ is manifest from the fact that at each site of the lattice each variable $\psi_+^a(t,x)$ is accompanied by a variable $\psi_-^b(t,x)$. Due to the relative minus sign in eq.~\eqref{S28} the products $\psi_+^a(t,x)\psi_-^b(t,x)$ are invariant under the transformation~\eqref{S28}. These transformations are local gauge transformations since they can be performed independently at every space-time point $(t,x)$.

The gauge transformations~\eqref{S28} are local Lorentz-transformations in $1+1$ dimensions, forming the group $SO(1,1)$. This can be seen by grouping the variables $\psi_+^a$ and $\psi_-^a$ into a two-component spinor
\bel{S29}
\psi^a(t,x)=\pvec{\psi_+^a(t,x)}{\psi_-^a(t,x)}\ .
\ee
Local infinitesimal Lorentz transformations act on $\psi^a$ as
\bel{S30}
\delta\psi^a=-\eta\otx\Sigma^{01}\psi^a\ ,\quad \Sigma^{mn}=\frac{1}{4}[\gamma^m,\gamma^n]\ .
\ee
The components $\psi_+$ and $\psi_-$ are Weyl spinors, which are irreducible representations of $SO(1,1)$. Since we have not introduced a complex structure for the Grassmann variables they are actually Majorana-Weyl spinors.

Local Lorentz symmetry is a central ingredient for gravitational theories with fermions. This is the reason why we have chosen examples with particle pairs (or hole pairs) at each site. The second symmetry ingredient of gravitational theories, namely diffeomorphism symmetry, will be seen for a naive continuum limit to be discussed below.

\subsection*{Lattice derivatives}

Lattive derivatives for Grassmann variables are defined by
\bel{S33}
\begin{split}
(\dt+\partial_x)\psi(t+\frac{\eps}{2},x+\frac{\eps}{2})=&\frac{1}{\eps}\big[\psi(t+\eps,x+\eps)-\psi(t,x)\big]\ ,\\
(\dt-\partial_x)\psi(t+\frac{\eps}{2},x-\frac{\eps}{2})=&\frac{1}{\eps}\big[\psi(t+\eps,x-\eps)-\psi(t,x)\big]\ .
\end{split}
\ee
In terms of these derivatives we write
\bel{S34}
\begin{split}
\psi(t+\eps,x+\eps)\psi(t,x)=&\eps\partial_+\psi(t+\frac{\eps}{2},x+\frac{\eps}{2})\psi(t,x)\\
\psi(t+\eps,x-\eps)\psi(t,x)=&\eps\partial_-\psi(t+\frac{\eps}{2},x-\frac{\eps}{2})\psi(t,x)\ ,
\end{split}
\ee
where
\bel{S35}
\partial_+=\dt+\partial_x\ ,\quad \partial_-=\dt-\partial_x\ .
\ee
This relation follows from the Grassmann identity $\psi(t,x)\psi(t,x)=0$.

The discrete fermionic model for the next-neighbor double-particle scattering automaton is given by eight terms
\bel{S36}
\eps^{-2}\big\{\Lba(t,x)+L(t+\eps,x)\big\}=\sum_{i=1}^8l_i\ .
\ee
The first two terms contain derivatives of $\psip^1$ and $\psim^1$,
\bel{S36a}
\begin{split}
l_1=&\phantom{-}\psip^1(\te,x)\dpl\psip^1(t+\efr{3},x+\frac{\eps}{2})\\
&\times\psim^2(\te,x+2\eps)\dm\psim^2(t+\efr{3},x+\efr{3})\\
&\times\psip^3(t+2\eps,x+3\eps)\psim^3(\te,x)\\
&\times\psip^4(\te,x+2\eps)\psim^4(t+2\eps,x+3\eps)
\end{split}
\ee
and
\bel{S36b}
\begin{split}
l_2=&-\psip^1(\te,x+3\eps)\dm\psip^1(t+\efr{3},x+\efr{5})\\
&\times\psim^2(\te,x+\eps)\dpl\psim^2(t+\efr{3},x+\efr{3})\\
&\times\psip^3(t+2\eps,x)\psim^3(\te,x+3\eps)\\
&\times\psip^4(\te,x+\eps)\psim^4(t+2\eps,x)\ .
\end{split}
\ee
The remaining six terms are displayed in appendix~\ref{app: C}. The expression~\eqref{S36} is the exact formulation of the discretization of spinor gravity which is described by a cellular automaton.

\subsection*{Continuum limit}

A naive continuum limit can be taken as $\eps\to0$, with lattice derivatives replaced by partial derivatives. We will discuss the conditions for the true continuum limit in more detail in later sections. In the naive continuum limit we can neglect the remaining differences in time- and space-points which are of higher order in an expansion in $\eps$. The continuum model based on eq.~\eqref{S36} simplifies considerably
\begin{align}\label{S37}
&\Lba(t,x)+L(t,x)=\eps^2\Big\{\psip^3\psim^3\psip^4\psim^4\nn\\
&\quad\quad\quad\times\gl\psip^1\dpl\psip^1\psim^2\dm\psim^2-\psip^1\dm\psip^1\psim^2\dpl\psim^2\gr\nn\\
&+\psip^1\psim^1\psip^2\psim^2\gl\psip^3\dpl\psip^3\psim^4\dm\psim^4-\psip^3\dm\psip^3\psim^4\dpl\psim^4\gr\nn\\
&\phantom{\Lba(t,x)+L(t,x)=}-\gl\psip\leftrightarrow\psim\gr\Big\}\ .
\end{align}
Here all Grassmann variables are taken at the same point $(t,x)$. We further use the identity ($\eps^{01}=-\eps^{10}=1$, $\eps^{00}=\eps^{11}=0$)
\bel{S38}
\begin{split}
\dpl\psi\g\dm\psi_\delta-\dm\psi\g\dpl\psi_\delta=&-2\gl\dt\psi\g\partial_x\psi_\delta-\partial_x\psi\g\dt\psi_\delta\gr\\
=&-2\eps^{\mu\nu}\partial_\mu\psi\g\partial_\nu\psi_\delta\ ,
\end{split}
\ee
and replace
\bel{S39}
\sum_t\sum_x=\eps^{-2}\int\text{d}t\text{d}x=\eps^{-2}\int_{t,x}\ .
\ee
This yields the continuum expression for the fermionic action
\bel{S40}
\begin{split}
S=&-2\int_{t,x}\eps^{\mu\nu}\Big\{\psip^1\partial_\mu\psip^1\psim^2\partial_\nu\psim^2\psip^3\psim^3\psip^4\psim^4\\
&+\psip^3\partial_\mu\psip^3\psim^4\partial_\nu\psim^4\psip^1\psim^1\psip^2\psim^2-\gl\psip\leftrightarrow\psim\gr\Big\}\ .
\end{split}
\ee
We use here the usual convention $x^\mu=(t,x)$ and $\partial_\mu=\partial/\partial x^\mu$. With respect to general coordinate transformations the Grassmann fields $\psi(t,x)$ transform as scalars, and $\partial_\mu\psi$ as covariant vectors. Due to the contraction of the derivatives with the antisymmetric $\eps$-tensor this continuum action is invariant under diffeomorphisms or general coordinate transformations. Together with the local Lorentz symmetry it has all ingredients for a model of spinor gravity.

In the opposite direction we consider eq.~\eqref{S36} as a discretization for the continuum model~\eqref{S40}. Such a discretization guarantees that the Grassmann functional integral is well defined. The particular discretization~\eqref{S36} preserves the gauge symmetry of local Lorentz-transformations. Diffeomorphism symmetry is not maintained by the discretization. Only lattice-diffeomorphism invariance~\cite{CWLDI} can be realized. Our particular discretization~\eqref{S36} results in a rather simple cellular automaton.

The local Lorentz-symmetry remains visible in the continuum limit. With
\bel{S41}
\psibar^a=\psi^{a\,T}\gamma^0=\gl\psim^a,-\psip^a\gr\ ,
\ee
we can construct Lorentz scalars
\bel{S42}
\begin{split}
\psibar^a\psi^b=&\psim^a\psip^b-\psip^a\psim^b\ ,\\
\psibar^a\bar\gamma\psi^b=&\psim^a\psip^b+\psip^a\psim^b\ ,
\end{split}
\ee
where $\bar\gamma$ is the analogue of $\gamma^5$ in two dimensions
\bel{S43}
\bar\gamma=-\gamma^0\gamma^1=\tau_3\ .
\ee
The factors $\psip^a\psim^a$ in eq.~\eqref{S40} involve the Lorentz scalars (no summation over the color index here)
\bel{S44}
\Htil^a=\psip^a\psim^a=-\frac{1}{2}\psibar^a\psi^a\ .
\ee

Furthermore, two-component Lorentz-vectors can be constructed as usual by
\bel{S45}
\begin{split}
V^{m,ab}=&\psibar^a\gamma^m\psi^b\ ,\\
V^{ab}=&\pvec{V^{0,ab}}{V^{1,ab}}=\pvec{-\gl\psim^a\psim^b+\psip^a\psip^b\gr}{\psim^a\psim^b-\psip^a\psip^b}\ .
\end{split}
\ee
Its infinitesimal Lorentz-transformation reads
\bel{S46}
\begin{split}
\delta V^{m,ab}&=\tilde\eps_{\ n}^mV^{n,ab}\ ,\quad \tilde\eps_{01}=-\tilde\eps_{10}=\eta\ ,\\
\tilde\eps_{\ n}^m=&\eta^{mp}\tilde\eps_{pn}\ ,\quad \tilde\eps_{\ 1}^0=-\eta\ ,\, \tilde\eps_{\ 0}^1=-\eta\ .
\end{split}
\ee
These objects exist in the discrete version of spinor gravity as well. We conclude that the continuum limit does not effect the local gauge symmetry connected to Lorentz transformations. Only the realization of diffeomorphism symmetry is non-trivial.

\subsection*{Zweibein}

In the continuum version we introduce fermion bilinears that have the same transformation properties as zweibeins (no sum over $a$)
\bel{S47}
\Etil_\mu^{\ m,a}=\psibar^a\gamma^m\partial_\mu\psi^a\ .
\ee
They transform as vectors under local Lorentz transformations since the inhomogeneous part vanishes due to $(\psi^a)^2=0$,
\bel{S48}
-\psibar^a\gamma^m\Sigma^{01}\psi^a\partial_\mu\eta=0\ .
\ee
With respect to diffeomorphisms the generalized zweibein-bilinears $\Etil_\mu^{\ m,a}$ transform as covariant vectors. The components of the zweibein-bilinears read explicitly
\bel{S49}
\begin{split}
\Etil_\mu^{\ 0,a}=&-\gl\psim^a\partial_\mu\psim^a+\psip^a\partial_\mu\psip^a\gr\ ,\\
\Etil_\mu^{\ 1,a}=&\psim^a\partial_\mu\psim^a-\psip^a\partial_\mu\psip^a\ .
\end{split}
\ee
What is new as compared to a standard definition of zweibeins is the additional color index $a$. The generalized zweibein transforms non-trivially with respect to additional color transformations - for a discussion of geometric consequences see refs.~\cite{CWLSG, CWSGDI, SG3, SG4}.

We can use the generalized zweibeins in order to build Lorentz-invariants by contraction with the invariant antisymmetric tensor $\eps_{mn}$,
\bel{S50}
\begin{split}
\eps_{mn}\Etil_\mu^{\ m,a}&\Etil_\nu^{\ n,b}=\Etil_\mu^{\ 0,a}\Etil_\nu^{\ 1,b}-\Etil_\mu^{\ 1,a}\Etil_\nu^{\ 0,b}\\
=&2\gl\psim^a\partial_\mu\psim^a\psip^b\partial_\nu\psip^b-\psip^a\partial_\mu\psip^a\psim^b\partial_\nu\psim^b\gr\ .
\end{split}
\ee
The objects
\bel{S51}
\begin{split}
\Etil^{ab}=&\frac{1}{2}\eps^{\mu\nu}\eps_{mn}\Etil_\mu^{\ m,a}\Etil_\nu^{\ n,b}\\
=&-\eps^{\mu\nu}\big\{\psip^a\partial_\mu\psip^a\psim^b\partial_\nu\psim^b-\gl\psip\leftrightarrow\psim\gr\big\}
\end{split}
\ee
transform under diffeomorphisms as scalar densities, similar to the determinant of the zweibein.

One concludes that terms of the form
\bel{S52}
S^{ab}=\int_{t,x}\Etil^{ab}f_s
\ee
are invariant both under local-Lorentz transformations and diffeomorphisms for $f_s$ an arbitrary function of Lorentz invariant scalars. The action~\eqref{S40} takes precisely this form
\bel{S53}
\begin{split}
S=&2\int_{t,x}\gl\Etil^{12}\Htil^3\Htil^4+\Etil^{34}\Htil^1\Htil^2\gr\\
=&\int_{t,x}\eps_{mn}\eps^{\mu\nu}\gl\Etil_\mu^{\ m,1}\Etil_\nu^{\ n,2}\Htil^3\Htil^4+\Etil_\mu^{\ m,3}\Etil_\nu^{\ n,4}\Htil^1\Htil^2\gr\ .
\end{split}
\ee

We can consider the fermion-bilinears $\Etil_\mu^{\ m,a}$ and $\Htil^a$ as composite bosons. In terms of these composite bosons the diffeomorphism- and Lorentz-invariant continuum action~\eqref{S53} takes a rather simple form. What is perhaps new is the appearance of \qq{colored} or \qq{flavored} zweibeins~\cite{CWLSG, CWSGDI} which carry a color index $a$ in addition to the world index $\mu$ and Lorentz index $m$. A composite metric can be introduced as a color-neutral object
\bel{222A}
g_{\mu\nu}=\Etil_\mu^{ma}\Etil_\nu^{na}\eta_{mn}\ .
\ee
As usual, it is invariant under local Lorentz transformations and transforms as a symmetric tensor under diffeomorphisms.


\section{Operators for observables}
\label{sec: OO}

We have presented a rather rich variety of cellular automata for which the fermion pictures are interesting two-dimensional quantum field theories. In the next part of this work we address the approach to the continuum limit. For this purpose we use the equivalence between automata and quantum field theories in order to demonstrate how many characteristic concepts of quantum mechanics find a useful application for cellular automata. This concerns notions as the Hamiltonian or possible order parameters built on fermionic bilinears which play an important role for the continuum limit of a very large number of cells.

In this section we address one more key concept of the quantum formalism that is realized by the use of wave functions, namely the association of operators to observables, and the quantum rule for expectation values. These operators do not need to commute. We will find useful operators for the description of probabilistic automata that do not commute with the operators which correspond to observables constructed from occupation numbers at a given time. In particular, we discuss the Hamilton operator. As familiar for quantum mechanics it is a key object for a description of the evolution in continuous time, and therefore essential for the understanding of the continuum limit.

\subsection*{Quantum rule for expectation values}

In quantum mechanics the expectation value of an observable $A$ at $t$ is given by the quantum rule, which reads in the real representation
\bel{P15}
\langle A(t)\rangle=\langle q(t)\hat{A}q(t)\rangle=q_\tau(t)\hat{A}\taur q_\rho(t)\ .
\ee
Here $\hat{A}$ is a suitable operator associated to $A$. In principle, it can depend on $t$, while it is $t$-independent for many simple observables. For the probabilistic cellular automaton the quantum rule can be derived from the general law for expectation values in classical statistics. For time-local observables this is given by
\bel{P16}
\exval{A(t)}=p_\tau(t)A_\tau\ ,
\ee
with $A_\tau$ the value of the observable for the bit-configuration $\tau$. Time-local observables are, for example, the occupation numbers $n\al(t)$ or $n\g(t,x)$. The value $A_\tau=(n\al)_\tau$ of such an observable for the bit-configuration $\tau$ equals one if the bit $\alpha$ in this configuration equals one, and zero if the bit-configuration hat $n\al=0$. Thus $(n\al)_\tau$ simply \qq{reads out} the corresponding bit of the configuration $\tau$.

We associate to the observable $n\al(t)$ the diagonal operator $\Nhat\al$ with elements
\bel{P17}
\gl\Nhat\al\gr\taur=(n\al)_\tau\delta\taur\ .
\ee
The equivalence of the quantum rule~\eqref{P15} with the classical statistical rule~\eqref{P16} is manifest
\bel{P18}
\begin{split}
\exval{n\al(t)}=&q_\tau(t)\gl\Nhat\al\gr\taur q_\rho(t)\\
=&\sum_{\tau,\rho}q_\tau(t)(n\al)_\tau\delta\taur q_\rho(t)\\
=&\sum_\tau(n\al)_\tau q_\tau^2(t)=\sum_\tau(n\al)_\tau p_\tau(t)\ .
\end{split}
\ee
This generalizes immediately to all observables that are functions of $n\al(t)$. Examples are the equal-time correlations $\exval{n\g(t,x)n_\delta(t,y)}$. The derivation of the quantum rule for expectation values is very simple, but may seem here somewhat trivial. We will extend the applicability of the quantum rule to a much larger set of observables, including associated operators that are no longer diagonal.

As a first set of observables beyond the time-local observables we take observables that involve occupation numbers at different times, $n\al(t_1)n\bet(t_2)$, whose expectation value is the unequal-time correlation function $\exval{n\al(t_1)n\bet(t_2)}$. We discuss in appendix~\ref{app: D} that expectation values can again be computed in terms of the wave function by the quantum rule~\eqref{P15}. For the particular case of cellular automata the operators associated to this type of observables remain diagonal and therefore commute with operators for time-local observables. This would not hold for a more general form of the step evolution operator.

\subsection*{Non-commuting operators}

The quantum rule for expectations values can also be used for observables that are represented by operators $\hat{B}$ with non-vanishing off-diagonal elements, $\hat{B}\taur\neq0$ for $\tau\neq\rho$. Examples are the momentum operator~\cite{CWNEW} or the Hamiltonian. We discuss the momentum operator for the particular case of one-particle states in sect.~\ref{sec: SPWF}, while the Hamilton operator is described in the present section. We also will discuss in the next section typical fermionic observables whose associated operators do not commute with the operators for time-local observables. For probabilistic automata the observables corresponding to non-diagonal operators typically do not have fixed values for a given bit-configuration. They rather describe properties of the probabilistic information, such as periodicity. In the present section we discuss the Hamiltonian and operators for time derivatives of observables, while further useful \qq{off-diagonal} operators will appear in subsequent sections.

\subsection*{Hamilton operator}

The Hamiltonian is known to be a very useful observable for the description of fermionic quantum systems. Having established an equivalence of probabilistic cellular automata with suitable fermionic quantum systems we may use the Hamiltonian as well for an understanding of properties of automata.

The orthogonal step evolution operator $\Shat$ can be written in an exponential form with hermitian Hamilton operator $H(t)$,
\bel{HH1}
\Shat(t)=\exp\big\{-i\eps H(t)\big\}\ ,\quad H\herm(t)=H(t)\ .
\ee
This general property is easily seen if we diagonalize $\Shat$ by a suitable unitary transformation $D$,
\bel{HH2}
\Shat'=D\Shat D^{-1}\ ,\quad D\herm D=1\ .
\ee
All eigenvalues of $\Shat$ have absolute value one, such that the diagonal form can be written as
\bel{HH3}
\Shat'=\text{diag}\gl\lambda_i\gr=\text{diag}\gl\exp(-i\eps H_i)\gr=\exp\gl-i\eps H'\gr\ ,
\ee
with real $H_i$ and
\bel{HH4}
H'=\text{diag}\gl H_i\gr\ .
\ee
The Hamilton operator $H$ obtains by transforming backwards
\bel{HH5}
H=D^{-1}H'D\ .
\ee
This proves the existence of $H$. For a concrete form for some particular step evolution operators, in particular the right-transport and left-transport operators, see ref.~\cite{CWNEW}.

Since $D$ may be a complex unitary matrix the eigenstates of $H$ are typically complex functions, given by
\bel{HH6}
H\vp_i=H_i\vp_i\ .
\ee
In the presence of a complex structure (see later sections) the real wave function $q$ is mapped to a complex wave function $\vp$ with half the number of components. The eigenstates of $H$ can then be realized directly in this complex language. They may be translated back to the real formulation if needed. For $H$ independent of $t$ the time-dependence of the eigenstates obeys
\bel{HH7}
\vp(t)=\exp\big\{-iH_i(t-t_0)\big\}\vp(t_0)=\exp\Big\{-2\pi i\frac{t-t_0}{\Delta_i}\Big\}\vp(t_0)\ ,
\ee
which agrees with the discrete evolution law for $t-t_0=m\eps$ with integer $m$. These are periodic oscillations with period $|\Delta_i|$ given by
\bel{HH8}
\Delta_i=P_i\eps=\frac{2\pi}{H_i}\ ,\quad P_i\in\mathbb{Z}\ ,
\ee
as appropriate for orthogonal step evolution operators. (See the clock automata in sect.~\ref{sec: FPGU} for a simple concrete realization.) The eigenstates of $H$ are the complete analogue to the solutions of the stationary Schrödinger equation for atoms. Knowing them explicitly, or their properties, can be an important step for the understanding of the time evolution of automata, in particular if the number of cells is very large.

If the eigenstate $\vp$ is a multicomponent vector with components $\vp\al$ the stationary solutions have time independent $|\vp\al|^2$. The periodic evolution involves only oscillations between the real and imaginary parts of $\vp$. One may want to associate $H$ with an energy observable $E$ that takes the sharp values $H_i$ for suitable probabilistic states (e.g. the periodic eigenstates). A suitable \qq{measurement} of $H_i$ could determine the period $|P_i|$ of a given periodic probability distribution or associated wave function. This determines $|H_i|=2\pi/(\eps|P_i|)$. The sign of $H_i$ is a more subtle issue to which we will turn later. We emphasize that the energy observable $H$ is not an observable that takes fixed values $H_\tau$ for a particular bit-configuration $\tau$. It may be called a \qq{statistical observable}~\cite{CWQPCS} whose values are determined by properties of the probability distribution or the wave function $\{q_\tau\}$, or more generally the whole time evolution of $\{q_\tau(t)\}$, such as periodicity.

Independently of the interpretation of energy $E$ as an observable the operator $H$ constitutes an important tool for the understanding of automata. In particular, it is a conserved quantity if $H$ does not depend on time, in the sense that the expectation value $\exval{E}$ computed according to the quantum rule
\bel{HH9}
\exval{E}=q^T(t)Hq(t)
\ee
is time-independent. The same holds for all powers of $E$ of functions $f(E)$, that are represented by operators $f(H)$. This follows from eq.~\eqref{HH1} since $H$ commutes with $\Shat$.

\subsection*{Energy spectrum}

The spectrum of eigenvalues of the Hamiltonian $H$ is called the energy spectrum. It is a characteristic feature of a given evolution law, rather than a property associated to particular bit configurations $\tau$. Up to the factor $\eps^{-1}$, which only sets units of energy or inverse time, the spectrum $\{H_i\}$ reflects directly the possible periods $P_i$ of the evolution,
\bel{HH1*}
H_i=\frac{2\pi}{P_i\eps}\ .
\ee
For long periods $P_i\gg1$ the energy eigenvalues can be much smaller than the microscopic energy scale $\eps^{-1}$. Energy differences can be even smaller,
\bel{HH2*}
\Delta E_{ij}=H_i-H_j=\frac{2\pi}{\eps}\frac{P_j-P_i}{P_iP_j}\ .
\ee

The time evolution of every finite invertible automaton is a periodic clock system~\cite{Wetterich:2020kqi}. For $MM_x$ bits and $N=2^{MM_x}$ bit-configurations the updating of a given configuration $\tau_1$ has to return to this configuration at the latest after $N$ updating steps. The maximal period is therefore given by $P\subt{max}=N$. If the system returns to $\tau_1$ after less steps, one has $P_1<N$. For the remaining $N-P_1$ configurations one can pick a particular configuration $\tau_2$. Again, the updating has to return to this configuration at the latest after $N-P_1$ updating steps, implying a second period $P_2\leq N-P_1$. Repeating this process we conclude that the automaton is a sum of clocks with periods $P_i$ obeying $\sum_iP_i=N$.

Generic periods grow proportional to $N$. In the continuum limit $M_x\to\infty$ the energy differences typically decrease exponentially
\bel{HH3*}
\eps E_{ij}\sim\exp\big\{-cM_x\big\}\ .
\ee
A generic energy spectrum becomes continuous. This does not contradict the possibility that certain subsystems, or even the whole system, can have a discrete energy spectrum. Also for some particular automata the characteristic energy differences decrease with an inverse power of $M_x$ instead of the exponential decrease,
\bel{HH4*}
\eps E_{ij}\sim M_x^{-b}\ .
\ee
Even though not realized necessarily in all cases the continuous energy spectrum is a characteristic feature for the continuum limit.

In the limit $M_x\to\infty$ we can discuss the behavior of certain \qq{equilibrium states} of the automaton from a thermodynamic point of view. It is the same as the thermodynamics of the associated fermionic quantum field theory. Equilibrium states are stationary in the sense that characteristic macroscopic expectation values and correlations do not depend on time. Such states typically involve only a few parameters, as temperature or chemical potentials. Those are related to the expectation values of the energy $\exval{E}$ and suitable conserved particle numbers.

\subsection*{Operators for time derivatives of observables}

For an observable $A(t)$ that is represented by a $t$-independent operator $\hat{A}$ we define a hermitian derivative operator 
\bel{HH10}
\partial_t\hat A=i\big[H,\hat A\big]\ .
\ee
Let us also define the discrete time-derivative observable
\bel{HH11}
\gl\partial_tA\gr_+=\frac1\eps\gl A(\te)-A(t)\gr\ .
\ee
The expectation value of this observable obeys
\begin{align}
\label{HH12}
\exval{\gl\partial_tA\gr_+(t)}=&\frac1\eps\gl q^T(\te)\hat Aq(\te)-q^T(t)\hat Aq(t)\gr\nn\\
=&\frac1\eps q^T(t)\gl\Shat^{-1}\hat A\Shat-\hat A\gr q(t)\nn\\
=&\frac1\eps q^T(t)\gl e^{i\eps H}\hat Ae^{-i\eps H}-\hat A\gr q(t)\nn\\
=&q^T(t)\gl i\big[H,\hat A\big]+\mathcal{O}(\eps)\gr q(t)\ .
\end{align}
For a sufficiently smooth wave function one can take the continuum limit and neglect the term $\mathcal{O}(\eps)$. In this continuum limit the expectation value of $\gl\partial_tA\gr_+(t)$ can be extracted from the quantum rule for the associated operator $\partial_t\hat A$.

We could define a different discrete derivative observable, as
\bel{HH13}
\gl\partial_tA\gr_s=\frac1{2\eps}\gl A(\te)-A(t-\eps)\gr\ .
\ee
In the continuum limit its expectation value coincides with the one for $\gl\partial_tA\gr_+$ and can therefore be represented by the same derivative operator $\partial_t\hat A$. Nevertheless, the two \qq{classical observables} $\gl\partial_tA\gr_+$ and $\gl\partial_tA\gr_s$ differ. These derivative observables do not take a fixed value for a given bit-configuration at $t$. Their possible measurement values involve the bit-configurations at two different times. The spectrum of possible measurement values is different for the two \qq{classical} derivative observables. If $A_\tau$ are the possible measurement values of $A(t)$, i.e. the spectrum of the diagonal operator $\hat A$, the possible measurement values of $\gl\partial_tA\gr_+$ are $(A_{\tau_1}-A_{\tau_2})/\eps$, while $\gl\partial_tA\gr_s$ has values $(A_{\tau_1}-A_{\tau_2})/(2\eps)$. This is a simple example that different classical observables can be represented by the same operator.

Unless $A$ is a conserved quantity the step evolution operator $\Shat$ does not commute with $A$, and we infer
\bel{HH14}
\big[H,\hat A\big]\neq0\ .
\ee
We may take $\hat A$ to be one of the diagonal operators for observables that can be constructed as functions of occupation numbers. Unless $\Shat$ is unity both $\Shat$ and $H$ have off-diagonal elements. In general, the derivative operator $\partial_t\hat A$ does not commute with $\hat A$ either
\bel{HH14}
\big[\partial_t\hat A,\hat A\big]=i\gl H\hat{A}^2+\hat{A}^2H-2\hat AH\hat A\gr\ .
\ee
We conclude that both $H$ and $\partial_t\hat A$ constitute simple examples that non-commuting operators play a useful role for the understanding of probabilistic cellular automata.

Under certain circumstances it may be possible to device a measurement procedure such that the only possible outcomes are the eigenvalues of the operator $\partial_t\hat A$. In this case this measurement procedure defines a genuine quantum observable. The expectation value of this quantum observable will again be given by the quantum rule for the operator $\partial_t\hat A$. The usefulness of the operator $\partial_t\hat A$ does not depend on the existence of a genuine quantum observable, however. This situation is the same as for quantum mechanics. There may be useful hermitian operators as $\partial_t\hat A$ for which a measurement procedure yielding only its eigenvalues cannot be found. As common in quantum mechanics, we sometimes call \qq{observable} a quantity whose expectation value is given by the quantum rule for the associated operator, being aware that this neither means a unique classical observable nor the existence of a genuine quantum observable in the sense of outcomes of individual measurements. Observable designates in this case rather an equivalence class of possible observables~\cite{Wetterich:2020kqi}. The common expectation value can be rather useful, as demonstrated for $\exval{\partial_t\hat A}$.

\subsection*{Conserved quantities}

Let us consider observables $A(t)$ that are represented for every $t$ by the same operator $\Ahat$. The time dependence of the expectation value $\exval{A(t)}$ arises then uniquely from the time evolution of the wave function. An observable corresponds to a conserved quantity if $\exval{A(t)}$ is independent of time for arbitrary wave functions. This is realized if $\Ahat$ commutes with the step evolution operator,
\bel{CQS1}
\big[\Ahat,\Shat(t)\big]=0\implies\exval{A(\te)}=\exval{A(t)}\ .
\ee
Similar to quantum mechanics, this relation is easily established by the quantum rule for expectation values,
\begin{align}
\label{CQS2}
\exval{A(\te)}=&q^T(\te)\Ahat q(\te)=q^T(t)\Shat^{-1}(t)\Ahat\Shat(t)q(t)\nn\\
=&q^T(t)\Ahat q(t)=\exval{A(t)}\ .
\end{align}

This property translates directly to the Hamiltonian. If an observable is associated to an operator $\Ahat$ which commutes with $H$, it is conserved,
\bel{CQA*}
\big[\Ahat,\hat H\big]=0\iff \partial_t\exval{A}=0\ .
\ee
The first direction follows from the definition~\eqref{HH1}: if $\Ahat$ commutes with $H$, it also commutes with the step evolution operator $\Shat$. The opposite direction follows from eq.~\eqref{HH12}: if $\exval{\partial_t A}\neq0$ the operator $\Ahat$ has necessarily to commute with $H$.

\subsection*{Grassmann operators}

So far we have defined operators acting on the wave function. They are identical for the automaton and the associated fermionic quantum field theory. For a Grassmann functional integral we can also define Grassmann operators $\cA$ as functions of the Grassmann variables. The expectation value of an observable $A$ can be evaluated by inserting the Grassmann operator $\cA$ into the Grassmann functional integral. Thus to a given observable one can associate both a quantum operator $\Ahat$ and a Grassmann observable $\cA$. The expectation value can be computed equivalently from the quantum rule~\eqref{P15}, or by insertion of $\cA$ into the functional integral. In appendix~\ref{app: E} we discuss this relation in detail. In particular, we establish the quantum operator $\Ahat$ that is associated to a given Grassmann operator $\cA$. Typically $\Ahat$ does not commute with the time local operators.



\section{Fermion operators}
\label{sec: FO}

We are used from fermionic quantum systems to employ operators that cannot be expressed in terms of occupation numbers. In view of the identical wave function for the probabilistic automaton and the associated fermionic quantum system it is our aim to show the usefulness of such operators for a description of cellular automata as well.

\subsection*{Switch operators}

Switch operators interchange two components of the wave function. We start with the extremely simple system of a single bit. The wave function has two components $q_1$ and $q_2$ for the occupied and the empty state respectively. The switch operator is represented by the Pauli matrix $\hat A_s=\tau_1$ and interchanges $q_1\leftrightarrow q_2$. It is a hermitian operator and thereby a potential observable. The eigenvalues are $\pm1$. We leave the question open under which circumstances a measurement procedure is possible for which a yes/no question decides between the two possible eigenvalues. This may require to embed the one-bit system into a larger system including some \qq{apparatus}, as exemplified by quantum mechanics where $\tau_1$ can be associated to the spin in a particular direction.

The eigenvectors of the two eigenvalues are only distinguished by the sign of a component of the wave function
\bel{SOO1}
A_sq_\pm=\pm q_\pm\ ,\quad q_\pm=\frac1{\sqrt{2}}\pvec{1}{\pm1}\ .
\ee
They lead to identical probabilities $p_1=p_2=1/2$. Thus the decision between the two eigenvalues cannot be based on the value of the occupation number. The expectation value
\bel{SOO2}
\exval{A_s}=q^T\tau_1q=2q_1q_2=\pm2q_1\sqrt{1-q_1^2}\ ,
\ee
contains information about the expectation value of the occupation number $\exval{n}=q_1^2$ according to
\bel{SOO3}
\exval{A_s}^2=4\exval{n}\gl1-\exval{n}\gr\ .
\ee
It is only the sign of $\exval{A_s}$ that cannot be determined by $\exval{n}$. In particular, if $A_s$ corresponds to a conserved quantity, the combinations~\eqref{SOO2},~\eqref{SOO3} are conserved. This implies that
\bel{SO4}
\gl2\exval{n}-1\gr^2=\gl1-\exval{A_s}^2\gr
\ee
is conserved, such that for each discrete step either $\exval{n}$ remains invariant or changes as $\exval{n}\to1-\exval{n}$. A conserved switch operator can be useful information for the understanding of the evolution of a probabilistic automaton.

The expectation value $\exval{A_s}$ changes if the sign of $q_1$ or $q_2$ changes. We have stated that the choice of signs of components of the wave function for a probabilistic automaton is arbitrary. One may therefore wonder if the expectation value $\exval{A_s}$ is a meaningful concept. We can perform a simultaneous change
\bel{SOO5}
\pvec{q_1}{q_2}\to\pvec{q_1}{-q_2}\ ,\quad \hat A_s\to-\hat A_s\ .
\ee
All relations remain invariant under this change. In this sense the choice of sign of $q_2$ remains indeed arbitrary. For the two different choices we have to use two different operator representations for the switch observable. This freedom of choice can also be seen as a freedom of choice of signs for the Grassmann elements $g_\tau$. The above simple setting is a particular case for a more general similarity transformation related to the choice of signs~\cite{CWQF}.

\subsection*{Annihilation and creation operators}

One often describes observables and the dynamics of fermionic quantum systems in terms of annihilation operators $a$ and creation operators $a\herm$,
\bel{AO1}
a=\pmat{0}{0}{1}{0}\ ,\quad a\herm=a^T=\pmat{0}{1}{0}{0}\ .
\ee
The switch operator for the one bit systems reads
\bel{AO2}
\hat A_s=a+a\herm\ .
\ee
For several species of fermions $\alpha$ this extends to fermionic annihilation operators $a\al$ and creation operators $a\al\herm$, which obey the usual anti-commutation relations
\bel{P47}
\{a\al\herm,a_\beta\}=\delta_{\alpha,\beta}\ ,\quad \{a\al,a_\beta\}=\{a\al\herm,a_\beta\herm\}=0\ ,
\ee
and allow us to express the particle number operators as
\bel{P48}
\Nhat\al=a\al\herm a\al\ .
\ee
For our real representation (see app.~\ref{app: A} and ref.~\cite{CWNEW} for detailed conventions) one has
\bel{P51}
a\al\herm=a\al^T\ .
\ee

For two species for fermions, $\alpha,\beta=1,2$, we may consider the hermitian operator
\bel{FO8}
\fhat=a_1\herm a_2+a_2\herm a_1\ ,\quad \fhat^2=\hat n_1\gl1-\hat n_2\gr+\hat n_2\gl(1-\hat n_1\gr\ .
\ee
This is again a type of switch operator. In the sector of one-particle states it switches between the two species. Applied to the zero- and two-particle states this switch operator yields zero. Again, the operator~\eqref{FO8} has off-diagonal elements and does not commute with the operators for the occupation numbers for the two species.
Explicit representations of the fermionic switch operators are displayed in appendix~\ref{app: F}.

\subsection*{Grassmann operators and conditional observables}

Simple Grassmann operators as $\psi\bet(\te)\psi\al(t)$ correspond often to conditional observables. Such observables have no fixed values for a given bit-configuration $\tau$. Their values typically involve properties of the step evolution operator $\Shat(t)$. For example, the observable $A_{\beta\alpha}(t)$ associated to the Grassmann operator $\psi\bet(\te)\psi\al(t)$ equals one if a particle $\alpha$ is present at $t$ and a particle $\beta$ is present at $\te$ only under the condition that the updating rule at $t$ changes a hole of type $\alpha$ to a hole of type $\beta$. Otherwise it vanishes. Expectation values $\exval{A_{\beta\alpha}(t)}$ are well defined for the cellular automaton and can be computed according to the quantum rule~\eqref{P15}. The associated quantum operator $\Ahat_{\beta\alpha}$ has, however, off-diagonal elements and does not commute with the time-local observables. We discuss this issue in detail in appendix~\ref{app: G}. In this appendix we also present the general rule which maps Grassmann operators $\cA$ to quantum operators $\Ahat$. For off-diagonal quantum operators the sign of the wave function matters. In appendix~\ref{app: G} we show that the choice of signs of the initial wave function is an arbitrary convention. Different choices of signs are a type of gauge transformation that does not affect the expectation values of observables.

\subsection*{Symmetry generators as conserved quantities}

The generators of continuous symmetries commute with the step evolution operator and therefore also with the Hamiltonian. They constitute conserved quantities. For non-abelian symmetries they do not commute among themselves. Symmetry generators are another important example for useful non-commuting operators.

As an example we may take the automata with global $SO(N)$-symmetry discussed in sect.~\ref{sec: ANA}. Infinitesimal symmetry transformations act on the Grassmann variables as ($\eta=L,R$)
\bel{GA1}
\delta\psi_{\eta a}\otx=\eps_zL_{ab}^z\psi_{\eta b}\otx\ ,\quad L_{ba}=-L_{ab}\ .
\ee
Here the matrices $L_{ab}^z$ are antisymmetric matrices in color space and $\eps_z$ are the infinitesimal transformation parameters. For $SO(N)$ there are $N(N-1)/2$ antisymmetric matrices. The generators of $SO(N)$ are linear combinations of $L^z$. The transformation acts at all positions $x$ and for all times $t$. In the language of sect.~\ref{sec: S} it has therefore the same action on $\psibar$ and $\psi$.

For a given $z$ the map $\psi\to L^z\psi$ translates to the Grassmann basis elements, defining the matrices $M^z$ similarly to eq.~\eqref{CSP11} by
\bel{GA2}
g_\tau[L^z\psi]=g_\rho\opsi M_{\rho\tau}^z\ .
\ee
The matrices $M^z$ commute with the step evolution operator, cf. eq.~\eqref{CSP9} for $D=\eps_z T^z$. They therefore also commute with the Hamiltonian. Let us assume next the existence of some matrix $\tilde I$ which commutes with $M^z$ and the step evolution operator or Hamiltonian, such that the products $T^z=\tilde I M^z$ are symmetric matrices. The \qq{expectation values},
\bel{GA3}
\exval{T^z}(t)=q_\tau(t)T\taur^zq_\rho(t)\ ,
\ee
are independent of $t$. We can associate $T^z$ to conserved quantities. The matrices $T^z$ are the generators of the symmetry group acting on the wave functions. The introduction of $\tilde I$ is necessary if the matrices $M^z$ are not symmetric. It typically reflects a multiplication with $i$ in the complex picture.

It is an interesting question if a measurement prescription can be devised which has as possible outcomes the discrete eigenvalues of the non-abelian charges $T^z$. In this case these charges are quantum observables in a strict sense. Independently of this issue the presence of conserved quantities~\eqref{GA3} is a rather useful tool for the understanding of the dynamics of the probabilistic automata. The matrices $M^z$ obey the same commutation relations as $L^z$,
\bel{GA4}
\big[L^x,L^y\big]=\tilde f^{xy}_{\ \ \ z}L^z\implies \big[M^x,M^y\big]=\tilde f^{xy}_{\ \ \ z}M^z\ ,
\ee
with $\tilde f^{xy}_{\ \ \ z}$ related to the structure constants of the Lie group. For a non-abelian symmetry group the non-abelian charges do not commute
\bel{GA5}
\big[T^x,T^y\big]=\tilde I\tilde f^{xy}_{\ \ \ z}T^z\ .
\ee
As familiar in fermionic quantum field theories one can express $T^z$ in terms of annihilation and creation operators.

\section{Vacua, order parameters and\\spontaneous symmetry breaking}
\label{sec: VOS}

Having translated the formal concepts of quantum mechanics and quantum field theory to the probabilistic cellular automata we will next see how key features of quantum field theory are realized for probabilistic automata. Our aim is to describe these automata in terms of propagating particles with scattering. In quantum field theory the properties of propagating particles depend on the vacuum state. We therefore discuss in the present section the notion of \qq{vacuum states} or \qq{ground states}.

A possible vacuum or ground state of a probabilistic cellular automaton is characterized by a time-translation invariant wave function. Depending on the precise setting this requires in a discrete formulation $q(t+2\eps)=q(t)$, or $q(t+4\eps)=q(t)$, or similar for more complex sequences of updatings. In case of chain automata we also require translation invariance in the position $x$, e.g. $q(t,x+4\eps)=q(t,x)$. These requirements are not strong enough to select a unique ground state. For given automata one may impose additional conditions as half-filling and particle-hole symmetry. More generally, one may impose stability requirements on possible ground states, but this will not be addressed systematically in the present paper.

In quantum field theory or many body quantum mechanics the notion of the ground state is usually also associated with the stability provided by the minimum of the Hamiltonian. Even with this condition the ground state may not be unique. A wave function of the ground state that has less symmetries than the symmetries of the Hamiltonian or the action indicates spontaneous symmetry breaking. The selection of the true state of a system may then depend on particular initial conditions or small disturbances (\qq{sources}). A ground state is often associated with the zero-temperature limit of thermal equilibrium. In this paper we will not address the question if and how thermodynamic equilibrium states are reached effectively by the dynamics of probabilistic automata. We rather discuss \qq{candidate ground states} which obey the restrictions of stationarity and homogeneity.

The discussion of the vacuum or ground state has to be carried out for each particular automaton separately. Since this key issue for the associated quantum field theory is typically a complex issue we only highlight here a few conceptual issues for one particular example, namely the spinor gravity automaton.

\subsection*{Half-filled vacua for the spinor gravity automaton}

As our example we investigate for the spinor gravity automaton given by the updating rules shown in Figs.~\ref{fig: C1} and~\ref{fig: S1} the possible half-filled vacua. At every position $x$ one has eight bits or occupation numbers $n_\pm^a(x)$. We require that on each site of the $(t,x)$-lattice four particles, and correspondingly also four holes, are present. We also require Lorentz invariance of the vacuum, which imposes two particles of the type $+$ and two particles of the type $-$. Thus the occupation number for the half-filled vacuum states obey
\bel{V1}
\sum_an_+^a(t,x)=\sum_an_-^a(t,x)=2\ .
\ee
Since left-movers and right-movers live on different sublattices we can discuss them separately.

Starting with the right-movers on the odd sublattice we recall that the updating rule always propagates two pairs of particles to the right, as displayed in Fig.~\ref{fig: S2}. Space-translation invariance by $4\eps$ is guaranteed if we repeat the structure of Fig.~\ref{fig: S2} in $x$. At this stage we have only a quarter filling, since $\sum_an_+^a(t,x)=\sum_an_-^a(t,x)=1$. The half-filled vacua are obtained if we add a second set of two pairs of particles. There are four possible sets of two pairs placed at $(t,x-\eps)$ and $(t,x+\eps)$ that are consistent with the updating rule,
\begin{align}
(A):&\ (12),\ (34)\ ,\quad (B):\ (34),\ (12)\ ,\nn\\
(C):&\ (21),\ (43)\ ,\quad (D):\ (43),\ (21)\ .
\end{align}

\begin{figure}

\includegraphics{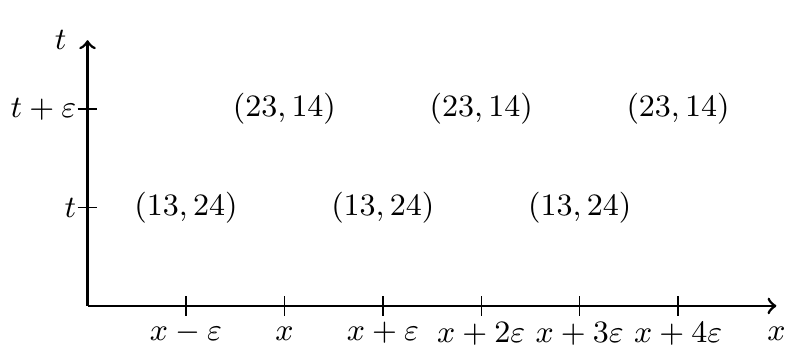}

\caption{Half-filled vacuum for right-movers of type $(AB)$.}
\label{fig: V1}

\end{figure}

For the combination of two of these sets there are six possibilities, corresponding to six possible vacua of this type for the right-moving sector. The two sets have to be different, such that the possible combinations of two sets are $(AB)$, $(AC)$, $(AD)$, $(BC)$, $(BD)$, $(CD)$. The combinations $(AB)$ and $(CD)$ are invariant under $t\to t+2\eps$ and $x\to x+2\eps$. The combination $(AB)$ yields the structure of the vacuum shown in Fig.~\ref{fig: V1}, where $(ab,cd)$ denotes $n_+^a=n_+^b=n_-^c=n_-^d=1$.
\begin{figure}

\includegraphics{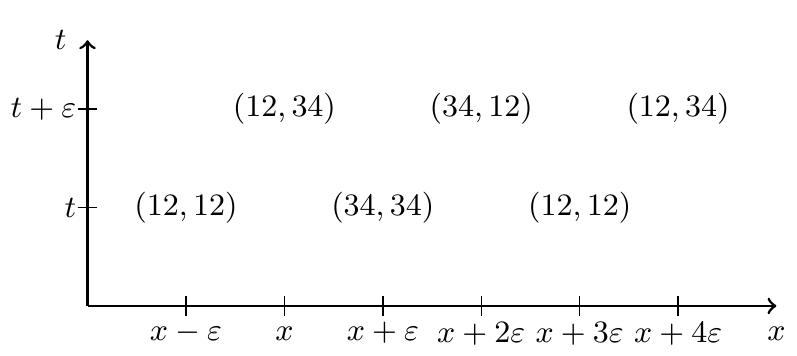}

\caption{Half-filled vacuum for right-movers of type $(AC)$.}
\label{fig: V2}

\end{figure}
\begin{figure}

\includegraphics{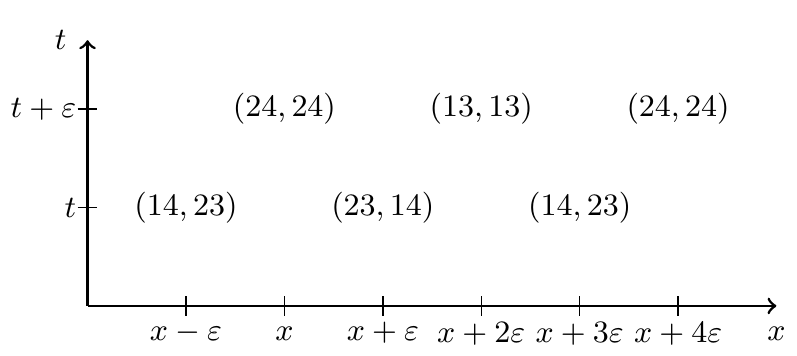}

\caption{Half-filled vacuum for right-movers of type $(AD)$.}
\label{fig: V3}

\end{figure}
For this vacuum the combinations $(12,34)$ at even $t$ and odd $x$ alternate with the combinations $(23,14)$ at odd $t$ and even $x$. The particle-hole transformation of the vacuum $(AB)$ replaces $(13,24)\to(24,13)$, $(23,14)\to(14,23)$. This corresponds to the combination $(CD)$. The combination $(AC)$ is invariant under the shift $t\to t+2\eps$, $x\to x+4\eps$. It is shown in fig.~\ref{fig: V2}, while the similar combination $(AD)$ is displayed in Fig.~\ref{fig: V3}. The particle-hole transform of $(AC)$ is $(BD)$, while $(AD)$ changes to $(BC)$ by an exchange of particles and holes. For these vacua the particle-hole transformation is equivalent to a shift in $x$ by $2\eps$.

Particle-hole symmetric vacua can be obtained by linear superpositions of wave functions. We denote by $q_{(AB)}$ the wave vector that equals unity precisely for the half-filled combination $(AB)$, and similar for the other combinations. This leaves us with three particle-hole symmetric vacua in the sector of right-movers
\bel{V3}
\begin{split}
q_R^{(1)}=&\frac{1}{\sqrt2}\gl q_{(AB)}+q_{(CD)}\gr\ ,\quad q_R^{(2)}=\frac{1}{\sqrt2}\gl q_{(AC)}+q_{(BD)}\gr\ ,\\
q_R^{(3)}=&\frac{1}{\sqrt2}\gl q_{(AD)}+q_{(BC)}\gr\ .
\end{split}
\ee
The treatment of the sector of left-movers can be done in complete analogy, resulting in particle-hole symmetric wave functions $q_L^{(i)}$. There are therefore nine different overall half-filled, Lorentz-invariant, and particle-hole symmetric vacuum states, given by the products
\bel{V4}
q^{(ij)}=q_R^{(i)}q_L^{(j)}\ .
\ee

At every site $(t,x)$ the probability to find a particle of a given type equals the probability to find a hole of this type. The different vacua are distinguished by different correlations to find different species at neighboring sites.

\subsection*{Order parameters}

We can characterize the different half-filled vacua by order parameters or vacuum expectation values. Among the possible order parameters are the expectation values of particle numbers of different colors. We restrict the discussion here to the right-movers, with obvious extension to the left-movers. For the vacuum $(AB)$ one has
\bel{V5}
\begin{split}
\exval{n_+^1}_1=&\exval{n_+^3}_1=\exval{n_-^2}_1=\exval{n_-^4}_1=1\ ,\\
\exval{n_+^2}_1=&\exval{n_+^4}_1=\exval{n_-^1}_1=\exval{n_-^3}_1=0\ ,
\end{split}
\ee
and
\bel{V6}
\begin{split}
\exval{n_+^2}_2=&\exval{n_+^3}_2=\exval{n_-^1}_2=\exval{n_-^4}_2=1\ ,\\
\exval{n_+^1}_2=&\exval{n_+^4}_2=\exval{n_-^2}_2=\exval{n_-^3}_2=0\ ,
\end{split}
\ee
where we use the shorthands (for $t$ and $x$ even and $n_t$, $n_x$ integer)
\bel{V7}
\begin{split}
\exval{n_\pm^a}_1=&\exval{n_\pm^a\gl t+2n_t\eps,x+(2n_x-1)\eps\gr}\ ,\\
\exval{n_\pm^a}_2=&\exval{n_\pm^a\gl t+(2n_t+1)\eps,x+2n_x\eps\gr}\ .
\end{split}
\ee
The particle-hole symmetric vacuum with wave function $q_R^{(1)}$ in eq.~\eqref{V3} has for all sites of the odd sublattice
\bel{V8}
\exval{n_+^a}=\exval{n_-^a}=\frac12\ .
\ee

For the half-filled vacuum $(AC)$ the mean particle numbers differ from the vacuum $(AB)$. Nevertheless, for the particle-hole symmetric combination $q_R^{(2)}$ in eq.~\eqref{V3} the equipartition~\eqref{V8} holds again. The different particle-hole symmetric vacua are not distinguished by the mean particle numbers for different species of particles. We can differentiate between them, however, by different correlation functions. We display a few correlation functions for diagonal neighbors at $(t,x-\eps)$ and $(t+\eps,x)$ in table~\ref{tab: X}.
\begin{table}

\centering
\renewcommand{\arraystretch}{1.5}

\begin{tabular}{C{0.08\textwidth}|*{3}{C{0.05\textwidth}}|*{3}{C{0.05\textwidth}}}

 & $AB$ & $AC$ & $AD$ & $BC$ & $BD$ & $CD$ \\
\noalign{\hrule height 1pt}
$\exval{n_+^{1\prime} n_+^1}$ & $0$ & $1$ & $0$ & $0$ & $0$ & $0$\\
$\exval{n_+^{2\prime} n_+^2}$ & $0$ & $1$ & $0$ & $0$ & $0$ & $0$\\
$\exval{n_+^{3\prime} n_+^3}$ & $1$ & $0$ & $0$ & $1$ & $1$ & $0$\\
$\exval{n_+^{4\prime} n_+^4}$ & $0$ & $0$ & $1$ & $0$ & $1$ & $1$\\

\hline

$\exval{n_-^{1\prime} n_-^1}$ & $0$ & $0$ & $0$ & $1$ & $0$ & $0$\\
$\exval{n_-^{2\prime} n_-^2}$ & $0$ & $0$ & $1$ & $0$ & $0$ & $0$\\
$\exval{n_-^{3\prime} n_-^3}$ & $0$ & $0$ & $0$ & $0$ & $0$ & $1$\\
$\exval{n_-^{4\prime} n_-^4}$ & $1$ & $0$ & $0$ & $0$ & $0$ & $0$\\

\hline

\end{tabular}

\caption{Correlation functions for diagonal neighbors at $(t+\eps,x)$ and $(t,x-\eps)$ for different vacua. Here $n_\pm^a$ stands for $n_\pm^a(t,x-\eps)$ and $n_\pm^{a\prime}$ stands for $n_\pm^a(t+\eps,x)$. These correlations are evaluated for even $t$ and even $x/\eps=0\ \text{mod}~4$. For $x/\eps=2\ \text{mod}~4$ the correlation functions change for the vacua $AB$, $AD$, $CD$, $BC$. Translation of $x$ by $2\eps$ switches $AB\leftrightarrow CD$, $AD\leftrightarrow BC$.}
\label{tab: X}

\end{table}
For the particle-hole symmetric vacuum with wave function $q_R^{(2)}$ this implies for all $a$
\bel{V9}
\begin{split}
q_R^{(2)}:\quad &\exval{n_+^a(t+\eps,x)n_+^a(t,x-\eps)}=\frac12\ ,\\
&\exval{n_-^a(t+\eps,x)n_-^a(t,x-\eps)}=0\ .
\end{split}
\ee
In contrast, for $q_R^{(1)}$ one finds
\bel{V10}
\begin{split}
q_R^{(1)}:\quad &\exval{n_+^{3,4}(t+\eps,x)n_+^{3,4}(t,x-\eps)}\\
=&\exval{n_-^{3,4}(t+\eps,x)n_-^{3,4}(t,x-\eps)}=\frac12\ ,\\
&\exval{n_+^{1,2}(t+\eps,x)n_+^{1,2}(t,x-\eps)}\\
=&\exval{n_-^{1,2}(t+\eps,x)n_-^{1,2}(t,x-\eps)}=0\ .
\end{split}
\ee
The correlation functions for these two vacua are different. We can use these correlations as \qq{order parameters} which distinguish the particle-hole symmetric vacua $q_R^{(1)}$ and $q_R^{(2)}$. We observe that the connected correlation functions differ from zero, e.g. for $q_R^{(1)}$
\bel{V11}
\exval{n_+^a(t+\eps,x)n_+^a(t,x-\eps)}-\exval{n_-^a(t+\eps,x)n_-^a(t,x-\eps)}=\frac14\ .
\ee

\subsection*{Half-filled random vacuum}

For the half-filled random vacuum the wave function is a product of a wave function $q_R$ for right-movers and a similar wave function $q_L$ for left-movers. We focus on $q_R$, which is a product of wave functions at positions $x$,
\bel{V12}
q_R(t)=\prod_xq_R(t,x)\ .
\ee
The product is over odd $x$ for $t$ even and over even $x$ for $t$ odd. For each position $x$ the local wave function $q_R(t,x)$ is a linear superposition of six wave functions, which correspond each to four particles and four holes.
\bel{V13}
q_R(t,x)=\frac{1}{\sqrt6}\sum_{A=1}^{6}q_R^{(A)}(t,x)\ .
\ee
We list the particles present for $q_R^{(A)}$ in table~\ref{tab: Y}, in the notation $(ab,cd)$ denoting $n_+^a=n_+^b=n_-^c=n_-^d=1$, with all other occupation numbers zero.
\begin{table}

\centering
\renewcommand{\arraystretch}{1.2}

\begin{tabular}{C{0.07\textwidth}|*{6}{C{0.06\textwidth}}}

$A$ & $1$ & $2$ & $3$ & $4$ & $5$ & $6$ \\
\noalign{\hrule height 1pt}
$t$ & $(12,12)$ & $(34,34)$ & $(13,24)$ & $(24,13)$ & $(14,23)$ & $(23,14)$\\
$t+\eps$ & $(12,34)$ & $(34,12)$ & $(23,14)$ & $(14,23)$ & $(24,24)$ & $(13,13)$\\
$t+2\eps$ & $(12,12)$ & $(34,34)$ & $(24,13)$ & $(13,24)$ & $(23,14)$ & $(14,23)$\\

\hline

\end{tabular}

\caption{Non-zero particle numbers for the local wave functions $q_R^{(A)}(t,x)$ for even $t$.}
\label{tab: Y}

\end{table}
The wave function is stationary in the sense $q_R(t+2\eps)=q_R(t)$. The evolution of $q_R^{(A)}$ can be followed from table~\ref{tab: Y} where $x$ is displaced by one unit at every time step.

For the half-filled random vacuum the particles are randomly distributed
\bel{V14}
\exval{n_+^a(t,x)}=\exval{n_-^a(t,x)}=\frac12\ .
\ee
The diagonal correlation functions obtain for even $t$ and $x$ as
\begin{align}
\label{V15}
&\exval{n_+^1(t+\eps,x)n_+^1(t,x-\eps)}=\exval{n_+^2(t+\eps,x)n_+^2(t,x-\eps)}\nn\\
=&\exval{n_-^1(t+\eps,x)n_-^1(t,x-\eps)}=\exval{n_-^2(t+\eps,x)n_-^2(t,x-\eps)}\nn\\
=&\exval{n_-^3(t+\eps,x)n_-^3(t,x-\eps)}=\exval{n_-^4(t+\eps,x)n_-^4(t,x-\eps)}\nn\\
=&\frac16\ ,
\end{align}
and
\bel{V16}
\exval{n_+^3(t+\eps,x)n_+^3(t,x-\eps)}=\exval{n_+^4(t+\eps,x)n_+^4(t,x-\eps)}=\frac12\ .
\ee
For the correlations $\exval{n_\pm^a(t+2\eps,x+\eps)n_\pm^a(t+\eps,x)}$ one finds a result similar to eqs.~\eqref{V15},~\eqref{V16} with colors $(1,2)$ exchanged with $(3,4)$. Finally, if one exchanges in these correlation functions $x+\eps$ by $x-\eps$ this exchanges the occupation numbers $n_+^a\leftrightarrow n_-^a$.
Averaging over four neighboring pairs yields
\bel{V17}
\begin{split}
\frac14\Big\langle& n_\pm^a(t+\eps,x)n_\pm^a(t,x-\eps)\\
+&n_\pm^a(t+2\eps,x+\eps)n_\pm^a(t+\eps,x)\\
+&n_\pm^a(t+\eps,x)n_\pm^a(t,x+\eps)\\
+&n_\pm^a(t+2\eps,x-\eps)n_\pm^a(t+\eps,x)\Big\rangle=\frac14\ .
\end{split}
\ee
The connected correlation function for these averages vanishes.

\subsection*{Spontaneous symmetry breaking}

Averaging over all $x$ and two neighboring $t$-layers the half-filled random vacuum obeys the relations~\eqref{V14} with vanishing diagonal connected two-point functions. It obeys the symmetries of color-exchange, exchange between $n_+^a\leftrightarrow n_-^a$, and particle-hole transformations. The corresponding symmetries of the action are preserved, and we may associate this wave function with the \qq{symmetric vacuum}.

One the other hand, the wave function $q_{(AB)}$ breaks some of these symmetries, for example particle-hole symmetry and some of the color-exchanges, cf. eqs.~\eqref{V5},~\eqref{V6}. Also the particle-hole symmetric vacuum $q_R^{(1)}$ breaks part of these symmetries for the correlation functions. We can associate these half-filled vacuum states with spontaneous symmetry breaking: the symmetry of the state is lower than the symmetry of the action.

We conclude from this simple example that characteristic features of quantum field theories as different vacua, order parameters and spontaneous symmetry breaking are visible for probabilistic automata. For the discrete spinor gravity automaton the vacuum solutions discussed here are all exact solutions for appropriate initial conditions. The vacuum structure in the continuum limit for some coarse grained formulation remains to be investigated. The discussion of particle-hole symmetric vacua highlights the importance of the probabilistic aspects for our setting.

The present section should be considered as only opening the discussion on the important aspects of order parameters and spontaneous symmetry breaking. For example, we have not yet addressed if there are vacua of the spinor gravity model for which the zweibein bilinear $\Etil_\mu^{m,a}$ introduced in sect.~\ref{sec: DSG} acquires an expectation value. This would permit the discussion of geometry for spinor gravity. As we will briefly discuss in sect.~\ref{sec: QMPP} such an expectation value gives rise to interesting simple fermion excitations obeying the Dirac equation.

\section{Single-particle wave functions}
\label{sec: SPWF}

Particles are local excitations of a given vacuum state. This means that the wave function differs in a region around the particle position $x_p$ from the vacuum wave function. For a single particle the wave function equals the vacuum wave function for all $x$ far enough away from the particle position. This basic concept of a particle makes it already clear that particle properties depend on the vacuum, as familiar in quantum field theory.

Sharp single-particle excitations of a given vacuum can be characterized by the sharp position $x_p$ of the particle. In a probabilistic setting this is replaced by a wave function $q(x)$ which has support in a region around $x_p$. It equals the vacuum wave function $q_0$ for $x$ far away from the particle's position. The one-particle wave function $q\g(x)$ may also carry labels for color etc., that can distinguish between different species of particles. The one-particle excitations can be of various types. For the Thirring-automaton we can add a right-moving or left-moving particle. They are described in ref.~\cite{CWNEW}. For the spinor gravity automaton, the four-particle state depicted in Fig.~\ref{fig: S2} can be taken as a composite single particle, with $x$ the position of the pair at the right. For the half-filled vacua in spinor gravity we can add or subtract right-moving or left-moving double pairs to or from the vacuum. We can also replace for all components of the vacuum wave function at $t\inn$ the two double-pairs at given neighboring positions by other fixed double-pairs. This default will propagate to the right or to the left. Single particles cannot interact. Without the presence of vacuum expectation values or, perhaps equivalently, effective disorder their evolution is given by the right-transport or left-transport operators.

For deterministic cellular automata the single-particle state is a given bit configuration that can be characterized for every $t$ by a sharp position $x(t)$. This defines a trajectory. For free non-interacting particles in our $1+1$-dimensional cases (without disorder) this is a straight line either to the right or to the left. For probabilistic cellular automata we specify the wave function $q(t,x)$ whose square defines the probability to find the particle at $t$ at the position $x$. Sharp deterministic wave functions involve a $\delta$-function $q(t,x)\sim\delta\gl x-\bar{x}(t)\gr$, with $\bar{x}(t)$ the trajectory of the particle. For probabilistic cellular automata the wave function is typically a smooth function centered around $\xbar(t)$. For our $1+1$-dimensional cellular automata there is no dispersion of the wave function in the absence of disorder or other forms of modified propagation.

For free massless particles the propagation of single particle states seems rather trivial. Nevertheless, important concepts of quantum field theory as a complex structure or the momentum representation and momentum operator find a direct application and simple interpretation for cellular automata. In the next section the one-particle states will also be a good example how the cellular automaton property of unique jump step evolution operators is lost once one proceeds to coarse graining.

We start without disorder and discuss the effects of disorder in the next section. For many forms of a modified propagation, in particular for disorder, a dispersion of the wave function is expected unless one starts initially with the sharp wave function of a deterministic automaton.

\subsection*{Discrete setting for single-particle system}

Single-particle quantum mechanics arises in a quantum field theory by restricting the wave function to one-particle states. The same holds for the probabilistic automaton. The momentum operator and the Fourier transform from a position representation to a momentum representation are standard tools in quantum mechanics. We want to discuss the emergence of these concepts in the context of probabilistic cellular automata. In order to be specific we work with a discrete setting, as appropriate for the basic formulation of the automaton. In consequence, we deal with a discrete version of momentum and discrete Fourier transforms, as familiar in solid state physics. The continuum limit can be taken at the end.

There are different versions to formulate the discrete setting. For the sake of concreteness and for later use we choose here one for which right-movers and left-movers are placed on different sublattices. This is not mandatory. A placement of all species on the same lattice is valid as well and used in other places of this paper. Consider a lattice with sites $(t,x)$ and lattice distances $\eps$, e.g. $x=m_x\eps$, $t=m_t\eps$. We assume periodicity in $x$, $m_x=0\dots2M_x-1$, with $(2M_x+m_x)\eps$ and $m_x\eps$ identified and integer $M_x$ equal to half the number of $x$-positions.
We place the right-movers on the odd lattice sites, $m_x+m_t$ odd, and the left-movers on the even lattice sites, $m_x+m_t$ even. For $t$ even the positions of the right-movers are specified by integers $m_R$, $x=(2m_R-1)\eps)$. The wave function $q_R(x)=q_R(m_R)$ has $M_x$ components. Similarly, for the left-movers the wave function $q_L(x)=q_L(m_L)$ with $x=2m_L\eps$ has also $M_x$ components, while a combined one-particle wave function $q(x)=\gl q_R(x),q_L(x)\gr$ has $2M_x$ components.

We will consider evolution steps in time from $t$ to $t+2\eps$, such that $t$ is always even. This has the advantage that the possible $x$-positions of the right-movers and left-movers are the same at every evolution step. We denote the combined step evolution operator by $S(t)=\Shat(\te)\Shat(t)$. The right-transport operator $S_R$ moves the particles to the right
\bel{SP1}
q_R(t+2\eps,x)=\sum_yS_R(x,y)q_R(t,y)=q_R(t,x-2\eps)\ ,
\ee
or
\bel{SP2}
q_R(t+2\eps,m_R)=\sum_{n_R}S_R(m_R,n_R)q_R(t,n_R)=q_R(t,m_R-1)\ .
\ee
The right-transport and left-transport operators
\bel{SP3}
\begin{split}
\gl S_R\gr_{mn}=&\delta_{m,n+1}=\delta_{m-1,n}\ ,\\
\gl S_L\gr_{mn}=&\delta_{m,n-1}=\delta_{m+1,n}\ ,
\end{split}
\ee
have non-zero elements neighboring the diagonal. (We omit indices $m_R$, $m_L$ etc. if the meaning is clear.)

\subsection*{Particle-hole transformation and complex\\structure}

Let us assume that the automaton has particle-hole symmetry and consider the single-particle states as excitations of a particle-hole symmetric vacuum. We can then introduce hole wave functions $q_R^c(t,x)$, $q_L^c(t,x)$ that follow the same evolution law~\eqref{SP1}-\eqref{SP3} as the particle wave functions $q_R(t,x)$, $q_L(t,x)$.
The particle-hole transformation maps $q_R^c\leftrightarrow q_R$, $q_L^c\leftrightarrow q_L$. We can employ this involution to define a complex structure. For the map $(q_R,q_R^c)\to\varphi_R$,
\bel{SP4}
\varphi_R\otx=\frac{1+i}{\sqrt2}q_R\otx+\frac{1-i}{\sqrt2}q_R^c\otx\ ,
\ee
and similar for $\varphi_L\otx$, the particle-hole transformation translates to complex conjugation for the complex wave function $\varphi\to\varphi^*$. The map $q\to q'=q^c$, $q^c\to {q^c}'=-q$ realizes in the complex formulation the multiplication with $i$, $\varphi\to i\varphi$. For more details on the complex structure based on the particle-hole transformation and generalization to multi-particle states see ref.~\cite{CWNEW}. In the following we will work with the complex one-particle wave functions $\varphi_R$ and $\varphi_L$. They describe a generalized one-particle concept comprising single (possibly composite) particles and holes. In sect.~\ref{sec: CSQM} we discuss a somewhat modified complex structure in more detail. The discussion in the present section takes over to other choices of a complex structure.

\subsection*{Momentum representation}

The right transport operator $S_R$ in eq.~\eqref{SP3} can be diagonalized by a discrete Fourier transformation,
\bel{SP5}
\begin{split}
\vp_R\otx\to\vp_R'(t,p)=&\sum_xD_R(p,x)\vp_R\otx\ ,\\
\vp_R(t,m)\to\vp_R'(t,q)=&\sum_mD_R(q,m)\vp_R(t,m)\ .\\
\end{split}
\ee
The $M_x\times M_x$ matrix $D$, 
\bel{SP6}
\begin{split}
D_R(p,x)=&\frac{1}{\sqrt{M_x}}\exp(-ipx)=\frac{1}{\sqrt{M_x}}\exp\big\{-i\eps p(2m-1)\big\}\ ,\\
D_R(q,m)=&\frac{1}{\sqrt{M_x}}\exp\big\{-\frac{i\pi}{M_x}q(2m-1)\big\}\ ,\\
\end{split}
\ee
is unitary, $D\herm D=1$. The values of the momentum $p$ are discrete, given by integers $q$
\bel{SP7}
p=\frac{\pi q}{\eps M_x}\ ,\quad |p|\leq\frac{\pi}{2\eps}\ ,\quad |q|\leq\frac{M_x}{2}\ ,
\ee
with periodic $q$ and $p$ identifying $M_x+q$ with $q$. The inverse of this similarity transformation is the inverse Fourier transform
\bel{SP8}
\vp_R\otx=\sum_p D_R^{-1}(x,p)\vp_R'(t,p)\ ,
\ee
with 
\bel{SP9}
D_R^{-1}(x,p)=D_R\herm(x,p)=D_R^*(p,x)=\frac{1}{\sqrt{M_x}}\exp(ipx)\ .
\ee

With respect to this similarity transformation the step evolution operator ~\eqref{SP3} transforms as
\bel{SP10}
S_R\to S_R'=D_RS_RD_R\herm\ ,
\ee
with
\bel{SP11}
\begin{split}
S_R'(q,q')=&\sum_{m,n}D_R(q,m)\gl S_R\gr_{mn}D_R^*(q',n)\\
=&\exp\gl-\frac{2\pi i}{M_x}q\gr\delta_{q,q'}\ .
\end{split}
\ee
(Here $\delta_{q,q'}$ is taken modulo $M_x$.) This is indeed a diagonal matrix, with absolute value of all eigenvalues equal to one.

The Fourier transform in the left-moving sector is analogous, with
\bel{SP12}
D_L(q,m)=\frac{1}{\sqrt{M_x}}\exp\big\{-\frac{2\pi iqm}{M_x}\big\}\ .
\ee
This yields the diagonal form of the left-transport operator
\bel{SP13}
S_L'(q,q')=D_L(q,m)D_L^*(q',n)\gl S_L\gr_{mn}=\exp\gl\frac{2\pi i}{M_x}q\gr\delta_{q,q'}\ .
\ee
The different sign of the angle in the phase factor as compared to $S_R'$ reflects the opposite direction of transport.

As familiar from quantum mechanics we can take the Fourier components as a new complex basis for the wave function, with
\bel{288}
\vp_R(t,m)=\sum_qD_R^{-1}(m,q)\vp'_R(t,q)\ .
\ee
The stepwise time evolution of the Fourier components is very simple, given by a phase rotation according to eq.~\eqref{SP11}
\bel{292A}
\vp_R'(t+2\eps,q)=\exp\left(-\frac{2\pi i}{M_x}q\right)\vp_R'(t,q)\ .
\ee
This demonstrates the applicability and usefulness of basis transformations for probabilistic automata. The probabilistic nature of the description is crucial for this possibility since sharp position states are not transformed to sharp momentum states, and vice versa. Basis transformations can only be performed in the formulation with wave functions. They do not exist on the level of the probability distribution.

\subsection*{Momentum operator}

In the discrete setting there are different ways to define the momentum operator. They will coincide in the limit of small momentum. A first definition uses the momentum basis in which $\tilde{P}'$ is defined as the diagonal operator
\bel{M1}
\tilde{P}'(q,q')=\frac{\pi q}{\varepsilon M_{x}}\delta_{q,q'}\ ,
\ee
with integer $q$ obeying the bound $|q|\leq M_{x}	/2$. (For the boundary of the interval we take $q=M_x/2$ in eq.~\eqref{M1}.)
In the position basis this operator reads for the left-movers
\begin{align}
\label{M2}
\tilde{P}(m,n) &= D_{L}\herm(m,q)\tilde{P}'(q,q') D_{L}(q',n) \nn\\
&={\sum_{q}}\frac{\pi q}{\varepsilon M_{x}}D_{L}^{*}(q,m)D_{L}(q,n)\nn\\
&={\sum_{q}}\frac{\pi q}{\varepsilon M_{x}^{2}} \exp\Big{\lbrace}\frac{2\pi i(m-n)q}{M_{x}}\Big{\rbrace}\ .
\end{align}
The expression for the right-movers is identical. Combining the contributions of $q$ and $-q$ one finds
\begin{align}\label{M3}
\tilde{P}(m,n)=&\frac{2\pi i}{\varepsilon M_{x}^{2}} \sum_{q=1}^{M_{\!x}/2\,-1}\bigg{\lbrace} q \sin \Big{(}\frac{2\pi (m-n) q}{M_{x}}\Big{)}\bigg{\rbrace}\nn\\
&+\frac\pi{2\eps M_x}(-1)^{m-n} \ .
\end{align}
The first part of this operator is antisymmetric and purely imaginary, and therefore hermitian as it should be. The second part is real and symmetric and therefore also hermitian. It vanishes in the continuum limit due to the opposite sign of two neighboring $n$ or $m$.

For large values of $\abs{m-n}$ the sine in the imaginary part of $\tilde P$ varies rapidly with $q$. The resulting contribution to $\tilde P$ is small due to the cancellation of contributions with opposite signs in the summation over $q$. Thus $\tilde P(m,n)$ is dominated by small $\abs{m-n}$. In particular, for $m=n\pm1$ one has
\bel{295X}
\tilde P(n\pm1,n)=\pm\frac{2\pi i}{\eps M_x^2}\sum_{q=1}^{M_x/2-1}q\sin\left(\frac{2\pi q}{M_x}\right)-\frac\pi{2\eps M_x}\ .
\ee
This yields for the difference
\bel{295Y}
\tilde P(n-1,n)-\tilde P(n+1,n)=-\frac{4\pi i}{\eps M_x^2}\sum_{q=1}^{M_x/2-1}q\sin\left(\frac{2\pi q}{M_x}\right)\ ,
\ee
which is a quantity of the order $-i/\eps$. In comparison, the diagonal part is suppressed by a factor $M_x^{-1}$,
\bel{295W}
\tilde P(n,n)=\frac{\pi}{2\eps M_x}\ .
\ee
For large $M_x$ we can identify $\tilde P$ with some type of a smoothened lattice derivative $\tilde\partial_x$,
\bel{295V}
\tilde P=\tilde\partial_x\ .
\ee

In the momentum basis the step evolution operator has a simple expression in terms of $\Ptil'$,
\bel{295A}
\vp'=\pvec{\vp_R'}{\vp_L'}\ ,\quad \Shat'=\pmat{S_R'}{0}{0}{S_L'}\ ,
\ee
with
\bel{295B}
\Shat'=\exp\gl-2i\eps H_f'\gr\ ,\quad H_f'=\Ptil'\tau_3\ .
\ee
The operator $H_f'$ is the Hamilton operator for a free single particle. Since the transition between the position and momentum basis is a similarity transformation the exponential form also holds in position space
\bel{295C}
\Shat=\exp\gl-2i\eps H_f\gr\ ,\quad H_f=\Ptil\tau_3\ .
\ee
While the Hamilton operator has a very simple expression in momentum space, its form in position space~\eqref{M3} is more complex.

An alternative definition of the momentum operator with a simpler expression in position space can be defined as
\bel{M4}
\widehat{P}'(q,q')=\frac{1}{2\varepsilon}\sin\Big{(}\frac{2\pi q}{M_{x}}\Big{)}\delta_{q,q'}\ .
\ee
For $|q|\ll M_{x} $ the definitions $\tilde{P}'$ and $\widehat{P}' $ coincide approximatively. In position space one finds the Fourier transform of $\widehat{P}'$ as
\bel{M5}
\widehat{P}(m,n)=\frac{1}{2\varepsilon M_{x}}\sum_{q}\bigg{\lbrace} \sin\Big{(}\frac{2\pi q}{M_{x}}\Big{)}\exp\Big{(}\frac{2\pi i (m-n) q}{M_{x}}\Big{)}\bigg{\rbrace}\ .
\ee
This momentum operator can be expressed in terms of the lattice derivative operator $\wpx$,
\bel{M6}
\widehat{P}(m,n)=-i\wpx(m,n)\; ,\quad \wpx(m,n)=\frac{1}{4\varepsilon}(\delta_{m,n-1}-\delta_{m,n+1})\ .
\ee
Indeed, one has (for left-movers and $x=2m\varepsilon$)
\begin{align}
\label{M7}
(\wpx \varphi)(x)=(\wpx \varphi)&(m)=\frac{1}{4\varepsilon}\Big{[}\varphi(m+1)-\varphi(m-1)\Big{]}\nn\\
&=\frac{1}{4\varepsilon}\Big{[}\varphi(x+2\varepsilon)-\varphi(x-2\varepsilon)\Big{]}\ .
\end{align}
We recall that the factor $i$ is replaced in the real formulation by the antisymmetric matrix $I$, such that $I\wpx$ is a symmetric matrix in the real formulation.

Neither $\tilde{P}$ nor $\widehat{P}$ commutes with the position operator $\widehat{X}$,
\bel{M8}
\widehat{X}(m,n)=2\varepsilon m\,\delta_{m,n}\ .
\ee
Indeed, for $\widehat{P}$ one obtains
\begin{align}
\label{M9}
\widehat{P}\widehat{X}&=-\frac{i}{2}\Big{[}(m+1)\delta_{m,n-1}-(m-1)\delta_{m,n+1}\Big{]}\ ,\nn\\
\widehat{X}\widehat{P}&=-\frac{i}{2}\Big{[}m\delta_{m,n-1}-m\delta_{m,n+1}\Big{]}\ ,
\end{align}
and therefore the commutator 
\bel{M10}
\Big{[}\widehat{P},\widehat{X}\Big{]}=-\tfrac{i}{2}(\delta_{m,n-1}+\delta_{m,n+1})\ .
\ee
In the continuum limit this yields the usual commutation relation $[\widehat{P},\widehat{X}]=-i$. The commutator $[\tilde{P},\widehat{X}]$ is more involved. In the continuum limit it reduces again to $[\tilde{P},\widehat{X}]=-i$.

We can relate $\widehat{P}$ to the step evolution operators~\eqref{SP3},
\bel{M11}
\widehat{P}=-\frac{i}{4\varepsilon}(S_{L}-S_{R})\ .
\ee
With $S_L^{-1}=S_R$, we obtain the evolution equation expressed in terms of the lattice derivative $\wpt$, 
\begin{align}
\label{M12}
&4\varepsilon\wpt \varphi_{L}(t,x)=\varphi_{L}(t+2 \varepsilon , x)-\varphi_{L}(t-2\varepsilon, x)\\
&=\big{(}S_{L}(x,y)-S_{L}^{-1}(x,y)\big{)}\varphi_{L}(t,y)=4\varepsilon i\widehat{P}(x,y)\varphi_{L}(t,y)\ .\nn
\end{align}
In the continuum limit this identifies
\bel{M13}
i\partial_{t}\varphi_{L}=H_{L}\varphi_{L}\quad , \quad H_{L}=-\widehat{P}\ .
\ee
Similarly, the Hamilton operator for the right-movers reads
\bel{M14}
H_{R}=\widehat{P}\ .
\ee
Comparing with eq.~\eqref{295C} shows that for the continuum limit $\widehat P$ and $\tilde P$ coincide.

\subsection*{Momentum observable}

For a fermionic quantum field theory the momentum is a central observable for the characterization of the one-particle states. This also has to hold for the associated probabilistic cellular automaton. We will find that for the automaton the momentum observable is a type of \qq{statistical observable} characterizing properties of the probability distribution rather than being a property of a given bit configuration.

In position space the eigenstates of the momentum operator are periodic functions. For the left-movers they read explicitly
\bel{M15}
\varphi_{p}(x)=\frac{1}{\sqrt{M_x}}\exp(ipx)=\frac{1}{\sqrt{M_x}}\exp\Big{(}\frac{2\pi imq}{M_x}\Big{)}\ ,
\ee
while for the right-movers one has an additional factor $\exp(-i\pi q/M_x)$ due to the different positioning. (For explicit expressions for the corresponding probability distribution in the real formulation see ref.~\citep{CWNEW}.)
The period of these eigenfunctions is $M_x/|q|$. 
If one can ``measure" the period of such an eigenfunction one can extract the value of $|q|$.
We may assign positive $q$ to the right-movers and negative $q$ to the left-movers, such that the Hamiltonian~\eqref{295C},~\eqref{M13},\eqref{M14} is positive.

With given $q$ we know the eigenvalues of $\tilde{P}$ or $\widehat{P}$ in the eigenstates, as given by the diagonal elements in eqs.~\eqref{M1},\eqref{M4}. We can interpret $\tilde P$ or $\widehat{P}$ as an operator for a momentum observable. This momentum observable takes the sharp value $\pi q/(\eps M_x)$ or $\sin(2\pi q/M_x)/(2\varepsilon)$ for an eigenstate with given $q$. It does not have a fixed value in anyone of the bit-configurations, however. The momentum observable is rather a ``statistical observable''~\cite{CWQPCS}, which measures properties of the probabilistic state, as the periodicity of wave functions and probability distribution. In this sense it has a comparable status to temperature or entropy in equilibrium systems, which do not have a well defined meaning for a given microstate. While the values of probabilistic observables can be determined by observation, they are not classical observables in the sense that they do not have fixed values for the bit-configurations, neither the time-local ones nor the overall ones. A classical correlation function for momentum and position observables does not exist. One can define ``measurement correlations''~\cite{CWQMCS, CWEM, CWPT} for sequences of ideal measurements. These correlation functions cannot be expressed by classical correlation functions. They therefore do not have to obey Bell's inequalities~\cite{BELL, CHSH}.

\subsection*{Continuum limit for free massless fermions}

The continuum limit for the simple quantum system of a right-moving or left-moving single particle is straightforward. We may keep a fixed length of the chain $L$ with discrete momentum spectrum $p=2\pi q/L$ according to eq.~\eqref{SP7}. The limit $M_x\to\infty$ corresponds to $\eps\to0$. A wave function whose Fourier components have support in a restricted momentum interval $|p|<p\subt{max}$ becomes smooth for $\eps\to0$. Lattice derivatives can be replaced by partial derivatives. The momentum operator becomes the continuum momentum operator $-i\partial_x$, with eigenvalues given by $p$. In the continuum limit the wave functions become complex vectors in a Hilbert space.

At this stage we have seen how all the familiar concepts of quantum mechanics have emerged very naturally for the one-particle states of the probabilistic automaton. This includes complex wave functions, non-commuting hermitian operators for observables, the quantum rule for expectation values, the probabilistic interpretation of quantum mechanics, change of basis and unitary time evolution. Still, our physical system is very simple, namely a free massless particle. We next investigate if a modified propagation, in particular disorder, can induce a mass for the particle and a potential, such that contact to the usual quantum mechanics in one space-dimension can be made.

\section{Quantum mechanics for particle in potential}\label{sec: QMPP}

In this section we propose a probabilistic automaton that realizes quantum mechanics for a particle in a potential in one space-dimension. This automaton is supposed to reproduce quantitatively all the simple textbook examples as tunneling through a potential barrier, the reflection and transmission laws etc. We want to establish a continuum limit for which the wave function of the probabilistic automaton obeys precisely the Schrödinger equation for a non-relativistic quantum particle. For this purpose we first derive the Dirac equation for a massive fermion in a potential, and take the non-relativistic limit subsequently.

Our proposal is based on the evolution in the presence of disorder discussed in sect.~\ref{sec: MP}. With the formalism of suitable quantum operators for single particle states discussed in the preceding section we can now make progress in the discussion of the continuum limit. The setting with randomly distributed disorder points may at first sight seem somewhat artificial. We will argue that precisely this type of scenario arises for fermionic models with interactions in the presence of suitable expectation values for fermion composites, as discussed in sect.~\ref{sec: VOS}. We believe that at the end a solution of the continuum limit for fermionic quantum field theories with interactions will be the most straightforward way to understand the quantum behavior of single particles. Nevertheless, the key issues can be understood from single-particle states of cellular automata with random disorder.

\subsection*{Disorder}

The propagation of a single particle gets more complicated in the presence of disorder. The particle changes direction whenever it hits a disorder point. If the disorder points are randomly distributed with a constant average number per site over a large enough volume one still expects a rather simple evolution for averaged quantities. We consider here generalized single particle excitations of a half-filled ground state. For simplicity we place for this discussion the right-movers and the left-movers on the same sublattice.

Starting from eq.~\eqref{DM1} a single disorder point at $(\tbar,\xbar)$ induces for the evolution at $\tbar$ an additional change
\bel{243A}
q'(\tbar,\xbar)=\Shat_{LR}(\tbar,\xbar)q(\tbar,\xbar)\ .
\ee
For the one-particle wave functions this amounts to
\bel{243B}
q'_R(\tbar,\xbar)=-q_L(\tbar,\xbar)\ ,\quad q'_L(\tbar,\xbar)=q_R(\tbar,\xbar)\\\\ ,
\ee
and similar for a single hole. In the complex language the additional jump at $\xbar$ reads
\bel{243C}
\vp'_R(\tbar,\xbar)=-\vp_L(\tbar,\xbar)\ ,\quad \vp'_L(\tbar,\xbar)=\vp_R(\tbar,\xbar)\ ,
\ee
which yields in the two-component notation
\bel{243D}
\vp'(\tbar,\xbar)=-i\tau_2\vp(\tbar,\xbar)=\exp\big\{-i\eps\Vbar(\xbar)\tau_2\big\}\vp(\tbar,\xbar)\ ,
\ee
where
\bel{243E}
\Vbar(\xbar)=\frac{\pi}{2\eps}\ ,\quad \vp(\tbar,\xbar)=\pvec{\vp_R(\tbar,\xbar)}{\vp_L(\tbar,\xbar)}\ .
\ee

Combining several possible disorder points $\xbar_i$ at a given $\tbar$, and a unit transformation for all positions $x$ different from the disorder points, one obtains
\bel{243F}
\vp'(\tbar,x)=\exp\big\{-i\eps H_V(x)\big\}\vp(\tbar,x)\ ,
\ee
where
\bel{243G}
H_V(x)=\Vtil(x)\tau_2\ ,\quad \Vtil(x)=\sum_i\frac{\pi}{2\eps}\delta_{x,\xbar_i}\ .
\ee
We may call a disorder point $\xbar_i(\tbar)$ at time $\tbar$ a \qq{disorder event} at the position $\xbar_i$ at time $\tbar$.

Writing also the propagation part of eq.~\eqref{DM1} in exponential form one arrives at
\bel{243H}
\vp(\tbar+\epstil,x)=\sum_y\exp\gl-i\eps H_f\gr(x,y)\exp\big\{-i\eps H_V(y)\big\}\vp(\tbar,y)\ ,
\ee
with
\begin{align}
\label{243I}
\exp\gl-i\eps &H_f\gr(x,y)=\pmat{\delta_{x,y+\eps}}{0}{0}{\delta_{x,y-\eps}}\nn\\
=&\sum_p D^{-1}(x,p)\exp\gl-i\eps\Htil_f(p)\gr D(p,y)\ ,
\end{align}
and
\begin{align}
\label{243J}
\Htil_f(p)=&p\tau_3\ ,\quad p=\frac{\pi q}{\eps M_x}\ ,\nn\\
D^{-1}(x,p)=&\frac1{\sqrt{M_x}}\exp(ipx)\ ,\ D(p,y)=\frac1{\sqrt{M_x}}\exp(-ipy)\ ,\nn\\
x=&m_x\eps\ ,\quad y=m_y\eps\ .
\end{align}
For simplicity we neglect a possible interaction term in the step from $t+\epstil$ to $t+2\epstil=\te$, such that $\vp(\tbar+\eps,x)=\vp(\tbar+\epstil,x)$. In position space one has
\bel{317A}
H_f=\Ptil\tau_3\ ,
\ee
with $\Ptil$ given by eq.~\eqref{M3}. In the absence of disorder ($\Vtil=0$) we recover after two steps eq.~\eqref{295C}.

\subsection*{Random Schrödinger equation}

The propagation part $H_f$ and the potential part $H_V$ do not commute. Nevertheless, since the step from $\vp\otx$ to $\vp\otex$ is a unitary transformation, we can write
\bel{243K}
\vp(\te)=\exp\big\{-i\eps H(t)\big\}\vp(t)\ ,
\ee
where $\vp$ is a $M_x$-component complex vector and $H(t)$ a hermitian $M_x\times M_x$ matrix. So far, we have taken position space with components of $\vp$ labeled by $x$ or $m_x$, but eq.~\eqref{243K} holds as well in momentum space or for any other basis. Here $H(t)$ is $t$-independent within a given time interval from $t$ to $\te$, but it may take different values in different $t$-intervals since the locations $\xbar_i(t)$ of the disorder points may be different for different evolution steps.

Eq.~\eqref{243K} is a solution of the continuous Schrödinger equation
\bel{243L}
i\partial_t\vp(t)=H(t)\vp(t)\ ,
\ee
with $H(t)$ piecewise constant for the intervals from $t$ to $\te$. For randomly distributed disorder points $\xbar_i(t)$ eq.~\eqref{243L} is a type of Schrödinger equation with random disorder. For every time interval the particle is reflected at different $\xbar_i(t)$, which may be interpreted as the action of a stochastic force.

Consider next a time interval $\Delta t$ much larger than $\eps$, $\Delta t/\eps\gg1$, and write the result of many evolution steps as
\bel{243M}
\vp(t+\Delta t)=\exp\big\{-i\Delta t\Hbar\big\}\vp(t)\ .
\ee
The hermitian matrix $\Hbar$ is a type of time-averaged Hamiltonian. This time-averaging entails an effective averaging over the stochastic potential. Unitarity of the evolution implies $\Hbar\herm=\Hbar$. For large enough $\Delta t/\eps$ one may assume that $\Hbar$ becomes time-independent and is no longer a stochastic quantity. The challenge consists in relating $\Hbar$ to $H_f$ and $H_V(t,y)$ in eq.~\eqref{243H}. We further assume that the wave function is sufficiently smooth in $x$ such that we can replace lattice-derivatives by partial derivatives $\partial_x$ acting on continuous functions $\vp\otx$.

\subsection*{Dirac fermion in a potential}

We want to show the existence of a continuum limit for which
\bel{243N}
\Hbar=Z_\vp\big[-i\tau_3\partial_x+\gl m+V(x)\gr\tau_2\big]\ .
\ee
This will lead to the Dirac equation for a particle with mass $m$ in a potential $V(x)$. We start from eq.~\eqref{243H} where $H_V$ depends on $\tbar$ since the disorder points $\xbar_i(\tbar)$ depend on $\tbar$. Performing a sequence of evolution steps one arrives at the exact defining relation for $\Hbar$,
\bel{CL1}
\exp\big\{-i\Delta t\Hbar\big\}=\prod_{n_t=0}^{N_t-1}\Big[\exp\gl-i\eps H_f\gr\exp\gl-i\eps H_V(t+n_t\eps)\gr\Big]\ ,
\ee
where
\bel{CL2}
N_t=\frac{\Delta t}{\eps}\ .
\ee
The order of the operators is with increasing $n_t$ to the left.

The time-averaged Hamilton operator $\Hbar$ can be written in the form
\bel{CL3}
\Hbar=H_0+\Delta H\ ,
\ee
where
\bel{CL4}
H_0=\frac1{N_t}\sum_{n_t=0}^{N_t-1}\Big(H_f+H_V(t+n_t\eps)\Big)\ .
\ee
The \qq{correction term} $\Delta H$ vanishes if the nonzero commutator of $H_f$ and $H_V$ can be neglected. It will be discussed below. Using the definition~\eqref{243G} for $H_V$ one obtains
\bel{CL5}
H_0=H_f+\Vbar(t,x)\tau_2\ ,
\ee
where
\bel{CL6}
\Vbar(t,x)=\frac{\pi}{2\eps}\nbar(t,x)\ .
\ee
Here
\bel{CL7}
\nbar(t,x)=\frac1{N_t}\sum_{n_t=0}^{N_t-1}\sum_i\delta_{x,\xbar_i(t+n_t\eps)}
\ee
is the time averaged number of disorder events at $x$ for the interval between $t$ and $t+\Delta t$. The intuitive interpretation of $H_0$ is
\bel{243O}
H_0=\exval{H_f}+\exval{H_V}\ ,
\ee
where the brackets stand for time averaging. For simplicity we will take a disorder distribution for which $\nbar(t,x)$ does not depend on time. The generalization to a time-dependent potential is straightforward.

We next restrict the discussion to sufficiently smooth wave functions for which $H_f$ can be expressed in terms of a partial derivative
\bel{CL8}
H_f=-i\partial_x\tau_3\ .
\ee
Such wave functions are dominated in momentum space by small $\abs{q}\ll M_x$, such that the two definitions of the momentum operator $\Ptil$ and $\widehat{P}$ effectively coincide. We further replace the average number $\nbar(x)$ of disorder events at the discrete points $x$ by the average number density $n(x)$ depending on a continuous variable $x$,
\bel{CL9}
n(x)=\frac{\exval{\nbar(x)}}{2\eps}\ ,
\ee
such that $\sum_m\nbar(m)=1/(2\eps)\int\text{d}x\,\nbar(m)=\int_xn(x)$. This yields the continuum version
\bel{CL10}
H_0=-i\partial_x\tau_3+\Vbar(x)\tau_2\ ,\quad \Vbar(x)=\pi n(x)\ .
\ee
For the identification $\Hbar=H_0$ one has $Z_\vp=1$, $m+V(x)=\pi n(x)$. Modified values of $Z_\vp$ and $m+V(x)$, as well as additional terms in eq.~\eqref{243N}, could arise from the so far neglected piece $\Delta H$ or from residual discretization effects.

\subsection*{Continuum Hamiltonian}

We next want to show that in the continuum limit $\eps\to0$ the correction term $\Delta H$ vanishes. This correction term arises from the non-zero commutator of $H_f$ and $H_V$. For a single time step one may use a formal expansion in $\eps$,
\begin{align}
\label{243Q}
&\exp\gl-i\eps H_f\gr\exp\gl-i\eps H_V\gr\nn\\
&=\exp\gl-i\eps(H_f+H_V)\gr+\mathcal{O}\gl\eps^2\big[H_f,H_V\big]\gr\ .
\end{align}
The commutator term can be neglected only if for the stochastic average the commutator $\big[H_f,H_V\big]$ remains constant for $\eps\to0$, or increases less fast than $\sim\eps^{-1}$. We will next investigate the conditions under which $\Delta H$ can be neglected.

Using the Hausdorff formula one has for a single time step
\bel{CL11}
\exp\gl-i\eps H_f\gr\exp\gl-i\eps H_V\gr=\exp\big\{-i\eps\gl H_f+H_V\gr-i\eps^2\Delta\big\}\ ,
\ee
with
\bel{CL12}
\Delta=-\frac i2\big[H_f,H_V\big]+\mathcal{O}(\eps)\ ,
\ee
where the term $\mathcal{O}(\eps)$ in $\Delta$ involves higher order commutators.

This generalizes to a chain of updating steps
\begin{align}
\label{PP1}
\exp\gl-i&\Delta t\Hbar\gr=\prod_{n_t=0}^{N_t-1}\Big[\exp\gl-i\eps H_f\gr\exp\gl-i\eps H_V(n_t)\gr\Big]\nn\\
=\exp\Big\{&-i\eps N_tH_0-\frac{\eps^2}{2}\Big[H_f,\sum_{n_t=0}^{N_t-1}\gl N_t-2n_t\gr H_V(n_t)\Big]\nn\\
&+\mathcal{O}(\eps^3)\Big\}\ .
\end{align}
The different $H_V(n_s)$ are distinguished by the different disorder events at times $t+n_s\eps$ and stand for mutually commuting diagonal matrices in position space. Comparison with eqs.~\eqref{CL1},~\eqref{CL3} defines $\Delta H$ in leading order in $\eps$.

In order to gain intuition we first consider a distribution of disorder events which is time-reflection invariant within the interval $\Delta t$. For $H_V(N_t-1-n_t)=H_V(n_t)$ one finds
\bel{PP2}
\sum_{n_t=0}^{N_t-1}\gl N_t-2n_t\gr H_V(n_t)=\sum_{n_t=0}^{N_t-1}H_V(n_t)\ ,
\ee
and therefore
\bel{PP3}
\Delta H=-\frac{i\eps}{2N_t}\Big[H_f,\sum_{n_t=0}^{N_t-1}H_V(n_t)\Big]+\mathcal{O}(\eps^2)\ .
\ee
This type of relation obtains actually under much wider circumstances. It is sufficient that time reflection symmetry in the interval $\Delta t$ holds in the average. Deviations of $\Delta H$ from eq.~\eqref{PP3} involve the commutator of $H_f$ with sums of differences of $H_V$ at different $n_t$. For sufficiently large $N_t$, and random distributions of disorder events the average value of these differences vanishes or is small. One may also switch the order of $H_f$ and $H_V$ between consecutive time steps. In this case $\Delta H$ only involves differences of $H_V$ at different $n_t$.

We base our estimate of $\Delta H$ on eq.~\eqref{PP3} and neglect the terms $\mathcal O(\eps^2)$,
\bel{PP4}
\Delta H=-\frac{i\eps}{2}\big[H_f,\Vbar(x)\tau_2\big]\ .
\ee
We next assume sufficiently smooth wave functions and smooth $\Vbar(x)$ such that $H_f$ can be approximated by eq.~\eqref{CL8},
\begin{align}
\label{PP5}
\Delta H=&-\frac\eps2\big[\tau_3\partial_x,\tau_2\Vbar(x)\big]\nn\\
=&\eps\tau_1\big\{\Vbar(x)i\partial_x+\frac i2\gl\partial_x\Vbar(x)\gr\big\}\ .
\end{align}
If $\Vbar(x)$ remains finite and differentiable for $\eps\to0$ at fixed $\Delta t$, i.e. $N_t\to\infty$, the term $\Delta H$ vanishes in this limit.

\subsection*{Conditions for continuum limit}

At this point we can state conditions for a continuum limit of the disordered one-particle automaton. First, the random disorder points should be sufficiently rare such that $\Vbar(x)$ remains finite in the limit $\eps\to0$ at fixed $N_t\eps$. For this limit the total number of disorder events $\hat n(x)$ per site $x$ or $m$ in the interval $\Delta t$ should be kept fixed to a finite value as $\eps\to0$, such that $\nbar(x)=\hat n(x)/N_t$ decreases proportional to $\eps\sim N_t^{-1}$ as $N_t$ diverges. Then $n(x)$ and $\Vbar(x)$ become independent of $\eps$.

Second, the number of disorder events should be sufficiently large and randomly distributed such that $n(x)$ and $\Vbar(x)$ can be approximated by sufficiently smooth functions of a continuous variable $x$. We may consider an interval $\Delta x$ for a typical resolution in space. A smooth description typically averages over $\Delta x$ and requires a large number of disorder events in the combined interval $\sim\Delta t\Delta x$. With $\Delta x=2N_x\eps$ we therefore require
\bel{PP6}
\sum_{x'\in I_x}\hat n(x')=\langle\hat n(x)\rangle N_x\gg1\ ,
\ee
where the sum is over all positions within the interval $I_x$ around $x$ with size $\Delta x$, and $\langle\hat n(x)\rangle$ is the space-averaged number of disorder events in the time interval $\Delta t$. For an almost homogeneous distribution of disorder events one has
\bel{PP7}
\Vbar(x)=\frac\pi{2\eps N_t}\hat n(x)\approx\frac\pi{2\Delta t}\langle\hat n(x)\rangle\approx m\ .
\ee
If we want to investigate time intervals $\Delta t$ with $m\Delta t\ll1$ this requires $\langle\hat n(x)\rangle\ll1$. For large enough $N_x$ there is no contradiction between the two constraints~\eqref{PP6} and~\eqref{PP7}. For $\langle\hat n(x)\rangle\ll1$ most sites in the interval $\Delta x$ have no disorder event for the whole time interval $\Delta t$. Smoothness is only realized by space-averaging due to the large number $N_x$ which diverges for $\eps\to0$ as $\Delta x/(2\eps)$. In contrast, for disorder distributions with $\langle\hat n(x)\rangle\gg1$ our treatment of a continuum limit is only valid for time intervals $\Delta t\gg m^{-1}$.

The combination of time-averaging and space-averaging in the combined interval $\Delta t\Delta x$ is of great help for guaranteeing the effective time reversal symmetry leading to the estimate~\eqref{PP3} and therefore to the relative suppression of $\Delta H$ proportional to $\eps$. It is sufficient that this effective symmetry holds in a space-averaged sense. More generally, it is sufficient that one can effectively replace in eq.~\eqref{PP1} $H_V(n_t)$ by a space averaged quantity $\exval{H_V(n_t)}$, and space averages $\langle H_V(t_1)-H_V(t_2)\rangle$ vanish approximately. The $x$-dependence of $\Vbar(x)$ is then due to different space averages for different $x$-intervals. In particular, if $\exval{H_V(n_t)}$ becomes independent of $n_t$ one obtains directly eq.~\eqref{PP2} and therefore the estimate~\eqref{PP5} for $\Delta H$.

For a complementary view one may consider an \qq{infinitesimal evolution step} by $\eps$ which acts on a space averaged wave function. If the action of $H_f$ and $H_V$ produces only a finite change of the averaged wave function one can expand
\begin{align}
\label{351X}
&\exp\gl-i\eps H_f\gr\exp\gl-i\eps H_V\gr\nn\\
=&\gl1-i\eps H_f\gr\gl1-i\eps H_V\gr+\mathcal{O}(\eps^2)\\
=&\gl1-i\eps(H_f+H_V)\gr+\mathcal{O}(\eps^2)=\exp\gl-i\eps H_0\gr+\mathcal{O}(\eps^2\gr\ .\nn
\end{align}
This implies that $\Delta H$ is indeed of the order $\eps$ and vanishes in the continuum limit $\eps\to0$.

The third condition for the continuum limit requires that the $x$-dependence of the wave function is sufficiently smooth. An averaging over $\Delta x$ is only appropriate if the wave function only undergoes small changes on this length scale. With typical momenta of the wave function $p\sim1/\Delta x$ one has for the momentum operator
\bel{PP8}
\abs q\sim\eps M_x\abs p\sim\frac{\eps M_x}{\Delta x}\sim\frac{M_x}{N_x}\ .
\ee
For large $N_x$ this value is much smaller than $M_x$, such that discretization effects become negligible. This implies that we can approximate $H_f=-i\partial_x\tau_3$. Taking things together, the double hierarchy,
\bel{PP9}
1\ll N_t,\ N_x\ll M_t,\ M_x\ ,
\ee
guarantees that we can combine a smooth evolution with time and a space resolution $\Delta t$ and $\Delta x$ much smaller than the total size of the system in time and space. As one may perhaps have expected the discretization effects are suppressed by $N_t^{-1}$ or $N_x^{-1}$.

\subsection*{Dirac equation}

We finally divide $\Vbar(x)$ into a homogeneous part $m$ and an inhomogeneous part $V(x)$,
\bel{PP10}
\Vbar(x)=m+V(x)\ .
\ee
According to the problem, there may be different criteria for this division. For example, if for large  $\abs x$ one has the same constant value of $\Vbar$, one may define $m$ by the condition $V(\abs x\to\infty)=0$. Another possibility is to take for $m$ the average of $\Vbar(x)$ over the whole space. The time evolution is independent of the split~\eqref{PP10}, but expansions in $V(x)/m$ may depend on the detailed definition.

For $\Hbar$ independent of $t$ the wave function with evolution~\eqref{243M} is a solution of the Schrödinger equation
\bel{243R}
i\partial_t\vp=\Hbar\vp\ .
\ee
A possible factor $Z_\vp$ in the somewhat more general form~\eqref{243M} can be absorbed into a rescaling of the time variable, resulting in
\bel{243S}
i\partial_t\vp=\big[-i\tau_3\partial_x+\gl m+V(x)\gr\tau_2\big]\vp\ .
\ee
This equation is the two-dimensional Dirac equation with a potential $V(x)$. Indeed, using the Dirac matrices $\gamma^0=-i\tau_2$, $\gamma^1=\tau_1$ it takes the form
\bel{SE3}
\big{(}\gamma^{\mu}\partial_{\mu}+m+V(x)\big{)}\varphi =0\ .
\ee
For $V(x)=0$ this evolution is compatible with Lorentz-symmetry, describing the free propagation of a Dirac-fermion with mass $m$. We will see that $V(x)$ corresponds to a potential in which the particle moves. In quantum field theory this type of equation is found for a Dirac fermion with Yukawa coupling to a scalar field, which has an $x$-dependent expectation value. This underlines the possible identification of effective disorder with a suitable vacuum expectation value. The scalar field may correspond to a composite of fermion fields.

\subsection*{Schrödinger equation for particle in a potential}

The Schrödinger equation for a particle in a potential $V(x)$ obtains as the non-relativistic limit of the evolution equation~\eqref{SE3} for the one-particle wave function. Let us assume $|V(x)/m|\ll 1$ and a weak dependence of $\varphi(t,x)$ on $x$. In this limit the approximate solution of eq.~\eqref{SE3} is obtained from a solution of the standard Schrödinger equation,
\bel{326A}
i\partial_t\chi=-\frac{\partial_x^2}{2m}\chi+V(x)\chi\ .
\ee

In order to see this we start from the Dirac equation with a potential~\eqref{SE3}
\bel{SE7}
\gamma^{\mu}\partial_{\mu}\varphi=-(m+V)\varphi\ .
\ee
Applying on both sides the operator $\gamma^{\mu}\partial_{\mu}$ one obtains a Klein-Gordon type equation
\bel{SE8}
(-\partial_{t}^{2}+\partial_{x}^{2})\varphi=(m+V)^{2}\varphi-\tau_{1}(\partial_{x}V)\varphi\ .
\ee
We write
\bel{SE9}
\varphi(t,x)=\exp(-imt)\tilde\chi (t,x)\ ,
\ee
resulting for $\chi (t,x)$ in the evolution equation
\bel{SE10}
i\partial_{t}\tilde\chi = \Big{(}-\frac{\partial_x^{2}}{2m}+V\Big{)}\tilde\chi +C\tilde\chi\ ,
\ee
with 
\bel{SE11}
C\tilde\chi = \Big{(}\frac{V^{2}}{2m}-\frac{\partial_{x}V}{2m}\tau_{1}+\frac{\partial_{t}^{2}}{2m}\Big{)}\tilde\chi \ .
\ee

The Schrödinger equation obtains in the limit where the term $C\chi$ can be neglected. It becomes valid for 
\bel{SE12}
\Big{|}\frac{V}{m}\Big{|}\ll 1 \, ,\quad |-\partial_{x}^{2}\chi|\ll|m^{2}\chi |\ ,
\ee
and either
\bel{SE13}
\Big{|}\frac{\partial_{x}\ln{V}}{m}\big{|}\ll 1
\ee
or
\bel{SE14}
|\partial_{x}V\tilde\chi|\ll |\partial_{x}^{2}\tilde\chi |\ .
\ee
If these relations hold we can compute $\partial_{t}^{2}$ iteratively from the Schrödinger equation
\bel{SE15}
\frac{1}{m}\partial_{t}^{2}\tilde\chi =\Big{(}\frac{\partial_x^{2}}{2m^{2}}-\frac{V}{m}\Big{)}\Big{(}-\frac{\partial_x^{2}}{2m}+V\Big{)}\tilde\chi\ ,
\ee
such that the last term in eq.~\eqref{SE11} is small as compared to the first term ot the r.h.s of eq.~\eqref{SE10} by virtue of the relations~\eqref{SE12}.

If the term $\sim\partial_x V\tau_1$ can be neglected the upper and lower component of $\tilde\chi$ obey the same equation. In order to determine the relative size and phase of the two components one has to reinsert into the Dirac equation. Indeed, with the approximations~\eqref{CL8}-~\eqref{CL10} the Dirac equation~\eqref{243S},~\eqref{SE3} is solved by
\bel{EMP1}
\vp_+=\frac1{\sqrt{2}}e^{-imt}\pvec{\chi_+-\frac{i\partial_x}{2m}\chi_+}{i\chi_+-\frac{\partial_x}{2m}\chi_+}\ ,
\ee
with $\chi_+$ a solution to the Schrödinger equation. This relates the two components of $\tilde\chi$, expressing them in terms of the single complex wave function $\chi_+$. The normalization of $\chi_+$ is given by
\bel{EMP2}
\int_x\chi_+^*\left(1-\frac{\partial_x^2}{4m^2}\right)\chi_+=1\ .
\ee
After multiplicative renormalization we identify $\chi_+$ with the non-relativistic wave function $\chi$ for a quantum particle in a potential.

Neglecting $i\partial_x$ and $V$ as compared to $m$ the wave function $\vp_+$ corresponds to the eigenvalue $m$ of the Hamiltonian. There exists a second solution $\vp_-\sim e^{imt}\chi_-$ which corresponds in this limit to the eigenvalue $-m$. The wave function $\vp_-^*$ describes the one-particle state for the antiparticle. As well known, the Dirac equation describes both particles and antiparticles.

We conclude that a rather simple probabilistic cellular automaton with disorder describes the evolution of a quantum particle in a potential in one space-dimension, provided that the continuum limit described before is a good approximation. Furthermore, the initial wave function has to be sufficiently smooth in $x$, and the parameters for the disorder need to be in a suitable range, such that eqs.~\eqref{SE12}-\eqref{SE14} hold. 
It would be an interesting (numerical) experiment on probabilistic cellular automata to reproduce the well known quantum features of the one-dimensional Schrödinger equation for a particle in a potential. 

\subsection*{Particle mass from spontaneous symmetry\\breaking}

For models with interactions the role of disorder can be played by expectation values of fermion composites. Order parameters can induce mass terms for the one-particle propagation. In the presence of interactions the motion of a single particle can be influenced by vacuum expectation values for fermion bilinears or more complex composites of an even number of fermions. This is similar to the Higgs mechanism for which the lepton and quark masses are induced by the vacuum expectation value of a scalar field. In our case the scalar would be a fermion composite. Scattering processes involving the fields constituting the order parameter can indeed change the direction of a given single particle. We can interpret $\nbar(x)$ and therefore $m$ as the average number of such direction-changing scatterings per site, caused by the presence of the particles represented by the expectation values. In this case the ``disorder'' reflects spontaneous symmetry breaking by a suitable expectation value. This mechanism offers the potential for explaining small parameters as $m\varepsilon$. They can arise if the corresponding expectation value is small. Furthermore, we may encounter states where the expectation value is position-dependent. This can induce the potential $V(x)$.

A simple example is the Gross-Neveu model and generalizations thereof. If one replaces in eq.~\eqref{MF7} the interaction term by a mean field expression
\bel{323A}
S=-\int_{t,x}\gl\bar\zeta_a\gamma^\mu\partial_\mu\zeta_a+\bar\zeta_a\zeta_a\exval{\bar\zeta_b\zeta_b}\gr
\ee
one obtains indeed the action of a massive Dirac fermion, with mass
\bel{323B}
m=\exval{\bar\zeta_b\zeta_b}\ .
\ee
The order parameter corresponds to spontaneous chiral symmetry breaking. This effect is well established in three or four dimensions in the corresponding Jona-Nambu-Lasinio models. We leave it open here for which type of two-dimensional models this type of spontaneous chiral symmetry breaking can be realized.

We briefly discuss here possible expectation values for the spinor gravity automaton. For our model of spinor gravity the insertion of a condensate
\bel{SE4}
\langle 
\tilde{E}_{\nu}{}^{n,2}\tilde{H}_{3}\tilde{H}_{4}\rangle = Z_{1}\delta_{\nu}^{n}
\ee
into the action\eqref{S53} would lead to the Lorentz invariant kinetic term
\bel{SE5}
S=-Z_{1}\int_{t,x}\overline{\psi}^{1}\gamma^{\mu}\partial_{\mu}\psi^{1}\ ,
\ee
and therefore to a well defined propagator for the fermion with $a=1$. Similar condensates can provide for propagators for the other fermions. A mass term for the fermions with $a=1$ could be generated by a condensate
\bel{SE6}
\varepsilon^{\mu\nu}\varepsilon_{mn}\langle \tilde{E}_{\mu}{}^{m,3}\tilde{E}_{\nu}{}^{n,4}\tilde{H}_{2}\rangle = 2 \overline{m}_{1}\ .
\ee
A propagation in a non-trivial geometric background becomes possible if the expectation value~\eqref{SE4} takes a more general $t$-and $x$-dependent form. 

\section{Continuum limit and coarse grained subsystems}
\label{sec: CLCGS}

The time-continuum limit can be taken if the time-dependence of the wave function or density matrix is sufficiently smooth on a scale given by $\eps$. In this case the limit $\eps\to0$ can be taken, and discrete lattice-time-derivatives become partial derivatives with respect to continuous time. Performing a continuum limit is not trivial, however. One typically averages over certain time intervals. Information can be lost by this averaging or \qq{coarse graining}. Discarding probabilistic information that is not needed for the continuum limit one arrives at a type of coarse grained subsystem.

For the coarse grained subsystems the evolution with $t$ will again be given by an evolution operator. One finds, however, that even for discrete time steps the evolution operator is no longer a unique jump operator. On the coarse grained or continuum level the probabilistic cellular automaton does no longer follow the deterministic evolution of a cellular automaton. A given coarse grained state can evolve with certain probabilities to different coarse grained states. The continuum evolution becomes probabilistic.

We have seen this feature explicitly for the computation of the averaged Hamiltonian in the preceding section. If the wave function is smooth enough such that the order of the propagation and the scattering at disorder points plays no longer a role, one effectively averages over two different evolutions with or without scattering. A given continuum wave function at $t$ evolves with certain probabilities to different continuum wave functions at $t+\Delta t$. One easily verifies that $\exp\gl-i\eps\Hbar\gr$ or $\exp\gl-i\Delta t\Hbar\gr$ are not unique jump matrices. The coarse grained evolution remains unitary but has replaced the automaton property by the characteristic quantum property of different possible paths.

Besides coarse graining in time a continuum limit also involves effective averaging in space. Again, the space-averaging leads to a loss of the unique jump property on the coarse grained level. This basic feature can be visualized by our automata with right-movers on odd lattice sites and left-movers on even lattice sites. Knowing only that a particle is within a given interval in $x$ does not tell if it is on an even or odd site, and therefore if it will move to the left or to the right. On the coarse grained level, it will move with a certain probability to the left, and with another probability to the right, or with still another probability stay within the interval.

In quantum field theories many highly developed methods are available for computing the continuum behavior of discrete lattice theories. They typically correspond to a simultaneous coarse graining in time and space. An efficient tool is the computation of the quantum effective action. This leads to macroscopic field equations in the continuum which include all effects of fluctuations. The most efficient methods are based directly on the functional integral and no not employ explicitly the quantum wave function or the density matrix.

For the conceptual understanding of what happens for probabilistic automata we present in this section the coarse graining procedure in the operator formalism. While in general much less efficient than the direct functional integral approach in practice, it nevertheless permits important insights on the mechanisms that lead to the characteristic probabilistic quantum features in the propagation of particles. A standard method employs the evolution of a quantum system in terms of the density matrix. Coarse graining in quantum systems proceeds by taking a subtrace of the density matrix. This has to be done in a basis where the evolution of the density matrix of the subsystem can be described by a suitably reduced step evolution operator. The evolution of the subsystem should only involve the probabilistic information encoded in the density matrix for the subsystem, being independent of the \qq{environment}. For a definition of suitable subsystems for the continuum limit of the probabilistic cellular automaton the quantum formalism is essential. It is also crucial that we treat with probabilistic cellular automata. No continuum limit exists for the sharp wave functions of a deterministic automaton. The quantum formalism with wave function of density matrix is a key element for these types of coarse graining.

\subsection*{Density matrix and subsystems}

For classical statistics in a real formulation we can define a density matrix $\rho$ in analogy to quantum mechanics. For pure states it reads 
\bel{D1}
\rho_{\tau\rho}(t)=q_{\tau}(t)q_{\rho}(t)\ .
\ee
It obeys
\bel{D2}
\rho^T=\rho\, \quad \tr\rho  =1
\ee
and has a positive semidefinite spectrum of eigenvalues. The evolution reads
\bel{D3}
\rho(t+\varepsilon)=\widehat{S}(t)\rho(t)\widehat{S}\,^T(t)\;,\quad\widehat{S}\,^T(t)\widehat{S}(t)=1\ ,
\ee
and expectation values of time-local observables are expressed in terms of the associated operator $\widehat{A}(t)$,
\bel{D4}
\langle A(t)\rangle=\tr \big{\lbrace}\rho(t)\widehat{A}(t)\big{\rbrace}\ .
\ee
These relations continue to hold beyond pure states for which $\rho$ is a more general positive hermitian and normalized matrix.

In the presence of a complex structure that is compatible with the evolution the pure state density matrix is a bilinear in the complex wave function $\varphi$,
\bel{D5}
\rho_{\tau\rho}=\varphi_{\tau}\varphi_{\rho}^{*}\, ,\quad \rho ^{\dagger}=\rho\, , \quad \tr \rho=1\ ,
\ee
with unitary evolution
\bel{D5}
\rho(t+\varepsilon)=U(t)\rho(t)U^{\dagger}(t)\,,\quad U^{\dagger}(t)U(t)=1\ .
\ee
Observables which are compatible with this complex structure are expressed by hermitian operators $\widehat{A}(t)$, with expectation values given by eq.~\eqref{D4}.
Again, general complex density matrices have a positive semidefinite spectrum of eigenvalues. In the following we use the complex formulation. The real formulation is a special case for which $\rho$, $U$ and $\widehat{A}$ are real matrices. The complex density matrix allows for a straightforward implementation of similarity transformations. For
\bel{D7}
\rho'(t)=D(t)\rho(t)D^{\dagger}(t)\, ,\quad D^{\dagger}(t)D(t)\ ,
\ee
and
\bel{D8}
U'(t)=D(t)U(t)D^{ \dagger }(t)\, , \quad \widehat{A}'(t)=U(t)\widehat{A}(t)U^{\dagger}(t)\, ,
\ee
all relations remain invariant.

The density matrix is a suitable object for defining general subsystems that are compatible with the evolution~\cite{Wetterich:2020kqi}. Assume that one can find a suitable basis such that $U'(t)$ becomes block diagonal
\bel{D9}
U'(t)=U_{E}(t)\otimes U_{S}(t)\ .
\ee
In this basis we can employ a double index notation \\ $\tau=(\alpha , \gamma)$, $\rho=(\beta ,\delta)$ such that
\bel{D10}
U_{\alpha\gamma , \beta\delta}'=(U_{E})_{\alpha\beta}(U_{S})_{\gamma\delta}\ .
\ee
In this notation the evolution equation reads
\begin{align}\label{D11}
\rho_{\alpha\gamma , \beta\delta}'&(t+\varepsilon)=\big{(}U_{E}\big{)}_{\alpha\alpha'}(t)\big{(}U_{S}\big{)}_{\gamma\gamma'}(t)\\&\times\rho_{\alpha'\gamma' , \beta' \delta' }'\big{(}U_{S}^{\dagger}\big{)}_{\delta'\delta}(t)\big{(}U_{E}^{\dagger}\big{)}_{\beta'\beta}(t)\ .\nn
\end{align}

We introduce the density matrix for the subsystem by a partial trace
\bel{D12}
\rho_{\gamma\delta}^{(S)}(t)=\rho_{\alpha\gamma , \beta\delta}'(t)\delta^{\alpha\beta}\ .
\ee
The evolution of this coarse grained density is independent of the ``environment", determined by the evolution operator $U_{S}$, as can be seen by taking a partial trace of eq~\eqref{D11}.
\bel{D13}
\rho_{\gamma\delta}^{(S)}(t+\varepsilon)=\rho_{\alpha\gamma , \beta\delta}'(t+\varepsilon)\delta^{\alpha\beta}=\big{(}U_{S}\big{)}_{\gamma\gamma'}(t)\rho_{\gamma'\delta'}^{S}(t)\big{(}U_{S}^{\dagger}\big{)}_{\delta'\delta}(t)\ .
\ee
We observe that even for real $\rho$ and $U=\widehat{S}$ the density matrix $\rho^{(S)}$ is, in general, a hermitian complex matrix and $U_{S}$ is a complex unitary matrix. The unitary matrix $D$ that brings $U=\widehat{S}$ to the block diagonal form~\eqref{D9} is, in general,  a complex unitary matrix, such that $\rho'$ and $U'$ are no longer real.

Not every local observable is compatible with the subsystem in the sense that its expectation value can be computed from $\rho^{(S)}$. Subsystem observables are independent of the environment. Their associated operators read
\bel{D14}
\widehat{A}'(t)=1_{E}\otimes\widehat{A}^{(S)}(t)\;,\quad
\widehat{A}_{\alpha\gamma , \beta\delta}'(t)=\widehat{A}_{\gamma\delta}^{(S)}(t)\delta_{\alpha\beta}\ ,
\ee
such that
\bel{D15}
\langle A(t)\rangle =\tr\big{\lbrace}\rho^{(S)}\widehat{A}^{(S)}\big{\rbrace}\ ,
\ee
where the trace is now over indices of the subsystem.

For cellular automata the subsystem is, in general, no longer described by a probabilistic cellular automaton. 
The evolution operator $U^{(S)}$ for the subsystem is unitary, but it is no longer a unique jump matrix. Even if we start with a deterministic automaton for which at $t$ a particular state $\overline{\tau}(t)$ is realized
\bel{D16}
\rho_{\tau\rho}(t)=\delta_{\tau , \overline{\tau}(t)}\delta_{\rho , \overline{\tau}(t)}\ ,
\ee
one has in the basis where $U'$ is block diagonal
\bel{D17}
\rho_{\tau\rho}'(t)=D_{\tau , \tau'}\rho_{\tau' \rho'}D_{\rho'\rho}^{\dagger}=D_{\tau\overline{\tau}}D_{\rho\overline{\tau}}^{*}\ .
\ee
This is a pure state matrix of the type~\eqref{D5}, with $\varphi_{\tau}=D_{\tau\overline{\tau}}$. The coarse grained density matrix is, in general, a mixed state density matrix
\bel{D18}
\rho_{\gamma\delta}^{(S)}=\sum_{\alpha} D_{\alpha\gamma , \overline{\tau}} D_{\alpha\delta , \overline{\tau}}^{*}\ .
\ee

We conclude that coarse graining leads rather generically to a probabilistic description. Some of the precise information needed to specify a sharp deterministic state is lost by the coarse graining. For an implementation of coarse graining a probabilistic description seems mandatory. The ``quantum formalism for classical statistics"~\cite{CWIT,CWQF} employed here yields a simple and appropriate description. We emphasize that coarse graining needs, in general, the notion of the density matrix. A description in terms of local probabilities alone is insufficient. This is seen directly for the evolution~\eqref{D13} for the subsystem: If $U_{S}$ is no longer a unique jump matrix, the evolution of the diagonal elements of $\rho^{(S)}$, which correspond to the local probabilities for the subsystem, involves on the r.h.s. also off-diagonal elements of $\rho^{(S)}$. There is no closed evolution equation for the local probabilities of the subsystem.

We finally state a necessary condition for the existence of a subsystem. The unitary transformation~\eqref{D8} does not change the spectrum of eigenvalues $\lbrace\lambda_{i}\rbrace$ of $U$ or $\widehat{S}$. In the ``direct product basis" the eigenvalues of $U'$ are products of eigenvalues of $U_{E}$ and $U_{S}$, denoted by $\lambda_{k}^{(E)}$ and $\lambda_{l}^{(S)}$. This implies that every eigenvalue of $U$ can be written as a product 
\bel{D19}
\lambda_{i}=\lambda_{k}^{(E)}\lambda_{l}^{(S)}\ ,
\ee
for suitable $k$ and $l$. On the other hand, every product $\lambda_{k}^{(E)}\lambda_{l}^{(S)}$ must be an eigenvalue $\lambda_{i}$ of $U$. Possible subsystems and their environments must have the property that the spectrum of $U$ coincides with the product spectrum of $U_{E}$ and $U_{S}$. This includes the degeneracies. This condition is also sufficient for the existence of corresponding subsystems. We can diagonalize $U$ and use the relation~\eqref{D19} in order to write the diagonal matrix in the direct product form for diagonal matrices $U_{E} $ and $U_{S}$. The direct product form will not be affected by separate similarity transformations of the subsystem and the environment. For many practical circumstances the division into subsystem and environment may not be perfect but constitute a good approximation.

In the remainder of this section we will demonstrate these general properties by simple examples. In particular, we address subsystems for the one-particle states of the cellular automata discussed in the preceding section. The matrix $D$ will typically be some type of Fourier transform.

\subsection*{Coarse grained subsystem}

Let us discuss the general issue of coarse grained subsystems~\cite{Wetterich:2020kqi} for a simple example. We consider a periodic space lattice with eight sites. Right-movers are placed at even $t$ on odd sites, and left-movers on even sites. We consider single-particle states, where the particle can be either a right-mover or a left-mover. The single-particle wave function $q(t)$ is an eight component vector, corresponding to the particle being placed on one of the eight positions. Since the right-movers and left-movers evolve independently we write the single-particle step evolution operator in a block diagonal form
\bel{SO1}
\Shat=\begin{pmatrix}S_R&0\\0&S_L\end{pmatrix}\ ,\ \ S_R=\begin{pmatrix}0&0&0&1\\1&0&0&0\\0&1&0&0\\0&0&1&0\end{pmatrix}\ ,\ \ S_R=\begin{pmatrix}0&1&0&0\\0&0&1&0\\0&0&0&1\\1&0&0&0\end{pmatrix}\ ,
\ee
noting
\bel{SO2}
S_L=S_R^T\ ,\quad S_LS_R=1\ ,\quad S_L^2=S_R^2\ ,\quad S_L^4=S_R^4=1\ .
\ee
Both $S_R$ and $S_L$ are orthogonal matrices.With
\bel{SO3}
S_R^2=S_L^2=\begin{pmatrix}0&1\\1&0\end{pmatrix}=\tau_1\otimes1\ ,
\ee
their eigenvalues are $(\pm1,\pm i)$. We could choose a different basis where the order of rows and columns follows the position on the lattice, but the block diagonal form~\eqref{SO1} is more convenient for our purpose.

We want to find a similarity transformation
\bel{SO4}
\Stil=D\Shat D\herm\ ,\quad D\herm D=1\ ,
\ee
such that $\Stil$ can be written in a direct product form
\bel{SO5}
\Stil=S_E\otimes S_S\ .
\ee
In this case $S_S$ can describe the evolution of the coarse grained subsystem while $S_E$ evolves the \qq{environment}. In this form we will be able to perform subtraces in order to define a subsystem whose evolution is independent of the environment~\cite{Wetterich:2020kqi}.
For our system this can be realized by
\bel{SO6}
S_S=U=\frac{1}{\sqrt{2}}\gl e^{-\frac{i\pi}{4}}S_R+e^{\frac{i\pi}{4}}S_L\gr\ ,\quad S_E=\begin{pmatrix}1&0\\0&i\end{pmatrix}\ .
\ee
With
\bel{SO7}
U=\frac{1}{\sqrt{2}}e^{-\frac{i\pi}{4}}\gl S_R+iS_L\gr\ ,\quad U^2=1\ ,
\ee
the eigenvalues of $U$ are $\pm1$. All eigenvalues of $\Shat$ can be represented as products of eigenvalues of $U$ and $S_E$, which is a necessary condition for the possibility of a direct product form~\eqref{SO5}.

The unitary matrix $D_U$ has to obey
\bel{SO8}
\begin{pmatrix}U&0\\0&iU\end{pmatrix}=D_U\begin{pmatrix}S_R&0\\0&S_L\end{pmatrix}D_U^{-1}\ .
\ee
This matrix has to exist because the eigenvalues of $\Shat$ and $\Stil$ coincide. It is obvious that the evolution operator $U$ for the subsystem is no longer a unique jump matrix, in contrast to $\Shat$. Also $U$ is not a real matrix anymore. This simple example demonstrates the emergence of more general unitary step evolution operators for subsystems.

\subsection*{Coarse graining in momentum space}

An example illustrating the change of character of the step evolution operator on the way towards the continuum limit of one-particle states is given by coarse graining in momentum space. Coarse graining in momentum space can be achieved by lowering the resolution or by eliminating high momentum modes. We give here examples in the context of the one-particle states. In the momentum basis the step evolution operator for the propagation of one-particle states (without disorder)~\eqref{SP11}~\eqref{SP13} reads
\bel{CG1}
\begin{split}
S'=&\begin{pmatrix}S_R'&0\\0&S_L'\end{pmatrix}\ ,\\S_R'=&\diag\gl\exp\{-\frac{2\pi iq}{M_x}\}\gr\ ,\ S_L'=\diag\gl\exp\{\frac{2\pi iq}{M_x}\}\gr\ .
\end{split}
\ee
The diagonal form of the step evolution operator in the momentum basis facilitates greatly the construction of subsystems since it permits directly to implement relations for eigenvalues~\eqref{D19}.

Our first example for coarse graining lowers the resolution in momentum space. We employ 
\bel{CG2}
S_R'=S_{RS}'\otimes S_{RE}'\ ,\quad S_L'=S_{LS}'\otimes S_{LE}'
\ee
with
\bel{CG3}
S_{RE}'=\begin{pmatrix}1&0\\0&\exp\gl-\frac{2\pi i}{M_x}\gr\end{pmatrix}\ ,\ S_{LE}'=\begin{pmatrix}1&0\\0&\exp\gl\frac{2\pi i}{M_x}\gr\end{pmatrix}\ .
\ee
The step evolution operators for the subsystem $S_{RS}'$ and $S_{LS}'$ remain in the form~\eqref{CG1}, but the range of $q$ only covers even $q$ such that we deal now with reduced $M_x/2\times M_x/2$ matrices.

We may translate the coarse grained subsystem back to position space. Since the wave function of the subsystem has only half the number of components as for the original system, the same has to hold for the subsystem in position space. We achieve this by restricting the Hilbert space in position space to functions that are periodic in $M_x/2$ instead of $M_x$. For this purpose we turn back to arbitrary integers $\qtil=q/2$, replacing the factors $2\pi iq/M_x$ in eqs.~\eqref{CG1} by $4\pi i\qtil/M_x$ or $2\pi i\qtil/(M_x/2)$, $|\qtil|\leq M_x/4$. A separate Fourier transform~\eqref{SP8} to position space for the right-movers and left-movers, with
\bel{CG4}
\begin{split}
D_R'=&\sqrt{\frac{2}{M_x}}\exp\gl-\frac{2\pi i}{M_x}\qtil(2m-1)\gr\\
=&\sqrt{\frac{2}{M_x}}\exp\gl-i\eps p\qtil(2m-1)\gr\ ,
\end{split}
\ee
and similar for $D_L'$ with $2m-1$ replaced by $2m$, leads again to the step evolution operators~\eqref{SP3}, now with periodic $m$, $n$ restricted to $|m|\leq M_x/4$. The only effect of this type of coarse graining is the identification of points $m$ and $m+M_x/2$. The subsystem discards the information contained in the part of the wave function that does not have the restricted periodicity. This way of coarse graining keeps right-movers and left-movers separate.

A second possibility of coarse graining in momentum space restricts the momentum range of the subsystem by taking
\bel{CG5}
S_L'=S_E'\otimes\Shat_L\ ,\quad S_R'=S_E'\otimes\Shat_R\ ,\quad S_E'=\begin{pmatrix}1&0\\0&-1\end{pmatrix}\ .
\ee
The $M_x/2\times M_x/2$ matrix $\Shat_L$ has the same diagonal form as $S_L'$ in eq.~\eqref{CG1}, with values of $q$ now restricted to the interval $[_+0,M_x/2]_-$, where the upper value is not included. The missing eigenvalues of $S_L'$ not contained in the spectrum of $\Shat_L$ obtains from the latter by multiplication with $(-1)$. For $\Shat_R$ we find the same eigenvalues as for $\Shat_L$ if we restrict the $q$-interval to the negative of the $q$-interval for $\Shat_L$. In view of the periodicity of $q$ we could have chosen for the coarse graining a different range of $q$-values characterizing the subsystem. For a range symmetric around zero the subsystem corresponds to a Hilbert space with less resolution. This type of subsystem averages effectively over neighboring points in position space.

We have chosen the ranges of $q$ for the subsystem in order to describe a subsystem that mixes right- and left-movers. Indeed, we can keep the resolution and cover all values of $q$ either by $\Shat_L$ or $\Shat_R$ by shifting $q$ for $\Shat_R$ by one unit, defining
\bel{CG6}
\Shat_R=\diag\gl\exp\big\{-\frac{2\pi i(q+1)}{M_x}\big\}\gr\ .
\ee
The interval for $q$ for $\Shat_R$ is then given by $[_+-\frac{M_x}{2},0]_-$. We can combine $\Shat_L$ and $\Shat_R$ into $S_S'$, such that
\bel{CG7}
S'=S_E'\otimes S_S'\ ,
\ee
with
\bel{CG8}
\begin{split}
S_S'(q,q')=\Big[&\exp\gl\frac{2\pi iq}{M_x}\gr\theta(q)\theta\gl\frac{M_x}{2}-1-q\gr\\
+&\exp\gl-\frac{2\pi i(q+1)}{M_x}\gr\theta\gl\frac{M_x}{2}+q\gr\theta(-1-q)\Big]\delta_{qq'}\ .
\end{split}
\ee
The range of $q$-values of $S_S'$ covers the whole periodic range of $M_x$ values with $|q|\leq M_x/2$.

An inverse Fourier transform yields the coarse grained evolution operator in a (periodic) position basis, $|m|\leq M_x/2$,
\bel{CG9}
S_S(m,n)=D\herm(m,q)S'(q,q')D(q',n)\ ,
\ee
where
\bel{CG10}
D(m,q)=\frac{1}{\sqrt{M_x}}\exp\big\{-\frac{2\pi iqm}{M_x}\big\}\ .
\ee
Performing the sums in the appropriate ranges yields
\bel{CG11}
\begin{split}
S_S(m,n&)=\frac{1}{M_x}\Big[\sum_{q=0}^{M_x/2-1}\exp\big\{-\frac{2\pi iq}{M_x}(m+1-n)\big\}\\
&+\exp\gl-\frac{2\pi i}{M_x}\gr\sum_{q=-M_x/2}^{-1}\exp\big\{\frac{2\pi iq}{M_x}(m-1-n)\big\}\Big]\ .
\end{split}
\ee
This evolution operator is no longer a unique jump operator. As expected, particles can move to the left or to the right with certain probabilities.

This can be seen explicitly for the case $M_x=4$ where one finds
\bel{CG12}
S_S=\frac12\begin{pmatrix}1+i&1&0&-i\\-i&1+i&1&0\\0&-i&1+i&1\\1&0&-i&1+i\end{pmatrix}\ ,
\ee
with $S_S^2=U$ as given by eq.~\eqref{SO7}. With eigenvalues $(1,i)$ identical to $S_S'$ one has $S_S^4=1$. This unitary evolution operator yields a certain probability that the particle stays at its position $(n=m)$, and somewhat smaller probabilities that it moves to the right $(m=n+1)$ or to the left $(m=n-1)$.

For general $M_x$ one can write $S_S(m,n)$ in the form
\begin{align}
\label{CG13}
S_S(m,n&)=\frac{1}{M_x}\sum_{q=0}^{M_x/2-1}\exp\big\{\frac{2\pi iq}{M_x}\big\}\Big[\exp\big\{\frac{2\pi iq}{M_x}(m-n)\big\}\nn\\
&+\exp\big\{-\frac{2\pi i}{M_x}\big\}\exp\big\{-\frac{2\pi iq}{M_x}(m-n)\big\}\Big]\ .
\end{align}
In particular, the diagonal elements read
\bel{CG14}
S_S(m,m)=\frac{2}{M_x}\sum_{q=0}^{M_x/2-1}\exp\gl\frac{2\pi iq}{M_x}\gr\ ,
\ee
and for the neighboring off-diagonal elements one finds
\bel{CG15}
\begin{split}
S_S(m,m+1)=&\frac{1}{M_x}\sum_{q=0}^{M_x/2-1}\Big[1+\exp\big\{\frac{2\pi i}{M_x}(2q+1)\big\}\Big]\ ,\\
S_S(m,m-1)=&\exp\gl-\frac{2\pi i}{M_x}\gr S_S(m,m+1)\ .
\end{split}
\ee
This demonstrates that the coarse grained step evolution operator $S_S$ is no longer a unique jump matrix.

\subsection*{Coarse graining and quantum formalism}

The coarse grained subsystem is no longer an automaton with a deterministic evolution. The step evolution operator remains unitary but is no longer a unique ump matrix. We can express the step evolution operator for the coarse grained subsystem again in the fermion language. The general bit-fermion map is defined for a general form of the step evolution operator. We will not do this exercise explicitly here. The general outcome coincides with our observation that the naive continuum limit in the fermionic picture loses the automaton property.

The quantum formalism for classical statistics, based on wave functions and density matrices, is a central ingredient for the construction of subsystems. No similar construction exists for probability distributions. For the coarse graining of the density matrix this becomes apparent since the time evolution of the coarse grained subsystem involves the off-diagonal elements of the density matrix. It cannot be formulated on the level of the diagonal elements, which encode a probability distribution on the coarse grained level. Similarly, the coarse graining in time by construction of an effective averaged Hamiltonian involves an object acting on wave functions. We conclude that an encoding of the time-local probabilistic information in wave functions or the density matrix becomes crucial for the construction of appropriate subsystems that lead towards the continuum limit.

\section{Complex structure and quantum mechanics}
\label{sec: CSQM}

We have expressed probabilistic cellular automata as discrete quantum field theories for fermions. These are quantum systems with all the formalism of quantum theory. The usual quantum mechanics for a single particle obtains from the quantum field theory as the special case of a one-particle excitation of some vacuum state. Furthermore, one deals with a continuum limit in space and time. The detailed dynamics of the one-particle quantum system will depend both on the updating rule for the automaton as reflected by the action in the Grassmann functional integral for fermions, and on the properties of the vacuum state. Usual quantum mechanics for a single particle is formulated in terms of a complex wave function or complex density matrix. We therefore have to specify the implementation of a complex structure.

There are different possible complex structures that are compatible with the evolution. For all of them the orthogonal step evolution operator of the automaton is mapped to a unitary matrix, guaranteeing the unitary evolution of the quantum system. We have encountered in sect.~\ref{sec: SPWF} a complex structure based on the particle-hole transformation, while for the Grassmann variables a complex structure has mapped two Majorana fermions to a Dirac fermion in sect.~\ref{sec: ANA}. These complex structures can be combined, and we will use the combined structure for a description of charged particles.

Let us start with the complex conjugation $K_c$ based on the particle-hole transformation. For every configuration $\tau'$ of the occupation numbers at $t$ we define the particle-hole conjugate configuration $\tau^c$ by exchanging occupied and non-occupied bits, $n\g(x)\leftrightarrow \gl1-n\g(x)\gr$. In a formulation with Ising spins $s\g(x)$ the particle-hole transformation flips the spin of all $s\g(x)$. The particle-hole transformation $K_c$ is an involution, $K_c^2=1$. Correspondingly, we can define the particle-hole transformation for the wave function of the probabilistic automaton $q(t)$,
\bel{CSQ1}
K_c:\ q(t)\mapsto q^c(t)\ ,\quad q^c_{\tau'}(t)=q_{\tau^c}(t)\ .
\ee
The $\tau'$-component of $q^c$ is given by the $\tau^c$-component of $q$.

We can split the configurations $\{\tau\}$ into two parts $\{\tau'\}$ and $\{\tau^c\}$ which are mapped onto each other by the particle-hole transformation~\cite{CWNEW}. Let us consider particle-hole invariant automata for which the particle number is conserved. We focus on a half-filled vacuum or ground state. If the configurations with one additional occupied bit belong to $\{\tau'\}$, the configurations with one missing occupied bit or additional hole belong to $\{\tau^c\}$. We will combine these configurations into a generalized one-particle sector. In this sector we can write the one-particle wave function in a two-component notation
\bel{CSQ2}
q_{\tau'}^{(1)}=\pvec{q_{\tau'}}{q_{\tau'}^c}\ ,\quad q^{(1)}=\pvec{q'}{q^c}\ ,
\ee
where $\tau'=(\gamma,x)$ denotes the different possibilities to have the generalized particle of type $\gamma$ at the position $x$. We can choose signs such that
\bel{CSQ3}
K_cq^{(1)}=\pmat{0}{1}{1}{0}q^{(1)}\ ,
\ee
with a block-diagonal step evolution operator
\bel{CSQ4}
\Shat^{(1)}q^{(1)}=\pmat{\Shat'}{0}{0}{\Shat'}q^{(1)}\ ,
\ee
given by the restriction of $\Shat$ to the one-particle sector.
The map of $q^{(1)}$ to a complex vector $\vp$ by
\bel{CSQ5}
\vp_{\tau'}=\frac1{\sqrt{2}}\big[q_{\tau'}+q^c_{\tau'}+i\gl q_{\tau'}-q^c_{\tau'}\gr\big]
\ee
identifies the multiplication with $i$ to a matrix multiplication with
\bel{CSQ6}
I_c=\pmat{0}{1}{-1}{0}\ ,\quad \vp\gl I_cq^{(1)}\gr=i\vp\gl q^{(1)}\gr\ .
\ee

A second possible complex structure associated to the combination of two Majorana spinors into a Dirac spinor can be formulated if the species of particles $\gamma$ can be divided into two classes, $\gamma=(\eta,\delta)$, $\eta=1,2$. Writing in the one particle sector
\bel{CSQ7}
q^{(1)}=\pvec{q_1}{q_2}=\pvec{q_{\eta=1,\delta}}{q_{\eta=2,\delta}}\ ,
\ee
the complex conjugation $K_M$ reverses the sign of $q_2$,
\bel{CSQ8}
K_Mq^{(1)}=\pmat{1}{0}{0}{-1}q^{(1)}=\pvec{q_1}{-q_2}\ .
\ee
Defining a complex wave function by
\bel{CSQ9}
\vp=q_1+iq_2
\ee
the multiplication by $i$ is realized by the map $I_M$,
\bel{CSQ10}
I_M=\pmat{0}{-1}{1}{0}\ ,\quad \vp\gl I_M q^{(1)}\gr=i\vp\gl q^{(1)}\gr\ .
\ee
We can identify $q_1$, $q_2$ with the real wave functions for two Majorana spinors, and $\vp$ with the complex wave function for a Dirac spinor. Both the pairs $(K_c,I_c)$ and $(K_M,I_M)$ obey the defining relations for a complex structure,
\bel{CSQ11}
K^2=1\ ,\quad I^2=-1\ ,\quad \big\{K,I\big\}=0\ .
\ee

A combined complex structure can be defined by
\bel{CSQ12}
K:\ q_1'\leftrightarrow q_1^c\ ,\quad q_2'\leftrightarrow-q_2^c\ .
\ee
This involves for $q_2$ an additional minus sign as compared to $K_c$ in eq.~\eqref{CSQ3}. Consistent with the complex conjugation we can define two complex wave functions
\begin{align}
\label{CSQ13}
\vp_+=&\frac1{\sqrt{2}}\gl q_1'+q_1^c+i(q_2'+q_2^c)\gr\nn\\
\vp_-=&\frac1{\sqrt{2}}\gl q_2'-q_2^c-i(q_1'-q_1^c)\gr\ .
\end{align}
The multiplication by $i$ is achieved by the map $I$
\begin{align}
\label{CSQ14}
I:\ &q_1'\mapsto-q_2'\ ,\quad q_2'\mapsto q_1'\ ,\nn\\
&q_1^c\mapsto-q_2^c\ ,\quad q_2^c\mapsto q_1^c\ .
\end{align}
In a direct product notation one has
\bel{CSQ15}
K=K_M\otimes K_c\ ,\quad I=I_M\otimes1\ .
\ee

One can impose a Majorana constraint
\bel{CSQ16}
q_1^c=q_1'\ ,\quad q_2^c=q_2'\ ,\quad \vp_-=0\ .
\ee
In this case only the Dirac spinors $\vp_+$ remain.
This choice corresponds to the identification $\psibar_a=\psi_a$ in eq.~\eqref{MF1}.  In the following we will focus on the complex structure~\eqref{CSQ15} with the constraint~\eqref{CSQ16}. In this case the one-particle wave function is particle-hole invariant, $q^c=q$. We can concentrate on $q$ for which the complex structure is given by $(K_M,I_M)$ once we keep in mind the identification $q^c=q$. The momentum eigenstates are now periodic oscillations between the two species.

As an example we take the Gross-Neveu model corresponding to $N=2$ in sect.~\ref{sec: ANA}. The one-particle wave function has two components $\vp_R$ and $\vp_L$ for the right- and left-movers. For the continuum limit in the absence of interactions the evolution is given by the Schrödinger equation
\bel{CSQ17}
i\partial_t\vp=H\vp=\widehat P\tau_3\vp\ ,\quad \widehat P=-i\partial_x\ .
\ee
The general solution reads
\begin{align}
\label{CSQ18}
\vp_R\otx=\int_p e^{-ip(t-x)}\vp_R(p)\ ,\nn\\
\vp_L\otx=\int_p e^{-ip(t+x)}\vp_L(p)\ .
\end{align}
We associate the Fouries components to particles with positive energy, $H\vp_{R,L}=p\vp_{R,L}$. For the components with negative $p$ we consider the complex conjugate wave functions $\vp^*_{R,L}\otx$. The corresponding components have again positive energy and are associated with antiparticles. We recall that complex conjugation is realized by the modified particle-hole transformation~\eqref{CSQ12}. The description of particles and antiparticles is related directly to the presence of particles and holes in the generalized one-particle wave function. We observe that the complex conjugation of eqs.~\eqref{CSQ17},~\eqref{CSQ18} can also be achieved by the simultaneous reversal of the time and space coordinate. This reflects the $CPT$-theorem of quantum field theory.

We can keep the same complex structure in the presence of interactions. Depending on the ground state or vacuum this may introduce a mass term and potential in the Schrödinger equation, replacing eq.~\eqref{CSQ17} by eq.~\eqref{243S}.

\section{Conclusions}
\label{sec: C}

This paper addresses basic properties of the continuum limit of reversible cellular automata for a very large number of cells. For this purpose we develop the fermion picture for large classes of cellular automata, or more general automata. This maps probabilistic automata to quantum field theories for fermions for which powerful methods exist for the computation of the continuum limit. The bit-fermion map from an automaton which processes bit-configurations to an equivalent fermion model is not limited to a particular dimension or a particular type of model. It exists whenever the global updating rule for the automaton is known. The equivalence with a fermionic quantum model helps to formulate the conceptual steps towards the continuum limit on the level of the propagation of the probabilistic information for the automaton. We have found that the quantum formalism for classical statistics, which introduces wave functions and the density matrix, is crucial for the definition of coarse grained subsystems on the way towards the continuum limit. On the coarse grained level the automaton property of an updating to a unique bit configuration is lost. It is replaced by a more general unitary evolution where a given coarse grained configuration can evolve with certain probabilities to many other configurations.

Our focus has been on rather simple reversible automata which allow for a straightforward interpretation in terms of propagating and interacting fermions. Despite the simplicity of the automata rather interesting discrete quantum field theories for interacting fermions have been found. This includes fermionic models with local gauge symmetries as, for example, local Lorentz symmetry. The very simple realization of local gauge symmetries by an automaton may be an interesting direction for the understanding of the fundamental interactions in nature~\cite{CWSLGT}. Several of our models have an interesting naive continuum limit, as a Lorentz-invariant two-dimensional model with Thirring- or Gross-Neveu-type interactions with abelian or non-abelian continuous symmetries, or two-dimensional spinor gravity with diffeomorphism symmetry and local Lorentz symmetry. The aim of this part of the investigation is to show that fermionic quantum field theories which admit a cellular automaton formulation are particular, but not limited to very special cases concerning the degrees of freedom and the symmetries of possible models.

The continuum limit requires a probabilistic setting for the cellular automaton. A continuous probability distribution or wave function for the bit-configurations may undergo only (\qq{infinitesimally}) small changes for an individual updating step, even though the change of the bit-configuration is discrete and cannot be taken as infinitesimally small. If this is the case, the discrete time evolution of the wave function by the step evolution operator turns to a differential time evolution equation. This type of continuum limit is not possible for deterministic cellular automata characterized by sharp bit-configurations at every discrete time step. For deterministic automata one could get a continuous evolution by interpolating between the discrete updating steps of the automaton. The discrete type of the evolution in fixed time intervals remains, however, in this case.

For typical continuous processes in nature no discrete time evolution of the type of an automaton is observed. For a description of this situation by an automaton the time step of the updating should be much smaller than the characteristic time for the change of observable quantities. Small changes of quantities during a single updating step are rather easily understood in terms of small changes of the probabilistic information. The expectation value for a particular occupation number can change continuously, while a sharp value can only jump between zero and one.

For the type of probabilistic automaton considered in this work the probabilistic aspect only enters by a probability distribution over initial bit-configurations. The updating remains the deterministic rule of the automaton. This contrasts the stochastic cellular automata for which the updating to different new configurations is done with certain probabilities. The latter are Markov chains, for which the initial information is diluted as the updating progresses with many steps. In our case the step evolution operator remains an orthogonal matrix such that the initial information is not lost. In the presence of a complex structure which is compatible with the evolution the time evolution of the complex wave function or density matrix is unitary. We may call this type of information preserving probabilistic setting a unitary probabilistic automaton. Of course, an orthogonal or unitary evolution does not forbid an approach to an \emph{effective} equilibrium. The issue is similar to the approach to thermal equilibrium of a pure many body quantum state. An approach to equilibrium is not mandatory, however, in contrast to many Markov chains for which the probability distribution converges to an equilibrium distribution.

The unitary probabilistic automata are discrete quantum systems, as reflected by the associated discrete fermionic quantum field theories. Quantum mechanics emerges from classical statistics~\cite{CWQMCS, CWEM}. A priori, we deal with real quantum mechanics, but a suitable complex structure which is appropriate for standard complex quantum mechanics is often found. In particular, we have indicated a rather general complex structure based on the particle-hole transformation. It is compatible with the evolution whenever the updating respects particle-hole symmetry.

A central issue for which we only have taken first steps concerns the continuum limit. A priori, it is not clear if the naive continuum limit reported here for several fermionic models corresponding to cellular automata reflects the main characteristics of the true continuum limit. Beyond the expected renormalization of couplings for models with interactions also qualitative questions arise. For typical continuum fermionic quantum field theories with interactions, including the ones for the naive continuum limit of our models, the time evolution for arbitrary discrete evolution steps is not of the type of an automaton. A given bit configuration does not evolve to a new fixed configuration. The step evolution operator remains unitary, but it is no longer a unique jump matrix. With more than one non-zero element in a column a given component of the wave function at $t$ contributes to more than one component at $t+\eps$.

The continuum limit has to account for this change of character of the step evolution operator. This will bring the cellular automata even closer to standard quantum mechanics. In generic quantum systems the step evolution operator is unitary, but not a unique jump matrix. The change of character of the step evolution operator can be seen by coarse graining. We have given several examples in section~\ref{sec: CLCGS}. For the coarse grained evolution we find indeed that the unitary evolution can be maintained, while the unique jump character of the step evolution operator is not preserved. The quantum formalism for classical statistics is a key ingredient for the construction of coarse grained subsystems.

Explicit coarse graining of the step evolution operator can become rather cumbersome in the presence of interactions. It is, fortunately, not the only way to perform a continuum limit for a discrete quantum field theory. In quantum field theories the relevant information can be extracted from expectation values and correlation functions, or the associated effective action as a generating functional. The continuum limit can be established by integrating out the short-distance fluctuations. Powerful renormalization group techniques exist for this purpose. They are typically much easier to handle than coarse graining of the Hamiltonian or step evolution operator.

For establishing a continuum limit for probabilistic cellular automata our formulation as discrete quantum field theories, either in a fermionic language for Grassmann variables or in a setting for occupation numbers or Ising spins, may prove very useful. This will reveal if all the interesting features of the naive continuum limit indeed emerge for the very simple automata discussed in this paper. Fortunately, probabilistic cellular automata can be explored numerically. This should allow for direct tests of our proposal.

\subsection*{Note added}

Since the first version of this paper a probabilistic cellular automaton for spinor gravity in four dimensions has been proposed in ref.~\cite{CWCASG}. The description of a quantum particle in a potential by a probabilistic cellular automaton has been developed further in ref.~\cite{CWPCAQP}.

\subsection*{Acknowledgment}

This work has been supported by the DFG collaborative research center SFB 1225 ISOQUANT and by the DFG excellence cluster \qq{STRUCTURES}.

\begin{appendices}

\appendix


\section{One-dimensional cellular automata, \\local neighbors and updating}
\label{app: A1}

In this appendix we establish that all chain automata are one-dimensional cellular automata. For this purpose we need to establish the neighbors of the cell $x$ in the sense of the cellular automaton and to determine the updating rule for each cell $x$. Our aim is to start from a given fermionic model encoded in $L(x)$ and to extract the corresponding updating rule. Indeed, the local elementary processes encoded in $L(x)$ determine the neighbors of a given cell and specify its updating rule.

For a start we discuss the particular simple case where $L(x)$ only involves Grassmann variables $\psi\g(x+\eps)$ and $\psibar_\delta(x)$. The same holds for
\bel{CA1}
K(x)=\exp\gl-L(x)\gr\ ,
\ee
and we can write
\bel{CA2}
K(x-\eps)=g_\sigma[\psi(x)]B_{\sigma\eta}(x)g_\eta'[\psibar(x-\eps)]\ .
\ee
The Grassmann elements $g_\sigma[\psi(x)]$ and $g_\eta'[\psibar(x-\eps)]$ are polynomials of the local Grassmann variables $\psi\g(x)$ and $\psibar_\delta(x-\eps)$, respectively. The overall local factor can be written as
\bel{CA3}
\Ktil=K(x-\eps)\cR(x)\ ,
\ee
where $\cR(x)$ does neither involve the variables $\psi\g(x)$ nor $\psibar_\delta(x-\eps)$.

We take the matrix $B$ as a restricted unique jump matrix, with a possible extension to a complete scattering automaton not discussed here explicitly. If for a given $\sigma$ all elements $B_{\sigma\eta}(x)$ vanish, no term in $\Ktil$ can contain the particular product of Grassmann variables $g_\sigma[\psi(x)]$. This implies that the bit-configuration corresponding to $g_\sigma[\psi(x)]$ cannot be generated in the updating step. All bit configurations that have in the cell $x$ the values of bits corresponding to $g_\sigma[\psi(x)]$ have to be excluded in this case, restricting the space of allowed states after the updating. On the other hand, if for a given $\eta$ all elements $B_{\sigma\eta}$ vanish, there is no bit-configuration to which the bit configuration $\eta$ in the cell $x-\eps$ could be mapped. Thus the states $\eta$ in the cell $x-\eps$ have to be excluded. The remaining (not excluded) states define the restricted matrix $B$. Invertibility requires that $B$ is a square matrix. Furthermore, for a fixed $\eta$ within the restricted set there can be only a single non-zero element $B_{\sigma\eta}$ with $\sigma=\sigma(\eta)$. Otherwise there would not be a unique updating. This has also to hold in the inverse direction and we conclude that $B$ is indeed a restricted unique jump matrix.

The cellular automaton structure is simple. The only neighbor of the cell $x$ (in the sense of the updating rule) is the cell $x-\eps$. If the cell $x-\eps$ has at $t$ the bit-configuration $\eta$, the cell $x$ has after the updating the bit configuration $\sigma(\eta)$, as specified by the single non-zero element of $B_{\sigma\eta}$ for a given $\eta$. We observe that the form of the remaining factor $\cR(x)$ in eq.~\eqref{CA3} is irrelevant for the updating of the cell $x$. It has no influence on the state of the cell $x$ after the updating.

For a more complex local chain there will be several factors $K(y_i)$ in which the variables $\psi\g(x)$ appear. The corresponding cells $y_i(x)$ are the neighbors of $x$ in the sense of the updating. We define
\bel{CA4}
K_N(x)=\prod_i K\gl(y_i(x)\gr\ ,\quad \Ktil=K_N(x)\cR(x)\ ,
\ee
where $\cR(x)$ does not involve the variables $\psi\g(x)$. We can write $K_N(x)$ in a form similar to eq.~\eqref{CA2}
\bel{CA5}
K_N(x)=\sum_{\sigma,\eta}g_\sigma[\psi(x)]B_{\sigma\eta}(x)g_\eta'\big[\psibar\gl y_i(x)\gr\big]C_\eta[\psi(z)]
\ee
where $\eta$ denotes now the combined bit-configuration in all cells $y_i(x)$. (Eq.~\eqref{CA2} is the special case of a single neighbor $y(x)=x-\eps$.) For more than one neighboring cell the number of bit-configurations $\eta$ is much larger than the number of bit-configurations $\sigma$. For example, for $M$ bits in a cell one has $2^M$ states $\sigma$, and $2^{2M}$ states $\eta$ in case of two neighbors. The factor $B_{\sigma\eta}$ can vanish for some of the configurations $\eta$. They can be connected to bit configurations in cells different from $x$. Furthermore, a given $g'_\eta$ may not only be multiplied by factors of $\psi(x)$, but also by Grassmann variables $\psi(z)$ in different cells. All this is accounted for by the factor $C_\eta[\psi(z)]$ which involves Grassmann variables in cells $z$ different from $x$.

We can now repeat similar steps as before. We first restrict the states $\sigma$ such that $B_{\sigma\eta}$ has at least one non-zero element. The unique jump property of the overall automaton implies that for every $\eta$ there cannot be more than one $\sigma$ for which $B_{\sigma\eta}$ differs from zero. Indeed, one cannot have for $\sigma'\neq\sigma$ a term in $K_N$ of the form
\bel{CA6}
a_\eta=\gl cg_\sigma[\psi(x)]+c'g_{\sigma'}[\psi(x)]\gr g_\eta'\big[\psibar\gl y_i(x)\gr\big]C_\eta[\psi(z)]\ ,
\ee
for both non-zero $c$ and $c'$. Employing the product from~\eqref{CA4} we would get a contribution where $a_\eta$ is multiplied by the same Grassmann element $g_\cR[\psibar\,]$ formed from the factors $\psibar$ in $\cR$. The product $(g_\eta'g_\cR[\psibar\,])$ defines the bit configuration before the updating. The overall unique jump property of the automaton requires that in $\Ktil$ this is multiplied by a unique Grassmann element $g[\psi]$, contradicting~\eqref{CA6}. For the configurations $\eta$ with non-zero $B_{\sigma\eta}$ one has a unique map $\eta\to\sigma(\eta)$ which defines the updating in the cellular automaton.

What is different from the first example of a single neighbor is that a given $\sigma$ does not need to be produced by the same $\eta$. Two different $\eta$, $\eta'$ can produce the same configuration in the cell $x$, $\sigma(\eta)=\sigma(\eta')$. This does not contradict overall invertibility if $C_\eta$ differs from $C_{\eta'}$. Knowing the updating map $\sigma(\eta)$ does not allow for a simple direct assessment if the cellular automaton is invertible or not. In our approach invertibility is guaranteed by the unique jump property of the step evolution operator $\Shat$ for the overall automaton.


\section{Conjugate and complex Grassmann variables}
\label{app: A}

In this appendix we briefly discuss a possible complex structure for Grassmann variables that is based on the notion of conjugate Grassmann variables. Here $\psibar\g(x)$ are considered as the variables conjugate to $\psi\g(x)$, as motivated by the modulo two property of Grassmann functional integrals. A particular complex structure brings certain symmetries which mix $\psi$ and $\psibar$ into the form of simple complex transformations.

We may combine the Grassmann variables $\psig$ and $\psibarg$ into a complex Grassmann variable
\bel{SY11}
\zetal\g(x)=\frac12\Big\{\psig+\psibarg+i\big[\psig-\psibarg\big]\Big\}\ ,
\ee
such that the exchange $\psi\leftrightarrow\psibar$ results in the complex conjugation $\zetal\leftrightarrow\zetal^*$. The transformation~\eqref{SY9} results in
\bel{SY12}
\delta\zetalg=i\eps\zetalg\ ,\quad \delta\zetalgg=-i\eps\zetalgg\ .
\ee
It is therefore compatible with this particular complex structure. We recognize in eq.~\eqref{SY12} a global phase rotation of all complex Grassmann variables $\zetalg$. Invariance of $\cL$ under this global $U(1)$-symmetry requires that reach term involves an equal number of factors $\zetal_\alpha=\zetalg$ and $\zetal_\beta^*=\zetal_\delta^*(y)$. With
\bel{SY13}
-i\zetal_\alpha\zetal_\beta^*=\frac12\gl\psi\al\psibar\bet+\psi\bet\psibar\al\gr-\frac i2\gl\psi\al\psi\bet+\psibar\al\psibar\bet\gr\ ,
\ee
\and\bel{SY14}
-i\gl\zetal\al\zetal\bet^*+\zetal\bet\zetal\al^*\gr=\psi\al\psibar\bet+\psi\bet\psibar\al\ ,
\ee
we recover eq.~\eqref{SY10}. For $\alpha=\beta$ one has (no sum over $\alpha$ here)
\bel{SY15}
-i\zetal\al\zetal\al^*=\psi\al\psibar\al\ .
\ee

We may also express the infinitesimal transformation~\eqref{SY6} in terms of $\zetal$, resulting in
\bel{SY16}
\delta\zetal\al=i\eps\zetal\al^*\ ,\quad\delta\zetal\al^*=-i\eps\zetal\al\ .
\ee
This transformation mixes $\zetal$ and $\zetal^*$. We may associate a conserved charge $\Qtil$ to the $U(1)$-symmetry~\eqref{SY9}~\eqref{SY12}. It differs from the total particle number which is associated to the symmetry~\eqref{SY6}. The combination~\eqref{SY14} is invariant under both symmetries~\eqref{SY12} and~\eqref{SY16}. The transformation~\eqref{SY9} and the associated charge $\Qtil$ will not play an important role in this paper. The reason is that the simple transport operators are not invariant. We typically will also employ complex structures for Grassmann variables that are different from eq.~\eqref{SY11}.


\section{Higher order terms for automata with global continuous symmetries}
\label{app: B}

This appendix gives an example how to construct the correction terms $\Delta L$ in the fermionic action which are needed in order to guarantee that the local factor $K$ generates a unique jump step evolution operator. A first example for the complete local scattering automata~\eqref{CS3} is given by eq.~\eqref{TS3}. In this appendix we discuss the SO(4)-invariant automaton~\eqref{NY1}-~\eqref{NY3}.

The \qq{correction term} $\Delta L$ in eq.~\eqref{NY1} contains terms with six or more Grassmann variables. We present here a few features of its $SO(4)$-invariant construction. One part of $K'$ in eq.~\eqref{NY13} is given by $K_2$
\bel{NY14}
K'=K_2+K''\ ,
\ee
where the part $K_2$ generates a unit step evolution operator in the sector where no pairs of right- and left-movers are present. These are the states with only right-movers or only left-movers. (The unit evolution of the totally empty state is not included in $K_2$.) We write
\bel{NY15}
K_2+K_2'=\exp\gl-L_0+\Delta K+\Delta K_2\gr-1\ ,
\ee
with
\bel{NY16}
\Delta K=-L\subt{int}-K_{2,2}=\psi\Ra\psi\Lb\psibar\Ra\psibar\Lb\ .
\ee
Similarly to $K_2$ the term $\Delta K_2$ does not contribute to scattering. It only serves to guarantee a unit evolution for all sectors with only right-movers or left-movers. The term $K_2'$ stands for contributions from the expansion of the exponential beyond the sectors of only right-movers or left-movers. It is part of $K''$ in eq.~\eqref{NY14}.

For example, the quadratic term in the expansion of $\exp\gl-L_0+\Delta K\gr$,
\bel{NY17}
\frac12L_0^2+\Delta K=-\frac12\gl\psi\Ra\psi\Rb\psibar\Ra\psibar\Rb+\Psi\La\psi\Lb\psibar\La\psibar\Lb\gr\ ,
\ee
yields in the two-particle sector the unit evolution of two right-movers or two left-movers. In the three-particle sector the same expansion yields a term
\begin{align}
\label{NY18}
-\gl\frac16L_0^3+L_0&\Delta K\gr=-\frac16\psi\Ra\psi\Rb\psi_{Rc}\psibar\Ra\psibar\Rb\psibar_{Rc}\nn\\
+&\frac12\psi\Ra\psi\Rb\psi_{Lc}\psibar\Ra\psibar\Rb\psibar_{Lc}+R\leftrightarrow L\ .
\end{align}
The first term generates the unit evolution in the sector with three right-movers or three left-movers and contributes to $K_2$. The second term corresponds to a unit evolution in the sectors with two right-movers and one left-mover, or two left-movers and one right-mover. It contributes to $K_2'$ and therefore to $K''$.

In the four-particle sector the expansion of $\exp\gl-L_0+\Delta K\gr$ contributes
\begin{align}
\label{NY19}
\frac1{24}&L_0^4+\frac12L_0^2\Delta K+\frac12\Delta K^2=\nn\\
\frac1{24}&\gl\psi\Ra\psi\Rb\psi\Rc\psi\Rd\psibar\Ra\psibar\Rb\psibar\Rc\psibar\Rd+R\leftrightarrow L\gr\nn\\
-\frac13&\gl\psi\Ra\psi\Rb\psi\Rc\psi\Ld+\psibar\Ra\psibar\Rb\psibar\Rc\psibar\Ld+R\leftrightarrow L\gr\nn\\
-\frac14&\psi\Ra\psi\Rb\psi\Lc\psi\Ld\psibar\Ra\psibar\Rb\psibar\Lc\psibar\Ld\ .
\end{align}
The first term yields the unit evolution in the sector of four right-movers or four left-movers, as appropriate for $K_2$. The second term contributes to the unit evolution in the sector with three right-movers and one left-mover, or right and left interchanged. The coefficient $-1/3$ is, however, twice the coefficient needed for a unit evolution, such that a corresponding correction is needed in $\Delta K_2$. Finally, the third term would yield a unit evolution in the sector with two right-movers and two left-movers. Such a term should not be present, since we have already the scattering contribution $K_{4,4}$ in this sector. The third term has therefore to cancel with other contributions, including from $\Delta K_2$.

Further contributions to $K(x)$ arise from mixed terms in the expansion ($\Delta L'=-\Delta K_2+\Delta L''$)
\begin{align}
\label{NY20}
K(x)=&\exp\big\{-L_0+K_{2,2}+\Delta K-\frac14K_{2,2}^2+\frac19K_{2,2}^3\nn\\
&\quad+\frac1{192}K_{2,2}^4+\Delta K_2-\Delta L''\big\}\nn\\
=&K_1+K_2+K_2'+K_3'=K_1+K_2+K_3\ .
\end{align}
For example, in the three particle sector the mixed term reads
\begin{align}
\label{NY21}
-&L_0K_{2,2}=-\psi\Rc\psi\Rb\psi\La\psibar\Ra\psibar\Rb\psibar\Lc+R\leftrightarrow L\nn\\
=&\frac12\psi\Rc\gl\psi\Ra\psi\Lb-\psi\Rb\psi\La\gr\psibar\Ra\psibar\Rb\psibar\Lc+R\leftrightarrow L\ .
\end{align}
The presence of such a term in $K(x)$ would contradict the unique jump character of the step evolution operator.  For a given pair $(a,b)$ of incoming right-movers there would be two different possibilities, corresponding to $\psi\Rb\psi\La$ and $\psi\Ra\psi\Lb$. We conclude that this term has to be canceled by $\Delta L''$.

A simple automaton arranges the correction terms $\Delta K_2-\Delta L''$ such that $K'=K_2+K_3$ yields the unit evolution in all sectors except the \qq{scattering sectors} covered by $K_1$ in eq.~\eqref{NY7}. The updating rule exchanges colors only if one, two or three pairs of right- and left-movers meet at $x$. For all other sectors one has a unit evolution for the interaction step. This updating rule fixes $\Delta L$ in eq.~\eqref{NY1} uniquely. Since the cancellations realized by $\Delta L$ only concern $SO(4)$-invariant terms, $\Delta L$ itself is $SO(4)$-invariant, as appropriate for an $SO(4)$-invariant local factor $K(x)$, and therefore for an $SO(4)$-invariant automaton.
As we have sketched here, the detailed construction of $\Delta L$ proceeds by a systematic expansion of $\exp(-L)$. In each sector of a given number of incoming particles one adjusts $\Delta L$ such that $K$ takes the form appropriate for the updating rule. We observe that $SO(4)$-invariance allows for different updating rules, that are realized by different forms of $\Delta L$.


\section{Discrete fermionic action for spinor gravity in two dimensions}
\label{app: C}

In this appendix we list the remaining six terms for the discrete formulation of spinor gravity in two dimensions. Similar to eqs.~\eqref{S36a},~\eqref{S36b} the next two terms involve lattice derivatives of $\psip^4$ and $\psim^3$,
\bel{S36c}
\begin{split}
l_3=&\phantom{-}\psip^4(t,x+\eps)\dpl\psip^4(t+\frac{\eps}{2},x+\efr{3})\\
&\times\psim^3(t,x+\eps)\dm\psim^3(t+\frac{\eps}{2},x+\frac{\eps}{2})\\
&\times\psip^1(\te,x)\psim^1(t,x-\eps)\\
&\times\psip^2(t,x-\eps)\psim^2(\te,x+2\eps)
\end{split}
\ee
and
\bel{S36d}
\begin{split}
l_4=&-\psip^4(t,x)\dm\psip^4(t+\frac{\eps}{2},x-\frac{\eps}{2})\\
&\times\psim^3(t,x)\dpl\psim^3(t+\frac{\eps}{2},x+\frac{\eps}{2})\\
&\times\psip^1(\te,x+\eps)\psim^1(t,x+2\eps)\\
&\times\psip^2(t,x+2\eps)\psim^2(\te,x-\eps)\ .
\end{split}
\ee
The next four terms $l_5$, $l_6$, $l_7$, $l_8$ are similar in structure to $l_1$, $l_2$, $l_3$, $l_4$, respectively. One exchanges $\psip$ and $\psim$ and observes an overall minus sign. Also the precise locations of the spinors differs, according to
\bel{S36e}
\begin{split}
l_5=&-\psim^1(\te,x+\eps)\dpl\psim^1(t+\efr{3},x+\efr{3})\\
&\times\psip^2(\te,x+3\eps)\dm\psip^2(t+\efr{3},x+\efr{5})\\
&\times\psip^3(\te,x+\eps)\psim^3(t+2\eps,x)\\
&\times\psip^4(t+2\eps,x)\psim^4(\te,x+3\eps)
\end{split}
\ee
and
\bel{S36f}
\begin{split}
l_6=&\phantom{-}\psim^1(\te,x+2\eps)\dm\psim^1(t+\efr{3},x+\efr{3})\\
&\times\psip^2(\te,x)\dpl\psip^2(t+\efr{3},x+\frac{\eps}{2})\\
&\times\psip^3(\te,x+2\eps)\psim^3(t+2\eps,x+3\eps)\\
&\times\psip^4(t+2\eps,x+3\eps)\psim^4(\te,x)\ .
\end{split}
\ee
Finally, the precise form of $l_7$ and $l_8$ reads
\bel{S36g}
\begin{split}
l_7=&-\psim^4(t,x)\dpl\psim^4(t+\frac{\eps}{2},x+\frac{\eps}{2})\\
&\times\psip^3(t,x)\dm\psip^3(t+\frac{\eps}{2},x-\frac{\eps}{2})\\
&\times\psip^1(t,x+2\eps)\psim^1(\te,x-\eps)\\
&\times\psip^2(\te,x+\eps)\psim^2(t,x+2\eps)
\end{split}
\ee
with
\bel{S36h}
\begin{split}
l_8=&\phantom{-}\psim^4(t,x+\eps)\dm\psim^4(t+\frac{\eps}{2},x+\frac{\eps}{2})\\
&\times\psip^3(t,x+\eps)\dpl\psip^3(t+\frac{\eps}{2},x+\efr{3})\\
&\times\psip^1(t,x-\eps)\psim^1(t+\eps,x+2\eps)\\
&\times\psip^2(\te,x)\psim^2(t,x-\eps)\ .
\end{split}
\ee
Here we have employed the freedom in the choice of signs for the Grassmann basis elements by taking a minus sign for the third and fourth entry in Figs.~\ref{fig: C1},~\ref{fig: S1}. The somewhat lengthy expression~\eqref{S36} is the exact translation of the updating rules in Figs.~\ref{fig: C1},~\ref{fig: S1} to a discrete model for fermions.


\section{Correlation functions at different times}
\label{app: D}

For the probabilistic cellular automaton we can evaluate correlations of bits or occupation numbers at different times. For the evolution of the expectation value $\exval{n\al(t_2)n_\beta(t_1)}$ we can compute for each initial configuration $\tau$ at $t\inn=0$ the value of $n\al(t_2)$ and $n_\beta(t_1)$ and form the product. This value $\gl n\al(t_2)n_\beta(t_1)\gr_\tau$ for the initial configuration $\tau$ has to be multiplied by the probability $p_\tau(0)$ for a given initial configuration $\tau$,
\bel{P19}
\exval{n\al(t_2)n_\beta(t_1)}=p_\tau(0)\gl n\al(t_2)n_\beta(t_1)\gr_\tau\ .
\ee
Instead of the initial time as a reference point we can take any other arbitrary time.
With $t_2-t_1=\Delta t>0$ one has
\bel{P20}
\exval{n\al(t+\Delta t)n_\beta(t)}=\sum_\tau p_\tau(t)\gl n\al(t+\Delta t)\gr_\tau \gl n_\beta(t)\gr_\tau\ .
\ee
Here $\gl n_\beta(t)\gr_\tau$ reads our directly the occupation number for the configuration $\tau$ at $t$. In contrast, for the determination of $\gl n\al(t+\Delta t)\gr_\tau$ we have to evolve the configuration $\tau$ at $t$ to $t+\Delta t$ by employing the updating rule for the automaton. Then $\gl n\al(t+\Delta t)\gr_\tau$ reads out the occupation number $n\al$ for the updated configuration originating from $\tau$.

We can equivalently evaluate this correlation from the quantum rule
\bel{P21}
\exval{n\al(t+\Delta t)n_\beta(t)}=q_\tau(t)\hat{A}\taur(\Delta t)q_\rho(t)\ ,
\ee
with operator $\hat{A}$ given by the operator product
\bel{P22}
\hat{A}(\Delta t)=\Nhat_\beta(\Delta t)\Nhat\al(0)\ .
\ee
Here we define
\bel{P23}
\Nhat_\beta(T)=U^{-1}(t+T,t)\Nhat_\beta U(t+T,t)\ ,
\ee
where the evolution operator $U(t+T,t)$ is given by the appropriate product of step evolution operators as described in the beginning of sect.~\ref{sec: SUG}, and $\Nhat_\beta$ is defined by eq.~\eqref{P17}. In consequence, one has $\Nhat\al(0)=\Nhat\al$.
In particular, one infers for a single time step
\bel{P24}
\Nhat_\beta(\eps)=\Shat^{-1}(t)\Nhat_\beta\Shat(t)\ ,
\ee
such that the matrix elements read for orthogonal $\Shat$
\bel{P25}
\begin{split}
\gl\Nhat_\beta(\eps)\gr\taur=&\gl\Shat^{-1}(t)\gr_{\tau\alpha}\gl\Nhat_\beta\gr_{\alpha\gamma}\gl\Shat(t)\gr_{\gamma\rho}\\
=&\sum\al\gl\Shat(t)\gr_{\alpha\tau}(n_\beta)\al\gl\Shat(t)\gr_{\alpha\rho}\ .
\end{split}
\ee
For general orthogonal $\Shat$ the off-diagonal elements of $\Nhat_\beta(\eps)$ do not vanish. For a unique jump matrix, however, $\Nhat_\beta(\eps)$ is again a diagonal matrix, since $\Shat_{\alpha\tau}\Shat_{\alpha\rho}=\delta_{\alpha,\bar\alpha(\tau)}\delta_{\alpha,\bar\alpha(\rho)}=\delta_{\alpha,\bar\alpha(\tau)}\delta_{\tau\rho}$. This extends to $\Nhat_\beta(T)$, which is a diagonal matrix.

In order to show that $\Nhat_\beta(\eps)$ is the correct operator associated to the observable $n_\beta(t+\eps)$ we first consider the expectation value of $n_\beta(t+\eps)$ according to the quantum rule
\bel{P26}
\begin{split}
\exval{n_\beta(t+\eps)}=&q_\tau(t)\gl\Nhat_\beta(\eps)\gr\taur q_\rho(t)\\
=&q^T(t)\Shat^T(t)\Nhat_\beta\Shat(t)q(t)=q^T(t+\eps)\Nhat_\beta q(t+\eps)\\
=&\gl\Nhat_\beta\gr_\tau p_\tau(t+\eps)\ ,
\end{split}
\ee
which coincides with the rule of classical statistics.
The construction of $\Nhat_\beta(\eps)$ is analogous to the Heisenberg operators in quantum mechanics. The step evolution operator $\Shat(t)$ can either be seen as evolving the wave function from $q(t)$ to $q(t+\eps)$. Equivalently, we may interpret its role as \qq{moving back} the operator $\Nhat_\beta$ to the reference point $t$.

Indeed, the diagonal elements of $\Nhat_\beta(\eps)$ correspond precisely to the values of $(n_\beta')_\tau$ for the configuration $\tau$ at $t$ which yield at $t+\eps$ configurations $\tau'$ with the values $(n_\beta)_{\tau'}$. If the updating rule changes for a given $\beta$ and given $\tau$ the occupation number zero at $t$ to one at $t+\eps$, the value $(n_\beta)_{\tau'}=1$ at $t+\eps$ corresponds to $(n_\beta')_\tau=0$ at $t$. Correspondingly, for this configuration with $n_\beta=1$ at $t+\eps$ one has a zero diagonal element of $\Nhat_\beta(\eps)$, $\gl\Nhat_\beta(\eps)\gr_{\tau\tau}=0$. This applies to all updatings of occupation numbers. In consequence, the diagonal value of the product $\gl\Nhat_\beta(\eps)\Nhat\al\gr_{\tau\tau}$ corresponds precisely to the value of $n_\beta(t+\eps)n\al(t)$ for the state $\tau$ at $t$. Weighing with the probabilities $p_\tau(t)$ yields the correlation function, demonstrating the validity of the quantum rule~\eqref{P21} for this correlation function.

The proof for a sequence of evolution steps proceeds iteratively, establishing eqs.~\eqref{P21}-\eqref{P23} for arbitrary $\Delta t$. We can also generalize these equations by replacing $n\al(t)$ and $n_\beta(t+\Delta t)$ by arbitrary time-local observables at $t$ and $t+\Delta t$. Furthermore, we can evaluate higher correlations for time-local observables at different times $t_n$. The use of wave functions, operators for observables and the quantum rule for expectation values is a rather convenient tool for the classical statistical setting of probabilistic cellular automata.
It allows us to evaluate correlations of observables at different times for a large number of probabilistic initial conditions. The associated operator has only to be computed once, and the quantum rule for the expectation value is then valid for arbitrary probability distributions or wave functions $q(t)$.


\section{Grassmann operators}
\label{app: E}

The notion of operators for observables translates to the fermionic formulation by a Grassmann functional integral. A general Grassmann operator $\cA\opsi$ is some Grassmann element formed from the Grassmann variables $\psi$. A time-local Grassmann operator $\cA(t)=\cA\opsit$ only depends on the Grassmann variables $\psi\al(t)$ at $t$. The expectation value associated to $\cA\opsi$ obtains by inserting the operator in the functional integral for the partition function $Z$,
\bel{P29}
Z=\int\cD\psi(t\inn\leq t'\leq\tf)\gbar_f(\tf)e^{-S}g\inn\ .
\ee
The partition function involves as a boundary term the \qq{conjugate final Grassmann wave function}
\bel{P29A}
\gbar_f(\tf)=q_\tau(\tf)\gbar_\tau(\tf)\ .
\ee
Here $q_\tau(\tf)$ can be extracted from the \qq{final Grassmann wave function} $g(\tf)$ at even $\tf$, which is given by
\bel{P27}
g_f(\tf)=\int\cD\psi(t\inn\leq t'\leq\tf-\eps)e^{-S}g\inn=q_\tau(\tf)g_\tau(\tf)\ .
\ee
With our normalization one has
\bel{P30}
Z=\int\cD\psi(\tf)\gbar_f(\tf)g(\tf)=q_\tau(\tf)q_\tau(\tf)=1\ .
\ee

The expectation value associated to the Grassmann operator $\cA\opsi$ is given by
\bel{P31}
\exval{A}=\int\cD\psi(t\inn\leq t'\leq\tf)\cA\opsi\gbar_f e^{-S}g\inn\ .
\ee
The general rule how to construct the Grassmann operator $\cA\opsi$ for a given time-local observable (given function of occupation numbers $n\al(t)$), as well as correlations thereof, is described in ref.~\cite{CWFGI}.

We are interested in the opposite direction: Given a simple Grassmann bilinear $\cA(t)=\psi_\beta(t+\eps)\psi\al(t)$, for which observable $A$ does the expectation value obey eq.~\eqref{P31}? And what is the associated quantum operator $\hat{A}$? We find that these quantum operators are typically non-diagonal matrices.
For $t$ even the insertion of $\cA(t)$ into the functional integral~\eqref{P31} replaces $\Kbar(t)$ by $\Khat(t)$,
\bel{P33}
\Kbar(t)\to\Khat(t)=\Kbar(t)\cA(t)\ .
\ee
For a general Grassmann operator $\cA(t)$ depending on $\psi_\beta(t)$ and $\psi\al(t+\eps)$ we can write eq.~\eqref{P31} as
\bel{P34}
\exval{A(t)}=\int\cD\psi(t+\eps)\cD\psi(t)\gbar(t+\eps)\Kbar(t)\cA(t)g(t)\ ,
\ee
where $\gbar(\te)$ is the conjugate Grassmann wave function
\bel{181A}
\gbar(\te)=q_\tau(\te)g'_\tau(\te)\ .
\ee
This expresses the expectation value $\exval{A(t)}$ in terms of the Grassmann wave functions $g(t)$ and $\gbar(\te)$ and the Grassmann observable $\cA$.

For a proof of eq.~\eqref{P34} we express $\gbar(\te)$ as a partial functional integral, similar to eq.~\eqref{P13},
\bel{P35}
\begin{split}
\gbar(t+\eps)&=\int\cD\psi(t'\geq t+2\eps)\gbar_f(\tf)e^{-S_>}\\
&=q_\tau(t+\eps)g_\tau'(t+\eps)\ ,
\end{split}
\ee
with
\bel{P36}
S_>=\sum_{t'=t+2\eps}^{\tf-2\eps}\gl\cL(t'+\eps)+\Lbar(t')\gr+\cL(t+\eps)\ .
\ee
Eq.~\eqref{P35} is established by evolving $\gbar(t)$ \qq{backwards} from $\gbar(\tf)$, using the evolution equations ($t$ even)
\bel{P37}
\begin{split}
\gbar(t+&\eps)=\int\cD\psi(t+2\eps)\gbar(t+2\eps)\Ktil(t+\eps)\\
=&q_\rho(t+2\eps)\Shat_{\rho\tau}(t+\eps)g_\tau'(t+\eps)=q_\tau(t+\eps)g_\tau'(t+\eps)\ ,
\end{split}
\ee
and
\bel{P38}
\begin{split}
\gbar(t)=&\int\cD\psi(t+\eps)\gbar(t+\eps)\Kbar(t)\\
=&q_\rho(t+\eps)\Shat_{\rho\tau}\gbar_\tau(t)=q_\tau(t)\gbar_\tau(t)\ .
\end{split}
\ee

Given the formula~\eqref{P34} we will be able to compute $\exval{A(t)}$ from the wave functions at $t$ and $\te$, and therefore make contact with the quantum rule for expectation values. We write
\bel{E13}
\exval{A(t)}=q_\tau(\te)\tilde A\taur q_\rho(t)\ ,
\ee
where the operator $\tilde A$ has the matrix elements
\bel{E14}
\tilde A\taur=\int\cD\psi(\te)\cD\psi(t)g'_\tau(\te)\Kbar(t)\cA g_\rho(t)\ .
\ee
This identifies the quantum operator $\Ahat$ as
\bel{E15}
\Ahat=\Shat^T(t)\tilde A\ .
\ee
In general, the off diagonal elements of $\Ahat$ do not vanish.

The conjugate Grassmann wave function of ref.~\cite{CWNEW}, app.~\ref{app: D}, obtains by performing the integral $\int\cD\psi(\te)g'_\tau(\te)\Kbar(t)$ with argument $\psibar(t)$ instead of $\psi(t)$ in $\Kbar(t)$.


\section{Explicit representation of fermionic switch operators}
\label{app: F}

In this appendix we briefly discuss switch operators of the type of eq.~\eqref{FO8}. These operators have non-zero off-diagonal elements in the occupation number basis. This is most easily seen by an explicit representation. The operator $\hat f$ in eq.~\eqref{FO8} is an off-diagonal $4\times4$ matrix.

With the explicit representation of app~\ref{app: A}, namely
\bel{FO9}
a_1=a\otimes1=\begin{pmatrix}0&0&0&0\\0&0&0&0\\1&0&0&0\\0&1&0&0\end{pmatrix}\ ,\quad a_2=\tau_3\otimes a=\begin{pmatrix}0&0&0&0\\1&0&0&0\\0&0&0&0\\0&0&-1&0\end{pmatrix}\ ,
\ee
and $a\herm\al=a^T\al$ one has
\bel{FO10}
\fhat=-\gl a\herm\otimes a+a\otimes a\herm\gr=-\begin{pmatrix}0&0&0&0\\0&0&1&0\\0&1&0&0\\0&0&0&0\end{pmatrix}\ .
\ee
This operator does not commute with the occupation number operators $\hat n_1=\hat n\otimes1$, $\hat n_2=1\otimes\hat n$, $\hat n=(1+\tau_3)/2$, according to
\bel{FO11}
\big[\fhat,\nhat_1\big]=-\big[\fhat,\nhat_2\big]=a\herm\otimes a-a\otimes a\herm\ .
\ee
Here we employ the standard relations
\bel{FO12}
a\herm\nhat=0\ ,\quad a\nhat=a\ ,\quad \nhat a\herm=a\herm\ ,\quad \nhat a=0\ .
\ee
We observe the relation
\bel{FO13}
\gl\big[\fhat,\nhat\al\big]\gr^2=-\fhat^2\ .
\ee

The operator $\fhat$ has a spectrum of eigenvalues $(1,0,0,-1)$. A suitable basis of eigenfunctions (in the order of the eigenvalues) is given by
\begin{align}
\label{FO14}
q^{(1)}=&\frac1{\sqrt{2}}\begin{pmatrix}0\\1\\-1\\0\end{pmatrix}\ ,\quad &&q^{(2)}=\begin{pmatrix}1\\0\\0\\0\end{pmatrix}\ ,\nn\\
q^{(3)}=&\begin{pmatrix}0\\0\\0\\1\end{pmatrix}\ , &&q^{(4)}=\frac1{\sqrt{2}}\begin{pmatrix}0\\1\\1\\0\end{pmatrix}\ .
\end{align}
We recognize that $\fhat^2$ acts as a projection on the single-particle states with wave functions $q_2=q_{(10)}$ and $q_3=q_{(01)}$. Within this subspace $\fhat$ acts as switch operator which interchanges the two single particles. We can view $\fhat$ as a product of the projection $\fhat^2$ and a switch operator $-\hat A_{s,23}$. We can also view it as a product of the exchange operator between the two particles $A_E=\tau_1\otimes\tau_1$ and a subsequent projection $\fhat^2$. These features generalize to arbitrary particle pairs $(\alpha,\beta)$ with
\bel{FO15}
\fhat_{\alpha\beta}=\fhat_{\beta\alpha}=a\herm\al a\bet+a\herm\bet a\al\ .
\ee


\section{Grassmann operators for fermionic \\observables}
\label{app: G}

In this appendix we explore simple bilinear Grassmann operators. We establish their relation to conditional fermionic observables and express them in terms of annihilation and creation operators. We also discuss in this context the choice of sign in the real wave function, which is common for the fermionic quantum field theory and the cellular automaton.

\subsection*{Conditional observables}

In order to understand the issue in its simplest form we will consider first a single Grassmann variable $\psi(t)$, with bilinear Grassmann operator
\bel{P32}
\cA(t)=\psi(t+\eps)\psi(t)\ ,
\ee
and $t$ even. We are interested in the expectation value of the associated observable which is given by eq.~\eqref{P31} or~\eqref{E13}. The most general from of $\Kbar(t)$ reads
\bel{P39}
\begin{split}
\Kbar(t)=&\phantom{-}\Shat_{00}(t)+\Shat_{01}(t)\psi(t)\\
&-\Shat_{10}(t)\psi(t+\eps)-\Shat_{11}(t)\psi(t+\eps)\psi(t)\ ,
\end{split}
\ee
and one has
\bel{P40}
\begin{split}
g(t)=&q_0(t)\psi(t)+q_1(t)\ ,\\
\gbar(t+\eps)=&q_0(t+\eps)\psi(t+\eps)+q_1(t+\eps)\ ,\\
g(t+\eps)=&\Shat_{00}(t)q_0(t)+\Shat_{01}(t)q_1(t)\\
&+\big[\Shat_{10}(t)q_0(t)+\Shat_{11}(t)q_1(t)\big]\psi(t+\eps)\ ,
\end{split}
\ee
with $q_0^2(t)$ the probability for a hole ($n(t)=0$) and $q_1^2(t)$ the probability for a particle ($n(t)=1$). (For an odd number $M$ of Grassmann variables at $t$ one needs additional minus signs in the relation between $\Kbar$ and $\Shat$.)
Insertion of $\cA(t)$ yields for $\Khat$ in eq.~\eqref{P33}
\bel{P41}
\Khat(t)=\Kbar(t)\cA(t)=\Shat_{00}(t)\psi(t+\eps)\psi(t)\ .
\ee
As a consequence, only configurations with $n(t)=n(t+\eps)=1$ contribute to the expectation value, which is therefore proportional to $\exval{n(t+\eps)n(t)}$. One infers from the expression~\eqref{P34}
\bel{P42}
\exval{A(t)}=\Shat_{00}(t)q_1(t+\eps)q_1(t)\ .
\ee
(For odd $M$ there are additional minus signs in eq.~\eqref{P34}.)
For a cellular automaton one has either $\Shat_{00}(t)=\Shat_{11}(t)=1$ and
\bel{P43}
\exval{A(t)}=\exval{n(t)}\ ,
\ee
or $\Shat_{01}=\Shat_{10}=1$, $\Shat_{00}=\Shat_{11}=0$, and $\exval{A(t)}=0$. We conclude that the Grassmann operator $\psi(t+\eps)\psi(t)$ is associated to a \qq{conditional observable}, which is given by the particle number if the updating rule at $t$ changes a hole to a hole ($\Shat_{00}(t)=1$), and zero otherwise ($\Shat_{00}(t)=0$).

This generalizes to the Grassmann operators
\bel{P44}
\cA_{\beta\alpha}(t)=\psi_\beta(t+\eps)\psi\al(t)\ .
\ee
The associated conditional observable $A_{\beta\alpha}$ is given by the product of occupation numbers $n_\beta(t+\eps)n\al(t)$ whenever the updating rule at $t$ changes a hole of type $\alpha$ into a hole of type $\beta$. For all other cases it takes the value zero. It is easy to see that $A_{\beta\alpha}$ vanishes for configurations for which at $t$ there is no particle of type $\alpha$. For these configurations $g_\tau(t)$ contains a factor $\psi\al(t)$, and the Grassmann integral over $\psi\al(t)$ vanishes for
\bel{P45}
\exval{A_{\beta\alpha}(t)}=\int\cD\psi(t+\eps)\cD\psi(t)\gbar(t+\eps)\psi_\beta(t+\eps)\Kbar(t)\psi\al(t)g(t)\ .
\ee
Thus $A_{\beta\alpha}(t)$ is proportional to $n\al(t)$. For the same reasons $A_{\beta\alpha}(t)$ vanishes for $n_\beta(t+\eps)=0$, since for these configurations $g_\tau'(t+\eps)$ in $\gbar(t+\eps)$ contains a factor $\psi_\beta(t+\eps)$, and we conclude $A_{\beta\alpha}(t)\sim n_\beta(t+\eps)n\al(t)$. Furthermore, the local factor $\Kbar(t)$ in eq.~\eqref{P45} should not contain $\psi_\beta(t+\eps)$ or $\psi\al(t)$. The product $\psi\al(t)g(t)$ involves only basis elements with a hole of type $\alpha$, and $\gbar(t+\eps)\psi_\beta(t+\eps)$ vanishes for basis elements with a particle of type $\beta$. The matrix elements of the step evolution operator which connect configurations with a hole of type $\alpha$ at $t$ to a hole of type $\beta$ at $t+\eps$ may be denoted as $\Shat_{\beta0,\tilde\tau;\alpha0,\tilde\rho}$, where $\tilde\tau$ denotes the configurations of occupation numbers $n_{\beta'}(t+\eps)$, $\beta'\neq\beta$, and similarly $\tilde\rho$ for the configurations of $n_{\alpha'}(t)$, $\alpha'\neq\alpha$. Neighboring configurations for which the hole to hole transition $\Shat_{\beta0,\tilde\tau;\alpha0,\tilde\rho}$ vanishes give no contribution to $\exval{A_{\beta\alpha}}$.
One concludes
\bel{P46}
\gl A_{\beta\alpha}(t)\gr_{\tilde\tau\tilde\rho}=\pm\Shat_{\beta0,\tilde\tau;\alpha0,\tilde\rho}(t)n_\beta(t+\eps)n\al(t)\ ,
\ee
where we recall that $\Shat_{\beta0,\tilde\tau;\alpha0,\tilde\rho}$ either takes the value $\pm1$ or $0$, depending on $\tilde\tau$, $\tilde\rho$. At this place we have paid no attention to the sign since the use of quantum operators below permits a simple bookkeeping.

Conditional observables of the type~\eqref{P46} are well defined for the cellular automaton. For a given automaton the updating rule defines the step evolution operator $\Shat(t)$, and the projection on the particular hole-hole transition $\Shat_{\beta0,\tilde\tau;\alpha0,\tilde\rho}$ can be inferred for every pair of configurations at $t+\eps$ and $t$. Again, it either vanishes or equals $\pm1$, and can be interpreted as a condition. The conditional observables $A_{\beta\alpha}(t)$ do, however, not take fixed values for a given configuration $\rho$ at $t$. They involve, in addition, the step evolution operator which transports a given configuration at $t$ to a new configuration at $t+\eps$. While being a well defined observable for the classical statistical system of the probabilistic cellular automaton, a conditional observable does not need to be represented by a diagonal quantum operator $\hat{A}_{\beta\alpha}$. We will see next that the associated quantum operator has indeed non-zero off-diagonal elements.

\subsection*{Non-commuting operators}

We want to find a representation of $A_{\beta\alpha}(t)$ by a quantum operator $\hat{A}_{\beta\alpha}$ in terms of the annihilation and creation operators. The Grassmann wave functions $g(t)$ have the property~\cite{CWNEW} that multiplication with $\psi\al$ at fixed quantum wave function $q(t)$ is equivalent to applying the annihilation operator $a\al$ on $q(t)$ at fixed Grassmann basis elements $g_\tau$,
\bel{P49}
\psi\al(t)g(t)=\psi\al(t)q_\tau(t)g_\tau(t)=(a\al)\taur q_\rho(t)g_\tau(t)\ .
\ee
The relation~\eqref{P49} holds for even $t$, and we have for odd $t+\eps$ a similar relation for the conjugate wave function
\bel{P50}
\begin{split}
\gbar(t+\eps)\psi_\beta(\te)=&q_\tau(t+\eps)g_\tau'(t+\eps)\psi_\beta(t+\eps)\\
=&(a_\beta)\taur q_\rho(t+\eps)g_\tau'(t+\eps)\ .
\end{split}
\ee
As a result, we can write the expectation value~\eqref{P45} as
\bel{P52}
\exval{A_{\beta\alpha}(t)}=q_\tau(t+\eps)\gl a_\beta\herm\Shat(t)a\al\gr\taur q_\rho(t)\ .
\ee
Employing further $q(t+\eps)=\Shat(t)q(t)$ we arrive at the quantum operator for the conditional observable $A_{\beta\alpha}(t)$
\bel{P53}
\hat{A}_{\beta\alpha}(t)=\Shat^T(t)a_\beta\herm\Shat(t)a\al\ .
\ee
With this quantum operator the expectation value $\exval{A_{\beta\alpha}}$ can be evaluated according to the quantum rule~\eqref{P15},
\bel{P54}
\exval{A_{\beta\alpha}(t)}=q_\tau(t)\gl\hat{A}_{\beta\alpha}(t)\gr\taur q_\rho(t)\ .
\ee
Only the symmetric part of the operator $\Ahat_{\beta\alpha}$ contributes in eq.~\eqref{P54}. We could symmetrize this operator. For simplicity of notation we keep the form~\eqref{P53}.

The operator~\eqref{P53} encodes the conditional structure discussed above. Since $a\al q(t)$ differs from zero only for the components for which a particle of type $\alpha$ is present, and similarly for $a_\beta q(t+\eps)$, one concludes $A_{\beta\alpha}\sim n_\beta(t+\eps)n\al(t)$. The structure $a\bet\herm\Shat(t)a\al$ projects $\Shat(t)$ on the subsector $\Shat_{\beta0,\tilde\tau;\alpha0,\tilde\rho}$. The use of the operator~\eqref{P53} is a rather economical way to keep track of the conditions. It also fixes the signs.

We may compare this operator to the operator for the condition-less product $n\bet(t+\eps)n\al(t)$,
\bel{P55}
\Nhat\bet(\eps)\Nhat\al=\Shat^T(t)a\bet\herm a\bet\Shat(t)a\al\herm a\al\ .
\ee
While eq.~\eqref{P55} results in a diagonal operator, this does not hold for $\hat{A}_{\beta\alpha}$. The two operators $\hat{A}_{\beta\alpha}$ and $\Nhat\bet(\eps)\Nhat\al$ do not commute. The simple rule for $A_{\beta\alpha}$ - annihilate a particle $\alpha$, move one step forwards, create a particle $\beta$, move one step backwards - does not commute with the measurement of $n\bet(\te)n\al(t)$. The operator $\hat{A}_{\beta\alpha}$ is an example for observables which are well defined for a classical statistical system and give rise to operators that do not commute with the diagonal operators for time-local observables.

\subsection*{Signs in the wave function}

If $\hat{A}\taur$ has non-zero elements for $\tau\neq\rho$, the expectation value~\eqref{P54} involves products $q_\tau(t)q_\rho(t)$ for $\tau\neq\rho$. They are no longer linear in the probabilities $p_\tau(t)=q_\tau^2(t)$, but rather involve roots $q_\tau=\pm\sqrt{p_\tau}$. The relative sign of $q_\tau$ and $q_\rho$ matters for eq.~\eqref{P54}. We can choose the sign of the components of the initial wave function $q_\tau(0)$ arbitrarily. For the computation of the expectation value~\eqref{P54} we need, however, to keep track of the signs in the evolution of the wave function.

A given sign convention fixes both the signs of $q_\tau(t)$ and the signs of the non-zero elements $\Shat\taur(t')$ of the step evolution operator. It therefore determines the signs of all matrix elements $\gl\hat{A}_{\beta\alpha}\gr\taur$. Switching to a different sign convention corresponds to the multiplication with diagonal sign matrices $D(t)$
\bel{P56}
q'(t)=D(t)q(t)\ ,\quad \Shat'(t)=D(t+\eps)\Shat(t)D(t)\ ,
\ee
where the diagonal elements of $D$ are given by signs, $D_{\tau\tau}=\pm1$. This is compatible with the evolution law~\eqref{P3}. The sign matrices $D(t')$ drop out for $t'>t$ in the definition of the operator $\Nhat\bet(T)$ in eq.~\eqref{P23}, or~\eqref{P55}. Therefore $\hat{A}(\Delta t)$ in eq.~\eqref{P21} transforms as
\bel{P57}
\hat{A}'(\Delta t)=D(t)\hat{A}(\Delta t)D(t)\ .
\ee
This compensates the change $q'(t)=D(t)q(t)$, and the expectation value is indeed independent of the choice of sign conventions. Similarly, $\hat{A}_{\beta\alpha}(t)$ is transformed to
\bel{P58}
\hat{A}'_{\beta\alpha}(t)=D(t)\hat{A}_{\beta\alpha}(t)D(t)\ ,
\ee
and the expectation value~\eqref{P54} does not depend on the sign convention.

\subsection*{General Grassmann operators}

The map between Grassmann operators $\cA$ and quantum operators $\hat{A}$ extends to a general Grassmann operator $\cA(t)$ that depends only on the variables $\psi\al(\te)$ and $\psi\al(t)$. The associated quantum operator $\hat{A}(t)$ can be constructed as
\bel{P59}
\hat{A}\taur(t)=\gl\Shat^T(t)\gr\taur\int\cD\psi(\te)\cD\psi(t)g_\sigma'(\te)\Kbar(t)\cA(t)g_\rho(t)\ .
\ee
Indeed, the quantum rule for the expectation value yields,
\bel{P60}
\begin{split}
&q_\tau(t)\hat{A}\taur(t)q_\rho(t)\\
=&\int\cD\psi(\te)\cD\psi(t)g_\sigma'(\te)\Kbar(t)\cA(t)q_\rho(t)g_\rho(t)\\
=&\int\cD\psi(\te)\cD\psi(t)\gbar(\te)\Kbar(t)\cA(t)g(t)\ ,
\end{split}
\ee
which agrees with eq.~\eqref{P34}. For $\cA(t)$ in a form where all factors $\psi(\te)$ are to the left of factors $\psi(t)$,
\bel{P61}
\cA(t)=\cA_+[\psi(\te)]\cA_-\opsit\ ,
\ee
(and $\Kbar(t)$ involving only terms with an even number of Grassmann variables) we can replace in $\cA_-$ the factors $\psi\al(t)$ by $a\al$, and for $\cA_+$ substitute $\psi\bet(\te)$ by $a\bet\herm$, such that
\bel{P62}
\hat{A}(t)=\Shat^T(t)\cA_+[a\bet\herm]\Shat(t)\cA_-[a\al]\ .
\ee

In the other direction, we can associate to the operator $\Bhat=\Shat\hat{A}$ the Grassmann expression
\bel{P63}
\Kbar(t)\cA(t)=\gbar_\tau'(\te)\Bhat\taur(t)\gbar_\rho(t)\ ,\quad \Bhat(t)=\Shat(t)\hat{A}(t)\ .
\ee
For a given step evolution operator $\Shat(t)$ this allows us to construct for every quantum observable $\hat{A}(t)$ the Grassmann factor $\Kbar(t)\cA(t)$. The expectation value $\exval{A(t)}$ obtains then by replacing in the Grassmann functional integral for $Z$ the local factor $\Kbar(t)$ by the factor $\Kbar(t)\cA(t)$. The construction~\eqref{P63} can be applied even if $\cA(t)$ is not available explicitly since only the product $\Kbar\cA$ is needed. This yields a general Grassmann expression~\cite{CWFGI} for all observables for which the expectation value can be computed from the quantum rule. We recall that the association between $\Shat\Ahat$ and $\Kbar\cA$ holds for even $t$, with a different formula for odd $t$ given in ref.~\cite{CWFGI}. The existence of this simple general construction relies on the modulo two property of Grassmann functional integrals.

An interesting special case for bilinear Grassmann operators \eqref{P44} arises if for a given $\alpha$ one can find $\beta$ such that
\bel{P64}
a_\beta\herm\Shat(t)a\al=\Shat(t)a\al\herm a\al\ .
\ee
In this case the operator $\hat{A}_{\beta\alpha}$ equals the particle number operator $\Nhat\al$. The simplest case where this is realized is $\Shat(t)=1$ for even $t$, with a possible non-trivial $\Shat(\te)$ for odd $\te$. In this case the particle number observables $n\al(t)$ are represented by the Grassmann operator $\psi\al(\te)\psi\al(t)$. The Grassmann variables $\psi\al(\te)$ play precisely the role of the conjugate Grassmann variables in ref.~\cite{CWFCB}. This generalizes to the automata~\eqref{GU17} for which $\cL$ is a fermion bilinear. In this case the index $\bar\beta(\alpha)$ for which eq.~\eqref{P64} is obeyed is given by
\bel{P65}
\bar\beta(\alpha)=F_{\bar\beta\alpha}\ .
\ee
In particular, if $t$ and $\te$ are related by right- or left-transport automata one has a simple Grassmann representation of the occupation number observable $N\al\otx$, namely
\bel{H26*}
\mathcal{N}\al\otx=\psi\al(\te,x\pm\eps)\psi\al\otx\ .
\ee
In the continuum limit this yields the familiar expression
\bel{H27*}
\mathcal{N}\al\otx=\psibar\al\otx\psi\al\otx\ .
\ee


\section{Conserved quantities and symmetries}

The familiar relation between symmetries and conserved quantities holds for the evolution of probabilistic cellular automata. For a global symmetry transformation $D$ one has $\Shat'=\Dbar\Shat D^{-1}=\Shat$ according to eq.~\eqref{CSP6}. Observables related by symmetry transformations have the same expectation value if the symmetry is not broken spontaneously.
For fixed $t$ we can perform a simultaneous symmetry transformation~\eqref{CSP7} on $q$ and $\Ahat$, such that the expectation value remains invariant ($D^TD=1$)
\bel{CQS3}
\exval{A}=\exval{q^TD^TD\Ahat D^TDq}={q'}^T\Ahat'q'\ ,
\ee
with
\bel{CQS4}
\Ahat'=D\Ahat D^T\ .
\ee
If the wave function is invariant under the symmetry, $q'=q$, one concludes that an observable $A'$ associated to $\Ahat'$ has the same expectation value as $A$,
\bel{CQS5}
\exval{A'}=q^T\Ahat'q={q'}^T\Ahat'q'=\exval{A}\ .
\ee
If further the evolution is compatible with the symmetry, $\Shat'=\Dbar\Shat D^T$, an invariant wave function remains invariant under the evolution
\begin{align}
\label{CQS6}
q'(\te)=&\Shat(t)q'(t)=\Shat'(t)q'(t)\nn\\
=&\Dbar(\te)\Shat(t)D^T(t)D(t)q(t)\nn\\
=&\Dbar(\te)\Shat(t)q(t)=\Dbar(\te)q(\te)\ .
\end{align}

These considerations are perhaps still somewhat abstract since the transformations act in the large space with arbitrary particle numbers. They show, however, that all standard quantum mechanical concepts for symmetries apply fully to probabilistic automata. Important practical simplifications occur if the symmetry transformations do not change particle numbers and the step evolution preserves particle number as well. For one-particle states the symmetry transformations are just what one is used to from one-particle quantum mechanics.

\end{appendices}

\nocite{*} 
\bibliography{refs}
\end{document}